\documentclass[superscriptaddress,prc,twocolumn,showpacs,preprintnumbers,amsmath,amssymb,bibnotes,secnumarabic,altaffilletter,floatfix]{revtex4}
\usepackage{graphicx}

\newcommand{\etal}{{\it et al.}}
\newcommand{\tltze}{$^{208}$Tl}
\newcommand{\bitof}{$^{214}$Bi}
\newcommand{\tij}{$\theta_{ij}$}
\newcommand{\hto}{H$_2$O}
\newcommand{\dto}{D$_2$O}
\newcommand{\mnox}{MnO$_{\mbox{x}}$}

\newcommand{\NS} {\chem{16}{N}}
\newcommand{\LI} {\chem{8}{Li}}
\newcommand{\chem}[2]{\mbox{$\rm ^{#1}{#2}$}}
\begin{document}

\title{Measurement of the $\nu_e$ and Total $^{8}$B Solar Neutrino Fluxes with
the  Sudbury Neutrino Observatory Phase I Data Set}

%
\newcommand{\ubc}{Department of Physics and Astronomy, University of 
British Columbia, Vancouver, BC V6T 1Z1, Canada}
\newcommand{\bnl}{Chemistry Department, Brookhaven National 
Laboratory,  Upton, NY 11973-5000}
\newcommand{\carleton}{Ottawa-Carleton Institute for Physics, Department of Physics, Carleton University, Ottawa, Ontario K1S 5B6, Canada}
\newcommand{\uog}{Physics Department, University of Guelph,  
Guelph, Ontario N1G 2W1, Canada}
\newcommand{\lu}{Department of Physics and Astronomy, Laurentian 
University, Sudbury, Ontario P3E 2C6, Canada}
\newcommand{\lbnl}{Institute for Nuclear and Particle Astrophysics and 
Nuclear Science Division, Lawrence Berkeley National Laboratory, Berkeley, CA 94720}
\newcommand{\lbla}{ Lawrence Berkeley National Laboratory, Berkeley, CA}
\newcommand{\lanl}{Los Alamos National Laboratory, Los Alamos, NM 87545}\newcommand{\llnl}{Lawrence Livermore National Laboratory, Livermore, CA}
\newcommand{\lanla}{Los Alamos National Laboratory, Los Alamos, NM 87545}\newcommand{\oxford}{Department of Physics, University of Oxford, 
Denys Wilkinson Building, Keble Road, Oxford OX1 3RH, UK}
\newcommand{\penn}{Department of Physics and Astronomy, University of 
Pennsylvania, Philadelphia, PA 19104-6396}
\newcommand{\queens}{Department of Physics, Queen's University, 
Kingston, Ontario K7L 3N6, Canada}
\newcommand{\uw}{Center for Experimental Nuclear Physics and Astrophysics, 
and Department of Physics, University of Washington, Seattle, WA 98195}
\newcommand{\uta}{Department of Physics, University of Texas at Austin, Austin, TX 78712-0264}\newcommand{\triumf}{TRIUMF, 4004 Wesbrook Mall, Vancouver, BC V6T 2A3, Canada}
\newcommand{\ralimp}{Rutherford Appleton Laboratory, Chilton, Didcot OX11 0QX, UK}
\newcommand{\iusb}{Department of Physics and Astronomy, Indiana University, South Bend, IN}
\newcommand{\fnal}{Fermilab, Batavia, IL}
\newcommand{\uo}{Department of Physics and Astronomy, University of Oregon, Eugene, OR}
\newcommand{\hu}{Department of Physics, Hiroshima University, Hiroshima, Japan}
\newcommand{\slac}{Stanford Linear Accelerator Center, Menlo Park, CA}
\newcommand{\mac}{Department of Physics, McMaster University, Hamilton, ON}\newcommand{\doe}{US Department of Energy, Germantown, MD}\newcommand{\lund}{Department of Physics, Lund University, Lund, Sweden}\newcommand{\mpi}{Max-Planck-Institut for Nuclear Physics, Heidelberg, Germany}\newcommand{\uom}{Ren\'{e} J.A. L\'{e}vesque Laboratory, Universit\'{e} de Montr\'{e}al, Montreal, PQ}\newcommand{\cwru}{Department of Physics, Case Western Reserve University, Cleveland, OH}\newcommand{\pnnl}{Pacific Northwest National Laboratory, Richland, WA}\newcommand{\uc}{Department of Physics, University of Chicago, Chicago, IL}
\newcommand{\mitt}{Laboratory for Nuclear Science, Massachusetts Institute of Technology, Cambridge, MA 02139}
\newcommand{\ucsd}{Department of Physics, University of California at San Diego, La Jolla, CA }
\newcommand{	\lsu	}{Department of Physics and Astronomy, Louisiana State University, Baton Rouge, LA 70803}\newcommand{\imp}{Imperial College, London SW7 2AZ, UK}
\newcommand{\uci}{Department of Physics, University of California, Irvine, CA 92717}
\newcommand{\ucia}{Department of Physics, University of California, Irvine, CA}
\newcommand{\suss}{Department of Physics and Astronomy, University of Sussex, Brighton  BN1 9QH, UK}\newcommand{	\lifep	}{Laborat\'{o}rio de Instrumenta\c{c}\~{a}o e F\'{\i}sica Experimental de
Part\'{\i}culas, Av. Elias Garcia 14, 1$^{\circ}$, 1000-149 Lisboa, Portugal}\newcommand{\hku}{Department of Physics, The University of Hong Kong, Hong Kong.}\newcommand{\aecl}{Atomic Energy of Canada, Limited, Chalk River Laboratories, Chalk River, ON K0J 1J0, Canada}
\newcommand{\nrc}{National Research Council of Canada, Ottawa, ON K1A 0R6, Canada}
\newcommand{\princeton}{Department of Physics, Princeton University, Princeton, NJ 08544}
\newcommand{\birkbeck}{Birkbeck College, University of London, Malet Road, London WC1E 7HX, UK}\newcommand{\snoi}{SNOLAB, Sudbury, ON P3Y 1M3, Canada}\newcommand{\uba}{University of Buenas Aires, Argentina}\newcommand{\hvd}{Department of Physics, Harvard University, Cambridge, MA}
\newcommand{\pny}{Goldman Sachs, 85 Broad Street, New York, NY}\newcommand{\pnv}{Remote Sensing Lab, PO Box 98521, Las Vegas, NV 89193}\newcommand{\psis}{Paul Schiffer Institute, Villigen, Switzerland}\newcommand{\liverpool}{Department of Physics, University of Liverpool, Liverpool, UK}\newcommand{\uto}{Department of Physics, University of Toronto, Toronto, ON, Canada}\newcommand{\uwisc}{Department of Physics, University of Wisconsin, Madison, WI}\newcommand{\psu}{Department of Physics, Pennsylvania State University,
     University Park, PA}\newcommand{\anl}{Deparment of Mathematics and Computer Science, Argonne
     National Laboratory, Lemont, IL}\newcommand{\cornell}{Department of Physics, Cornell University, Ithaca, NY}\newcommand{\tufts}{Department of Physics and Astronomy, Tufts University, Medford, MA}

\affiliation{\aecl}
\affiliation{\ubc}
\affiliation{\bnl}
\affiliation{\uci}
\affiliation{\carleton}
\affiliation{\uog}
\affiliation{\lu}
\affiliation{\lbnl}
\affiliation{\lifep}\affiliation{\lanl}\affiliation{\lsu}\affiliation{\mitt}
\affiliation{\nrc}
\affiliation{\oxford}
\affiliation{\penn}
\affiliation{\princeton}
\affiliation{\queens}
\affiliation{\ralimp}\affiliation{\snoi}\affiliation{\uta}\affiliation{\triumf}
\affiliation{\uw}

\author{B.~Aharmim}\affiliation{\lu}
\author{Q.R.~Ahmad}\affiliation{\uw}
\author{S.N.~Ahmed}\affiliation{\queens}
\author{R.C.~Allen}\affiliation{\uci}
\author{T.C.~Andersen}\affiliation{\uog}
\author{J.D.~Anglin}\affiliation{\nrc}
\author{G.~B\"uhler}\affiliation{\uci}
\author{J.C.~Barton}\altaffiliation{Deceased}\affiliation{\oxford}
\author{E.W.~Beier}\affiliation{\penn}
\author{M.~Bercovitch}\affiliation{\nrc}
% ...
\collaboration{SNO Collaboration}
\noaffiliation

\date{\today}

\begin{abstract}
	This article provides the complete description of results from the
Phase I data set of the Sudbury Neutrino Observatory (SNO).  The Phase I data set is based
on a 0.65 kt-year exposure of heavy water to the solar $^8$B neutrino flux. 
Included here are details of the SNO physics and detector model, 
evaluations of systematic uncertainties, and estimates of backgrounds.
Also discussed are SNO's approach to statistical extraction of the signals from
the three neutrino reactions (charged current, neutral current, and elastic
scattering) and the results of a search for a day-night asymmetry in the $\nu_e$ flux.
Under the assumption that the $^8$B spectrum is undistorted, the measurements
from this phase yield a solar $\nu_e$ flux of $\phi(\nu_e) =
  1.76^{+0.05}_{-0.05}\mbox{(stat.)}^{+0.09}_{-0.09}~\mbox{(syst.)}
  \times 10^{6}$~cm$^{-2}$~s$^{-1}$, and a non-$\nu_e$ component
$\phi(\nu_{\mu\tau}) =
  3.41^{+0.45}_{-0.45}\mbox{(stat.)}^{+0.48}_{-0.45}~\mbox{(syst.)}
  \times 10^{6}$~cm$^{-2}$~s$^{-1}$.  The sum of these components provides a
total flux in excellent agreement with the predictions of Standard Solar
Models.  The day-night asymmetry in the $\nu_e$ flux is found to be $A_{e} = 7.0
\pm 4.9~\mathrm{(stat.)^{+1.3}_{-1.2}}\%~\mathrm{(sys.)}$, when the asymmetry
in the total flux is constrained to be zero.
\end{abstract}


\maketitle


\section{Introduction \label{sec:intro}}

	More than thirty years of solar neutrino 
experiments~\cite{cl,kam,sage,gallex,SK,gno} indicated that the total flux
of neutrinos from the Sun was significantly smaller 
than predicted by models of the Sun's energy generating 
mechanisms~\cite{BP01,TC}.  The
deficit was not only universally observed but had an energy dependence which
was difficult to attribute to astrophysical sources. The data were consistent
with a negligible flux of neutrinos from solar $^7$Be~\cite{hata,hamish}, 
though neutrinos
from $^8$B (a product of solar $^7$Be reactions) were observed.  A
natural explanation for the observations was that neutrinos born as
$\nu_e$s change flavor on their way to the Earth, thus producing an 
apparent deficit in experiments detecting primarily $\nu_e$s.  Neutrino
oscillations---either in vacuum~\cite{pontecorvo,mns} or
matter~\cite{wolf,ms}---provide a mechanism both for the flavor change and the
observed energy variations.

	While these deficits argued strongly for  neutrino
flavor change through oscillation, it was clear that a far more compelling
demonstration would not resort to model predictions but look directly
for neutrino flavors other than the $\nu_e$ emitted by the 
Sun.  The Sudbury Neutrino Observatory 
(SNO) was designed to do just that: provide direct evidence of solar
neutrino flavor change through observation of non-electron 
neutrino flavors by making a flavor-independent measurement of the total
$^8$B neutrino flux from the Sun~\cite{hhchen}.   As a real-time detector, SNO was
also designed to look for specific signatures of the oscillation mechanism, such as
energy- or time-dependent survival probabilities.  For example, depending upon the
values of the mixing parameters, the matter (MSW) effect leads to  different $\nu_e$
fluxes during the day and the night and to a distortion in the expected energy
spectrum of $^8$B solar neutrinos.

We present in this article the details of the analyses presented in previous SNO
publications~\cite{snocc, snonc, snodn}, including the exclusive $\nu_e$ and
inclusive active neutrino fluxes, a measurement of the $\nu_e$ spectrum, the
difference in the neutrino fluxes between day and night, and determination of the
neutrino mixing parameters.  We will concentrate here on the low-energy threshold 
measurements of Refs.~\cite{snonc,snodn} which included the first measurements of
the total $^8$B flux, but will describe the differences between these analyses
and the high-threshold measurement presented in Ref.~\cite{snocc}.  

We begin in Section~\ref{sec:overview} with an overview of the SNO
detector and data analysis.
In Section~\ref{sec:dataset} we describe the data set used for
the measurements made in the initial phase (hereafter Phase I) of SNO using pure D$_2$O as
the target-detector.  Section~\ref{sec:model}
describes the detector model which is ultimately used both
to calibrate the neutrino data and to provide distributions used to fit our
data. Section~\ref{sec:dataproc} describes the processing
of the data, including all cuts applied, reconstruction of position
and direction, and estimations of effective kinetic energy for
each event.  Section~\ref{sec:sysunc} details the systematic uncertainties
in the model, which translate into uncertainties in the
neutrino fluxes.  Section~\ref{sec:bkds} describes the measurement
of backgrounds remaining in the data set, including neutrons from
photodisintegration, the tails of low energy radioactivity, and cosmogenic
sources.  Section~\ref{sec:sigex} details the methods used to fit for
the neutrino rates, and Section~\ref{sec:norm} the ingredients
which go into normalization of the rates.  Sections~\ref{sec:results}
and~\ref{sec:daynight} present the flux results and results of a search for an
asymmetry in the day and night fluxes.  Appendix~\ref{sec:physint}
describes the methods used to calculate mixing parameters from these data, and
Appendix~\ref{sec:apdxa} gives details of the cuts we used to remove instrumental
backgrounds.

We will refer in this article to Ref.~\cite{snocc} as the `ES-CC paper',
Ref.~\cite{snonc} as the `NC paper', Ref.~\cite{snodn} as the `Day-Night paper',
and collectively we call them the `Phase I publications'. 

\section{Overview of SNO \label{sec:overview}}

\subsection{The SNO Detector}

	SNO is an imaging Cherenkov detector using heavy water (D$_2$O) as both the
interaction and detection medium~\cite{NIM}.  SNO is located in Inco's
Creighton Mine, at $46^{\circ} 28^{'} 30^{''}$ N latitude, $81^{\circ} 12^{'}
04^{''}$ W longitude.  The detector resides 1730 m below sea level with an
overburden of 6020 meters water equivalent, deep enough that the rate of cosmic ray
muons passing through the entire active volume is just 3 per hour.   
\begin{figure}
\begin{center}
\includegraphics[height=0.55\textwidth]{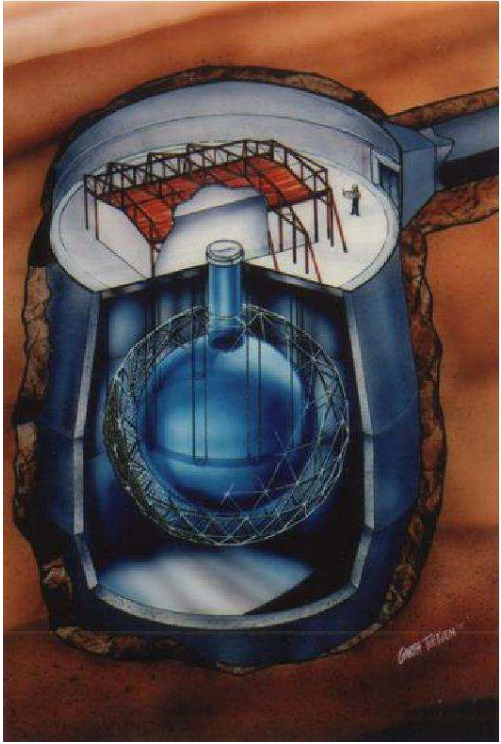}
\caption{Schematic of SNO Detector. \label{fig:detector}}
\end{center}
\end{figure}

Figure~\ref{fig:detector} is a schematic of  the detector.  One thousand metric
tons of heavy water are contained in a 12-m diameter transparent acrylic vessel
(AV).  Cherenkov light produced by neutrino interactions and radioactive
backgrounds is detected by an array of 9456 Hammamatsu model R1408 8-inch
photomultiplier tubes (PMTs), supported by a stainless steel geodesic sphere
(the PMT support sphere or PSUP).  Each PMT is surrounded by a light
concentrator (`reflector'), which increases the photocathode coverage to
nearly $55$\%.  The channel discriminator thresholds are set to fire on 1/4
of a photoelectron of charge.  Over seven kilotons of light water shield the
heavy water from external radioactive backgrounds: 1.7~kT between the acrylic
vessel and the PMT support sphere, and 5.7~kT between the PMT support sphere and
the surrounding rock. The 5.7~kT of light water outside the PMT support sphere
is viewed by 91 outward-facing 8-inch PMTs that are used for identification of
cosmic-ray muons.  An additional 23 PMTs, arranged in a rectangular array, are
suspended in the outer light water region. These 23 PMTs view the neck of the
acrylic vessel and are used primarily in the rejection of instrumentally generated
light.

	The detector is equipped with a versatile calibration deployment system
which can place radioactive and optical sources over a large range of the $x$-$z$
and $y$-$z$ planes in the AV.  Sources that can be deployed include a diffuse
multi-wavelength laser for measurements of PMT timing and optical
parameters~\cite{laserball}, a $^{16}$N source which provides a triggered sample of
6.13~MeV $\gamma$s~\cite{n16}, and a $^8$Li source that delivers tagged $\beta$s with an
endpoint near 14~MeV~\cite{li8}.  In addition, high energy (19.8~MeV) $\gamma$s are
provided by a $^3{\rm H}(p,\gamma)^4{\rm He}$ (`pT') source~\cite{poon} and neutrons
by a $^{252}$Cf source.  Some of the sources can also be deployed on vertical axes
within the light water volume between the acrylic vessel and PMT support sphere.

\subsection{Physics Processes in SNO}

	SNO was designed to provide direct evidence of solar neutrino
flavor change through comparisons of the interaction rates of three
different processes:

\begin{center}
  \begin{tabular}{ll}
     $ \nu_x + e^- \rightarrow \nu_x + e^-$  & (ES)\\
     $\nu_e + d \rightarrow p + p + e^-$\hspace{0.5in} & (CC)\\
     $ \nu_x + d \rightarrow p + n + \nu_x$ & (NC)\\  \\
  \end{tabular}
 \end{center}

	The first reaction, elastic scattering (ES) of electrons,
has been used to detect solar neutrinos in other water Cherenkov experiments.
It has the great advantage that the recoil electron direction is strongly
correlated with the direction of the incident neutrino, and hence the
direction to the Sun $(\cos \theta_{\odot})$.   This ES  reaction is sensitive
to all neutrino flavors.  For $\nu_e$s, the elastic scattering reaction has 
both charged and neutral current components,  making the cross section for 
$\nu_e$s $\sim$ 6.5 times larger than that for $\nu_{\mu}$s  or $\nu_{\tau}$s.

Deuterium in the
heavy water provides loosely bound neutron targets 
for an exclusively charged
current (CC) reaction which, at solar neutrino 
energies,  occurs only for $\nu_e$s.  In addition to providing exclusive
sensitivity to $\nu_e$s, this reaction has the advantage that the recoil
electron energy is strongly correlated with the incident neutrino energy,
and thus can provide a precise measurement of the $^8$B neutrino energy spectrum. The
CC reaction also has an angular correlation with the Sun which falls as
$(1-0.340\cos \theta_{\odot})$~\cite{vogel}, and has a cross
section roughly ten times larger than the ES reaction for neutrinos within SNO's
energy acceptance window.

	The third reaction, also unique to heavy water, is a purely
neutral current (NC) process.  This has the advantage that it
is equally sensitive to all neutrino flavors, and thus provides a direct measurement
of the total active flux of $^8$B neutrinos from the Sun.  Like the CC
reaction, the NC reaction has a cross section nearly ten times as large as the ES
reaction. 

	For both the ES and CC reactions, the recoil electrons are 
detected directly through their production of Cherenkov light.  For the NC
reaction, the neutrons are not seen directly, but are detected in a 
multi-step process.  When a neutrino liberates a neutron from a deuteron,
the neutron thermalizes in the D$_2$O and may eventually be captured by another
deuteron, releasing a 6.25~MeV $\gamma$ ray.  The $\gamma$ ray either Compton 
scatters an electron or produces an $e^+e^-$ pair, and the Cherenkov radiation of
these secondaries is detected.

	To determine whether neutrinos that start out as $\nu_e$s in the
solar core convert to another flavor before detection on Earth, we have
two methods: comparison of the CC reaction rate to the NC reaction rate,
or comparison of the CC rate to the ES rate.  The NC-CC comparison has the advantage
of high sensitivity. When we compare the total flux to the $\nu_e$
flux, we expect the former to be roughly three times the latter if both
solar neutrino experiments and standard solar models are correct.  In
addition, many uncertainties in the cross sections for the two processes will
largely cancel.   

	The comparison of CC to ES has the advantage that 
recoil electrons from both reactions provide neutrino spectral
information.  The spectral 
information can ultimately be used to show that any excess in the ES
reaction over the CC reaction is not caused by a difference in the effective
neutrino energy thresholds used to analyze the two reactions~\cite{fogli01,vill99}.  The CC-ES comparison
also has the advantage that the strong angular correlation of the ES electrons with
the direction to the Sun demonstrates that any excess seen is not due to some
unexpected non-solar background.  Lastly, the CC-ES comparison can be made using
both SNO's ES measurement and  the high precision ES measurement made by the
Super-Kamiokande collaboration~\cite{SK}.  This provides a high sensitivity cross
check for the CC-NC comparison with different backgrounds and systematic
uncertainties.

	The goal of the SNO experiment is to determine the
relative sizes of the three signals (CC, ES, and NC) and to
compare their rates.  We  
cannot separate the signals on an event-by-event basis;  instead, we
`extract' the signals statistically by using the fact that they are
distributed distinctly in the following three derived quantities: the
effective kinetic energy $T_{\rm eff}$ of the $\gamma$ ray resulting from the
capture of a neutron produced by the NC reaction or the
recoil electron from the CC or ES reactions,  the 
reconstructed radial position of the interaction ($R^3$)
and the
reconstructed  
direction of the event relative to the expected direction of a neutrino arriving
from the Sun ($\cos \theta_{\odot}$).  We measure the radial positions in units of
AV radii, so that $R^3 \equiv (R_{\rm fit}/R_{AV})^3 = 1.0$ when an event
reconstructs at the edge of the heavy-water volume.
\begin{figure}
\begin{center}
\includegraphics[height=0.3\textheight]{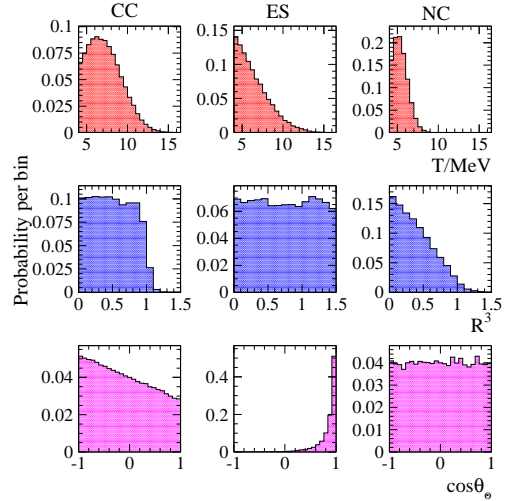}
\caption{The energy (top row), radial (middle row), and directional
(bottom row) 
distributions used to build pdfs to fit the SNO signal
data. $T_{\rm eff}$ is the effective kinetic energy of the $\gamma$ from neutron
capture or the electron from the ES or CC reactions, and $R$ is the
reconstructed event radius, normalized to the 600~cm radius of the acrylic vessel. \label{fig:pdfs}} 
\end{center}
\end{figure}

	Figure~\ref{fig:pdfs} shows simulated distributions for each of
the signals.   The top row  shows the energy distributions for each of
the three signals.  The strong correlation between the electron energy and
the incident neutrino energy for the CC interaction produces a spectrum
which resembles the initial $^8$B neutrino spectrum, while the recoil
spectrum for the ES reaction is much softer.  The NC reaction is, within
the smearing of the Compton scattering process and the resolution of the
detector, essentially a line spectrum, because the $\gamma$ produced
by the neutron capture on deuterium always has an energy of 6.25~MeV.

	The distributions of reconstructed event positions $R^3$,
normalized to the radius of the acrylic vessel $R_{AV}$, are shown in the
middle row of Fig.~\ref{fig:pdfs}.  We see here that the CC reaction, which
occurs only on deuterons, produces events distributed uniformly within the heavy
water, while the ES reaction, which can occur on any electron, produces events
distributed uniformly well beyond the heavy-water volume.  The small leakage of
events just outside the heavy-water volume (just outside $R^3=1$)
for the CC reaction is due to the resolution tail of the reconstruction algorithm.

The NC signal, however, does not have a uniform distribution inside
the heavy water, but instead decreases monotonically from the
central region to the edge of the acrylic vessel.  
The reason for this is the long ($\sim$ 120~cm) thermal diffusion length for
neutrons in D$_2$O.  Neutrons produced near the edge of the heavy-water volume 
have a high probability of wandering outside it, at which point they can
be captured on hydrogen either in the acrylic vessel or the H$_2$O
surrounding the vessel.  The capture cross section on hydrogen is nearly 600 times
larger than on deuterium, and therefore these hydrogen captures occur almost
immediately, leaving no opportunity for the neutrons to diffuse back into the
fiducial volume.  Further, 
such hydrogen captures
produce a 2.2~MeV $\gamma$ ray which is well below the analysis
threshold, and therefore events from these captures do not appear in the NC $R^3$
pdf shown in Figure~\ref{fig:pdfs}.  

	The bottom row of Fig.~\ref{fig:pdfs} shows the reconstructed
direction distribution of the events.  In the middle of that row we see
the peaking of the ES reaction, pointing away from the Sun.  The $\sim$
$1-1/3\cos \theta_{\odot}$ distribution of the CC reaction is also clear
in the left-most plot.	The NC reaction shows no correlation with the solar
direction---the $\gamma$ ray from the captured neutron carries no directional
information about the incident neutrino.

	One last point needs to be made regarding the distributions
labelled `NC' in Fig.~\ref{fig:pdfs}: they represent equally well
the detector response to any neutrons, not just those produced by 
neutral current interactions, as long as the neutrons are distributed
uniformly in the detector.  For example, neutrons produced through  
photodisintegration by $\gamma$ rays emitted by radioactivity
inside the D$_2$O will have the same distributions of energy, radial
position, and direction as those produced by solar neutrinos.  These
neutrons are an irreducible background in the data analysis, and must
be kept small through purification of detector materials.

\subsection{Analysis Strategy}

	To determine the sizes of the CC, ES, and NC signals we use the nine
distributions of Fig.~\ref{fig:pdfs} to create probability density functions
(pdfs) and perform a generalized maximum likelihood fit of the data to the same
distributions.  There are, however, three principal prerequisites before we can
begin this `signal extraction' process:  we must process the data so that we can
create distributions of event energies, positions, and directions; we need to build
a model of the detector so that we can create the pdfs like those in
Fig.~\ref{fig:pdfs}; and we need to provide measurements of any residual
backgrounds.

	Data processing begins with the calibration of the raw data, converting ADC
values into PMT charges and times.  The calibrated charges and times allow us to
reconstruct each event's position and direction, as well as estimate event energy.
We also apply cuts to the data set during processing to remove as many background
events as possible without sacrificing a substantial number of neutrino signal
events.

	The signal extraction process described above implicitly assumes that the
pdfs used in the fit are built from a complete and  accurate representation of the
detector's true response.  The model we use to create the pdfs must therefore
describe everything from the physics of neutrino interactions, to the propagation of
particles and optical photons through the detector media, to the behavior of the
data acquisition system.  The model needs to reproduce the response to signal events
at all places in the detector, for all neutrino directions, for all neutrino
energies, and for all times.  It must also track changes in the detector over time,
such as failed PMTs or electronics channels.

	Although our suite of cuts is very efficient at removing background events,
we nevertheless must demonstrate that the residual background levels are negligible
or we must produce measurements of their size.  The latter is particularly important
for the photodisintegration neutrons---because they look identical to the NC signal,
they cannot be removed, and must be measured and subtracted from the total neutron
count resulting from the maximum likelihood fit.

	Signal extraction estimates the numbers of CC, NC, and ES events; conversion
to fluxes requires acceptance corrections for each of the signals and, for the NC
signal, adjustments for the capture efficiency of neutrons on deuterons.  The final
normalization also includes neutrino interaction cross sections, detector livetime,
and the number of available targets.

	For our first publications we performed three independent
analyses of the data presented 
in this article~\cite{the:mgb,the:kh,the:msn}.  Prior to final
processing, we chose from these three analyses two
independent approaches for each major analysis component (cut sets,
reconstruction algorithms, energy calibration, etc.).  Comparisons of the
results of the independent approaches were used to validate every component
of the analysis---one approach was designated `primary' and used for the
Phase I published results, and one was designated
`secondary' and used as the verification check.  Table~\ref{tbl:analcomps}
lists the approaches for each of the analysis components. In this
article, we describe both the primary and secondary approaches
used.

\section{Data Set \label{sec:dataset}}

     The data set used in the analysis we describe here was acquired between
November 2, 1999 and May 31, 2001, and represents a total of 306.4 live days.  
Although the SNO detector is live to neutrinos during nearly all calibrations, 
data taken during the calibration periods---roughly 10\% of the time the
detector is running---is not used for solar neutrino analysis.  Other losses
of livetime result from mine power outages, detector maintenance periods and 
the loss of underground laboratory communication or environmental systems.

     The SNO data set is divided into `runs', a new run being started either at a
change in detector conditions (such as the insertion of a calibration source) or
after a maximum duration has been exceeded (in Phase I, no more than four days).
The runs used for the final analysis were selected based upon criteria external to
the data themselves. Selected runs were those for which calibration sources were not
present in the detector, no major electronics systems were off-line,  no maintenance
was being performed, and no circulation of the D$_2$O that caused light to be
produced inside the detector was being undertaken. 

	The SNO detector responds to several triggers, the primary one being a
coincidence of 18 or more PMTs firing within a period of $\sim$ 93~ns (the
threshold was lowered to 16 or more PMTs after December 20, 2000). The 
rate of such triggers averaged roughly 5~Hz.  
The detector also triggered if the total charge collected in all PMTs
exceeded 150 photoelectrons. A `random' trigger pulsed the detector at 5~Hz
throughout the data set, and a pre-scaled trigger fired after every thousandth
11-PMT threshold crossing.  Information about which condition caused the trigger for
a given event was saved as part of the primary data stream.  The
overall trigger rate was between 15 and 20~Hz.

	Although the overall detector configuration was kept stable during the
data taking period, we performed two fixes worthy of comment.  The first was a
change to the charge- and time-digitizing analog-to-digital converters (ADCs).
Soon after the start of production running, it was discovered that the ADCs were
developing non-linearities well beyond their specification.  During most of the
data taking period, bad ADCs were periodically replaced or repaired, but on August
18, 2000, a permanent fix was implemented.	In addition, roughly halfway through
the data taking period, we discovered a small rate dependence to the PMT timing
measurements. Although small, the rate dependence did affect our position
reconstruction. We developed a hardware solution to mitigate the effect, and also
created an off-line calibration to remove it.  The hardware change was completed in
December, 2000, and the off-line calibration was applied to the entire data set.

	Other minor changes---failure of individual PMTs (at an average rate of
about 1\% per year), alteration of front-end discriminator thresholds, or
repair of broken electronics channels---were tracked and the status of every
channel was stored in the SNO database at the beginning of each run for use in
the offline data analysis.  In addition, the front-end electronics timing and
charge responses were calibrated twice each week, much more frequently than the
observed variations of pedestals or slopes.  Calibration of phototube gain,
timing, and risetime response was done roughly monthly.

	To provide a final check against statistical bias, the data
set was divided in two, an `open' data set to which all analysis
procedures and methods were applied, and a `blind' data set upon 
which no analysis within the signal region (between 40 and 200 hit phototubes)
was performed until the full analysis program had been finalized.  The blind 
data set began at the end of June 2000, at which point we began analyzing just 10\%
of the data set, leaving the remaining 90\% blind.  The total size of the blind data
set thus corresponded to roughly 30\% of the total livetime.

\section{Physics and Detector Model \label{sec:model}}

	Both reconstruction of event kinetic energy and construction
of the distributions shown in Fig.~\ref{fig:pdfs} require a model of
the detector's response to Cherenkov light created by neutrino
interactions. For energy reconstruction, the model we use for the response
is analytical, and for the creation of the pdfs in Fig.~\ref{fig:pdfs}
the model is a Monte Carlo simulation.	Most of the required inputs
are the same for both models: the physics of the passage of electrons
and $\gamma$-rays through the various detector media and the associated
production of Cherenkov light, the optical properties of the detector,
and the state and response of the detector PMTs, electronics, and trigger.
In addition, for the Monte Carlo simulation to correctly predict the energy
spectra and direction distributions, it must include
the total and differential cross sections for the CC, ES, and NC neutrino
interactions, as well as the incident $^{8}$B neutrino spectrum.  Lastly,
to produce the correct radial distributions for the neutrons from the NC
reaction, the Monte Carlo model also simulates the transport and capture
of low energy ($<$20~MeV) neutrons.

	In the following section, we describe the details of each component of
the models and the calibrations applied.  As will be seen here and in subsequent
sections of this article, the Monte Carlo simulation reproduced nearly all
the distributions of interest we measured with our calibration sources to a high
degree of accuracy.

\subsection{Neutrino Spectrum and Interactions \label{sec:nuspecint}}

	  In the Monte Carlo model, neutrino energies are picked by weighting
the $^8$B neutrino energy spectrum by the neutrino interaction cross
sections, $\sigma(E_{\nu})$, for each of the three reactions (ES,CC, and NC).
The energies and directions of the secondary electrons and neutrons are generated
through a convolution of the $^{8}$B spectrum measured by
Ortiz~\etal~\cite{ortiz} with the corresponding normalized double differential
cross sections $d^2N/dE d\Omega$.  For the ES reaction, the simulation used the
cross sections as presented by Bahcall~\cite{jnb_na}, which do not include
radiative corrections (a roughly 2\% correction that was later applied to the
extracted ES rate---see Section~\ref{sec:results}).  For the CC and NC
reactions we used the calculations by Butler, Chen and Kong (BCK)~\cite{BCK},
with an $L_{1,A}$ scale factor of 5.6~fm$^3$, but then rescaled the overall
cross-sections to the values found by Nakamura {\em et al.}~\cite{nsa} and
applied correction factors to account for the radiative corrections as
determined by Kurylov {\em et al.}~\cite{kmv}.
As a general verification check, we also ran
the simulation with several other cross section
calculations~\cite{KN,NSGK}, which show agreement at the 1-2\% level.
In addition, the simulation did not include variation in the fluxes due to the
eccentricity of the Earth's orbit---this variation (and its uncertainty) were
included at a later stage in the analysis (see Section~\ref{sec:results}).  

\subsection{Background Processes}

	Radioactive backgrounds are also modeled through Monte Carlo
simulation.  The simulation includes the branching fractions into $\beta$s and
$\gamma$s of each nuclide known to be present in the detector, and includes
angular correlations between decay $\gamma$-rays if appropriate.  The
background events can be generated within any of the media represented in the
Monte Carlo simulation, including the D$_2$O, H$_2$O, acrylic, Vectran support
ropes, PMT glass and related components, PMT support structure, etc.

\subsection{Cherenkov Light from Electrons and $\gamma$-ray Interactions}

	The Monte Carlo simulation of the neutrino interactions and
backgrounds produces electrons and  $\gamma$-rays whose initial energy and 
angular distributions depend only upon neutrino and nuclear physics. We have
compared the output of the simulation at this stage to analytic calculations of
these distributions and find excellent agreement.

	To go from the initial energy and angular distributions to
the photons seen by the photomultiplier tubes, the Monte Carlo model
simulates both the propagation and interaction of electrons, neutrons,
and $\gamma$-rays within the detector media, and the consequent production
of Cherenkov light.

We used the EGS4~\cite{egs4} (Electron Gamma Shower) code to simulate the
interactions of electrons and $\gamma$-rays.  EGS4 provides some critical
pieces of physics: conversion of $\gamma$-rays into electrons through
Compton scattering, pair production, and the photoelectric effect; and
energy loss and multiple scattering of electrons~\cite{egsnote}.  At solar neutrino
energies, multiple scattering of the electrons as they propagate severely distorts
the Cherenkov cone, and we therefore simulate the production of Cherenkov light by
adding Cherenkov photons along each electron's entire trajectory.  


The EGS4 code simulates individual tracks by a series of straight segments,
with a small fractional change in the kinetic energy in each step arising from
energy loss in the medium. At the end of each step an angular deflection is
generated, drawn from the Moli\`{e}re distribution, to simulate multiple
scattering.  If all Cherenkov photons from a given step are produced at the
Cherenkov angle $\theta_c$ relative to the direction of the straight track
segment, the final pattern will be a series of cones. If the step size is
doubled the number of cones is halved; the angular distribution of the
Cherenkov light is thus sensitive to the step size. This artifact is removed by
linearly interpolating, for each photon generated, the local direction cosines
of the track between successive steps.

To choose the optimal EGS4 step size, we compared the output
of our implementation of the EGS4 code to data on electron scattering,
and found that energy step sizes in the range of 0.001~MeV to
0.05~MeV  reproduced the data best~\cite{the:lay}.  We verified the EGS4
treatment of multiple scattering by comparing output Cherenkov distributions
averaged over many electron trajectories with those from an independent
Goudsmit-Sanderson treatment of multiple scattering. With a step size of 1\% in
energy loss, we found very good agreement when the interpolation of direction
cosines is included, even at energies as low as 1~MeV.

For generating Cherenkov light on each segment of an electron's path,
we use the asymptotic formula for light yield:
\begin{equation}
\frac{dI}{d\omega}=\frac{\omega e^2 L sin^2\theta_c}{c^2}
\label{eq:cherenkov}
\end{equation}
In Eqn.~\ref{eq:cherenkov}, the yield {\it I} (with dimensions of energy per unit
frequency interval), is given as a function of angular frequency, {\it $\omega$},
and is proportional to path length {\it L}.  We have verified  the use of this
asymptotic formula by calculating the interference between two unaligned segments,
and have found that the interference does not produce significant lowering of light
yield.

The number of photons produced is then sampled from a Poisson
distribution and the creation points of these photons are positioned
randomly along the segment.  Photons are emitted at an angle $\theta_c$
to the electron track direction, which is interpolated as described above, and
is kept fixed within each step of the track. 

\subsection{Neutron Transport}
\label{sec:mcnp}

	In addition to electrons and $\gamma$-rays, the Monte Carlo model must
account for the propagation and capture of neutrons throughout the detector
media.  The most important of these neutrons are those which result from 
disintegration of deuterons through neutrino neutral current
interactions, and those produced through photodisintegration of 
the deuterons by $\gamma$-rays.  

For neutron propagation,  we use the MCNP~\cite{mcnp} neutron transport code
developed at Los Alamos National Laboratory, but restrict its use to the
propagation of neutrons, ignoring additional particles (e.g. $\alpha$s)
which may be created by neutron interactions.  The creation of additional
particles is recorded, but the particles are not propagated, with the
exception of $\gamma$-rays and electrons which are handled by EGS4.
MCNP was chosen because of its widespread verification and usage, and
because of its sophisticated handling of thermal neutron transport
in general and molecular effects in H$_2$O and D$_2$O in particular, without
which accurate simulation of neutron transport in the SNO detector
could not be carried out.

MCNP is primarily intended as a non-analog code, which uses weighted sampling
techniques to study rare processes. It has a set of physics-related routines
that form the core of its simulated neutron transport, and it is these that are
used in the Monte Carlo simulation.  The MCNP code uses extensive data tables
to provide partial and total interaction cross sections as a function of
neutron energy, the energy-angle spectrum of the emergent neutrons, and other
interaction data.

To verify our implementation of MCNP, we compared many of the
low-level simulation parameters in several different media, such as the
neutron step length, the emitted neutron energy, and the directions of
initial and final trajectories for each interaction.  We performed these
tests for neutron energies from $10^{-3}$ eV to 10~MeV, and in over a
thousand comparisons of distributions between MCNP and our simulation,
none were found to be anomalous.

We also checked that our simulation could reproduce representative cross
sections at thermal energies, and match the diffusion equation closely
in the limit $\Sigma \ll \Sigma_a$, where $\Sigma$ and $\Sigma_a$ are the
macroscopic interaction and absorption cross sections, respectively. MCNP
(and hence our simulation) has been shown by Wang {\it et al.}~\cite{wang}
to predict the absolute number of neutrons captured in an experiment
involving neutron thermalization with an accuracy of at worst 3\%. At
the same time, Wang {\it et al.} have shown that the ratio of the numbers
of captured neutrons predicted by MCNP in related experimental setups is
accurate to within 0.3\%. Based on our studies, we believe these numbers
apply to the SNO detector as well.

\subsection{High Energy Processes}

	To simulate muon events and any other lepton above 2~GeV, the SNO Monte
Carlo simulation relies on the CERN package LEPTO 6.3~\cite{pkgs,lepto}. The
lower-energy electromagnetic components of the resultant muon showers are then
passed to the EGS4 code and the rest of the SNO simulation, as described above.
Hadrons produced by the interaction of these muons are handled by the FLUKA and
GCALOR packages.

\subsection{Detector Geometry}

	The Monte Carlo simulation includes a detailed model of the
detector geometry, including the position and orientation of the PMT support
sphere and its resident PMTs, the position and thickness of the acrylic vessel
including support plates and ropes, the size and position of the acrylic
vessel `neck', and a full model of the structure of the PMTs and their associated
light concentrators.  The values were based primarily upon surveys and measurements taken
before the elements were installed in the detector. The positions of the acrylic
sphere and PMT support sphere were updated after the detector was filled with water,
to account for the effects of buoyancy.  For the work we describe in this article,
all simulations assumed that the acrylic vessel and PMT support sphere were
concentric, though small adjustments to this were made at a later stage in the
analysis (see Section~\ref{sec:sigex}).  

The orientation of the PMT array with respect to true North was determined
on the cavity deck after the detector  was constructed and filled with
water, by surveying chords between the PMT array suspension points with a
commercial marine gyrocompass.  Multiple chords were surveyed and averaged
and coupled to detailed deck surveys, PMT array construction drawings, and
field tests of the geodesic sphere's rigidity.  The absolute orientation
of the array was determined to ~0.5 degrees. This survey was in reasonable
agreement ($2.5^\circ$) with the original Inco mine surveys.  The coordinate system
used for the Monte Carlo model and for data analysis put $z$ along the detector's
vertical axis, and $x$ along true North.

\subsection{Detector and PMT Optics}
\label{sec:optics}

	By far the most important parts of the detector model 
are the optical properties of the detector media and the
photomultiplier tubes.	SNO is optically more complex than previous water
Cherenkov detectors: photons traverse multiple optical media from
the fiducial volume to the PMTs, and the light concentrators surrounding
the PMTs have their own optical properties.  Therefore the energy response
of the SNO detector varies significantly with radial position and event
direction---an event near the edge of the volume and pointing outward produces
a very different ($\sim$ 5\%)  number of hits than an event pointing inward,
which is yet different from an event near the center. For more detailed
descriptions of the optical measurements, see Refs.~\cite{the:moffat,the:ford}.

	Although we extensively calibrated the detector with Cherenkov
sources of different energies and characteristics that were deployed at many
different positions, the optical model provides a way of predicting the response at
positions, energies, directions, and times (of year) not sampled by the sources.
The model is used both in a Monte Carlo simulation of the detector's response to
neutrino and background events, and in an analytic form to estimate the energy of
each event (see Section~\ref{sec:enecal}).

	In principle, there are many optical parameters which must be measured:
attenuation and scattering lengths of D$_2$O, acrylic, and H$_2$O, the reflection
coefficients at the D$_2$O-acrylic interface, the acrylic-H$_2$O interface, and of
the PMTs, light concentrators, and PMT support sphere.  For the optical measurements
we describe in this article, we considered only light in a narrow ($\pm 4$~ns)
timing window, called the `prompt time window'.  The prompt time window allows us to
characterize scattering as an additional attenuation, and allows us to accurately
calculate a response without requiring detailed knowledge of the geometry and
parameters of reflections.

\begin{figure}
\begin{center}
\includegraphics[height=0.25\textheight]{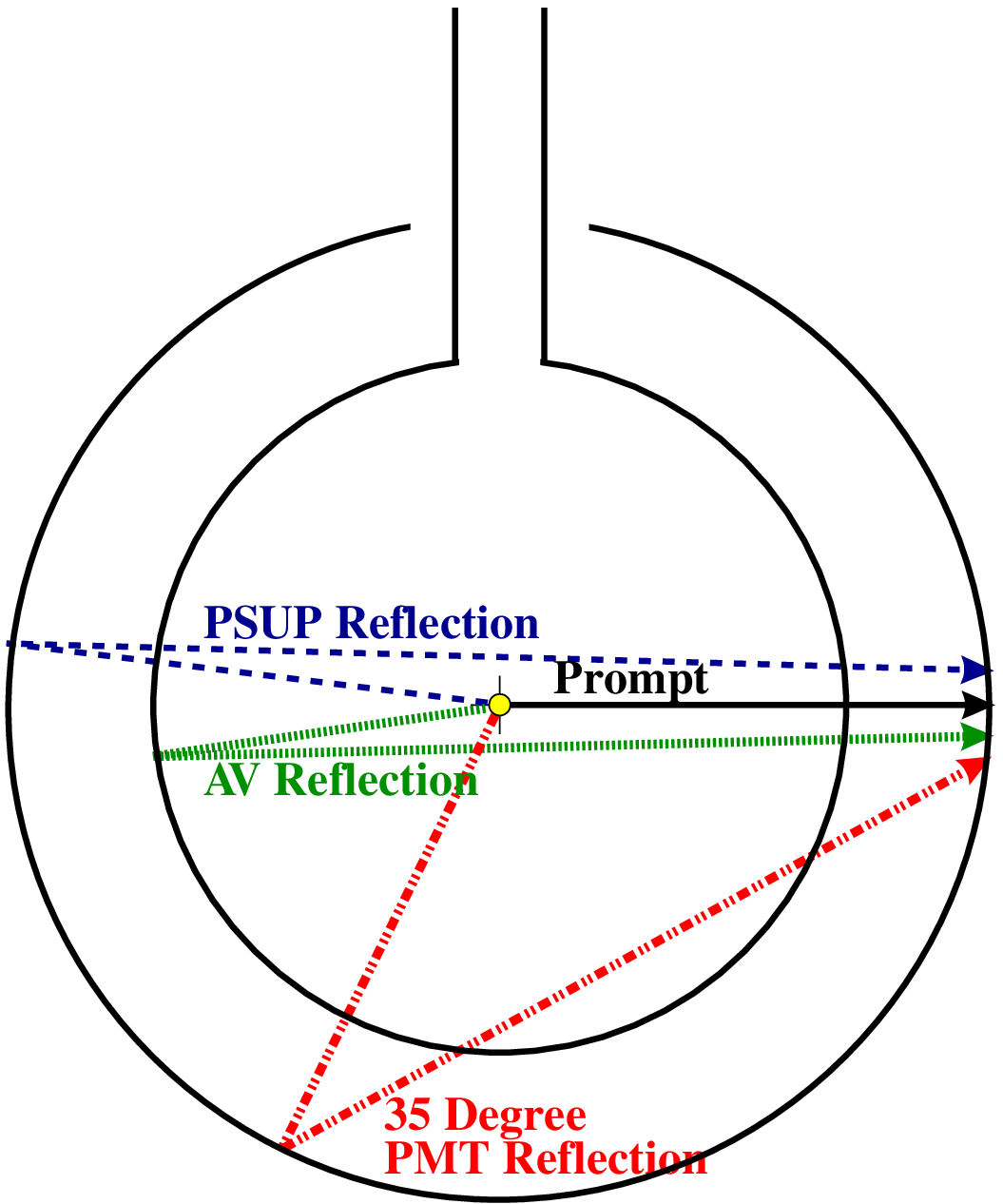}
\vspace{0.1in}
\includegraphics[height=0.25\textheight]{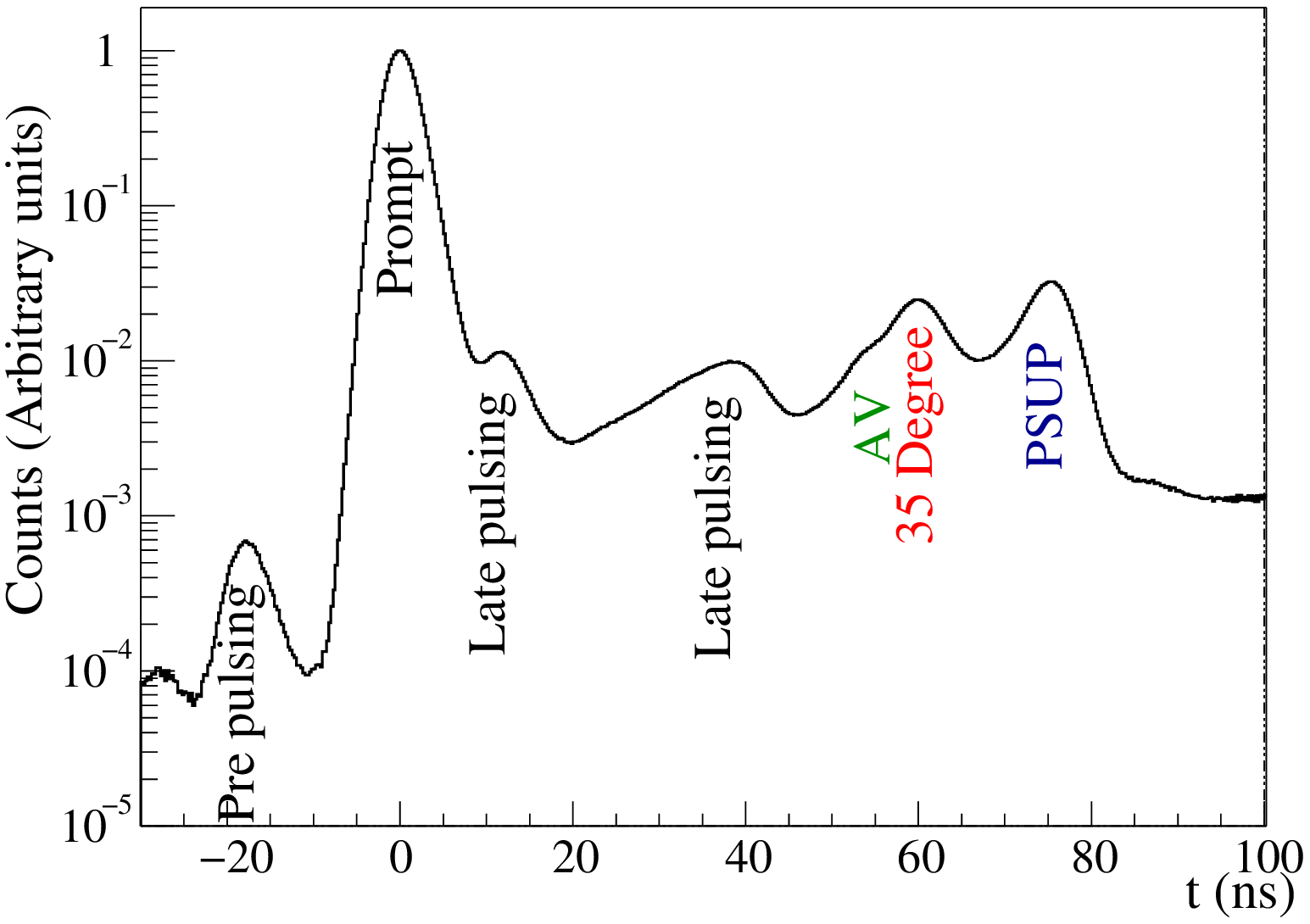}
\caption{{\it Top}: Optical light paths within the detector.  {\it Bottom}: PMT time residual distribution for laser data.\label{fig:lastim}}
\end{center}
\end{figure}

	We measured the optical parameters using a pulsed nitrogen laser
source (the `laserball') whose light was transmitted into the detector through an optical fiber and diffused in a small sphere containing 50~$\mu$m diameter glass
beads suspended in a silicon gel.  In addition to the primary wavelength of 337.1~nm, a
series of dyes provided additional wavelengths of 365, 386, 420, 500, and
620~nm. These values were chosen to provide good coverage over the range of
detectable Cherenkov wavelengths.  The top panel of Figure~\ref{fig:lastim}
illustrates the various optical paths taken by the light for the source at the
center of the detector, and the bottom
panel the measured distribution of the differences between PMT hit times and
the laserball trigger time, corrected for photon time-of-flight (the
`time-residual distribution').
As the figure shows, the prompt window of the time-residuals is centered on the peak at $t=0$, and several other peaks
including the reflections off the acrylic and the PMT array are indicated.

As with nearly all SNO calibration sources, the laserball can be deployed
almost anywhere in two orthogonal planes within the acrylic vessel, as well as
outside the vessel along a few vertical axes.  For the data scans used to
determine the optical parameters, we collected data four times with the
laserball at the center and 18 times off-center at radii between 100~cm and
500~cm.  Each of the central-position data collections was done with four
different azimuthal orientations of the laserball, to help understand anisotropies
in its light output.  We kept the laser intensity relatively low (typically
only about 5\% of the PMTs registered hits for each laser pulse) so that
the corrections that we applied to  account for multiple photons hitting a
single tube were small.

The optical model used to predict the number of prompt counts
$N_{ij}$ observed
in PMT $j$ in a given run $i$, within the $\pm4$~ns window, is parameterized
as follows:

\begin{equation}
N_{ij}=N_{i}\Omega_{ij}R_{ij}T_{ij}L_{ij}\varepsilon_{j}e^{-\left(d_{d}\alpha_{d}+d_{a}\alpha_{a}+d_{h}\alpha_{h}\right)}. 
\label{eq:model}
\end{equation}

$N_{i}$ is a normalization parameter, proportional to the number of photons
emitted by the laserball in run $i$ that can be detected within the prompt
time window at each PMT. $\Omega_{ij}$ is the solid angle subtended by PMT
$j$ with respect to the source position for run $i$. $R_{ij}$
is the PMT and concentrator assembly response aside from solid angle considerations,
parameterized as function of the incidence angle on the PMT. $T_{ij}$ is the
product of the
Fresnel transmission coefficients for the heavy-water/acrylic/light-water
interfaces. $L_{ij}$ is the laserball light intensity distribution,
parameterized as a function of the polar and azimuthal angles of the light ray
relative to the laserball center. The $\varepsilon_{j}$ are the relative 
PMT efficiencies for normally incident light, combining concentrator, PMT and
electronics effects.  $d_{d}$, $d_{a}$ and $d_{h}$ are the distances of the light
paths through the D$_{2}$O, acrylic, and H$_{2}$O respectively.  The
$\alpha$s are the attenuation coefficients of the respective media, including the
effects of both bulk absorption and Rayleigh scattering.

The parameters $\Omega_{ij}$, $T_{ij}$, $d_{a}$, $d_{d}$
and $d_{h}$ can be calculated from the source position and
detector geometry, but the normalization $N_{i}$
and laserball intensity distribution $L_{ij}$ must be determined from
the source data, together with the parameters required for the optical response
model, $R_{ij}$, $\alpha_{d}$, and $\alpha_{h}$.
The acrylic attenuation coefficient, $\alpha_{a}$, is fixed to
\emph{ex-situ} measurements performed as described
in~\cite{av}. To take into account the probability of multiple
photoelectron (MPE) hits, the number of prompt counts $N_{ij}$ is
corrected by inverting the expected Poisson distribution of the hit counts,
\begin{equation}
N_{ij}^{MPE}=-N_{pulses}\ln(1-N_{ij}/N_{pulses}), 
\end{equation}
where $N_{pulses}$ is the total number of laser pulses in the run.

To remove the dependence on the imprecisely known PMT efficiencies $\varepsilon_{j}$,
instead of $N_{ij}^{MPE}$ for each PMT we use an `occupancy ratio' $O_{ij}$ of the
MPE-corrected number of counts in PMT $j$ for run $i$ to the MPE-corrected number of
counts for a run with the laserball in the center of the detector, $O_{0j}$.

The terms that can be calculated purely from source-PMT geometry are the solid
angle $\Omega_{ij}$ and the product $T_{ij}$ of the Fresnel transmission
coefficients.  These two terms are used to correct the occupancy ratios measured
with calibration data:

\begin{equation}
 O^{data}_{ij} = \frac{N_{ij}^{MPE}}{N_{0j}^{MPE}}( \frac{ \Omega_{0j}T_{0j} } { \Omega_{ij}T_{ij} }) 
\end{equation}
The occupancy ratio calculated from the optical model is: 
\begin{equation}
 O^{model}_{ij} = \left( \frac{N_{i}^{MPE}}{N_{0}^{MPE}} \right) \frac{R_{ij}L_{ij}}{R_{0j}L_{0j}} e^{ \delta d_{d} \alpha_{d} + \delta d_{a} \alpha_{a} + \delta d_{h}
\alpha_{h}} 
\end{equation}
Here $\delta d_x= d_{ij} - d_{i0}$ is the difference in path length
between run $i$ and a run with the laserball in the center for light
traveling from the laserball to the $j$th PMT through each of the
three  modeled media (heavy water, acrylic, and 
light water) .
We then derive the optical parameters by minimization of the $\chi^2$
between the data and the model: 
\begin{equation}
 \chi^{2} = \sum^{N_{runs}}_{i}
\sum^{N_{PMT}}_{j} \frac{(O^{data}_{ij} - O^{model}_{ij})^{2}} {
(\Delta O_{ij})^{2} + \sigma^{2}_{PMT ij} } 
\label{eq:optchi2}
\end{equation}

The parameters over which $\chi^2$ is minimized are the attenuation
coefficients, the average angular response $R_{ij}$ (assumed to be the same for
every PMT) as a function of the incident angle of the light, the normalization
constant $N_i$, and the laserball anisotropy $L$ as a function of solid angle.
In Equation~\ref{eq:optchi2},  $\Delta O_{ij} $ is the statistical uncertainty
in the occupancy ratio due to counting statistics and $\sigma^{2}_{PMT ij}$ is
an additional uncertainty introduced to account for tube-by-tube variations
in the PMT angular response as a function of the incidence angle of the light. 

\begin{figure}
\begin{center}
\includegraphics[height=0.2\textheight]{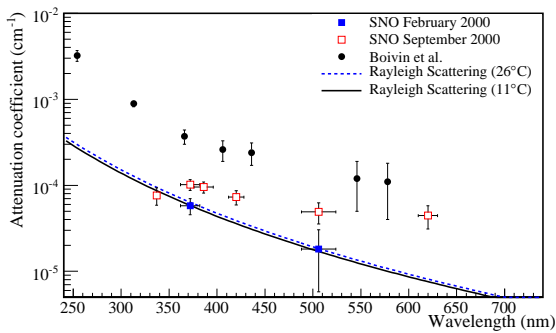}
\caption{Measured D$_2$O attenuation lengths, compared to the data of
Boivin~\etal~\cite{boivin}.  \label{fig:d2oatten}}
\end{center}
\end{figure}

\begin{figure}
\begin{center}
\includegraphics[height=0.2\textheight]{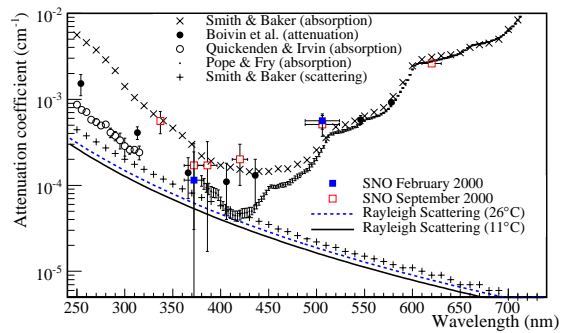}
\caption{Measured H$_2$O attenuation lengths compared to the data of Smith and
Baker~\cite{smbaker}, Boivin~\etal~\cite{boivin}, Quickenden and Irvin~\cite{quick}, and Pope and Fry~\cite{pope}.  \label{fig:h2oatten}}
\end{center}
\end{figure}

	Figure~\ref{fig:d2oatten} shows the D$_2$O attenuation lengths measured
in SNO for two different data sets, compared to previous measurements and the
Rayleigh scattering limit.  We see that the SNO heavy water is the clearest
large sample ever measured.  Figure~\ref{fig:h2oatten} shows the attenuation
lengths for the light water surrounding the heavy water volume.

	In addition to the attenuation lengths, minimization of the
$\chi^2$ shown in Equation~\ref{eq:optchi2} also returns the response
of the photomultiplier tubes and light concentrators as a function of
incidence angle.  The form of this response is one of the biggest sources
of the position-dependence to the overall detector response.

	Within the fit to the optical model, we parameterize the angular
dependence as a simple binned response function, with 40 bins ranging from
normal incidence to the highest angle possible from sources inside the heavy
water volume (roughly $40^\circ$).  Here, normal incidence is defined as normal
to the front plane of the PMT and concentrator assembly (the face of the
concentrator `bucket'), or, in other words, parallel to the PMT axis of
symmetry.  For the detector response used in the energy calibration (see
Section~\ref{sec:ecalibrator}), it is this binned form which is used.  

\begin{figure}
\begin{center}
\includegraphics[height=0.40\textheight]{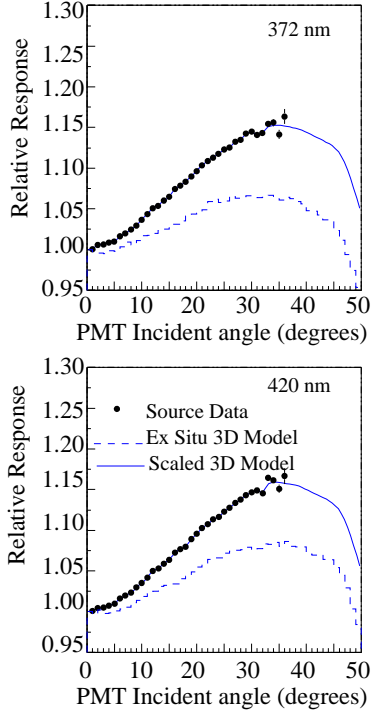}
\caption{PMT angular response model compared to laser source data and to the hybrid model that adjusts the {\it ex-situ} measurements to agree with the data, for two
wavelengths. Note that the zero has been suppressed. \label{fig:pmtang}}
\end{center}
\end{figure}

	Within the Monte Carlo simulation, however, the Cherenkov photons are
tracked through a complete three dimensional model of the PMT geometry.
The model was based entirely on {\it ex-situ} measurements of the 
photocathode and concentrator assembly~\cite{the:lay}. By including the
full geometry, the Monte Carlo model has the advantage 
that it correctly reproduces
the timing of reflected photons, in particular the important `$35^\circ$
reflections' shown in Fig.~\ref{fig:lastim} that occur when a photon bounces off 
the photocathode and then the PMT concentrator~\cite{the:moffat}. These reflected photons ultimately affect the
accuracy of event position reconstruction, which depends upon the timing
of the PMT hits.  Rather than using the optical fit of
Equation~\ref{eq:optchi2} to extract all the microscopic parameters
associated with the three dimensional PMT model,
we created a hybrid model in which a
small number of the three dimensional parameters were tuned in order to reproduce the
binned angular response derived from the optical fit.  These parameters altered
the {\it ex-situ}-measured PMT photocathode efficiency as a function of radial distance from the PMT central
axis.  Light that strikes the concentrators at normal incidence (defined
the same way as above) is reflected to the
edge of the photocathode, and thus with the tuned photocathode efficiency the 
overall hit probability for these photons was reduced.  Figure~\ref{fig:pmtang}
shows the comparison between the resultant modeled response and the 
measurement.  With the hybrid model, we
correctly reproduce both the PMT timing and angular response,
at the cost of a somewhat phenomenological (rather than an entirely
physical) basis for the Monte Carlo model.

	We have studied the sensitivity of our optical measurements to
laserball position uncertainties, data selection criteria, laserball isotropy
and acrylic vessel position.  The dominant systematic uncertainties
associated with the optical parameters arise from uncertainties in the position
of the laser source relative to the PMTs, and enter primarily through
calculation of the PMT solid angle used in Equation~\ref{eq:model}.  We
estimated these uncertainties in several different ways, including making
independent measurements of the positioning of the sources by touching the
walls of the acrylic vessel, and timing the reflections of the laser light off
the PMT array.  

\subsection{Energy Scale \label{sec:enescale}}

The calibrated optical parameters are used as input to the Monte Carlo model.  The
model accounts for photon scattering and absorption, tracking through the region of
the PMT concentrators, to the PMT face, and ending with absorption in the photocathode
and photoelectron emission.   Electron optics in the PMT and subsequent charge
collection and discrimination are not modeled, but an overall efficiency for these
processes is included as a probability for a given photo-electron to produce a PMT
hit.  This probability is defined as
\begin{equation}
P_{hit} = \epsilon_{c} \epsilon_{t}
\label{eq:phit}
\end{equation}
\noindent
where $\epsilon_{c}$ is the efficiency for collecting photoelectrons produced
at the photocathode onto the first PMT dynode ($\sim$ 70 \%),
and $\epsilon_{t}$ is the fraction of PMT pulses which generate a hit after passing through the electronics
chain and discriminator (approximately 80\% for 1/4 p.e. threshold), so that $P_{\rm hit} \approx 0.56 $.  $P_{\rm hit}$ thus sets the detector's `energy
scale', and allows the model to correctly predict the number of detected PMT 
hits per MeV given an event's location and direction.  Phenomenologically,
the determination of $P_{\rm hit}$ corresponds to determining the average quantum
and detection efficiency of the PMT array, though in practice it includes
other effects such as incompletely modeled optical responses and 
the efficiencies of the instrumentation. 

As is described later in Section~\ref{sec:enecal}, we used two estimators of an
event's energy: an estimation based on the raw number of total hits in the event (the
`$N_{\rm hit}$' estimator), and an estimation based on hits in a narrow $\pm 10$~ns time
window, corrected for position- and direction-dependent effects (the energy
`reconstructor').  The energy reconstructor was used to produce the initial
Phase I results~\cite{snocc,snonc,snodn}, and the $N_{\rm hit}$ estimator, which has different sensitivities to
systematic effects, was used as a verification check.  The energy reconstructor's
$\pm 10$~ns window was chosen to be wider than the $\pm 4$~ns optical calibration
prompt-time window to maximize the number of hits available for reconstruction,
without needing to include significant corrections for scattered or reflected
photons.

To determine the absolute energy scale for both estimators, we compared
Cherenkov events from the ${}^{16}$N calibration source to Monte Carlo 
predictions of the detector's response to the source.  The code used to make
the predictions simulated the production and emission of $\gamma$-rays, and
included a model of the source geometry and optics. The state of the detector
(for example, the average PMT noise rate and off-line or inoperative PMTs and
electronics channels) at the time of the calibration run was taken into
account.

$P_{\rm hit}$ is determined using $^{16}$N data with the source deployed at the
detector center.  For the energy reconstructor, we found the peak of the in-time hit
distribution occured at 36.06 hits, for $^{16}$N runs taken mid-way through the
D$_2$O phase.  Based on this number, the value of $P_{\rm hit}$ which correctly
scaled the Monte-Carlo was 0.566, a correction of approximately $5\%$ to the value
of $P_{\rm hit}$ determined from {\it ex-situ} estimates of the PMT collection
efficiency and hardware thresholds.  The energy scale was sampled by many ${}^{16}$N
calibration runs made throughout the running period. As shown in
Fig.~\ref{fig:edrift},  we found a small energy scale drift which appeared to be
caused by small changes in detector optics or PMT characteristics to which the
optical calibration was not sensitive, such as the global PMT quantum efficiency.
To correctly model the response as a function of time, we therefore applied a
correction to event energy using a piecewise linear fit to Fig.~\ref{fig:edrift}
(described further in Section~\ref{sec:enecal}).  In the Monte Carlo model, we used
a fixed energy scale for all simulations, set to reproduce data taken during the
middle of the data acquisition period. Note that the absolute calibration of $P_{\rm
hit}$ and the drift correction function are the only corrections applied to the
simulation, after the inputs from the optical model.

\subsection{Electronics and Trigger \label{sec:mcdaq}}

	The Monte Carlo model includes many of the details of the detector
instrumentation.  We tracked the detector state run-by-run, saving in the SNO
database information such as the number of electronics channels online, number
of working PMTs, and number of working trigger signals.  This 
information was fed into the Monte Carlo simulation, so that each data run
was simulated with the correct detector configuration.  Although the thresholds
and gains of the individual PMTs were also tracked, we did not use this
information to simulate individual PMT responses, but set all PMTs
to the average (see Section~\ref{sec:enescale}).

	The PMT noise rate was also tracked in every run using the
pulsed trigger described in Section~\ref{sec:dataset}.  The average noise
rate for each run is used in a simple Poisson model to add noise hits
to the simulated events.

	The PMT hit timing was simulated using test-bench timing
measurements, and included a nearly Gaussian prompt peak whose width was
1.6~ns, as well as the pre-pulsing and latepulsing structure seen in
Fig.~\ref{fig:lastim}.  We simulated the PMT single photoelectron charge
spectrum also using distributions drawn from test-bench measurements, with each
PMT assumed to have the same gain.  We did not simulate tube-by-tube
efficiencies due to different PMT thresholds and gains.

An `event' within the simulation is subject to the same trigger
criterion as events in the SNO detector, using a model of the analog
trigger signals themselves~\cite{the:msn,elec}.  Although the model can include the measured
trigger efficiencies, the SNO trigger threshold is set so low and the
trigger efficiency is so high that the difference between using a `perfect'
trigger and the true trigger efficiencies in the model was negligible.
We therefore simulated events with perfect efficiency.

	After an event is triggered in the simulation, the PMT times are
calculated relative to the trigger time and stored along with the simulated PMT
charges.  We did not digitize the PMT times and charges in the simulation
because studies of the effects of the digitization showed only negligible effects
on the analysis.  The final simulated data thus looked like calibrated PMT
times and charges.

%

\section{Data Processing \label{sec:dataproc}}
	
	In this section we describe the data processing used to calibrate, filter,
and reconstruct the data set.  As discussed in Section~\ref{sec:overview} and shown
in Table~\ref{tbl:analcomps}, we created multiple distinct methods for all major
analysis components.  In the following we discuss the multiple methods used for 
identification and removal of instrumental backgrounds, position and direction
reconstruction, and energy estimation.  We leave the estimation of the numbers of 
residual background events to Section~\ref{sec:bkds}.

\subsection{Raw Data}

	Each event recorded by the SNO detector contains several items of `header'
information: the trigger ID number, a word specifying which trigger or triggers
fired in the event, the master clock time, and an absolute clock time synchronized
to the GPS system.  The GPS system provides time with resolution of 100~ns and an
accuracy of $\sim 300$~ns. For each hit channel three digitized charges
(a high gain, short integration-time charge; a high gain, long integration-time
charge; and a low gain, long integration-time charge) and one time are recorded.
All hit times are relative to the time-of-arrival of the global trigger.  

\subsection{Charge and Timing Calibrations}
\label{sec:ecapca}

	To convert the digitized charges and times to values that can be
used in reconstruction and energy calibration, we subtracted pedestal values
and converted the times from ADC counts to nanoseconds.  The time conversion
was done by linearly interpolating between 10 precisely measured pulser
calibration times.  The digital resolution for the times was
approximately 0.1 ns, more than 10 times smaller than the intrinsic PMT time
resolution.  The charges were not converted into picocoulombs or
photoelectrons, but left as pedestal-subtracted ADC count values.

       The pedestals and timing slopes were measured twice weekly, and during
data processing we applied the most recently measured set of calibrations.  The
pedestals were extremely stable---the variations from calibration to
calibration were typically as small as could be measured (below one ADC count).  The
output of the pedestal and time calibration included quality control flags
that we used to reject channels which were noticeably bad, or came from boards
that had been replaced but not yet calibrated.

	In addition to the pedestals and slopes applied to the digitized times,
we also measured and subtracted the global channel-to-channel timing offsets
(caused by differences in PMT transit times and small variations in signal path
lengths) using data from the laserball source described in
Section~\ref{sec:optics}.  The laserball data also provided us with a
charge-dependent correction to the measured PMT times, necessary to account for the
variation due to the risetime of the PMT pulses. 

	As was discussed in Section~\ref{sec:dataset}, during the data
acquisition period we discovered two problems with the charge and timing
calibrations. The first problem was the slow development of non-linearities in
the time- and charge-digitizing ADCs. Although we ultimately developed a
hardware fix for the ADCs, for data taken before the fix was implemented we
applied the quality control flags discussed above to reject affected channels.

	The second problem was the small rate dependence of the time and charge
pedestal values---the pedestal calibrations were typically taken at high rate
while the actual neutrino data was low rate, and therefore the `true' pedestal
needed for the neutrino data could be a few counts different from the
calibrated pedestal value. We developed a hardware solution to mitigate this
problem, too, but also adjusted the time pedestal of each channel offline based
upon the time since it last recorded a hit.  This adjustment removed most of
the problem, but for nearly all important calibrations (such as energy scale or
the reconstruction of event position) we used radioactive source data taken at
both high and low rates to ensure there were no residual effects.  The rate
dependence of the charge measurement was not corrected, but, as described later
in Section~\ref{sec:ecalibrator}, the overall analysis was designed to 
depend only weakly on the charge measurement.

\begin{figure}[h]
\begin{center}
\includegraphics[ width = 3.0in ] {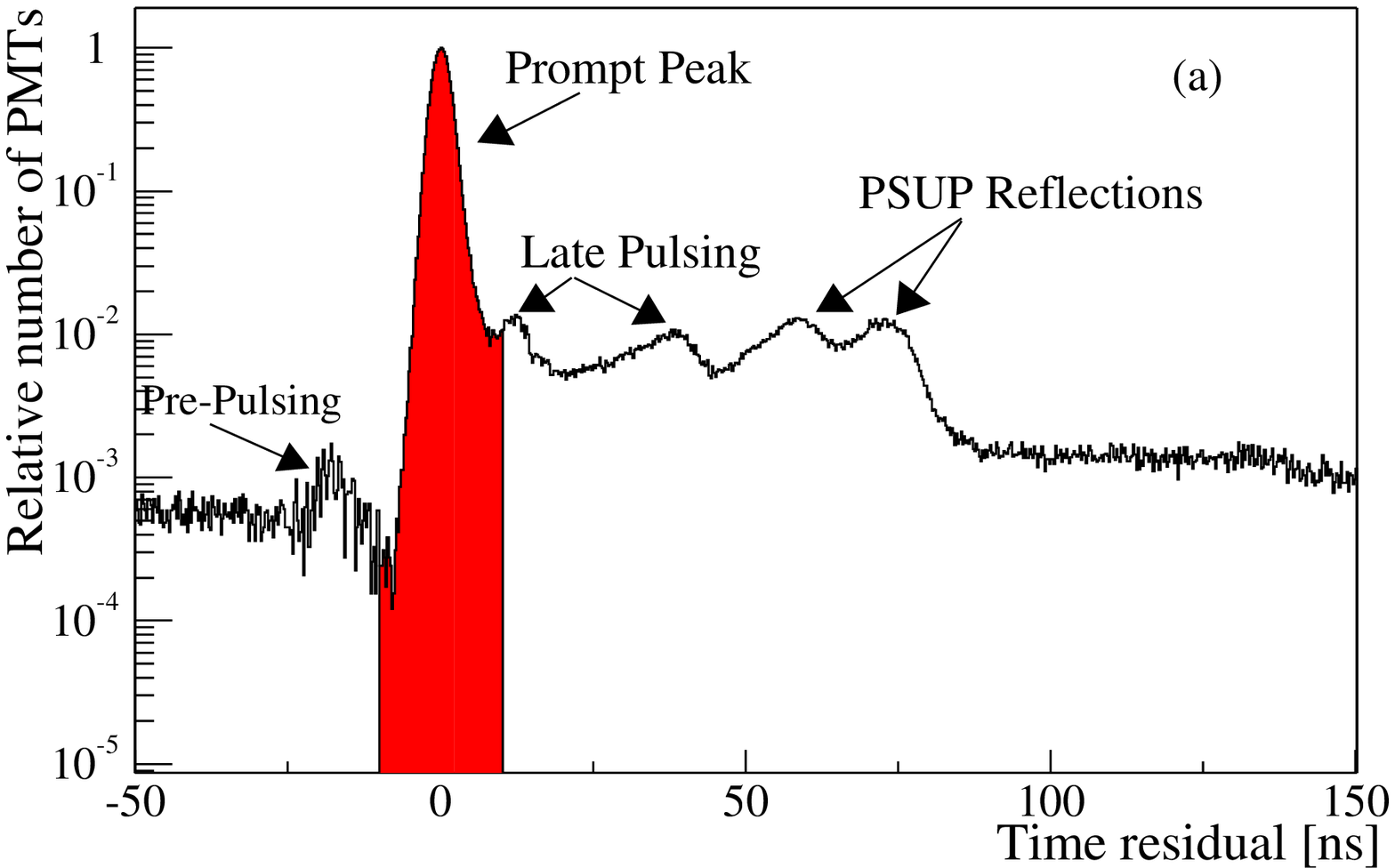}
\includegraphics[ width = 3.0in ] {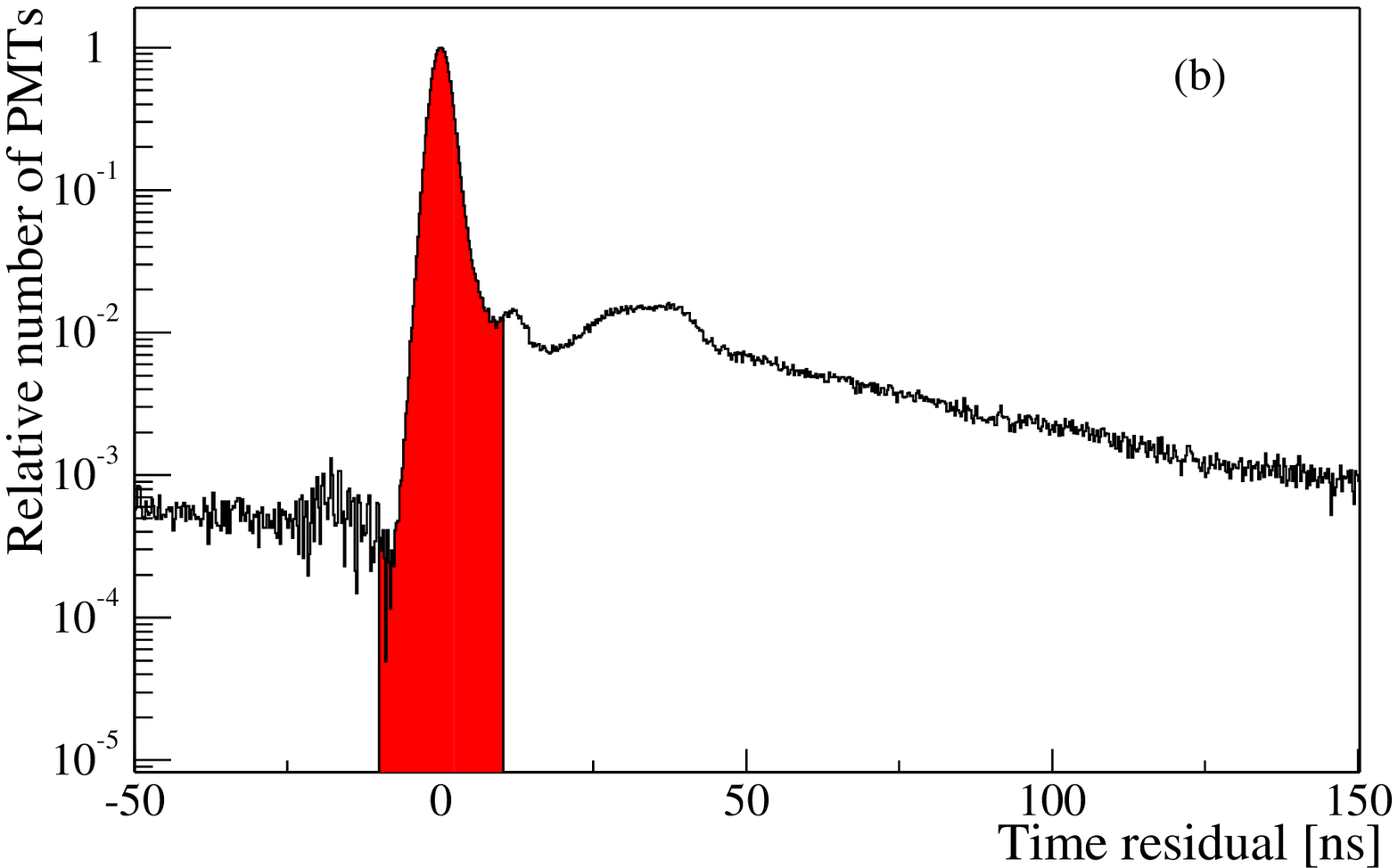}
\caption[$^{16}$N time residual histograms from calibration data]{Time residual histograms from ${}^{16}$N calibration data at (a) center of the detector and
(b) $r=500$ cm.  The pre-pulsing and late pulsing peaks are properties of the PMTs and do not depend on the source position, while the reflection peaks
from the PMT support sphere vary with source position.  The shape and fraction of light in the main peak used for energy calibration (shaded) is reasonably 
insensitive to source position.
\label{fig:tres_data}}
\end{center}
\end{figure}

	Figure~\ref{fig:tres_data} shows the width of the `prompt' peak of the
time residuals for the $^{16}$N calibration source deployed at the center of
the detector.  The 1.5~ns width is slightly better than what we had anticipated
based on benchtop measurements.

\subsection{Instrumental Background Cuts \label{sec:inst}}

In addition to neutrino interactions, cosmic rays, and radioactive decays,
the SNO detector also collects and records many background
events produced by the detector instrumentation itself.  They have several
sources and span the energy range of interest for solar neutrino
analysis.  Although these events are relatively easy to distinguish
from neutrino events, because of their much higher frequency a high
rejection fraction is needed to ensure they do not contaminate the final data
sample.  More information on the instrumental backgrounds and the cuts used to
remove them can be found in Appendix~\ref{sec:apdxa} and
Refs.~\cite{the:mccauley,fistdoc}.

There are four distinct classes of instrumental background sources:
\begin{itemize}
\item Photomultiplier Tubes

	Small discharges within a PMT can produce detectable light.
Although for a single PMT this occurs rarely (roughly once each week),
integrated over the entire array we see roughly one such `flasher' event each
minute.  Further, seismic activity within  the mine---either natural or
mining-related---can cause thousands of PMTs to flash within several tens of
milliseconds.

	The PMTs can also produce light due to high voltage breakdown in
their connectors or bases.  Such events light up nearly the entire PMT array,
and are accompanied by electronic pickup in neighboring electronic channels and
crates. 

\item External Light

Light outside the PMT array can generate detectable hits by entering
through the neck region of the acrylic vessel or through the backs of the
photomultiplier PMTs.  For example, static discharges in the neck of the
acrylic vessel, and at the boundary of the acrylic, nitrogen cover gas,
and the water surface, can produce hits at the bottom of the PMT array.

\item Electronic Pickup

	Activity near the electronics racks causing electronic noise can
produce radiative pickup in many channels at once.  Readout of a
crate can occasionally produce hits confined to a single card in an
electronics crate.

\item Acrylic Backgrounds.

The acrylic vessel itself sometimes emits isotropically distributed
light at several locations; this light does not appear to be associated with
any radioactivity.

\end{itemize}

\begin{figure}[h]
\begin{center}
\includegraphics[ width = 2.8in ] {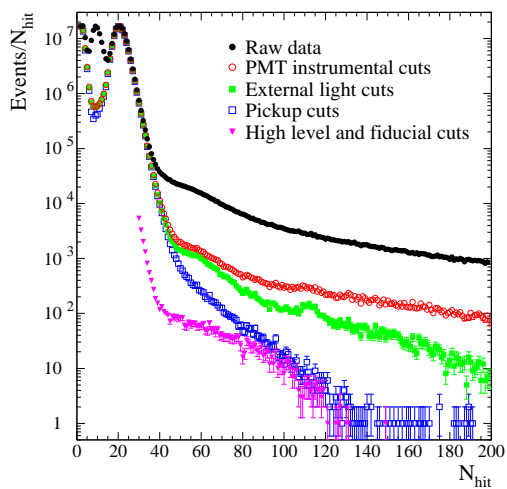}
\caption{The reduction in the total number of events, as a function of the
number of hit PMTs, for successive applications of the instrumental background
cuts.  In the figure, 1~MeV corresponds to roughly 8.5 hit PMTs. \label{fig:dcreduce}}
\end{center}
\end{figure}

	To remove the vast majority of these events efficiently, we developed a
suite of `low-level' cuts which are applied to the data set before
reconstruction (see Appendix~\ref{sec:apdxa}). The cuts are based on information
such as the distribution of PMT charge measurements, the total integrated
charge, the time distribution of PMT hits, the inter-event timing, the spatial
distribution of PMT hits, and the firing of veto PMTs installed in the neck
region and outside the PMT support sphere.  `Flasher' events, for
example, are characterized by a high charge in the offending PMT; electronic
pickup events have many channels whose integrated charge is near the pedestal
level.  The cuts were designed individually as coarse filters to remove the
most obvious background events, but the combination of the cuts 
removed nearly all the instrumental backgrounds (see
Section~\ref{sec:contam}) before the more sophisticated stages of the analysis.
Figure~\ref{fig:dcreduce} illustrates the removal of instrumental backgrounds from
the raw PMT data as successive groups of cuts are applied. Each group of cuts
primarily targets a different source of instrumental background.  The figure also
shows the effects of the high level (`Cherenkov Box') cuts described in 
Section~\ref{sec:cerbox} and the fiducial volume cut which restricts events in the
final signal sample to have a reconstructed radial position $R_{\rm fit}<550$~cm.  We see that
in the region of interest for solar neutrinos (40--120 hit PMTs) the cuts
reduce the number of events in the data set by several orders of magnitude.

	Each of the cuts returns a simple binary decision.  The results are
saved as tags for each event, and the actual elimination of events based on
the tags is done at the end of the analysis. 

\begin{figure}[h]
\begin{center}
\includegraphics[ height = 3.0in ] {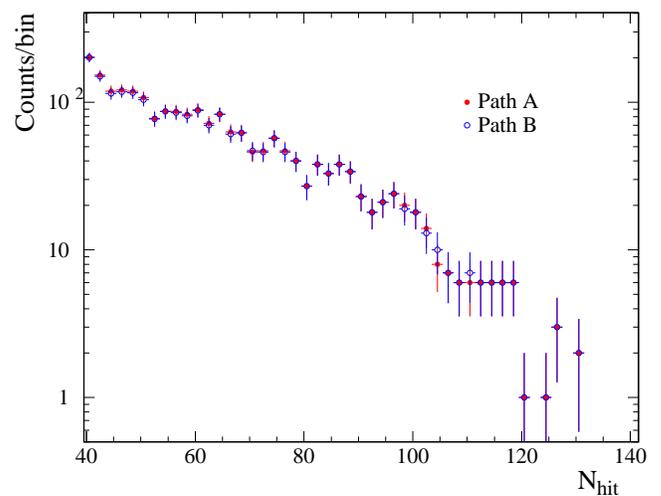}
\caption{Comparison of the number of events remaining after application of the
two separate sets of instrumental background cuts, as a function of the number
of hit PMTs. As can be seen, the differences between the data sets are very 
minor.  \label{fig:damnfist}}
\end{center}
\end{figure}

	With such a large reduction in the number of events, we were
particularly cautious in developing the cuts and measuring their signal
acceptance.  Nearly all the cuts were developed on a small subset of the total
data set, primarily data taken during detector commissioning and the collection
of radioactive source calibration data.  Unbiased data sets containing instrumental
backgrounds (such as bursts of flasher events caused by seismic activity) were also
used in the creation of the cuts. We developed two separate sets of
cuts, created by groups working independently, and performed extensive
comparisons between them.  Figure~\ref{fig:damnfist} compares the energy
spectra (as measured by the number of hit PMTs) for a set of neutrino
data that has been been subjected to both sets of cuts.  As can be seen in the 
plot, the differences between the numbers of accepted events is extremely 
small, and our measurements showed that this difference 
is consistent with the difference in the signal acceptances of the two sets of 
cuts.

	As described in Section~\ref{sec:insteff}, the acceptance of signal
events for the final suite of low-level cuts was measured to be greater than 
99.5\%.

\subsection{Position and Direction Reconstruction \label{sec:recon}}

	We use reconstructed position and direction both to produce the pdfs
shown in Fig.~\ref{fig:pdfs} as well as to reject background events
originating in the light water and PMTs.  Further, as described in
Section~\ref{sec:enecal}, estimation of an event's energy requires
knowledge of its position and direction.  We used two different position
reconstruction algorithms.  For the final analysis presented here and in
our initial Phase I publications, we used one to provide the starting position
and direction (the `seed') for the other, thus ultimately obtaining a more
accurate fit than either algorithm would have produced alone.

	Both reconstruction algorithms use time-of-arrival of photons at
the PMTs as the primary basis for determining event position. The algorithms
treat photons as being created at a point at a single instant,  
and then calculate the arrival times using straight-path trajectories from the
point source to a hit PMT. A likelihood is then calculated through comparison
of the actual hit times to the hypothesized distribution of times.  The second
of the two algorithms also uses the angular distribution of PMT hits relative
to an hypothesized electron direction.  A likelihood is calculated by comparing
the measured angular distribution of hits to the hypothesis that the event
begins as a single 5~MeV Cherenkov electron.

The first step in the fitting procedure of the first algorithm is to
search a coarse three-dimensional grid of 1.5 meter spacing across the entire
detector volume.  At each grid point a likelihood function is maximized with
respect to time, the only remaining free parameter. 
The 20 grid points with the highest
likelihoods are used as starting points for maximizing the same likelihood
function, but this time in four parameters, $x,y,z,$ and $t$.  The highest
likelihood value found determines the best fit vertex~\cite{gridfit}.

The probability density function (pdf) used to calculate the likelihood in this
stage of reconstruction depends solely on the PMT time-of-flight
residuals $t^{\rm res}_i$ relative to the hypothesized fit vertex position. For the
$i$th PMT, $t^{\rm res}_i$ is defined as
\begin{equation}
\label{eqn:ftg-tresid} 
t^{\rm res}_i = t_{i} - t_{\rm e} - |\vec{r}_{\rm e} - \vec{r}_{i}|{n^*}/c 
\end{equation} 
where $t_i$ is the hit time of the $i$th PMT, $t_{\rm e}$ is the time
being fit, $\vec{r}_{\rm e}$ is the event position being fit and  $\vec{r}_{i}$ is
the PMT position.  The photons are assumed to travel at a group velocity
$c/{n^*}$ with ${n^*}$ an effective index of refraction averaged over the media
in the detector. For this stage of the fitting, the pdf $P(t^{\rm res})$
was generated by Monte Carlo simulation of low energy background events
in the light water region. 
The fit for vertex position and time amounts to shifting $\vec{r}_{\rm
e}$ and $t_{\rm e}$ until the largest number of PMT hit times lie
underneath the peak of the in-time distribution.
The logarithm of the likelihood function used to do the fit at this stage is:
\begin{equation}
\label{eqn:ftg-log}
{\log\cal L} = \Sigma_{i=1}^{\rm N_{\rm hit}} \log(P(t^{\rm res}_i)).
\end{equation}

Once the vertex location has been determined, the direction is fit by a maximum
likelihood method based on a pdf of the angular distribution of photons
relative to the initial direction of a simulated 5~MeV electron.

	The vertex and direction obtained thus far are passed to the second
reconstruction algorithm which differs primarily in that it simultaneously fits
the event position, time, and direction using both timing and angular
information.  The log-likelihood function maximized as a function
of $\vec{r}_e$, $t_e$, and $\vec{v}_e$ is:
\begin{equation}
{\log\cal L} = \sum_{i=1}^{N_{hit}} \log {\cal P}(\vec{r}_e,\vec{v}_e,t_e;t_i,
\vec{r}_i)
\end{equation}
where $t_i$ is the measured PMT hit time and $\vec{r}_i$ is the PMT position;
$\vec{r}_e$ is the event vertex, $\vec{v}_e$ is the event direction,
and $t_e$ is the event time.  As before, the angular part of the pdf is based on the
assumption that the event begins as a single Cherenkov electron.

The probability ${\cal P}$ contains two terms,
to allow for the possibilities that the detected photon arrives directly
from the event vertex (${\cal P}_{\rm direct}$), or results from
reflections, scattering, or random PMT noise (${\cal P}_{\rm other}$).
These probabilities are weighted based on 
data collected in the laserball calibration runs: ${\cal P}
= f_{\rm direct}{\cal P}_{\rm direct} + f_{\rm other}{\cal P}_{\rm
other}$, with $f_{\rm direct} = 0.879$, and $f_{\rm other}= 0.121$.

\begin{figure}
\centerline{
\includegraphics[height=0.3\textheight]{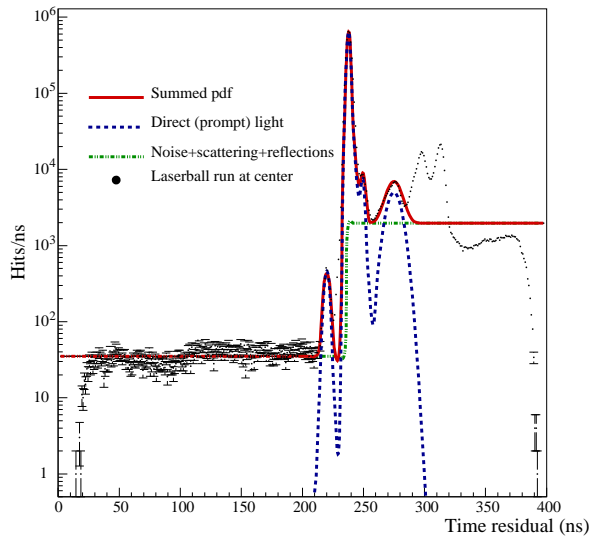}
}
\caption{\label{fig:path_time_pdf} A plot of the distribution of PMT
hit times from a laser calibration run.  Overlaid on the data are the
pdfs used in the fitter to characterize the direct light
(dashed) and the pdf describing the non-direct light (dotted).  The
summed pdf is also displayed (solid line).  As only the relative times
of the PMT hits are relevant in the event reconstruction, the offset of
the ``prompt'' peak from zero is unimportant.}
\end{figure}

${\cal P_{\rm direct}}$ and ${\cal P_{\rm other}}$ are further
broken down into separate time and angle factors:${\cal P_{\rm direct} =
P^{\rm TIM}P^{\rm ANG}}$, for example.  The time factor was
based on the time residual distributions determined from the laserball
calibration data with the source at the center of the detector.
(The time residual is defined in Equation~\ref{eqn:ftg-tresid}).
We characterized the direct light time distribution with a sum of four Gaussians
corresponding to prompt, pre-pulse, late-pulse and
after-pulse PMT hits.  Non-direct light was characterized by
a step function with the value for $t^{\rm res}<0$ corresponding to random PMT
noise, and for $t^{\rm res}>0$ corresponding to random noise plus an average
contribution from reflected and scattered light.  Figure~\ref{fig:path_time_pdf}
displays the PMT time distribution from the laser calibration data
along with the functions used to describe the distribution.

The angle factor is the Poisson probability for a single photon
hit in a PMT.
\begin{equation}
{\cal P}^{\rm ANG} = N_\gamma\rho_ie^{-N_\gamma\rho_i}
\end{equation}
where $N_\gamma = N_{\rm hits}/P_{hit} \equiv N_{\rm hits}/0.55$~is an estimate of
the number of photons which strike PMTs (see Equation~\ref{eq:phit}) and
\begin{equation}
\rho_i = \frac{1}{2\pi} g(\cos\alpha_i)\Omega_i
\end{equation}
where $g(\cos\alpha)$ is the angular distribution of the photons
relative to the initial electron direction, $\alpha_i$ is the angle of
the $i$th PMT relative to the hypothesized electron direction as measured from the
vertex, and $\Omega_i$ is the solid angle of the $i$th PMT as
viewed from the vertex:
\begin{equation}
\label{eq:sang}
\Omega_i = \frac{\pi r_c^2}{d_i^2}\hat{d}_i\cdot\hat{r}_i .
\end{equation}
In Equation~\ref{eq:sang},  $r_c$ is the radius of the PMT concentrator 
`bucket' (see Section~\ref{sec:optics}),
$\vec{d}_i$ is the vector from the event vertex to the center of the
face of the concentrator bucket, and $\vec{r}_i$ is the position of the 
front face of the PMT in detector coordinates.

\begin{figure}
\centerline{
\includegraphics[height=0.3\textheight]{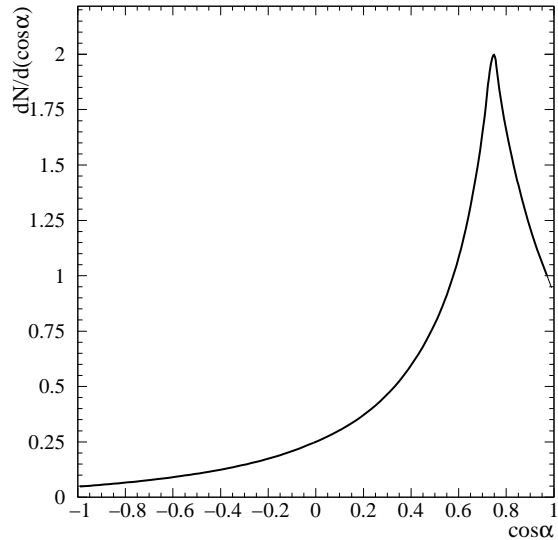}
}
\caption{\label{fig:path_angle_pdf} Parameterized 
angular distribution of Cherenkov photons relative to the initial
direction of a Monte Carlo 5 MeV electron.  This distribution is
used as the pdf for the direct PMT hits in reconstruction.}
\end{figure}

Figure~\ref{fig:path_angle_pdf} shows the angular distribution assumed for the
direct photons.  The non-direct photons are assumed to be isotropic relative to
the event vertex and hence to have a flat distribution in $\cos\alpha$.

	The azimuthal symmetry of Cherenkov light about the event direction
dilutes the precision of reconstruction along the event direction.  
Scattering of photons out of the Cherenkov cone thus systematically tends to
drive the reconstructed event vertex downstream of the true event position.  To
compensate for the systematic drive, after initial estimates of 
position and direction are obtained, a correction is applied to shift the
vertex back along the direction of the event, varying with the distance of the
event from the PMT sphere as measured along its direction.

In the final stage of the fit, the hypothesis that the event was
a single electron is tested.  We do this using two figure-of-merit
criteria calculated from the angular distribution of the PMT hits
relative to the event vertex and direction.  The first of these 
is a Kolmogorov-Smirnov test of the uniformity of the azimuthal distribution of PMT
hits around the event direction.  
The second is a two-dimensional Kolmogorov-Smirnov test of 
the distribution of hit PMT directions azimuthally and in 
$\cos{\alpha}$ relative to the reconstructed event direction.

\begin{figure}
\includegraphics[height=0.3\textheight]{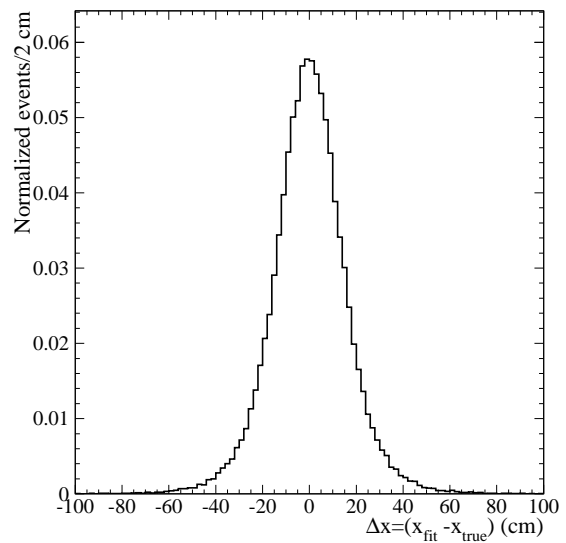}
\caption{\label{fig:xres_CC} The distribution of the difference 
between the reconstructed $x$ position ($x_{\rm fit}$) and the true position
$x_{\rm true}$ for a sample of Monte Carlo simulated CC electrons.}
\end{figure}

Figure~\ref{fig:xres_CC} shows the $x$-coordinate resolution of vertex reconstruction
for events for a Monte Carlo simulated sample of CC electrons. The performance
of the reconstruction algorithm on data and its associated uncertainties will
be presented in Section~\ref{sec:posdir}.

\subsection{Energy Calibration}
\label{sec:enecal}

	We used two different estimators of event energy as assurance against
unexpected systematic errors.  One was simply the total number of
hit PMTs (`$N_{\rm hit}$'), without any adjustment for the position dependence of
the energy scale within the detector.  For this estimator, the energy spectra 
in the top row
of Fig.~\ref{fig:pdfs} were replaced by `$N_{\rm hit}$ spectra'.
The second estimator, the energy
`reconstructor', used the fitted event position and direction and the analytical
form of the optical model described in Section~\ref{sec:optics}.  The energy
reconstructor was used to produce the results reported in the intial Phase I
publications, and the $N_{\rm hit}$ estimator used for validation of those results.
In this section, we briefly discuss the $N_{\rm hit}$ estimator (for more details,
see Refs.~\cite{the:msn,the:kh,the:pw}) and give a more complete description of the
energy reconstructor.

\subsubsection{`$N_{\rm hit}$' Energy Estimator}
\label{sec:nhit}

Using  the total number of hit PMTs in an event ($N_{\rm hit}$) as an
energy estimator has the advantage that it is simple: it uses no cuts on charge
or time to define good and bad hits, it integrates over uncertainties in the
time distribution of reflected and scattered light, and it applies no
corrections to the data itself.  Also, the additional statistics gained by
including scattered and reflected light can lead to a narrower energy
resolution overall.  Although the calibrations of our optical model have
explicitly been done only for prompt light (see Section~\ref{sec:optics}), as
Fig.~\ref{fig:tres_data} shows the fraction of late light in an event is only
$\sim$12\%. We can therefore include reflected and scattered light even if our
knowledge of the optical parameters which govern its generation and propagation
are somewhat worse than for direct light.  Most importantly, the use of total
$N_{\rm hit}$ is sensitive to different systematic effects from the
prompt-light energy reconstructor described in the next section.

	To use total $N_{\rm hit}$ to extract signal fluxes, we employ the
Monte Carlo simulation to generate pdfs like those shown in Fig.~\ref{fig:pdfs}, with
the top row replaced by $N_{\rm hit}$ spectra.  With the data untouched by any
correction or calibration, one must ensure that the Monte Carlo simulation takes into
account the variations in detector state  over the data collection livetime.
For example, the number of working channels as a function of time and the
change in PMT noise rates must be tracked and fed either into the Monte Carlo
simulation (as described in Section~\ref{sec:mcdaq}) or applied as subsequent corrections.

\begin{figure}[h]
\begin{center}
\leavevmode
\includegraphics[height=0.30\textheight]{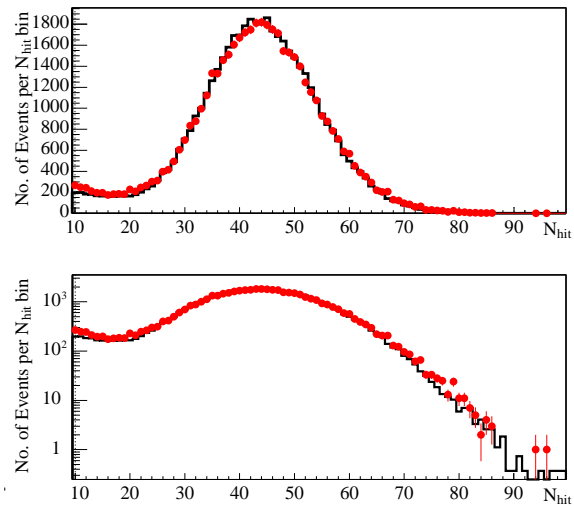} 
\caption{$N_{\rm hit}$ distributions for Data (dots) and Monte Carlo simulation (line) for a $^{16}$N calibration at the detector center, on both linear and logarithmic scales. \label{fig:n16mccomp}}
\end{center}
\end{figure}

	The only calibration necessary here is therefore that described in
Section~\ref{sec:enescale}---the initial calibration of the Monte Carlo model
to ensure that the predicted number of hits per event agrees with the
measurements using sources.  The uncertainty of this calibration will be
discussed in Section~\ref{sec:sysuncert_ene}. Figure~\ref{fig:n16mccomp} shows
a comparison of the Monte Carlo model's prediction of the distribution of
$N_{\rm hit}$ for the $^{16}$N source to an actual source run.

\subsubsection{Energy Reconstructor}
\label{sec:ecalibrator}

Unlike the `$N_{\rm hit}$' energy estimator, the energy reconstructor 
corrects for detector optical, temporal, and spatial effects to
assign a most probable energy
to each event.  Given an event's position, direction, and number of hit
PMTs, the energy reconstructor uses the analytic form of the optical model
described in Section~\ref{sec:optics} to estimate the number of PMT hits the
event would have produced had it been created at the center of the detector.
A scale factor is then applied to convert the number of hits to an equivalent 
electron energy.

This reconstruction has several advantages over the simple `$N_{\rm hit}$'
estimation.  First, it allows us to produce energy spectra labeled in MeV,
rather than the detector-specific $N_{\rm hit}$.  Also, by correcting for the
detector's point-to-point variation in response we can choose to use a single
analytic function to map true energy to reconstructed energy,
rather than relying on the entire Monte Carlo model to provide the detector
response. With such an analytic function, we and others wishing to fit our data
set can create pdfs in energy which do not require the entire detector
simulation.

	As described in Section~\ref{sec:optics}, measuring the optical
parameters that characterize late hits, such as the degree of
scattering and various reflection coefficients, can be difficult.
In addition, in a particular event, there is no way to uniquely
determine the flight paths of such out-of-time photons.  The energy
reconstructor therefore begins by eliminating out-of-time hits,
restricting PMT times to be within $\pm$10~ns of the prompt
time peak.  The remaining hits are treated as if they came directly from
the reconstructed event vertex.  The $\pm 10$~ns window is applied to the PMT
time residuals defined by:

\begin{equation}
t_{\rm res} = t_{\rm pmt} - t_{\rm fit} - t_{\rm travel} - t_{\rm shift}, 
\label{eq:tresrsp}
\end{equation}

\noindent{where}
\begin{eqnarray*}
t_{\rm pmt} & = & \mbox{calibrated PMT hit time} \\
t_{\rm fit} & = & \mbox{fitted event time} \\
t_{\rm travel} &  = &  \mbox{travel time from vertex to PMT} \\
t_{\rm shift} &  = & \mbox{average risetime correction shift.}
\end{eqnarray*}

$t_{\rm shift}$ is necessary because, as described in Section~\ref{sec:ecapca}, we
discovered rate dependencies to the charge and time pedestal values.  Although
the effect was small, it meant that the measured PMT times, which nominally
were corrected for PMT pulse risetime based on the integrated event charge,
could vary as a function of event rate. By removing the risetime correction
from the energy calibration, this variation was no longer an important
source of systematic uncertainty, and with the prompt time cut used here, the
loss of PMT timing precision is not critical.  The value of $t_{\rm shift}$ was
picked to center the uncorrected PMT timing residuals at $t_{\rm res}=0$.

Time residual histograms for $^{16}$N source runs at radii of $0.0$~cm
and $500$~cm are shown in Fig.~\ref{fig:tres_data}.  One can clearly see
the effects of scattering at the higher radius.  The PMT reflection peaks,
which are more than 50~ns from the prompt peak with the source at
the center, move closer as the source is moved toward the PMT array.

With the `prompt' PMTs in an event identified, we define an 
effective number of PMTs hit as 

\begin{displaymath}
N_{\rm eff} = N_{\rm win} - N_{\rm dark} 
\end{displaymath}
\noindent with
\begin{eqnarray*}
N_{\rm win} & = & \mbox{ number of in-time hits ($\pm 10 $ ns)} \\
\end{eqnarray*}
and
\begin{eqnarray*}
N_{\rm dark} & = & \mbox{expected number of in-time noise hits.}
\end{eqnarray*}

The average number of PMT noise hits, measured using the pulsed trigger
described in Section~\ref{sec:intro}, was found to be 2.1 in the 440~ns event
timing window.  (This is equivalent to an average dark noise rate for each
photomultiplier tube of $\sim$ 593~Hz). Since the dark noise hits are 
uniformly distributed throughout the 440 ns window, the expected number of
noise hits $N_{\rm dark}$ within the energy reconstructor's 20~ns timing window
is just 0.1.  This number is small enough (equivalent to roughly 10 keV) that
accounting for variations from run-to-run would have had a negligible impact.

We then apply optical and gain corrections to determine the equivalent
$N_{\rm eff}$ at the detector center to produce a `corrected $N_{\rm hit}$':

\begin{equation}
N_{\rm cor}  =  N_{\rm eff} \times \frac{1}{\epsilon_{\rm response}/\epsilon_0} \times \frac{1}{\epsilon_{\rm hardware}} \times \frac{1}{\epsilon_{\rm drift}}.
\label{eq:neff_eqn}
\end{equation}

In Equation~\ref{eq:neff_eqn}, $\epsilon_0$ is the detector's optical response for an event at the detector center, and  $\epsilon_{\rm response}$ represents the detector's 
optical response for  events at a given position ($\vec{r}$)
and direction ($\vec{u}$):
\begin{eqnarray} 
\epsilon_{\rm response} & = \sum_{\theta^{\prime}} \sum_{\phi^{\prime}} \sum_{\lambda} \frac{\epsilon_{\rm PMT}(\lambda)}{\lambda^2} g(\theta^{\prime},\phi^{\prime}) \nonumber \\ 
& \times e^{-\mu_1 d_1-\mu_2 d_2-\mu_3 d_3} R(\theta^{\prime},\phi^{\prime} ) M(r,\theta^{\prime},\phi^{\prime}) &  
\label{eq:epsresp}
\end{eqnarray}
In Equation~\ref{eq:epsresp}, the sums are over 10 polar ($\theta^{\prime}$) and 10 azimuthal
($\phi^{\prime}$) angle bins relative to the reconstructed event
vertex and direction ($\theta^{\prime}$=0), and wavelengths $\lambda$
in a range (220-710 nm) that span the wavelengths to which the
detector is sensitive. 
The factor $\epsilon_{\rm PMT}(\lambda)$ represents the efficiency of the PMT
as a function of wavelength, and $g(\theta^{\prime}, \phi^{\prime})$ is 
the angular distribution of Cherenkov light about the event direction. The
$\mu_i$ are the inverse of the wavelength-dependent attenuation lengths for
each medium ($i=1,2,3$ corresponding to D$_2$O, acrylic, and H$_2$O), and
the $d_i$ are the distances through each medium that photons travel from the
event vertex to the PMT array in each ($\theta^{\prime}$,$\phi^{\prime}$) bin.
$R(\theta^{\prime},\phi^{\prime})$ is the PMT angular response, and
$M(r, \theta,\theta^{\prime},\phi^{\prime})$ is a correction for the multiple
hit probability (which depends on the event position, $r$). 
The largest variation in $\epsilon_{\rm response}$ within the
D$_2$O as a function of source radius is about 7\%, with its largest values
occurring near $R=450$~cm.  

The efficiency $\epsilon_{\rm hardware}$ is applied to correct for the number of
PMTs
available in a given event, which is tracked run-by-run and logged in the SNO
analysis database.  In addition, PMTs which are known to have poor response are
flagged during the PMT calibrations described in Section~\ref{sec:ecapca};
their effect is then included as a reduction in $\epsilon_{\rm hardware}$.  

\begin{figure}[htb]
\begin{center}
\includegraphics[height=0.25\textheight] {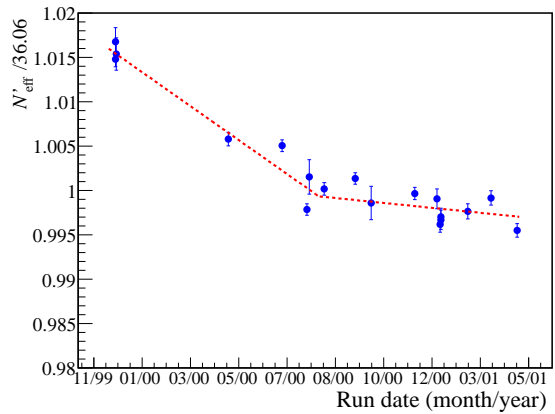}
\caption[Photon collection efficiency versus time]{Photon collection efficiency ($P_{\rm hit}$) versus time.  Shown is the value of the peak of the $N_{\rm eff}^{\prime}$ distribution from $^{16}$N calibration runs at the detector center, relative to the average over the data set. The dashed line is used as $\epsilon_{\rm drift}$, the  correction to the absolute energy scale.\label{fig:edrift}}
\end{center}
\end{figure}

We apply $\epsilon_{\rm drift}$ only to data (not Monte Carlo events), and we use it to 
correct for small changes in the overall
photon collection efficiency of the detector over time. 
Figure~\ref{fig:edrift} shows the time-dependent behavior 
of $N_{\rm eff}^\prime$, defined by  
\begin{equation}
N_{\rm eff}^{\prime} = N_{\rm eff} \times \frac{1}{\epsilon_{\rm hardware}},
\label{eq:neffprime}
\end{equation}
and we can see that, as discussed in Section~\ref{sec:enescale}, there was a
drop in overall detector gain of about $1.8\%$ during the first
several months of production running followed by slower drop for the remainder
of the running period.  The dashed line in
Fig.~\ref{fig:edrift} is used as a correction to the energy scale as a
function of date, and is given by
\begin{eqnarray}
\epsilon_{\rm drift} &  = & 1.595 - [6.315\times  10^{-5}\times {\rm JDY}] \\ \nonumber
                     &    & \mbox{ for JDY $<$ 9356} \\
                     &  = &   1.004 - [9.170\times 10^{-6}\times ({\rm JDY}-9356)] \\
\nonumber
                     &    &  \mbox{ for JDY $\ge$ 9356}
\end{eqnarray}
where JDY is `SNO Julian Date'. SNO Julian Day 9356 corresponds to
midnight UTC, on August 12, 2000. 
As Section~\ref{sec:enescale}
describes, the Monte Carlo model's energy scale was left fixed to the level
determined in the middle of the data acquisition period, and so no
$\epsilon_{\rm drift}$ correction is applied to simulated events.  

\begin{figure}[htb]
\begin{center}
\includegraphics[height=0.2\textheight] {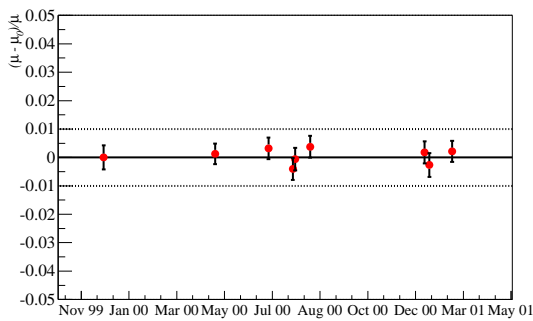}
\caption[${}^{16}$N stability runs]{Fractional deviation of mean $N_{\rm win}$ 
for ${}^{16}$N source calibration runs after drift correction 
applied. Only a subset of the data in Fig.~\ref{fig:edrift} is shown. \label{n16_cor}}
\end{center}
\end{figure}

Figure ~\ref{n16_cor} shows the fractional deviation of the mean $N_{\rm win}$ after
applying the drift correction.  The mean deviation of this value from zero is about
0.25\% which is consistent with statistical variation.

To map the corrected number of hit PMTs ($N_{\rm cor}$) to electron
energy, sets of Monte Carlo calculations are performed for mono-energetic
electrons at the detector center, at  different electron energies.  For
each electron energy, we fit a Gaussian to the resultant $N_{\rm eff}$ spectrum 
to obtain a mean value.  This is done for event energies covering our region of 
interest for solar neutrino analysis, from about 2-30~MeV, resulting in a 
linear relationship between $N_{\rm eff}$ and energy in MeV.

Using Monte Carlo events we calculate $N_{\rm cor}$ from Eq.~\ref{eq:neff_eqn}
and use the generated linear map to produce a calibrated energy
spectrum.  
For reference, Table
~\ref{mu0s} shows the predicted $E_{\rm eff}=T_{\rm eff}+0.511{~\rm MeV}$ peaks for various calibration sources.

\begin{table}[htb]
\begin{center}
\begin{tabular}{lc}  \hline \hline
Source & Peak $E_{\rm eff}$ [MeV] \\ \hline
${}^{16}$N     & $5.486$ \\ 
pT             & $19.2$ \\ 
n(d,t)$\gamma$ & $5.59$ \\ \hline \hline 
\end{tabular}
\end{center}
\caption[Predicted $E_{\rm eff}$ peaks for calibration sources]{Predicted $E_{\rm eff}$ peaks for calibration sources.\label{mu0s}}
\end{table}

\begin{figure}[htb]
\begin{center}
\includegraphics[height=0.25\textheight] {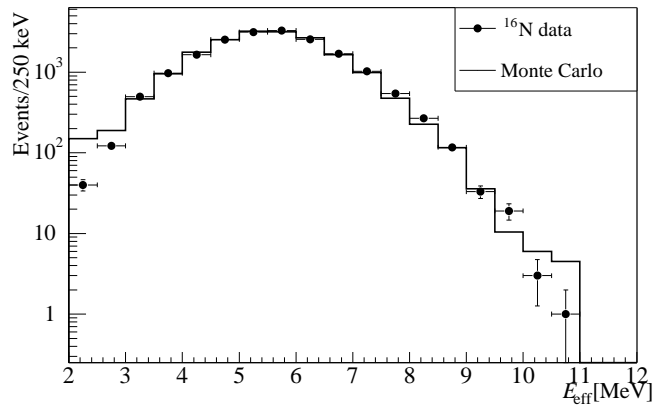}
\caption[$^{16}$N $E_{\rm eff}$ spectra for data and Monte Carlo]{$^{16}$N $E_{\rm eff}$ spectra for calibration data and a Monte Carlo calculation, for a deployment near the center of the detector.\label{fig:eeff_compare}}
\end{center}
\end{figure}

Figure ~\ref{fig:eeff_compare} shows the $E_{\rm eff}$ spectra for $^{16}$N data
and Monte Carlo, showing good agreement between energies in the region of
interest for the solar neutrino analysis ($T_{\rm eff}>5$~MeV).

\subsection{`Cherenkov Box' Cuts}
\label{sec:cerbox}

	Although the `low-level' instrumental background cuts described
in Section~\ref{sec:inst} are very efficient at removing backgrounds
with specific characteristics (high charge in one or more PMTs, poor timing
distributions, etc.) we still want to ensure that the final data set contains
no events which are inconsistent with Cherenkov light.  The defining characteristics of
Cherenkov light are that it has a very narrow time distribution, and a hit pattern
consistent with a Cherenkov cone.  We therefore formulated two cuts, one based on
timing and the other on hit pattern, which define a `Cherenkov box' in which we
expect only neutrino events and background events due to radioactivity to lie.  These cuts used
derived information---such as the reconstructed position of each event---as opposed
to the low-level information used in the cuts described in Section~\ref{sec:inst}
and Appendix~\ref{sec:apdxa}. We therefore refer to them as `high-level' cuts.

	Our measure of Cherenkov timing is simply the ratio of
in-time hits to the total number of hits in an event, where `in-time'
is defined using reconstructed time-of-flight residuals like those of
Equation~\ref{eq:tresrsp}. Unlike in Eq.~\ref{eq:tresrsp}, however, here we
use the risetime-corrected hit times.  The in-time window for this
ratio is $-2.5 \rightarrow +5.0$~ns relative to the prompt timing peak, and
we restrict neutrino candidate events to have an  in-time ratio (ITR) $> 0.55$.

	For the hit pattern cut, we reject events for which the mean angle
between all pair-wise combinations of hit PMTs ($\theta_{ij}$) is either too large
($>1.45$ radians) or too small ($<0.75$ radians).  The PMT pair angles are calculated as viewed
from the reconstructed event vertex, and only PMTs within a small time window 
(within $\sim \pm 9$~ns of the prompt peak) are used. Events with mean pair angles
greater than 1.45 radians are `too isotropic' to be Cherenkov light; those with pair
angles below 0.75 radians are `too narrow' compared to a Cherenkov ring.

\begin{figure}[h]
\begin{center}
\includegraphics[height=0.3\textheight]{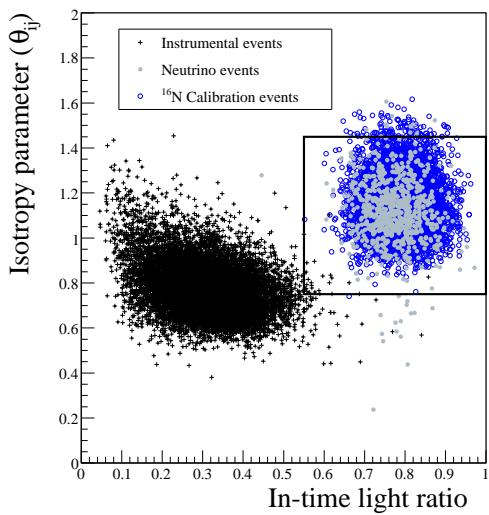}
\caption{Fraction of in-time light for an event versus its mean PMT pair angle ($\theta_{ij}$). Events which pass the low-level cuts (grey) and those which fail
(black) plotted on the high-level cut axes.  $^{16}$N calibration source events
are plotted in blue.
\label{fig:bifurcate}}
\end{center}
\end{figure}

	Figure~\ref{fig:bifurcate} shows events plotted in these two
characteristics.  In black are events which have been tagged by the low-level
instrumental background cuts, in grey are neutrino candidate events, and in 
blue are events from the $^{16}$N calibration source.  As we see,
most of the candidate events lie within the same Cherenkov box as do the 
$^{16}$N calibration source events.  In fact, many candidate
events which lie outside are mis-reconstructed events rather than instrumental
backgrounds, as determined by calibration source data.

\subsection{Muon and Atmospheric Neutrino Follower Cuts \label{sec:follcuts}}

	    While the rates in SNO of muons and atmospheric neutrino
interactions are just 3 per hour, their products can be a dangerous source of
background.  Spallation by cosmic ray muons produces neutrons as well as
long-lived radioisotopes.  Atmospheric neutrino interactions can also produce
neutrons, through both neutral current and charged current processes.

	We remove cosmic-ray-muon spallation products by cutting
all events occurring within 20~s of a muon event. The muon
identification criteria require more than five hits in the outward-looking
veto PMTs and more than 150  hits in inward-looking PMTs.  To avoid
large detector deadtime from this cut, we do not impose a deadtime
following an event which satisfies these criteria but is also tagged by the
low-level cuts as arising from a discharge in the neck of the acrylic vessel.

	To remove the products of atmospheric neutrinos and muons missed by the muon
tag, we also cut all events within a 250~ms interval following any event that has
more than 60 PMT hits.  This cut removed 53 events from the final neutrino candidate
sample.  In Section~\ref{sec:cosmons} we describe our estimate of the number of
background events in the final data set passed by these cuts, and in
Section~\ref{sec:livetime} the associated livetime loss.

\subsection{Fiducial Volume and Energy Threshold}
\label{sec:fvolecut}

	The last set of cuts applied to the data set are intended primarily to
remove backgrounds associated with low energy radioactivity within the detector.
Radioactive decays within the heavy water volume typically produce much lower 
energy events ($\sim 2$~MeV) than interactions by the $^{8}$B solar neutrinos (up to
15~MeV), and we therefore remove the vast majority of backgrounds by imposing an
energy threshold of $T_{\rm eff}= E_{\rm eff} - 0.511~{\rm MeV} > 5.0$~MeV.  

	Events originating from radioactivity in the regions outside the
heavy water---from the light water shield, the acrylic vessel, the
PMTs and associated support structure or the cavity walls---can remain
in the final data sample only if they are both above the energy threshold
and have misreconstructed positions.  Nevertheless, these regions have
far higher radioactivity levels than the D$_2$O (see Sections~\ref{sec:pd_bgd}
and~\ref{sec:lowext}), and we therefore restrict the fiducial volume of the
final sample to avoid these backgrounds.  Our requirement that
the final events reconstruct within 550~cm of the detector
center also has the advantage that backgrounds from misreconstruction
of light produced by the acrylic vessel are minimal (see Section~\ref{sec:inst}).
Further, our understanding of the detector optics and response is best within
the 550~cm fiducial volume.

\begin{figure}[h]
\begin{center}
\leavevmode
\includegraphics[height=0.3\textheight]{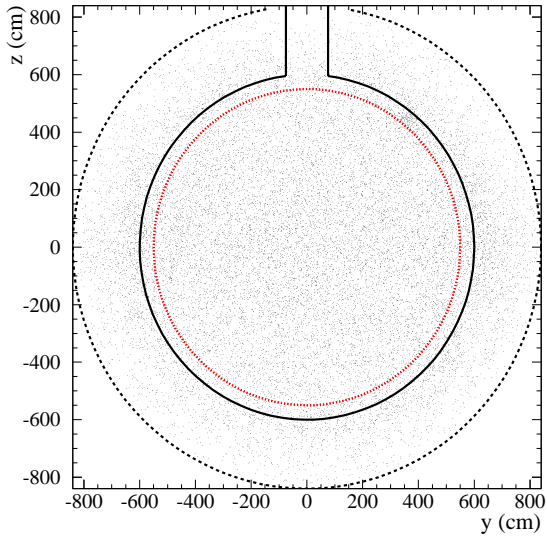} 
\caption{Projection of reconstructed positions of events within the SNO 
detector onto the $z$-$y$ plane, for an energy threshold of $T_{\rm eff} > 5.0$~MeV. The dashed line shows
the radius of the PMT support sphere (PSUP), the solid line the
acrylic vessel, and the dotted line the fiducial volume cut used by
SNO. \label{fig:zvx}}
\end{center}
\end{figure}

	Figure~\ref{fig:zvx} shows a $z$ vs. $y$
projection for events above a threshold of $T_{\rm eff} = 5.0$ that pass all
cuts except the fiducial volume restriction.  The fiducial volume used in this 
analysis indicated by the red line.  As can be seen in the figure, there is a region
of higher activity, a `hot spot', near $z=450$~cm and $y=400$~cm.   Although the
origin of this hot spot is unknown, the characteristics of events which reconstruct
there  are consistent with decays in the natural radioactive chains.  We discuss the
hot spot as a source of background in Section~\ref{sec:av_pd} and as an {\it
in-situ} `calibration' source in Section~\ref{sec:diurnalE}.

\subsection{Data Processing Cut Summary}

	Table~\ref{tbl:cutsum} details the number of events remaining after
each cut applied during data processing.
\begin{table}[h!]
\caption{Number of events remaining in data set after each step in the
data processing described in Section~\ref{sec:dataproc}. \label{tbl:cutsum}}.
\begin{center}
\begin{tabular}{lr}
\hline\hline
Data Processing Step & Events Remaining \\ \hline
Total event triggers &  450188649 \\
Neutrino triggers (hit multiplicity)  &  191312560 \\
Analysis $N_{\rm hit}$ cut ($N_{\rm hit} > 21$) & 10088842 \\
Low-level cuts &  7805238 \\
`Cherenkov Box' cuts &   3418439\\
Fiducial volume cut &  67343 \\
Energy threshold ($T_{\rm eff}>5$~MeV)& 3440  \\
Muon follower cut &  2981 \\
Atmospheric $\nu$ followers  &  2928 \\ \hline
Total $\nu$ candidates  &  2928 \\ \hline \hline
\end{tabular}
\end{center}
\end{table}

\section{Systematic Uncertainties on the Model \label{sec:sysunc}}

	The pdfs shown in Fig.~\ref{fig:pdfs}, created by the model described in
Section~\ref{sec:model}, represent our best estimates of the true distributions of
neutrino event energies, directions, and radial positions.  Before we fit the
processed data set, we need to evaluate the uncertainties on the pdf shapes.  
We rely on the model to generate the pdfs, rather than on calibration source data,
because the sources themselves do not identically reproduce the neutrino signal
data. Calibration source events differ from neutrino events in many ways: the source
and deployment hardware affect the detected energy; sources like $^{16}$N are
$\gamma$ sources but the detected products of the CC and ES reactions are electrons;
the calibrations were performed at discrete points in the detector, while the
neutrino events occur throughout the volume; and the calibrations were done at
particular times while the neutrino data is distributed over the entire data
acquisition livetime.  As described in Section~\ref{sec:model}, the only two direct
inputs to the model from the calibration source data are the optical properties of
the detector and the overall energy scale.

What the calibration source data do give us is a powerful way of determining
the systematic uncertainties on the model predictions of detector response.
Rather than determining these uncertainties by varying (and covarying) each of
the relevant parameters in the model (such as optical attenuation lengths or
PMT efficiencies),  we made direct comparisons of source data to model
predictions of the response for each source.  The differences between the model
predictions and the source data were then used as estimates of the systematic
uncertainty on the model's ability to reproduce the detector behavior.  As
explained later in Section~\ref{sec:sigex},  we determine the effects of these
uncertainties on the neutrino flux measurements by varying the pdfs of
Fig.~\ref{fig:pdfs} by amounts consistent with the uncertainties, and then
re-fitting for the fluxes.  

SNO's extensive array of calibration sources and the ability to place them at
many positions within the detector allowed us to explore the dependence of the
uncertainties on nearly every way in which the simulation and the calibration
data differed. Different source types allowed checks of the dependence on
particle species, particle energies, and  calibration-source apparatus;
position dependence was provided by scans of sources throughout two orthogonal
detector planes; rate dependence was explored by varying calibration source
rate; time dependence was determined through periodic deployments of sources
throughout the data acquisition period.

In addition to the Monte Carlo model, we also developed a set of analytic pdfs that
described the response of the detector and used them to do a similar signal
extraction (see Section~\ref{sec:sigex}).  The determination of the systematic
uncertainties for the analytic pdfs were derived from direct fits to the calibration
source data.  

   We describe below our determination of the uncertainties on the pdf shapes
through model-data comparisons.  Systematic uncertainties affecting the overall
normalization of the fluxes such as those associated with livetime, knowledge of
the Earth's orbital eccentricity, and uncertainties on the acceptance of the cuts
applied to the data set, are presented in Section~\ref{sec:results}.

\subsection{Position and Direction Reconstruction \label{sec:posdir}}

	As discussed in Section~\ref{sec:recon}, we compared the results
of two separate reconstruction algorithms to verify their performance.
Ultimately, one algorithm, which used PMT timing information
alone, was used to provide the `seed vertex' for the second algorithm,
which simultaneously fits event position and direction using both the timing
and angular distribution of the hit PMTs.  As there was not a significant
difference in the uncertainties of the two algorithms, we describe below how
the uncertainties were estimated for the final hybrid method.

	The algorithm characteristics for which we need to determine
uncertainties fall into three classes:
\begin{itemize}

\item {\it Vertex Accuracy}: The average distance between the true interaction
position (in $x$, $y$, and $z$) and the reconstructed position.  Many effects
that can produce a systematic shift (such as a shift along the event direction
due to the azimuthal symmetry of Cherenkov light) are already accounted for
in the model, and what we are interested in here is the uncertainty on the
model's prediction of these shifts.

\item {\it Vertex Resolution}: The width of the distribution of reconstructed
event positions relative to their true positions.  The resolution itself
is well-modeled, but we need to determine the uncertainty on the model
prediction.

\item {\it Angular Resolution}: The distribution of reconstructed directions
relative to the initial electron direction.
\end{itemize}

	Ultimately, reconstruction uncertainties affect our
flux measurement uncertainty in two ways.  First, we need to know
the uncertainty in our prediction of geometric acceptance---how many
events we expect to reconstruct inside our fiducial volume.  This acceptance
uncertainty depends both on uncertainty in vertex 
resolution (if, say, the true resolution is broader or narrower than we
believe, then we will over- or underestimate the number of events) and upon
the possibility of systematic shifts in the mean fit position (outward
or inward, upward or downward, etc.).  

	The second way in which these uncertainties affect our final
answer is in the shapes of the pdfs we use for signal extraction.  An
error in the response function used to model the detector (either through
Monte Carlo simulation or with an analytical model) will alter the number of events derived
from our fits to the data.  For this, reconstruction of both direction
and position is important.

\subsubsection{Vertex Accuracy}
\label{sec:vshift}

A systematic shift inward or outward in mean reconstructed position is the most
dangerous of the reconstruction-related uncertainties.  Such a shift
effectively shrinks or grows the fiducial volume.  A +1\% uncertainty in
scaling on the radial coordinate, for example, produces a 3\% uncertainty in
the number of accepted events within the fiducial volume.

	In estimating the uncertainty in vertex accuracy, we examine both \NS~and
\LI~data.  We take our primary estimate of the uncertainty from the \NS~data, and 
check for effects which depend on event energy or source type with
the \LI~data.
Both sources generate electrons with known position distributions (in the
case of the \NS~source, the electron position distribution includes the effects of
Compton scattering by the $\gamma$ ray). To estimate the shift in the mean
reconstructed vertex and the width of the resolution function, we convolve these
known position distributions with a hypothetical resolution function, and then fit
the resultant convolution to the data by allowing the mean and width of the resolution function to vary~\cite{the:mgb,the:msn}.  That is, a 
function $\xi(x_{fit}; \sigma, \mu)$ is fit to the one-dimensional reconstructed position 
distribution (here shown in $x$), 
\begin{equation}
  \xi(x_{fit}; \sigma, \mu) = \int\limits_{-\infty}^{\infty} F(x_{fit}, \sigma,
  \mu; x_{src}) S(x_{src}) dx_{src}
\end{equation}
where $S$ is the electron source distribution and $F$ is the reconstruction
resolution function for electrons.  $F$ includes both the width of the resolution
($\sigma$) and a shift in the mean ($\mu$). The one-dimensional form chosen for $F$ is a simple Gaussian,
\begin{equation}
  F(x_{fit}, \sigma, \mu; x_{src}) = \frac{1}{\sqrt{2\pi}\sigma}
  e^{-\frac{[(x_{fit}-x_{src})-\mu]^2}{2\sigma^2}}
\end{equation}                                      
motivated by Monte Carlo studies of reconstructed electron position
distributions.

Although a better fit to the Monte Carlo distributions is obtained using the sum of a Gaussian
and an exponential (the data suggest exponential rather than Gaussian
tails~\cite{the:mgb}), for signal extraction using Monte Carlo signal pdfs we need
only the Gaussian, since we are just trying to characterize differences between
Monte Carlo distributions and calibration data distributions.  As described above,
in signal extraction we use these differences to evaluate the systematic
uncertainties on the fitted event rates by convolving the Monte Carlo generated pdfs
with smearing functions designed to broaden and shift the Monte Carlo simulated
position distributions so that they look like those we have obtained with the data.
In other words, we fit for the function $F$ for both Monte Carlo simulation and
calibration data, and then find the Gaussian which smears the Monte Carlo-derived
$F$ to yield the $F$ we measure for the data.  This `smearing' Gaussian is then
convolved with the Monte Carlo-generated signal position pdfs (the second row of
Fig.~\ref{fig:pdfs}) and the signal extraction procedure repeated. The resultant
change in the fiducial volume and the number of extracted neutrino events yields the
uncertainty on the neutrino fluxes.

For our secondary signal extraction method (using analytical pdfs), one needs to
include the exponential tails.  In this case, the goal is to produce pdfs by
convolving the expected true position distribution for events inside the detector
volume with a resolution function derived primarily from data.  To correctly
reproduce the event position distributions without using the Monte Carlo model,
the more complete distribution is therefore needed.

For \chem{16}{N}, the form of $S(x_{src})$ is the one-dimensional projection of the
three-dimensional 
Compton scattering distribution, $S(r)\sim \exp^{\frac{-r}{\lambda}}/r^2$, 
with $\lambda=37$ cm. The \chem{8}{Li} source is
approximated as a source of electrons on a shell 10.7 cm in diameter. 

\begin{figure}
\begin{center}
\includegraphics[height=0.28\textheight, angle=-90]{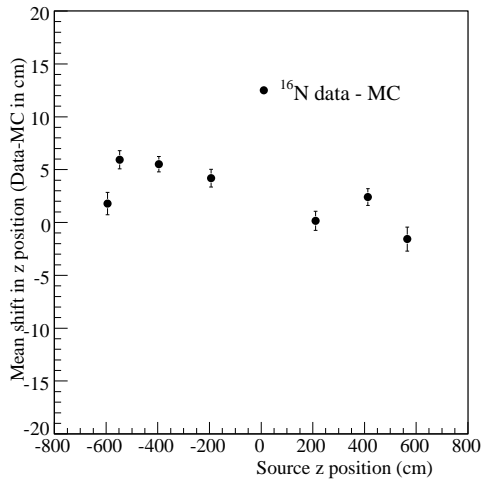}
\caption{Difference between \NS~source data and \NS~Monte Carlo simulations of
the vertex accuracy, using a precision $z$-axis source scan. Vertex accuracy
is defined here as the difference between expected source position
(based on the measurements by the source-positioning mechanism for data
and the true source position for Monte Carlo simulated events) and 
reconstructed position.  
The shift between the Monte Carlo prediction and the
data are consistent with a model that scales each event's
reconstructed radial position by $\pm$1\%.
\label{fig:zscan}}
\end{center}
\end{figure}

The derived values of $\sigma$ and $\mu$ in our resolution function $F$ are in
general functions of position, energy, and source type.  
We look first at position dependence.
The comparison between Monte Carlo simulation and source data is done
by first deriving the mean fit position as described above and comparing it to 
the measured source position, based on the information from the 
source-positioning mechanism.  The precision of the source-positioning
is the limit to our overall uncertainty, and hence the primary measurements of 
position-dependent shifts are done using scans along the $z$-axis,  where the 
source position uncertainty is expected to 
be smallest ($\sim$2~cm).  Figure~\ref{fig:zscan} compares the mean reconstructed 
vertex for an \NS~$z$-axis scan taken in October 2000 to results from a Monte Carlo 
simulation of the same scan.

	One approach to assigning a systematic uncertainty on the vertex accuracy
based on measurements like those in Fig.~\ref{fig:zscan} would be to use the
maximum difference between the Monte Carlo predicted shift and the measured shift
as the `worst case' systematic shift, and treat that as the uncertainty.  Such
an approach would overestimate the uncertainty, however, because the plot shows
that there is a distribution of differences and not a simple overall offset. A
second approach would be to use the RMS of the distribution of (data-Monte Carlo
simulation) residuals as the uncertainty.  This would be appropriate if the
residuals were normally distributed about zero, indicating that the remaining
differences between the model and the data came from many (small) contributions.  It
is clear in the figure, however, that there is some indication of a systematic
variation of the residuals with source position---the largest residuals occur near
the bottom of the AV.  With so few calibration points (relative to the volume of the
detector) we must therefore create a model for a position-dependent systematic shift
in the mean reconstructed position that is consistent with the data.

	For this uncertainty, the best we can do is construct a plausible
worst-case which is consistent with the data we have.  Such a worst-case is
actually easy to create: as described above only a systematic shift
in reconstructed position inward or outward can have a significant effect
on the overall flux measurement.  The data shown in Fig.~\ref{fig:zscan},
which were taken with the $^{16}$N source on the $z$-axis so that $R\approx z$,
can be fit with a roughly linear shift as function of $R$.  For the lower half
of the AV the slope of such a linear shift is $\sim$~6~cm/600~cm=0.010, while
for the upper half the shift is smaller.  We therefore used a
$\pm$1\% scaling of reconstructed event radial position as our
systematic uncertainty on vertex accuracy.

	We looked at more than just this one scan, since
we need to explore all the ways in which the \NS~data used so far is not 
representative of neutrino events.  Among the ways we know it is not
representative is in data rate---the \NS~source is typically run at 
$\sim$ 100~Hz or so with events averaging $\sim 40$ hit tubes, while
physics data (including all backgrounds) is typically in the regime of $\sim$ 15-20~Hz with an
average of 12 hit tubes or so.  To understand the differences between this
relatively high rate environment and the low rate neutrino data (especially
given the known rate dependences discussed in Section~\ref{sec:dataset}),
we also took \NS~scans for which the source rate was lower than the typical physics
data acquisition rate (below 15 Hz or so).  During these scans (and for nearly all
other calibration runs), the standard physics triggers remained enabled, and
therefore the overall trigger rate rate was very similar to that for a typical neutrino run.
The scans were done 
along the $z$ axis, and the event vertices fit both before and after the 
rate-dependent correction described in Section~\ref{sec:dataset}.  

	The assumption of a systematic shift as a function of $R$, based on the
$z$-axis $^{16}$N scans, gives us a conservative estimate of the effects of
reconstruction uncertainties on fiducial volume uncertainties.  Nevertheless, we
also examined off-axis scans to ensure that there was no large axis-dependent
uncertainty.  In all cases, the uncertainties were consistent with the
radius-dependent $\pm$1\% shift described above.

\begin{figure}
\begin{center}
\includegraphics[height=0.28\textheight, angle=-90]{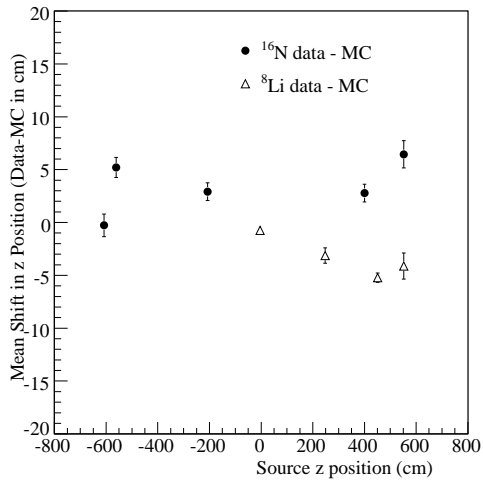}
\caption{Comparison of vertex accuracy in for \NS~and \LI~data for the
October 2000 \NS~$z$ scan and the December 2000 \LI~scan.  Vertex accuracy
is defined here as the difference between expected source position
(based on the measurements by the source-positioning mechanism for data
and the true source position for Monte Carlo simulated events) and 
reconstructed position.  The shifts between the Monte Carlo prediction and the 
data for both sources are consistent with a model that scales each event's
reconstructed radial position by $\pm$1\%.
\label{fig:n16li8shift}}
\end{center}
\end{figure}

	Having explored the position dependence of the uncertainty on vertex 
accuracy, we need to turn now to the source and energy (or $N_{\rm hit}$) dependence.  
For this we
compared the $z$ scan data for \NS~with that for \LI.  
Figure~\ref{fig:n16li8shift} shows the comparison of the difference
between the expected and reconstructed vertex positions for 
these sources compared to the Monte Carlo prediction of the difference.
As we
can see, there is no major difference in the vertex accuracy
of the two sources, despite the fact that the $^8$Li data produces electrons
up to 15~MeV and the $^{16}$N source produces monoenergetic 6.13~MeV $\gamma$-rays.  

	The \LI~-\NS~comparison also serves as a check of the time
dependence of the vertex accuracy, since these data were taken at times
separated by a few months.  We additionally looked at data separated by
over a year, using other scans, and this, too, was consistent with the
simple $\pm$1\% radial scaling model.

\subsubsection{Vertex Resolution}
\label{sec:vres}

\begin{figure}
\begin{center}
\includegraphics[height=0.28\textheight, angle=-90]{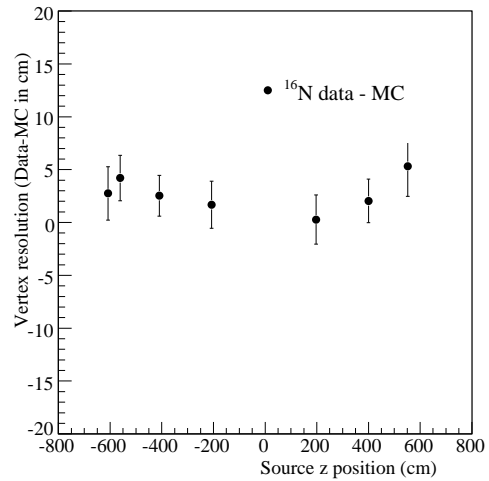}
\caption{Difference between vertex resolution for October 2000 high rate 
$z$ scan as a function of $z$ and Monte Carlo prediction.
\label{fig:zscan_sig}}
\end{center}
\end{figure}

	We measured the uncertainty on the vertex resolution in the same way 
as the vertex accuracy,
through comparisons of \NS~and \LI~data to Monte Carlo simulation.  The resolution is
obtained using the Gaussian convolution described in Section~\ref{sec:vshift}
for the different source distributions (Compton scatters for the \NS, a
spherical shell for the \LI).  Figure~\ref{fig:zscan_sig} compares the resolution 
obtained this way for the October 2000 $z$ scan to the
Monte Carlo simulation.  Here we see differences in resolution between the data and the 
Monte Carlo simulation of 1-5~cm, with the data having a systematically broader resolution than
the simulation.  Such a systematic broadening is not unexpected, as there are many
effects in real data which will worsen the resolution relative to the Monte Carlo
simulation (shifts in timing calibrations, knowledge of the source position during
calibration, knowledge of the true angular distribution of PMT hits around the event
direction, etc.) but few if any that will make it better.  We nevertheless
treat the systematic difference between data and the Monte Carlo simulation as a
double-sided uncertainty.

\begin{figure}
\begin{center}
\includegraphics[height=0.3\textheight]{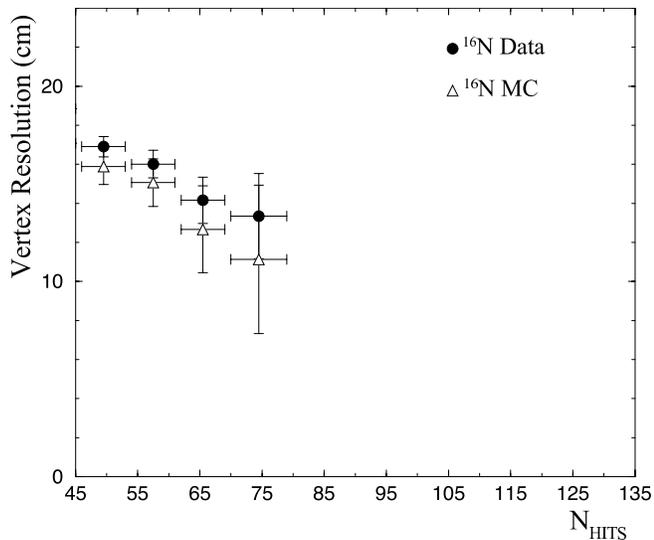}
\caption{Vertex resolution for high rate \NS~data as a function of 
$N_{\rm hit}$ compared to Monte Carlo prediction.
\label{fig:n16nhitdep}}
\end{center}
\end{figure}

\begin{figure}
\begin{center}
\includegraphics[height=0.3\textheight]{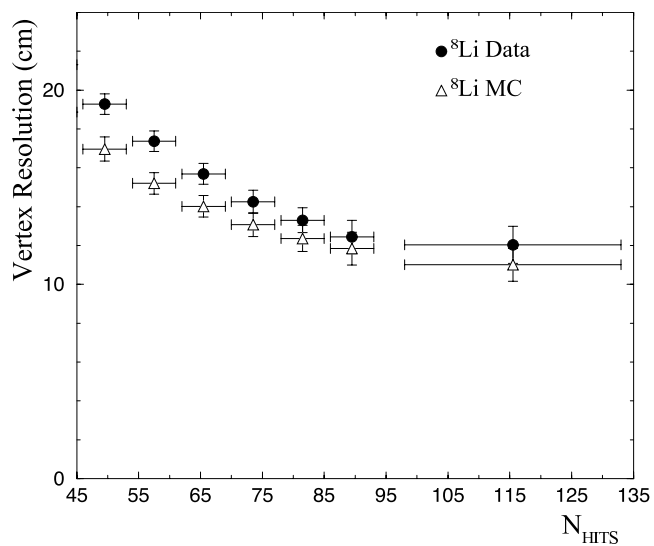}
\caption{Vertex resolution for \LI~data as a function of 
$N_{\rm hit}$ compared to Monte Carlo prediction.
\label{fig:li8nhitdep}}
\end{center}
\end{figure}

We also explored the energy ($N_{\rm hit}$) dependence of the resolution, since we expect 
the resolution to depend on energy ($N_{\rm hit}$) both through the increase in
the number of hits available at higher energies as well as the sharper angular
distribution of the Cherenkov cone.  Figure~\ref{fig:n16nhitdep} compares the Monte
Carlo prediction of the $N_{\rm hit}$ dependence of the vertex resolution for \NS~events to
source data and Figure~\ref{fig:li8nhitdep} does the same for \LI~data.  Both show
reasonably good agreement on the magnitude of the resolution (to a few cm) as well as
its slope with energy.  While the \LI~source data is somewhat suspect because
of the blockage of backward light by the source chamber, it is the only data
available for
testing the Monte Carlo predictions at high energies. While we may be willing to
accept the Monte Carlo simulation's handling of higher energy physics (the scaling
of Cherenkov photon production, for example) the effects of more photons
(such as crosstalk or a timing bias for multi-photoelectron hits) are not necessarily well-modeled.
Figure~\ref{fig:li8nhitdep} demonstrates that those
uncertainties are not large enough to matter.  The fact that the \LI~and the
\NS~data agree well where they overlap in $N_{\rm hit}$ also suggests that
source effects are not significant.

	Rate-dependent effects were checked using low-rate source data, and we tested
time-dependence using source runs taken along different detector axes at times
separated by more than a year.  In none of these comparisons did we see any effects
beyond those shown in the previous figures.

	We therefore take as our overall systematic uncertainty on the resolution
the rms of the differences between the data and the Monte Carlo simulation shown in
Figure~\ref{fig:zscan_sig}, which is about 2.5~cm.  As mentioned above, we do not
shift the resolution in the Monte Carlo simulation to agree with the data, but treat this
uncertainty as a double-sided uncertainty (that is, $\pm 2.5$~cm).  We estimate the
effects of this resolution uncertainty on the neutrino fluxes by convolving the
Monte Carlo predicted position pdfs with a Gaussian designed to broaden the Monte Carlo simulation's
resolution function by 2.5~cm.  The Monte Carlo prediction for the width of our
resolution is 15~cm for \NS~events, and we therefore convolved the position pdfs
with a Gaussian whose width was 9~cm.

\subsubsection{Summary of Vertex Uncertainties}

	For the uncertainty in vertex accuracy, we have found that a $\pm$1\%
radial scaling of the fit position is a reasonable worst-case model for the
differences between Monte Carlo simulation and \NS~source data.  We have further explored
the dependence on position, source type, energy, and time and found that in
none of these cases is the uncertainty worse than this.  For vertex resolution,
we have done a similar study, and find that the uncertainty in the resolution
is roughly 2.5~cm which is equivalent to convolving a Gaussian of width 9~cm
with the Monte Carlo predicted resolution response.

\subsubsection{Angular Resolution}
\label{sec:angres}

An ideal calibration source for measuring angular resolution would be a
directed source of single electrons with tunable
energies.  The angular resolution function (for a given electron position, 
direction, and energy) in the detector would then be the distribution of 
$\theta$, the angle between the reconstructed and the known initial electron 
directions. While the \chem{8}{Li} source does provide a source
of tagged electrons, we do not know the 
initial directions of individual electrons.
Instead, we developed a method for determining the 
angular resolution and uncertainties using $\gamma$-rays from the
$^{16}$N source~\cite{the:mgb,the:msn,the:cj}.

\begin{figure}
\includegraphics[height=0.3\textheight]{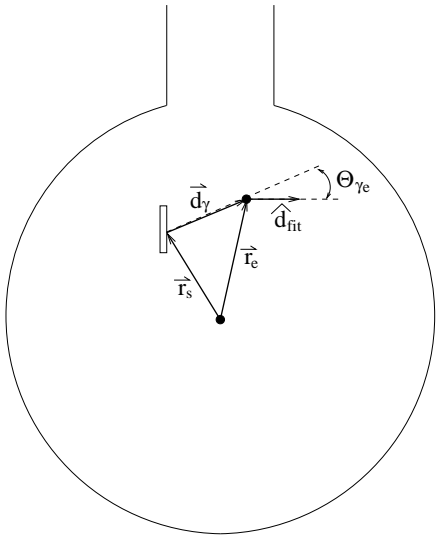}
\caption[Diagram showing vectors involved in measurement of angular
  resolution using the \chem{16}{N} $\gamma$-ray calibration source]{Diagram
    showing vectors involved in measurement of angular resolution using the
    \chem{16}{N} $\gamma$-ray calibration source.
\label{fig:angres}}
\end{figure}

The \chem{16}{N} calibration source data can be used to 
determine angular resolution uncertainty, by relying on the 
colinearity of Compton scattered electrons with the $\gamma$ direction, when the
$\gamma$ loses the majority of its energy.
If the scattering vertex, $\vec{r}_e$, is known, the $\gamma$-ray direction,
$\hat{d}_{\gamma}$, is
related to the \chem{16}{N} source position, $\vec{r}_s$, by the simple vector
relation (See Figure~\ref{fig:angres})
\begin{equation}
\hat{d}_\gamma = \frac{\vec{r}_e-\vec{r}_s}{\mid \vec{r}_e-\vec{r}_s \mid}
\end{equation}
The dot product of this unit vector with the
reconstructed event direction gives the cosine of the angle $\theta_{\gamma e}$:
\begin{equation}
\cos \theta_{\gamma e} = \hat{d}_\gamma \cdot \hat{d}_{fit}
\label{eq:angres_true}
\end{equation}

With the fit vertex $\vec{r_{\rm fit}}$ used as an estimate of the Compton
scattering vertex (${\vec{r}_e}$) 
Equation~\ref{eq:angres_true} becomes 
\begin{equation}
\cos \theta = \frac{\vec{r}_{fit}-\vec{r}_s}{\mid \vec{r}_{fit}-\vec{r}_s
  \mid} \cdot \hat{d}_{fit} 
\end{equation}
Note that this manner of determining the angular resolution depends on vertex
reconstruction uncertainties, since the vertex is used to calculate the
direction of the Compton scattered electron relative to the
incident $\gamma$-ray. In order to minimize the
effect of vertex reconstruction errors on the angular resolution measurement, we
only used events reconstructing a large distance from the \chem{16}{N} source
as compared to the vertex resolution.

\begin{figure}
\includegraphics[height=0.3\textheight]{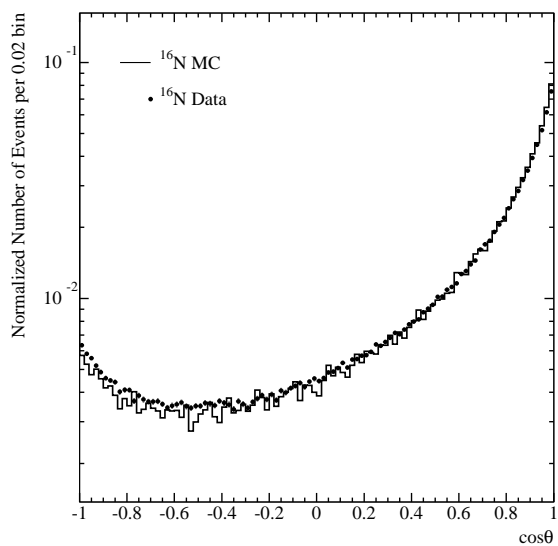}
\caption[The
  angular resolution from \chem{16}{N} is in good agreement with
  Monte Carlo simulation]{The angular resolution tail from \chem{16}{N} is in 
  good agreement with Monte Carlo simulation.}
\label{fig:angres_10372_log}
\end{figure}

Figure~\ref{fig:angres_10372_log} shows a comparison of the $\cos{\theta}$
distributions between real and simulated \chem{16}{N} calibration data with the
source at the center of the detector. The data plotted
are restricted to events that were reconstructed more than 1.5 m from the source
position.  We see that
the Monte Carlo model predictions are in good agreement with the measurements
for this particular location. 

\begin{figure}
\includegraphics[height=0.3\textheight]{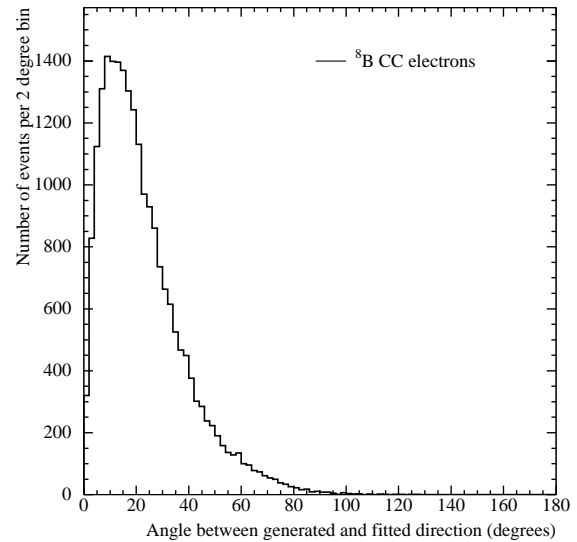}
\caption{Angular 
resolution for Monte Carlo simulated CC electrons. Shown is the distribution 
of angle between the Monte Carlo predicted initial electron direction and fit
direction for $N_{\rm hit}>$65 and $R_{\rm fit}<550$ cm. Roughly 68\% of the distribution
is contained within 26.7 degrees.
\label{fig:angres_cc}}
\end{figure}

To characterize the angular resolution 
we define a measure which is the angle between the initial electron 
direction and the fit direction that contains 68\% of the angular distribution. 
Notice from Figure~\ref{fig:angres_cc} that this
is determined from Monte Carlo simulation to be 26.7$^\circ$ for charged current electrons 
at energies near that of the $^{16}$N source and within the 550~cm fiducial volume.

	The systematic uncertainty on angular resolution is somewhat harder to
define than that for the uncertainty on position resolution. The angular
resolution is a complicated function as Fig.~\ref{fig:angres_cc} indicates.  As will
be discussed later in more detail in Section~\ref{sec:anal_response}, for our secondary
analysis which uses analytic response functions to build pdfs, we fit a
parameterized function (Eq.~\ref{aresp}) to distributions like that in
Figs.~\ref{fig:angres_10372_log} and~\ref{fig:angres_cc}. We then created new pdfs
with the parameters on the angular response function varied over their $\pm
1\sigma$ uncertainty range, and used the changes measured in the extracted numbers
of events as our $\pm 1\sigma$ angular resolution systematic uncertainty.
For our primary analysis, in which we used Monte Carlo-generated pdfs, we
used a perturbation function to `smear' the pdf.  The perturbation function was
chosen so that it that reproduced the differences seen in the comparison between
data and Monte Carlo simulation like that shown in Fig.~\ref{fig:angres_10372_log}: a
narrowing of the forward peak and the addition of an isotropic component that puts
up to 2\% of events into the tail.  The effects of this smearing are similar to the
variations of the analytic angular response function discussed in
Section~\ref{sec:anal_response}.

\subsection{Energy Response}
\label{sec:sysuncert_ene}

	For the integral flux measurements of our initial Phase I
results~\cite{snocc,snonc,snodn}, the dominant uncertainty on our measurements
derives from the uncertainty on the detector's energy scale.
The reason is that at the energy
threshold used, a small variation in response leads to a large variation in the
number of accepted events.  The natural covariance between the charged current and
neutral current signals---the fact that the differences between the pdfs shown in
Fig.~\ref{fig:pdfs} are predominantly in the energy distributions---makes the
problem significantly worse.  We therefore needed to be particularly careful in
evaluating these uncertainties.

\subsubsection{Energy Scale}

	As discussed in Section~\ref{sec:model}, the energy scale---the number of
PMT hits per MeV of electron energy, or the adjustment of reconstructed electron
energy to agree with physical electron energy---was determined through deployment of
the $^{16}$N source at the detector center.  In addition to the center deployment,
we also made two extensive scans, covering two orthogonal planes within the
detector.  The scans were performed in December 1999 and January 2001.  The primary
estimate of our systematic uncertainty on the energy scale is the volume-weighted
average difference between the Monte Carlo model prediction of the detector response
to the source at each point and the source data itself. There are many contributors
to non-zero differences: the statistics of the calibration source data, small errors
in the optical calibrations which are input to the Monte Carlo detector model, and
unmodeled or incompletely modeled detector effects such as crosstalk between
electronics channels and PMT-to-PMT variations.  To account for both the non-zero
mean of the volume-weighted difference distribution and its width, we add them 
linearly and take the sum as our uncertainty on the position-dependence of the
energy scale.

\begin{figure}[ht]
\begin{center}
\includegraphics[width=0.3\textheight]{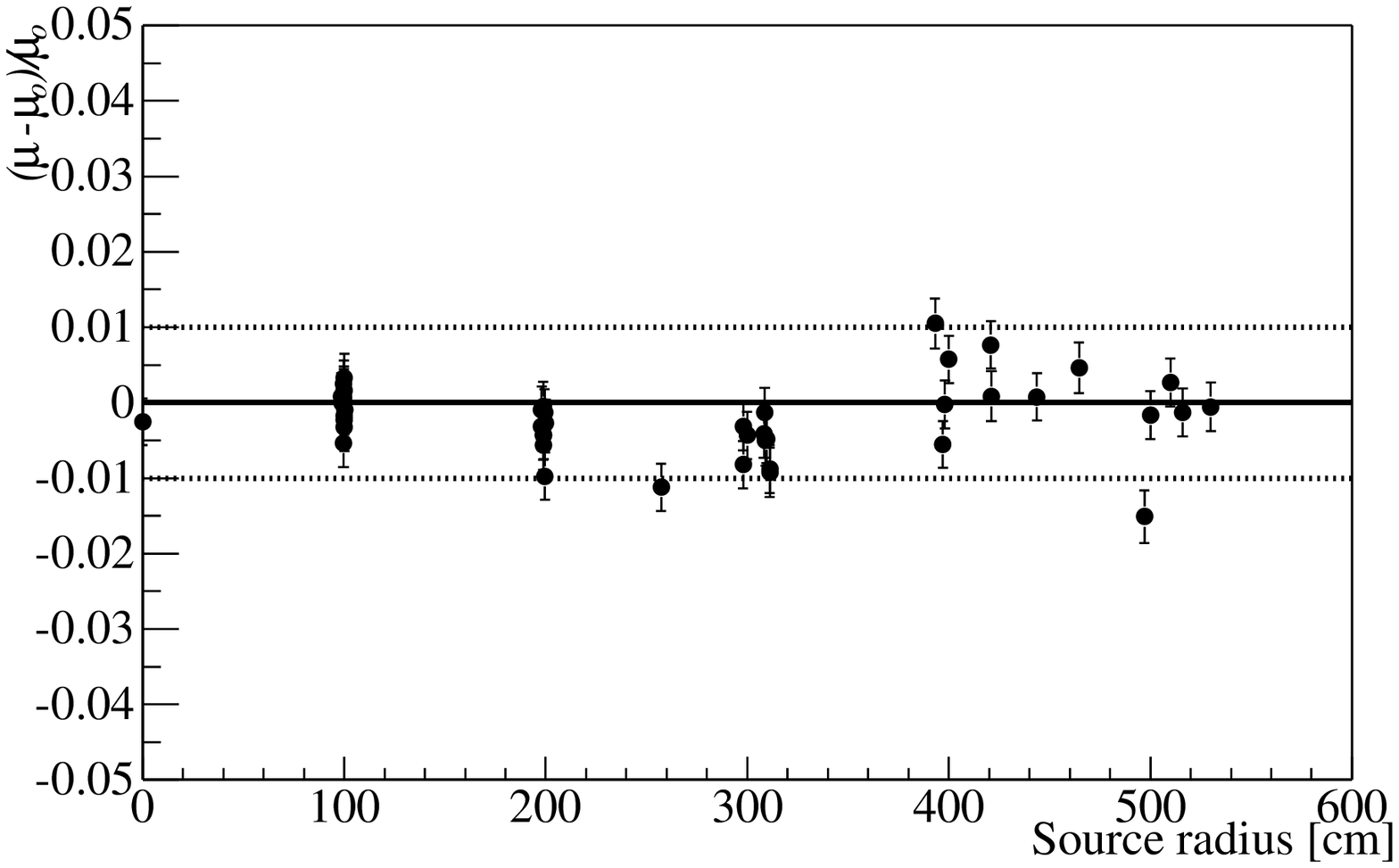}
\includegraphics[width=0.3\textheight]{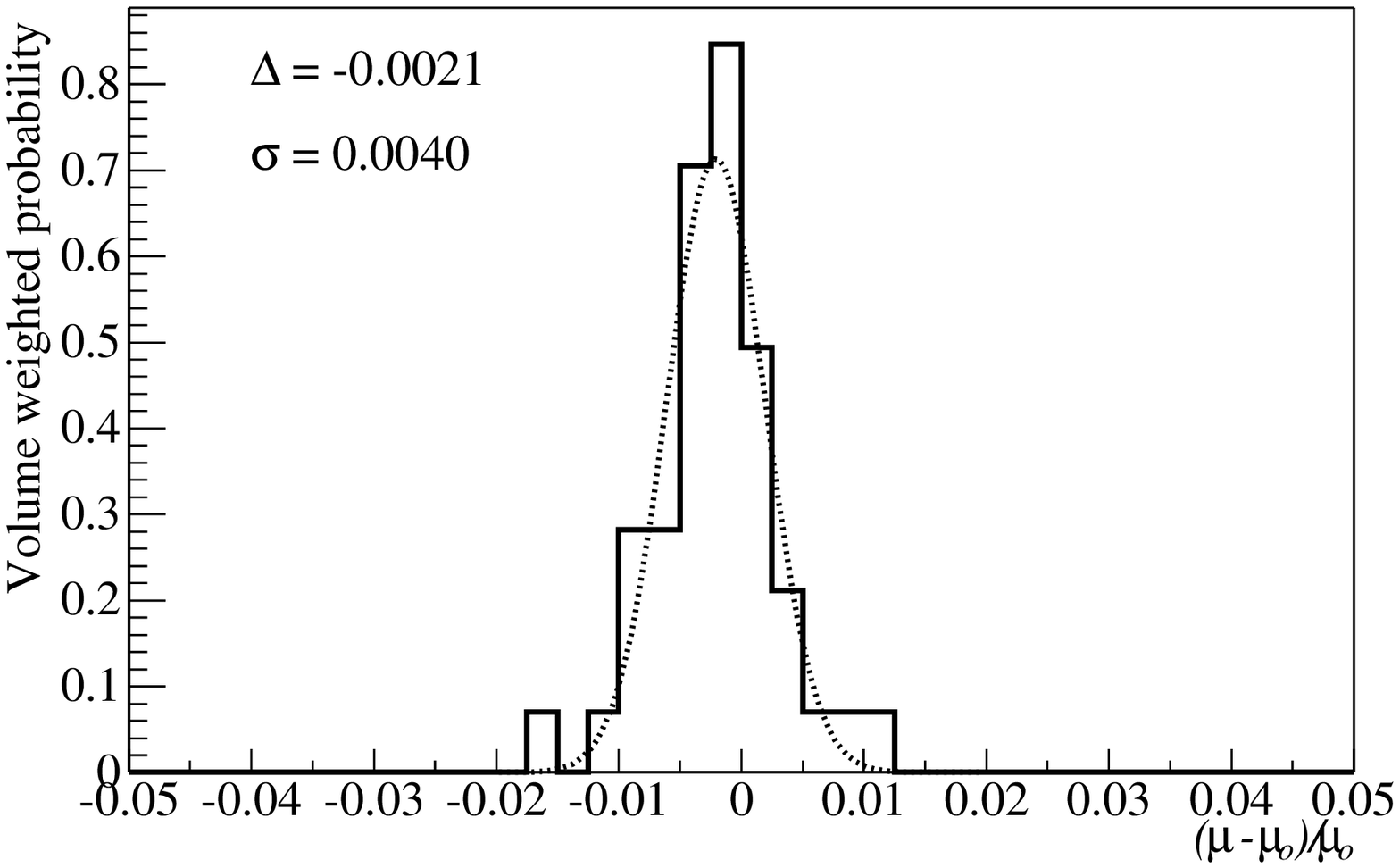}
\caption{Fractional deviation in effective kinetic energy peak $\mu$ from the Monte Carlo prediction $\mu_0$ for December 1999 position scans.  Shown in the top frame is the deviation versus source position.  The bottom frame shows the distribution of deviations weighted by volume.  A conservative $1\sigma$ limit of $0.72\%$ is obtained by adding the mean offset and distribution width linearly. The scatter in the top frame for points at the same radius is due to the fact that these source locations had different $(x,y,z)$ coordinates which gave the same radial positions. \label{plot_dec_scans}}
\end{center}
\end{figure}

\begin{figure}[ht]
\begin{center}
\includegraphics[width=0.3\textheight]{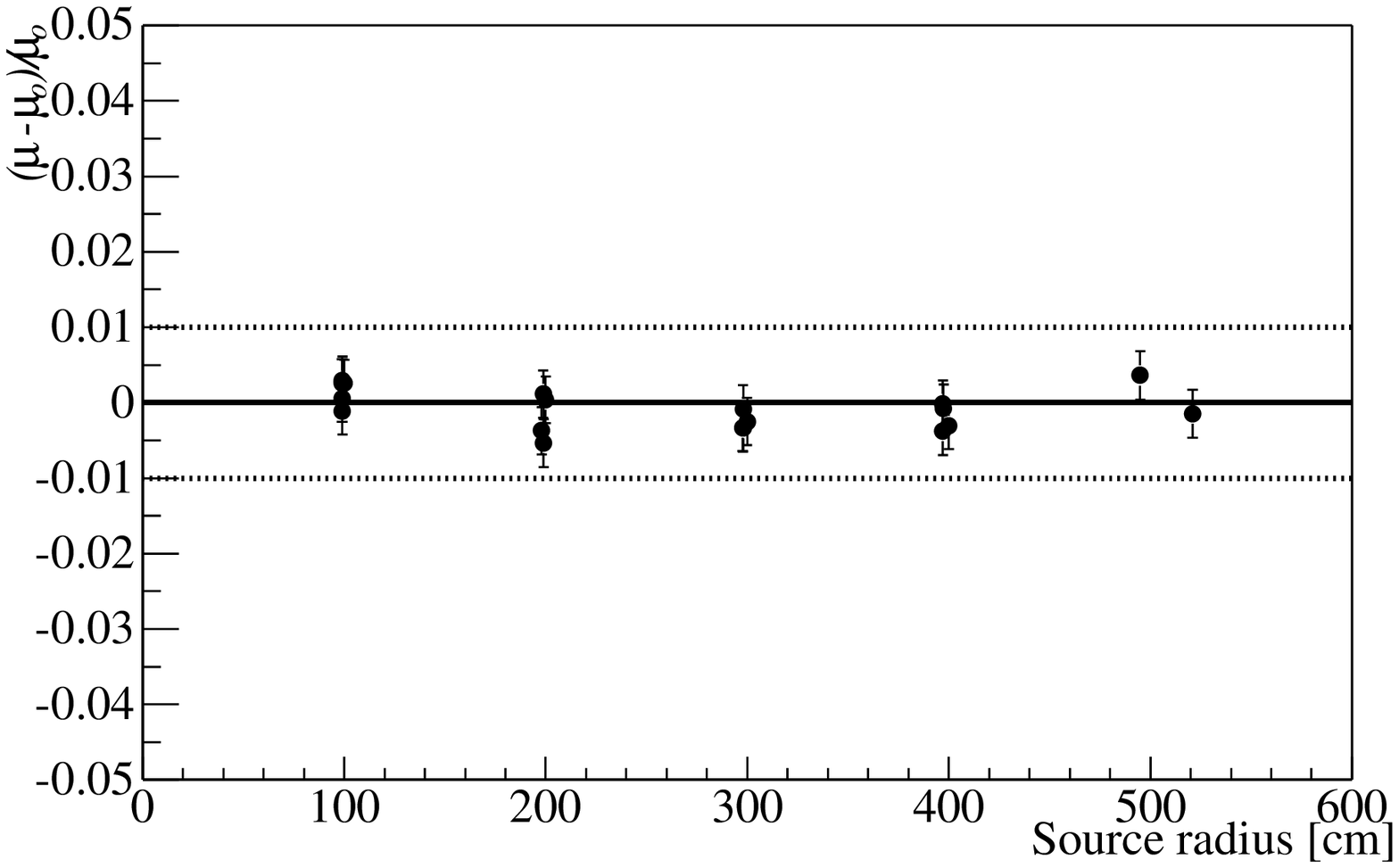}
\includegraphics[width=0.3\textheight]{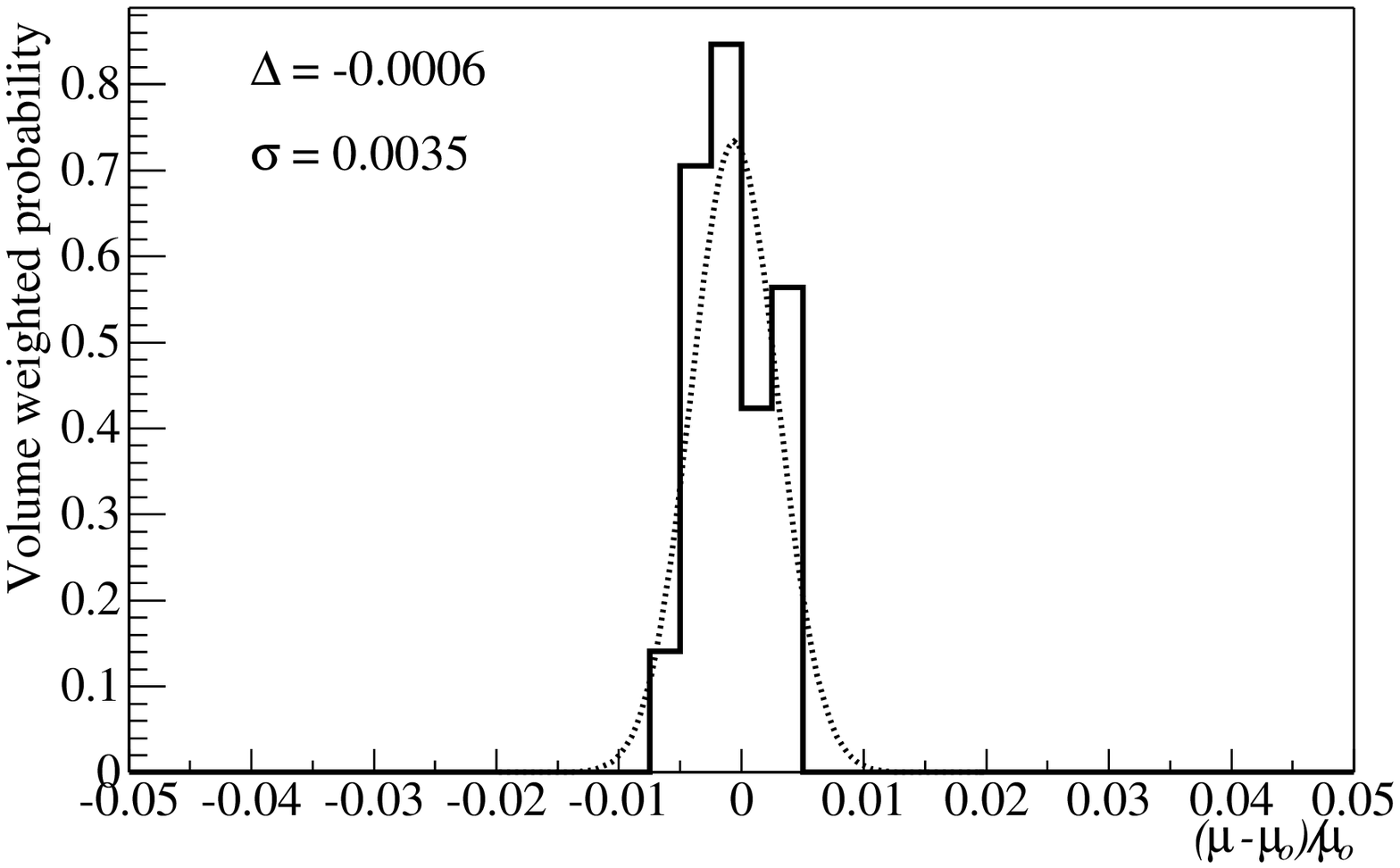}
\caption{Fractional deviation in effective kinetic energy peak $\mu$ from Monte Carlo prediction $\mu_0$ for the January 2001 position scans.  Shown in the top frame is the deviation versus source position.  The bottom frame shows the distribution of deviations weighted by volume.\label{plot_jan_scans}}
\end{center}
\end{figure}

Figures ~\ref{plot_dec_scans} and ~\ref{plot_jan_scans} show the fractional
differences as a
function of the source radial position as well as the volume-weighted
distribution of those differences, for one set of position scans.  Data taken along the
$+z$-axis are excluded from the Figures because the effects of the acrylic
vessel neck shift the energy peak substantially. This shift is a small effect
on neutrino data because there are so few events which occur in the neck
region.  The scatter in the points of Fig.~\ref{plot_dec_scans} that occur
for deployments at the same radial position is due to the fact that the source
was deployed at different $(x,y,z)$ coordinates for these radii.  Based on the
December 1999 and January 2001 scans, our $1\sigma$ estimate on the
position-dependent energy scale uncertainty for the energy reconstructor is
0.72~\%.  For the total light ($N_{\rm hit}$) energy estimator, the uncertainty
is 1.03\%.

We determined the uncertainties associated with source modeling by varying the
relevant source description parameters.  Details of how this was done can be
found in Refs.~\cite{the:ham,n16}.  The contribution to the uncertainties
from source modeling are less than $0.3 \%$, and the total source-related
uncertainty including uncertainties in the $^{16}$N decay scheme and
uncertainty in the tracking of $\gamma$-rays and electrons by EGS4 is 
$\sim 0.5 \%$.

The rate dependence of the calibrated times and charges described in 
Section~\ref{sec:ecapca} implies that the high rate calibration data 
($\sim 200$ Hz) may not correctly characterize the energy scale for 
neutrino data (typically 20 Hz).  Although a rate-dependent correction
was applied to the PMT hit times (Section~\ref{sec:ecapca}), and the energy calibrator
did not use any charge-dependent timing corrections 
(Section~\ref{sec:ecalibrator}), we nevertheless included a small systematic
uncertainty to account for residual rate-dependent effects.  This uncertainty
was determined through comparisons of high and low rate calibration data
taken with the $^{16}$N source and the $^{252}$Cf neutron source.
These comparisons showed no statistically significant 
rate-dependent effects, and the resultant uncertainty associated with
rate-dependent effects (driven by the statistical sensitivity of the 
comparisons) was  0.39\%.

	Variations in channel thresholds can also lead to unexpected
changes to the energy scale for neutrino data  that are not completely
represented by calibration data.  The probability of crosstalk between
adjacent channels is a very sensitive measure of the channel thresholds,
and by monitoring this probability we were able to limit the uncertainty
on the energy scale from such variations throughout the neutrino data set to 
0.45\%. As a verification that the calibration data was not significantly
different from the neutrino data, we compared the mean number of noise
hits measured with the pulsed trigger (see Section~\ref{sec:ecalibrator}) during
neutrino data collection to that measured during calibration source runs, and
found no significant differences. 

	In addition to threshold, the gains of the PMTs
may also vary and lead to energy scale variations.  To measure gain
stability, we compared the high edge of the single photoelectron charge peak
for neutrino data and $^{16}$N data, and found that the gain was stable
to 1.25\%.  This variation translates into an uncertainty of 0.28\% in
efficiency, and thus in energy scale.

The complete list of the contributions to the energy scale systematic
uncertainties appears in Table~\ref{tbl:uncnew} where a 0.39\% uncertainty is
attributed to rate dependence, 0.45\% to threshold variations, and 0.28\% to
gain variations. With the suite of uncertainties added in quadrature, the
energy scale uncertainty for the energy calibrator is 1.21\%, and for the total
light estimator ($N_{\rm hit}$) is 1.39\%. 

\begin{table}[htb]
\caption{Breakdown of systematic uncertainties on energy scale for the
total light ($N_{\rm hit}$) energy estimator, and the energy reconstructor.  \label{tbl:uncnew}}
\begin{center}
\begin{tabular}{lcc}
\hline \hline
Contributing factor &  $N_{\rm hit}$ & $T_{\rm eff}$\\ 
&& (MeV) \\ \hline
Scale including Time Drift & 0.46\% & 0.25\% \\ 
Position dependence & 1.03\% & 0.72\% \\ 
Source & 0.46\% & 0.46\% \\ 
Rate dependence & 0.39\% & 0.39\% \\ 
Gain variation  & 0.28\% & 0.28\% \\ 
Threshold variations - XTalk & 0.45\% & 0.45\% \\ 
Channel accounting & 0.1\% & negligible \\ 
Background noise & 0.1\% & negligible \\ 
Time calibration & negligible & 0.5\% \\ \hline
Total & 1.39\% & 1.21\% \\ \hline \hline
\end{tabular}
\end{center}
\end{table}

As a cross-check, energy calibration computations have been applied to
n(d,t)$\gamma$-ray event data from the ${}^{252}$Cf source (high and low rate),
low rate ${}^{16}$N data, and pT data.  

\begin{figure*}[t]
\begin{center}
\includegraphics[ angle=-90, width=0.5\textheight] {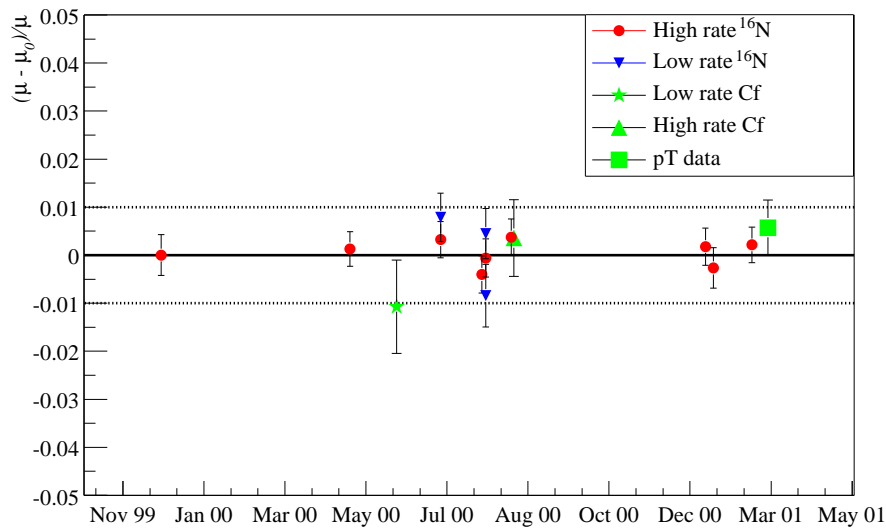}
\caption{Fractional deviation in kinetic energy peak for all calibration sources including low rate neutron and ${}^{16}$N data. Here, $\mu$ is the value at the peak for the data, and $\mu_0$ is the Monte Carlo simulation's prediction for the peak. \label{all_sources}}
\end{center}
\end{figure*}

Figure ~\ref{all_sources} shows the deviations in the $E_{\rm eff}$ peak for all
sources.

\subsubsection{Differential Energy Scale Uncertainty}

   For integral flux measurements, the most important uncertainty on the
energy scale is near threshold, where small shifts in the scale can lead
to large shifts in the fluxes.  Most of SNO's calibration source data for
Phase I (primarily $^{16}$N) have a central value of energy near
this threshold.  Differential non-linearities in the energy scale can affect
the integral measurements, however, because they alter the shapes of the pdfs
used for signal extraction. For a spectral measurement, in which each recoil
electron energy bin is treated independently, such non-linearities matter more.

	The primary sources of potential non-linearity are small errors in
the modeling of the PMT hit efficiencies as a function of the number of
incident photons (the `multiphoton effect'), and detector artifacts that
vary with the number of photons such as channel-to-channel crosstalk.  
For $^{8}$B solar neutrino events within SNO's
550~cm fiducial volume the probability of more than one photon hitting
a PMT is small, and errors on the modeling of these efficiencies
are negligible.

	The probability of crosstalk is also small, but it can still lead to
one or two additional hits in an event which if ignored could produce
a noticeably non-linear scale.  The prompt time cut of the energy calibrator
removes roughly 2/3 of the crosstalk hits, because their times are delayed 
slightly. Using 19.8~MeV $\gamma$ rays from the pT source to measure of the shift in energy
scale with energy, and interpolating between the $^{16}$N and pT results to the
$^8$B spectrum, we limit the additional shift in energy scale from
non-linearities to 0.23\% at the pT source energy ($T_{\rm eff}=19.1$~MeV), 
decreasing linearly to zero at the $^{16}$N energy ($T_{\rm eff}=4.98$~MeV).

	The functional form for the non-linear piece of the shift in kinetic
energy $\delta T_{\rm eff}$ as a function of reconstructed effective energy is
\begin{equation}
\delta T_{\rm eff} =  \alpha \times 19.1 \times \frac{T_{\rm eff}-4.98}{13.61}
\end{equation}
where $\alpha$ is limited to be $\pm 0.0023$.  The shift is measured in MeV.

\subsubsection{Energy Resolution}

	Uncertainties on the detector's energy resolution have a smaller effect
on the measured neutrino fluxes than energy scale uncertainties.  Small differences
between the true pdfs and the models of those pdfs used for signal extraction do not
have a big effect on the overall acceptance.  Resolution uncertainties have a much
bigger effect on measurements of the backgrounds from low energy radioactivity, as
described in Section~\ref{sec:certail}.

	To measure the uncertainty on energy resolution, we compared the
reconstructed energy distributions for calibration data to Monte Carlo simulations
of that data, for the 6.13~MeV $\gamma$-ray $^{16}$N source and the 19.8~MeV
$\gamma$-ray pT source.   For the $^{16}$N source, the measurements were made at
many locations throughout the detector volume along the two planes allowed by the
calibration system, and for the pT source at several locations along the $z$-axis
including positions out to $R \sim 450$~cm.  

	Figure~\ref{fig:eeff_compare} compares the distribution of reconstructed
energy for both data and Monte Carlo simulations of an $^{16}$N deployment near the
detector center.  To measure the resolution, we fit a Gaussian between 4 and 7~MeV
to distributions like those in Fig.~\ref{fig:eeff_compare}.  We found that on
average, the resolution for the data was $\sim$ 2.5\% broader than for the Monte
Carlo, and that the variations from point to point between the two was also of order
2\%.  We conservatively added these two measurements linearly, for a combined
resolution uncertainty of 4.5\%.  For the analytic parameterization of the
resolution function given in Section~\ref{sec:anal_response}, there are additional
uncertainties associated with the extraction of the parameters from the Monte Carlo
simulation.  Those uncertainties are given in that section.

	For the pT source, we found a much bigger difference between the data and
Monte Carlo simulation's resolutions---roughly 10\%, primarily due to the fact that the many
neutrons produced by the source affect our ability to measure the resolution in the
data, but are not modeled in the Monte Carlo simulation.  A linear function was
used to interpolate the uncertainties between the $^{16}$N and pT energies:
\begin{equation}
\label{eq:sigTunc}
\frac{\Delta \sigma_T}{\sigma_T} = 0.045 + 0.00401\times(T_{\rm eff}-4.98) 
\end{equation}

\section{Background Measurement \label{sec:bkds}}

	After processing, the events remaining above the analysis
threshold and within the 550~cm fiducial volume are primarily recoil electrons
and $\gamma$-rays produced in association with neutrino interactions, but may
also include instrumental, radioactive, and cosmogenic backgrounds.  In this
section we describe measurements to determine the residual contamination from
each background source.  

\subsection{Instrumental Contamination \label{sec:contam}}

	Although the suite of low-level cuts described in
Section~\ref{sec:inst} is highly effective at removing instrumental
backgrounds, and the subsequent reconstruction and `high-level' cuts reduce
residual contamination still further, we must estimate how many
events from instrumental sources remain in the final data set.

	Instrumental backgrounds are a particularly difficult problem, because
it is not possible to model every possible ill-understood non-Cherenkov
background source.   Instead, we need a method that can determine the
background level irrespective of its source.
	
	The method we adopted for this analysis combined the low-level cuts and
the high-level cuts in what is sometimes referred to as a `bifurcated
analysis'~\cite{E787}.  For more detail than we give here, see
Ref.~\cite{the:vrusu}.

In a bifurcated analysis one picks two cuts (or
two sets of cuts as we have done) and counts the numbers of events in the
data set rejected by either cut, both cuts, or neither cut.  We assume that the
data set consists of just two classes of events, signal events $\nu$ and
background events $\beta$, so that the total number of events in the data set
is just $S=\beta + \nu$.  The background contamination in the final signal
sample is just the fraction of $\beta$ that passes both sets of cuts.  If the
acceptance for background events by cut set $i$ is $y_i$, the final background
contamination is $K=y_1 y_2 \beta$.  If the acceptance for signal events by cut
set $i$ is $x_i$, the final number of signal events is $x_1 x_2 \nu$.

We start with three separate event totals:
the number of events that pass both cuts ($a$), the number
that fail cut 1 but pass cut 2 ($b$), and the number that pass cut 1 but fail 
cut 2 ($c$).  We then relate all of these with a linear system of equations:
\begin{eqnarray}
a+c=x_1 \nu+y_1 \beta\\
\label{eq1}
a+b=x_2 \nu+y_2 \beta\\
\label{eq2}
a=x_1 x_2\nu+y_1 y_2 \beta\\
\label{eq3}
\beta+\nu=S
\label{eq3prime}
\end{eqnarray}
\noindent
which we solve analytically to determine the remaining background contamination
$K= y_1 y_2 \beta$.  The values for the cut acceptances will be discussed
later, in Section~\ref{sec:accept}.

We illustrate the general approach in Fig.~\ref{fig:bifurcate}, which shows
$^{16}$N events events in blue, neutrino candidate events in grey, and
instrumental background events in black on a `high-level' cut plot.  The
in-time ratio (ITR) is on the horizontal axis and the average PMT pair angle
$\theta_{ij}$ on the vertical axis.  The $^{16}$N data define a `Cherenkov box'
(see Section~\ref{sec:cerbox}) that contains most of the neutrino event
candidates.  Most instrumental background events, defined as events which fail
the `low level' cuts, lie outside the Cherenkov box.  Our bifurcated analysis
measures the ratio of the number of events failing the low-level cuts that lie
within the Cherenkov box to those that lie outside, and also measures the
number of events that pass the low-level cuts but lie outside the Cherenkov
box. The number of background events within the Cherenkov box (that is, which
pass both the low-level and high-level cuts) is then the product of these two
numbers.  For the final Phase I data set, we find using this technique that the
overall contamination has a $95\%$ confidence level upper limit of $K\leq3$
events.  

The dominant systematic uncertainties in this analysis are the uncertainties
on the cut acceptances (see Sections~\ref{sec:insteff} and~\ref{sec:hlcuteff}) and the possibility
of variations in the efficiency of the cuts for removing backgrounds.
For the latter, we looked at the stability of each cut as a function of
time using calibration source data.

For this analysis to work, the two sets of cuts we use must be orthogonal to
one another---we must be sure that the probability of passing the low-level
cuts does not increase the probability of passing the high-level cuts.
To demonstrate orthogonality, we loosened the cuts (essentially opening the
final `signal box' defined by those events which pass both sets of cuts)
and measured the increase in the number of background events.  With the
looser cuts, we found the increase in the number of background events agreed
well with what would be expected for orthogonal cuts.

To ensure that there were no instrumental backgrounds missed by this analysis,
we also examined many different distributions of events and hits in `detector
coordinates'---the number of hits as a function of electronics channel (rather
than PMT), the distribution of event directions relative to the detector's
zenith (rather than the solar direction), and the general PMT-by-PMT occupancy.
We found no evidence of any remaining non-Cherenkov-light background.

	In addition, we repeated the bifurcated analysis using different sets
of cuts---for example replacing the cut on the mean PMT pair angle with a cut
on an event `isotropy' parameter derived from the full two-point PMT-PMT correlation
function, or using only a subset of the low-level cut suite. All
differences in the results were very small and within our 
expectations.

\subsection{Photodisintegration Background \label{sec:pd_bgd}}

By far the most dangerous background to the NC measurement are
the neutrons produced through photodisintegration of deuterons by low
energy radioactivity.  In particular, $^{232}$Th 
and $^{238}$U have $\gamma$-rays at the end of their decay chains (2.61~MeV and
2.44~MeV, respectively) that are
above the 2.22~MeV deuteron binding energy. Low levels of these nuclei can
be found in all the components of the detector: the heavy water, the acrylic
vessel, the light water, PMT support structure, as well as the PMT glass and 
base hardware.  The neutrons produced by photodisintegration
are indistinguishable from those produced by the NC reaction, and therefore
measurements of the background levels inside the detector are crucial for
correct normalization of the total $^{8}$B flux.  It is critical to measure the
levels of $^{232}$Th and $^{238}$U separately as the fraction of decays that
lead to $\gamma$-rays above 2.2~MeV are very different, 36\% and 2\%,
respectively. Additionally, the photodisintegration cross-section depends
strongly on the decay $\gamma$-ray energy.

	The first step in dealing with these backgrounds was to build
the detector with very stringent radio-purity targets for all components.
Table~\ref{tbl:radio_target} lists the $^{232}$Th and $^{238}$U  target
levels for the D$_2$O, AV, and H$_2$O.  (At these radiopurity levels,
the background to the NC signal is approximately 1 neutron produced per day
or $\sim$10\% of the NC signal).  In this section, we describe the
techniques developed to measure the $^{232}$Th and $^{238}$U
concentrations in different detector regions, and the resultant numbers of
background neutrons which these measurements imply.

\begin{table}[h!]
\caption[The target maximum radio-purity levels for different components in the SNO detector]{The target radio-purity levels for different components in the SNO detector.}
\protect\label{tbl:radio_target}
\begin{center}
\begin{tabular}{lcc} \hline \hline
Component & $^{232}$Th  &  $^{238}$U \\ 
                      &  (g/g)   & (g/g)   \\ \hline
D$_2$O  &  3.7$\times$10$^{-15}$  & 4.5$\times$10$^{-14}$ \\ 
H$_2$O & 3.7$\times$10$^{-14}$  & 4.5$\times$10$^{-13}$ \\ 
AV & 1.9$\times$10$^{-12}$  & 3.6$\times$10$^{-12}$ \\ \hline \hline
\end{tabular}
\end{center}
\end{table}

As in the rest of the analysis, we used two independent approaches to measuring
backgrounds within the H$_2$O and D$_2$O.  Methods that
remove water from the detector and perform direct radioassays to determine
the concentration of impurities are called {\it ex-situ} techniques, and 
methods that measure background levels using the Cherenkov light observed 
within the SNO detector are called {\it in situ}.

\subsubsection{{\it Ex-situ} Techniques for Determining Water Radioactivity}   
\label{sec:exsitu}

The {\it ex-situ} techniques circulate large samples of water from the detector
volumes, extract background isotopes from the samples, and count the
number of decays using instrumentation external to the SNO detector.
We developed three such {\it ex-situ} techniques: extraction of
Ra isotopes using manganese oxide (\mnox) beads~\cite{bib:mnox_nim},
extraction of  Ra, Th and Pb isotopes using hydrous titanium oxide
(HTiO)-loaded membranes~\cite{bib:htio_nim}, and degassing the $^{222}$Rn
from the $^{238}$U chain (the ``Rn assay'', Ref.~\cite{bib:rn_nim}).

In the \mnox\ technique, \dto\ or \hto\ is passed through polypropylene
columns that contain beads coated with a manganese oxide compound,
which extracts Ra from the flowing water.  After a large volume of water
has passed through the columns, they are removed and dried.  The dried
columns are then attached to a gas flow loop on an electrostatic counter.
The Rn produced from Ra decay is swept from the columns into the
electrostatic counter where it decays.  The charged Po ions from the decay of Rn are carried
by the electric field onto an $\alpha$ counter where the decays of the Po
are detected, and their $\alpha$ energy spectra are collected.  For the
$^{232}$Th chain, the relevant Po $\alpha$ decays are $^{216}$Po (6.8~MeV
$\alpha$) and $^{212}$Po (8.8~MeV $\alpha$), whereas the relevant ones
for the U chain are $^{218}$Po (6.0~MeV $\alpha$) and $^{214}$Po (7.7~MeV
$\alpha$).  A number of \mnox\ assays were carried out for both the D$_2$O
and H$_2$O.  
The $^{232}$Th value for the MnOx data, averaged over the
neutrino livetime, for Phase I of the experiment is
\begin{eqnarray}\nonumber
\mbox{\dto}:& 2.15 \, ^{+0.90}_{-0.94} \,\times\, 10^{-15} \ \mbox{g Th/g D}_2\mbox{O} \\ \nonumber
\mbox{\hto}:& 8.1 \, ^{+2.7}_{-2.3} \,\times\, 10^{-14} \ \mbox{g Th/g H}_2\mbox{O} \\ \nonumber
\end{eqnarray}
where the statistical and systematic uncertainties have been combined in
quadrature.  Ref.~\cite{bib:mnox_nim} provides a more
detailed discussion of the evaluation of the systematic uncertainties.

In the HTiO technique~\cite{bib:htio_nim}, \dto\ or \hto\ is passed through
hydrous titanium oxide (HTiO) trapped on filtration fibers.   The HTiO
ion-exchanger is first deposited onto a microfiltration membrane.  Then 
columns containing the loaded filters are used to extract $^{224}$Ra (from the
Th chain) and $^{226}$Ra (from the U chain) from a large volume of \dto\ or
\hto.  After extraction, the Ra is eluted with nitric acid, and subsequently
concentrated to $\sim$10~m$l$ of eluate.  This is then mixed with liquid
scintillator and counted using $\beta$-$\alpha$ delayed coincidence
counters~\cite{the:taplin}.  For the $^{232}$Th chain, the coincidences
of the $\beta$-decay of $^{212}$Bi and the $\alpha$-decay of $^{212}$Po are
counted, whereas the coincidences of the $\beta$-decay of $^{214}$Bi and the
$\alpha$-decay of $^{214}$Po are counted for the $^{238}$U chain.  The
HTiO and MnOx measurements were in good agreement, but the MnOx result
above was used as the final {\it ex-situ} measurement of the $^{232}$Th 
concentration because the measurements were made more regularly.

The measurements of $^{226}$Ra concentration in the \dto\ and the \hto\ by the
\mnox\ and the HTiO techniques described above are not, however, sufficient to
determine the total radioactive background from the $^{238}$U chain.  Even a
small ingress of underground laboratory air (with its $\sim$3~pCi/$l$ of $^{222}$Rn)
can lead to significant disequilibrium between $^{226}$Ra and $^{214}$Bi.  To
tackle this problem, we developed a Rn assay technique~\cite{bib:rn_nim}. Water
drawn from discrete sample points in the detector is flowed through a degasser
to liberate Rn.  The Rn is purified and collected in a cryogenic collector.
The subsequent $\alpha$ decays are counted in a Lucas cell scintillator (ZnS)
chamber on a 2.54-cm diameter photomultiplier tube.  Since there is a delay of
many $^{220}$Rn lifetimes between the preparation of the Lucas cells and their
subsequent counting, this method is sensitive only to $^{222}$Rn decays.

\begin{figure*}[h]
\begin{center}
    \includegraphics[height=0.75\textheight]{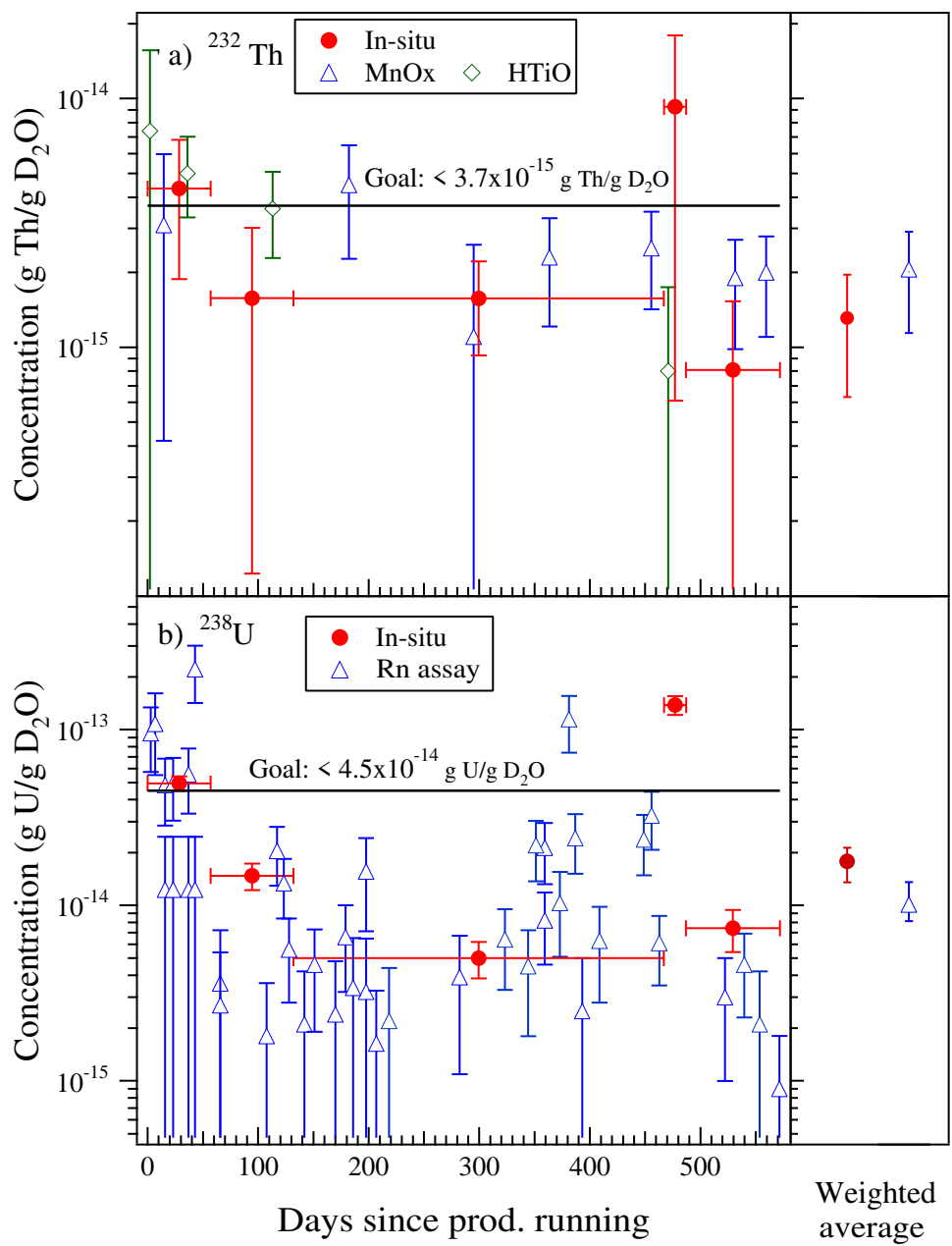}
    \caption[Comparing the {\it ex-situ} and {\it in-situ} results as a function of time]{Comparing the {\it ex-situ} and {\it in-situ} results as a function of time.}
    \protect\label{fig:inex_cmp}
\end{center}
\end{figure*}

The Rn assay results for different sampling points in the \dto\ and the
\hto\ as a function of time and systematic uncertainties in the results
are discussed further in Ref.~\cite{bib:rn_nim}.  It can be seen from the
measurements presented there and here in Fig.~\ref{fig:inex_cmp} that during 
the early phase of the production running, the Rn level in the detector was 
much higher than our target level. After a few months the levels dropped, and remained better than the target levels,  with the
exception of some excursions for short intervals.

\subsubsection{{\it In-situ} Technique for Determining Water Radioactivity}  
\label{sec:dto_insitu}

The {\it in-situ} technique identifies and measures the different radioactive
backgrounds using the Cherenkov light produced by the events within the SNO
detector itself.  The goal of the {\it in-situ} analysis is twofold: first to
separate decays from the $^{238}$U chain from those of the $^{232}$Th chain,
and second to determine the corresponding radioactivity levels based on the
total numbers counted.  We applied this analysis to both the D$_2$O and
H$_2$O. Unlike the {\it ex-situ} analysis, the {\it in-situ} analysis
is integrated over the same livetime as the neutrino data, rather than being
sampled at discrete times.  Moreover, it measures the amounts of the
radionucludies, $^{208}$Tl and $^{214}$Bi (from the $^{232}$Th and $^{238}$U
decays chains, respectively), that give rise to $\gamma$ rays above 2.2~MeV.
The {\it in-situ} technique measures the isotopes that produce
photo-disintegration backgrounds directly and does not assume secular
equilibrium in the decay chain.  As in the {\it ex-situ} analysis described
above, we are interested in this analysis in measuring the overall detector
radioactivity, and from that measurement calculating the number of neutrons
produced in the decays of the associated daughters.  We therefore used a lower
energy threshold than our nominal signal analysis threshold of $T_{\rm
eff}=5.0$~MeV, to ensure that we had enough background statistics to make a
meaningful measurement.  Although the $Q$ values of many of the radioactive
decays we are studying are below even this lower threshold, the broad energy
resolution of the detector leads to a substantial number that reconstruct 
above threshold.  For more detail, see Refs.~\cite{the:chen,the:mcgregor}.

The \tltze\ decay  has a $Q$ value of $\sim$5.0~MeV,
and the \bitof\ decay a $Q$ value of 3.27~MeV.
Almost every \tltze\ decay emits a 2.614~MeV $\gamma$, one or more low energy
$\gamma$s and a $\beta$ with an endpoint of $\sim$1--1.8~MeV, whereas there is a
unique branch in the \bitof\ decay to the ground state of $^{214}$Pb that
produces a single $\beta$ with an endpoint energy of 3.27~MeV.  Above an
analysis threshold of $T_{\rm eff} \sim 3.8$~MeV, the detected events from
$^{214}$Bi decays are dominated by the 3.27~MeV endpoint $\beta$-decay electrons, while
those from $^{208}$Tl decays may have multiple energetic electrons produced by
Compton scattering as well as $\beta$-decay. The $^{214}$Bi decays will
therefore have a PMT hit pattern resembling that of a single electron, while
$^{208}$Tl decays appear more isotropic.

The different hit patterns of $^{214}$Bi and $^{208}$Tl events allowed us to use the
distribution of event `isotropy' (characterized by the mean angle between PMT pairs,
$\theta_{ij}$) to separate the \tltze\  and  \bitof\ decays statistically.  (The
energy spectra from these decays are too similar above 3.8~MeV to allow separation
using pdfs in event energy).  The parameter \tij~is calculated by taking the average
angle relative to the reconstructed event vertex for all hit PMT pairs within a
prompt light time window in an event.  It is the same variable as was used as one of
our `Cherenkov box' cuts, as discussed in Sections~\ref{sec:inst}
and~\ref{sec:contam}.

\begin{figure}[h]
\begin{center}
   \includegraphics[height=0.3\textheight]{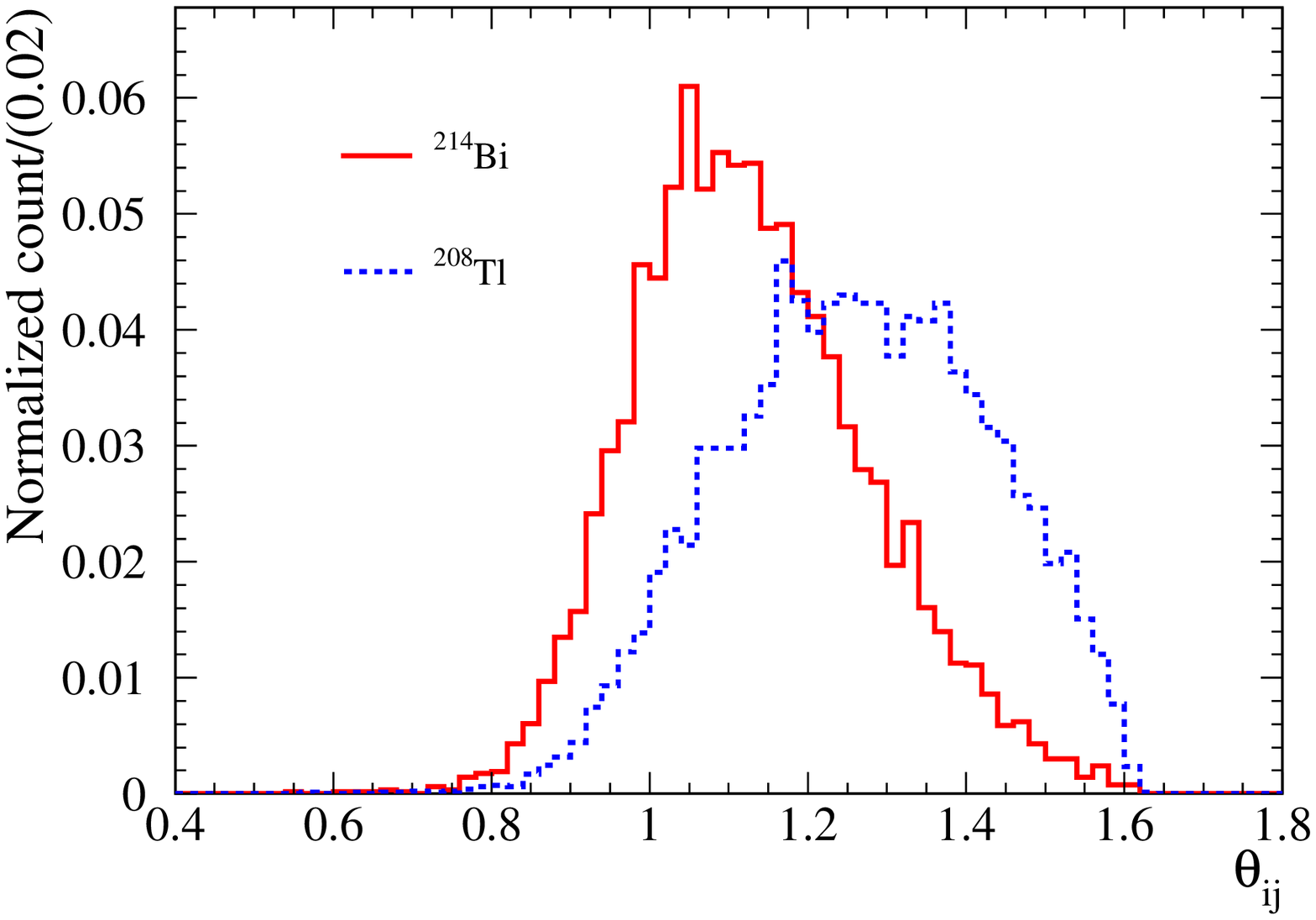}
   \caption[Difference in event light isotropy  between \tltze\ and \bitof\ decays]{Monte Carlo-predicted event isotropy distributions for \tltze\ and \bitof\ decays.  The isotropy parameter shown here on the abscissa, \tij, is the average opening angle between all fired photomultiplier tube pairs centered at the reconstructed event vertex.  More isotropic light distribution in an event results in a higher \tij\ value. \label{fig:thetaij}}
\end{center}
\end{figure}

Figure~\ref{fig:thetaij} shows the Monte Carlo model's prediction of the difference in the distribution of \tij\ between \tltze\
and \bitof\ decays.  
Statistical separation of the \tltze\ and \bitof\ events is obtained
by a maximum likelihood fit to the \tij\ distribution of the Cherenkov events.

The {\it in-situ} analysis of background radioactivity in the D$_2$O has its
own background---events from the H$_2$O region which 
misreconstruct into the D$_2$O volume and therefore look like D$_2$O radioactivity. To
avoid this `background to the background', the {\it in-situ} analysis was done
using a smaller fiducial volume ($R_{\rm fit}<450$~cm) than the 550~cm fiducial
volume used for the neutrino analysis.  Ultimately, the concentrations of
radioactivity determined from the {\it in-situ} analysis are scaled to the full 
volume.  The second background in the {\it in-situ} analysis is the neutrino
events themselves, and to avoid these a narrow `monitoring' window in energy
is chosen, $N_{\rm eff}=33-36$, which corresponds to the energy range $T_{\rm eff}\approx
3.8$~MeV to $T_{\rm eff} \approx 4.2$~MeV (see Section~\ref{sec:ecalibrator} for a
discussion of the relationship between $N_{eff}$ and energy).

In the {\it in-situ} analysis of the \hto\ background, a slightly different
energy window was used, from $T_{\rm eff} \sim$4.0~MeV to $T_{\rm eff} \sim
4.5$~MeV.  The higher energy window was used because of increased
contamination from other background sources (e.g., $\beta-\gamma$ decays from
the PMT array).  The fiducial volume for the H$_2$O analysis was chosen to be
far from the acrylic vessel and D$_2$O volume but well within the angular
acceptance of the PMTs and light concentrators, 650~cm$<R<$680~cm.  
Selection of events with an outward-going reconstructed direction
further reduced contamination.

From the {\it in-situ} analysis the equivalent $^{232}$Th and  
$^{238}$U concentrations in the \dto\ are: 
\begin{center}
\begin{tabular}{ll}
     $^{232}$Th:  &  1.34 $\pm$ 0.62 $^{+0.33}_{-0.38}$ $\times$ 10$^{-15}$ g Th/g D$_2$O \\
     $^{238}$U:  &  17.8 $\pm$ 1.4 $^{+3.2}_{-4.1}$ $\times$ 10$^{-15}$ g U/g D$_2$O.
\end{tabular}
\end{center}
where the first uncertainty is statistical and the second systematic.

The dominant systematic uncertainties in the {\it in-situ} study of \dto\
radioactivity are in the energy scale and in the 
\tij\ pdfs.  As is true for the neutrino pdfs, these \tij\ pdfs 
were derived from the Monte Carlo simulation. We verified their shapes
by comparing them to the distributions obtained during periods of 
Rn ingress into the target volume.  

Temporal variation of the detector energy scale was modeled to study its effect
on the extracted $^{232}$Th and $^{238}$U concentrations.  We have included in
the systematic uncertainties those due to contamination from other background
sources in the monitoring window.  In addition, we have included uncertainties
due to potential non-uniformities of the spatial distributions, and thus the
numbers represent the estimate of the total radioactivity in the D$_2$O, 
not just that within $R<450$~cm.

Similarly, the equivalent  $^{232}$Th and  $^{238}$U concentrations in the \hto\  determined from the {\it in-situ} analysis are:
\begin{center}
\begin{tabular}{ll}
     $^{232}$Th:  &  14.2 $\pm$ 0.6 $\pm$ 6.6 $\times$ 10$^{-14}$ g Th/g H$_2$O \\
     $^{238}$U:  &  75.5 $\pm$ 1.2 $\pm$ 32.9 $\times$ 10$^{-14}$ g U/g H$_2$O.
\end{tabular}
\end{center}
where again the statistical uncertainty is listed first.

The systematic uncertainties in the \hto\ analysis are considerably larger than
those in the \dto\ analysis,  with the largest component 
in the \hto\ analysis being the contribution from the energy scale
uncertainty, whose magnitude is 42\% of the measured $^{232}$Th and $^{238}$U
concentrations.  The large uncertainty is due, in part, to the fact that 
the optics of the outer regions of the detector are difficult to model (particularly
the optics of the PMT and concentrator assembly), and that we calibrated these outer
regions less frequently than the inner fiducial region.

\subsubsection{Overall $^{232}$Th and $^{238}$U Concentration Determined for the Water}

The {\it in-situ} and {\it ex-situ} techniques are independent,
and their systematic uncertainties have been independently
assessed.   Fig.~\ref{fig:inex_cmp} shows good agreement 
between {\it ex-situ} ($^{232}$Th: MnOx, $^{238}$U: Rn assay) and 
{\it in-situ} measurements.  
For the $^{232}$Th chain, we have therefore used the weighted mean of the
results, including additional uncertainties associated with the {\it ex-situ}
sampling.
The $^{238}$U chain activity is dominated by Rn ingress, which
is highly time dependent, and we have therefore used the {\textit{in-situ}} 
determination for this activity as it includes the appropriate weighting of
neutrino live time.  For the present data set, we find the  equivalent equilibrium
$^{238}$U and $^{232}$Th concentrations in the \dto\ to be
\begin{eqnarray}
^{232}\mbox{Th} &:& 1.61 \pm 0.58\times 10^{-15} \mbox{g Th/g \dto} \nonumber \\
^{238}\mbox{U} &:& 17.8 ^{+3.5}_{-4.3} \times 10^{-15} \mbox{ g U/g \dto},  \nonumber
\end{eqnarray} 
where we have added the statistical and systematic uncertainties in quadrature.  The concentrations in the \hto\ are
\begin{eqnarray}
^{232}\mbox{Th} &:&  9.1 \pm 2.7 \times 10^{-14}\mbox{g Th/g \hto} \nonumber \\
^{238}\mbox{U} &:& 75.5 \pm 33.0 \times 10^{-14} \mbox{ g U/g \hto}.  \nonumber
\end{eqnarray}

\subsubsection{Acrylic Vessel Radioactivity}
\label{sec:av_pd}

To determine the photodisintegration background due to radioactivity
in the walls of the acrylic vessel, we first need to establish
the vessel's radioactivity load.  It is difficult to apply the {\it in-situ}
technique, primarily because the vessel is very clean and its Cherenkov
signals are masked by the dominant H$_2$O background.  The approach here is
therefore to first determine the radioactivity load of the acrylic vessel from
radioassay results, and then to use Monte Carlo simulations to deduce the
photodisintegration background.  We discuss below contributions to the
radioactivity from the acrylic vessel panels and bonds, from surface activity
caused by mine dust, and from a `hot spot' of unknown origin.

During the production of the AV panels, acrylic samples were analyzed for internal $^{232}$Th and
$^{238}$U radioactivities by neutron activation analysis.  The $^{232}$Th
concentration in the thermoformed acrylic panels was found to be
0.25$\pm$0.04~ppt  $^{232}$Th.

Additional radioactivity was presumably introduced 
during bonding of the acrylic panels, possibly from the glue, environmental 
dust, or plating of radioactive isotopes.  It is difficult to determine this
background, as dust might be embedded in the bond during the construction.  
The surface area and volume of the bonds are much smaller than those of the
vessel as a whole, and therefore we estimate an uncertainty of
$^{+1}_{-0}$~$\mu$g  $^{232}$Th ($\sim$ total amount of Th from dust on the
inner surface of the vessel) due to embedded dust in the bonds.  Adding this in
quadrature to the uncertainty of the Th concentration in the thermoformed
panels, we estimate 7.5$^{+1.7}_{-1.3}$~$\mu$g of $^{232}$Th for the full vessel.
This represents an expected 6.2$^{+1.4}_{-1.1}$ detected photodisintegration
neutrons in the full Phase I data set.

Since the U contribution to the backgrounds is less than $^{232}$Th for a given
concentration and the U-to-Th ratio in materials is normally less than 1, the
$^{238}$U concentration in the vessel did not pose as significant a problem
as $^{232}$Th.  Neutron activation of virgin acrylic samples gave 2$\sigma$
upper limits ranging from 0.1~ppt to 1~ppt $^{238}$U.  We therefore estimate
0.5$\pm$0.5 ppt $^{238}$U as the total contamination, under the
assumptions that the thermoforming process introduced the same amount of
$^{238}$U into the panels as Th ($\sim$0.2~ppt) and that the embedded dust in
the bonds has the same U-to-Th ratio as mine dust.  This translates to
15$\pm$15~$\mu$g $^{238}$U in the vessel.

During its construction and after final cleaning, the areal density of $^{232}$Th deposited on the
surface of the acrylic vessel was determined from X-ray fluorescence (XRF)
analysis  of dust samples lifted off the vessel's surface by adhesive tapes.
The amount of $^{232}$Th on the acrylic vessel determined from the XRF
analysis after its final cleaning was found to be:
\begin{center}
\begin{tabular}{rc}
     Inner AV surface:  &  0.87 $\pm$ 0.17 $\mu$g $^{232}$Th \\ 
     Outer AV surface:  &   0.96 $\pm$ 0.19 $\mu$g $^{232}$Th. 
\end{tabular}
\end{center}
The $^{238}$U load could not be determined directly from the dust sample because of the limited sensitivity of the XRF.  The dust sample was assumed to have the same composition as mine dust---a $^{238}$U/$^{232}$Th ratio of 0.187$\pm$0.024.  The amount of $^{238}$U on the acrylic vessel is then:
\begin{center}
\begin{tabular}{rc}
     Inner AV surface:  &  0.16 $\pm$ 0.04 $\mu$g $^{238}$U \\ 
     Outer AV surface:  &   0.18 $\pm$ 0.04 $\mu$g $^{238}$U. 
\end{tabular}
\end{center}

As discussed in Section~\ref{sec:fvolecut} and shown in Fig.~\ref{fig:zvx},
an anomalous `hot spot', which appears to be radioactivity embedded in the acrylic
vessel, was identified during analysis of Cherenkov events near the edge of the
fiducial volume.
We derived an estimate of the radioactivity level of the hot spot using
data from low energy calibration sources (for example, $^{232}$Th embedded
within acrylic) as well as extensive Monte Carlo simulations that included
variations of the optical properties of the vessel.
Based on these analyses, we find that under the hypothesis that the
radioactivity $m_{hs}$ of the hot spot is all Th (the worst case), its level
is:
\begin{displaymath}
  m_{hs} \;=\; 10\,\pm\,1\,\mbox{(stat.)}\,^{+8.5}_{-3.5}\,\mbox{(sys.)}\,\mu\,\mbox{g Th equivalent}.
\end{displaymath}
We assumed in these analyses that the hot spot was located on the outer
surface of the acrylic vessel.  The dominant systematic uncertainty 
was the uncertainty of the energy scale at the acrylic vessel.
Because of the complicated light propagation in the
acrylic vessel, the systematic uncertainty associated with the energy
response was estimated at $\sim$30\%.  
Although we assumed the hot spot was comprised solely of Th-chain
radioactivity, our studies indicated that there are compensating effects
among Q values, detector efficiencies, neutron propagation, and
photo-disintegration rates for $\gamma$s that cause the estimated
photodisintegration neutron rate to be relatively constant regardless of the
relative $^{238}$U and $^{232}$Th composition.

\subsection{Determining the Total Photodisintegration Background \label{sec:pdtot}}

Monte Carlo calculations were performed to determine the equivalent
$^{232}$Th and $^{238}$U quantities in different detector regions that would
produce one photodisintegration neutron in the \dto~per day.
Table~\ref{tbl:pdconv} summarizes these results.

\begin{table}
\caption{Equivalent $^{232}$Th and $^{238}$U masses that each produce a photodisintegration neutron in the D$_2$O target per day.  Radioactivities are assumed to be in secular equilibrium and evenly distributed in the respective detector regions.  The uncertainties shown here are statistical.  Systematic uncertainties are dominated by the uncertainty in the cross section, which is $\sim$1\%.}
\protect\label{tbl:pdconv}
\begin{center}
\begin{tabular}{lcc}\hline \hline
 & $^{232}$Th & $^{238}$U  \\
 & ($\mu$g) & ($\mu$g) \\ \hline
\dto & 3.79$\pm$0.01& 29.8$\pm$0.76 \\ 
AV & 10.83$\pm$0.04 & 82.92$\pm$1.75 \\ 
\hto & 278.3$\pm$5.2 & 2325$\pm$111 \\ \hline \hline
\end{tabular}
\end{center}
\end{table}

Using the $^{232}$Th and $^{238}$U concentrations, the equivalent masses in
Table~\ref{tbl:pdconv}, and the neutron detection efficiency for $T_{\rm eff}>5$~MeV
and a fiducial volume of $R_{\rm fit}<550$~cm, we find the expected total number of
detected photodisintegration neutrons arising from internal radioactivities
from different detector components to be 71.3$^{+11.6}_{-11.9}$~counts, which
is 12\% of the expected neutral-current signal.

Contributions from different regions are summarized in
Table~\ref{tbl:pd_summary}.  Calibrated neutron detection efficiencies
(see Section~\ref{sec:neutrons}) were used in these calculations.
\begin{table}
\caption[Summary of photodisintegration neutron background ($T_e>$5~MeV) in the fiducial volume ($R_{\rm fit}<$550~cm.) for Phase I of the experiment]{Summary of the estimated number detected photodisintegration neutrons ($T_{\rm eff}>5$~MeV) in the fiducial volume ($R_{\rm fit}<$550~cm.) for Phase I of the experiment}
\protect\label{tbl:pd_summary}
\begin{center}
\begin{tabular}{cccc} \\ \hline \hline 
               &   $^{232}$Th   &  $^{238}$U &  Total  \\
               &   (counts)    &   (counts) &  \\ \hline 
D$_2$O    &   18.4$\pm$6.5 &   25.9$^{+5.0}_{-6.3}$ & 44.3$^{+8.2}_{-9.1}$ \\ 
AV     &   14.2$^{+5.8}_{-6.6}$ &  1.6$\pm$1.6  &  15.8$^{+6.0}_{-6.8}$ \\ 
H$_2$O    &  5.6$^{+3.6}_{-2.2}$ & 5.6$^{+4.2}_{-2.9}$ & 11.2$^{+5.5}_{-3.6}$ \\ \hline
 Total  &  38.2$^{+9.4}_{-9.5}$ & 33.1$^{+6.7}_{-7.1}$  &  71.3$^{+11.6}_{-11.9}$ \\ \hline \hline  
\end{tabular}
\end{center}
\end{table}

\subsubsection{Other Possible Sources of Photodisintegration}

In the following we briefly discuss other possible sources of photodisintegration. 
\begin{enumerate}

\item {\em $\beta-\gamma$ from the PMTs and PSUP structure} \\ 
For U and Th decays in the PMT/PSUP region to photodisintegrate a
deuteron, the $\gamma$s emitted must travel a very long
distance ($>$10 attenuation lengths).  A Monte Carlo study was performed
to estimate the photodisintegration background due to these decays in
the PMT/PSUP region.  Based upon this study, we estimate an upper limit of
0.009 neutron captures per day in the fiducial volume, corresponding to 
$<$1.4 neutrons detected for the full Phase I data set.

\item {\em Outer H$_2$O $\beta-\gamma$} \\ Radioassay results demonstrate
that the  H$_2$O outside the photomultiplier tube support structure 
has an average $^{238}$U concentration very similar to that in the
inner H$_2$O (i.e. between the acrylic vessel and the PSUP).  Because of
the large radial attenuation of neutrons produced in the outer region of the
detector, we concluded that contributions to the total photodisintegration
background from the cavity H$_2$O are negligible.

\item {\em Sources other than Th/U}  \\ An extensive literature search was made
for long-lived isotopes with high energy $\gamma$ decays that could be present
in the heavy water.  The only possibilities found were those that could have
been produced had the water been used in a reactor.  As the SNO heavy water was
never used this way, there are no isotopes known to us other than $^{232}$Th
and $^{238}$U that are capable of producing photodisintegrating $\gamma$-rays.

\end{enumerate}

\subsection{Low Energy $\beta$-$\gamma$ Backgrounds}
\label{sec:certail}
	
	The number of events originating within the D$_2$O volume that appear
above threshold is kept small primarily by ensuring that the radioactivity
levels in the heavy water are low.  In addition to the neutrons produced
through photodisintegration, the primary particles from decays of U and Th
daughters (low energy $\gamma$s and $\beta$s) can also lead to events in the
final data sample.  Although nearly all decays in these chains have Q values
lower than the $T_{\rm eff} = 5.0$~MeV analysis threshold, the broad energy
resolution of the detector at low energies allows a small fraction of these
decays to appear above threshold.  We refer to backgrounds in the D$_2$O as
$\beta$-$\gamma$ backgrounds to distinguish them from the neutron backgrounds
described above.

Outside the heavy water volume, however, the acrylic vessel, the light water,
and in particular the PMT array and support structure have relatively high
levels of radioactivity.  The vast majority of these events (as well as of high
energy $\gamma$-rays coming from the cavity walls) are
removed by the 550~cm fiducial volume cut (see Section~\ref{sec:fvolecut}).

	We therefore have two distinct approaches to these two classes of
backgrounds: for events originating within the heavy water the dominant issue
is how well we understand the energy response of the detector, while for events
originating outside we must know the reconstruction accuracy well.

\subsubsection{Internal to D$_2$O Volume \label{sec:lowint}}

	As described in Sections~\ref{sec:model} and~\ref{sec:sysunc},
the Monte Carlo model is well calibrated within the fiducial volume,
reproducing the measured energy spectra of different sources over a
range of energies covering nearly the entire solar neutrino energy regime.
With the exception of the energy scale itself, the model parameters were
derived independently from the calibration sources---thus the successful
simulation of the source data is the result of the physical basis of the
model itself.  We therefore can reasonably expect that the model will
accurately simulate the characteristics of other radioactive decays that
differ only in the physical particles they produce.

	Our approach to estimation of these low energy $\beta$-$\gamma$
backgrounds was to simulate Tl- and Bi- chain decays for
each run in the SNO data set, and to apply the analysis chain described
in Section~\ref{sec:dataproc} to these simulated data.	To minimize
uncertainties associated with analysis efficiencies, we do not use
the Monte Carlo to make an absolute prediction of the number of events above
threshold, but to predict the ratio of the number of detected $\beta$-$\gamma$
events to the number of detected photodisintegration neutrons.  This ratio is
then normalized by using the predictions for the number of photodisintegration
neutrons from the {\it ex-situ} radioassay and {\it in-situ} Cherenkov analyses
described in Sections~\ref{sec:exsitu} and~\ref{sec:dto_insitu}.

Based upon the Monte Carlo simulation, the energy spectra 
for $\beta$-$\gamma$ events are well represented by simple
Gaussians in the energy range $4.5<E<6.5$~MeV.  The mean ($\mu$)
and width ($\sigma$) from these fits are 2.019 MeV and 0.8773 MeV for $^{208}$Tl decays,
and 2.588 MeV and 0.7828 MeV for $^{214}$Bi decays.  

	To determine the systematic uncertainties on the ratio
of the numbers of $\beta$-$\gamma$ events to photodisintegration neutrons,
we began with the uncertainties on the Monte Carlo model described
in Section~\ref{sec:sysunc} and on all applied cuts
(described later in Sections~\ref{sec:insteff} and~\ref{sec:hlcuteff}).
We then created 10000 `hypothetical' SNO experiments whose energy scale,
resolution, vertex accuracy, etc., were slightly different from the baseline
Monte Carlo prediction by amounts consistent with the measured
uncertainties on each of those quantities.  For each hypothetical experiment
the ratio of $\beta$-$\gamma$ events to photodisintegration neutrons was calculated
for each decay chain, and a 1$\sigma$ confidence interval was determined from
the distribution of the ratio over the 10000 trials.

With $T_{\rm eff} > $5.0~MeV and $R_{\rm fit}<$550~cm, the ratios
between the numbers of detected $\beta$-$\gamma$ events to 
detected photodisintegration neutrons are:
\begin{center}
\begin{tabular}{ll}
     $^{208}$Tl:  &  0.162  $^{+0.092}_{-0.030}$  \\
     $^{214}$Bi:  & 0.670  $^{+0.460}_{-0.125}$.
\end{tabular}
\end{center}

Given the estimated numbers of detected photodisintegration neutrons (Th: 18.4$\pm$6.5, U: 25.9$^{+5.0}_{-6.3}$, see Section~\ref{sec:pdtot}), the expected numbers of 
$\beta$-$\gamma$ events from these decays in the final data set are:
\begin{center}
\begin{tabular}{ll}
     $^{232}$Th:  &  3.0 $^{+2.0}_{-1.3}$~counts  \\
     $^{238}$U:  & 17.4  $^{+12.4}_{-5.3}$~counts \\
    Total: & 20.4 $^{+12.6}_{-5.5}$~counts.
\end{tabular}
\end{center}

As a test of this method, we used data taken during two periods in which the radon
levels in the detector were 1-2 orders of magnitude higher than their nominal
levels.  As can be seen in Fig.~\ref{fig:inex_cmp}, the first of these 
`high radon' periods occurred near the start of data taking, while the initial radon
load was decaying away, and the second period occurred roughly 90\% through the
run, when a pump failed and allowed radon to enter the D$_2$O volume.  Using the
method described above, we predicted the excess number of events as a function of 
energy during these periods and found good agreement with the data.

We also compared the Monte Carlo predictions and uncertainties 
to data taken with shielded low energy Th sources.   The shield
was intended to allow only $\gamma$-rays from the source to be seen by
the detector, so that uncertainties associated with the optics of 
$\beta$-originated Cherenkov light within the source itself could be ignored.

\begin{figure}
\centering
\includegraphics[height=0.30\textheight]{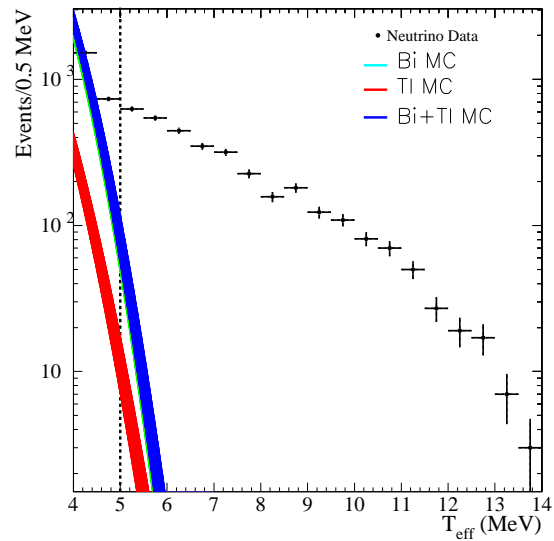}
\caption[Comparison of Monte Carlo prediction with systematic uncertainties to the the total D$_2$O neutrino data]{Comparison of Monte Carlo predictions of
$\beta$-$\gamma$ background energy spectrum within the D$_2$O to the total 
neutrino data set. The curves are not fit to the data, they are normalized by
the {\it in-situ} and {\it ex-situ} estimates and simply overlaid on the
neutrino energy spectrum.\protect\label{fig:d2o_tail_nu_sup}}
\end{figure}

Figure~\ref{fig:d2o_tail_nu_sup} shows the final estimate for the 
number of $\beta$-$\gamma$ decays that make it into the
final neutrino data set.  The curves shown are not a fit to the data set---they
are normalized by the {\it in-situ} and {\it ex-situ} background estimates and
simply overlaid on the data. The widths of the bands indicate the 
uncertainties on the estimates.

\subsubsection{External to D$_2$O Volume \label{sec:lowext}}

	Radioactive decays within the acrylic vessel itself, the light water
region, and the photomultipliers and associated support structure can also
produce events above the analysis energy threshold and within the fiducial
volume.  Events leak into the fiducial volume in two ways: $\gamma$-rays 
can travel unscattered from their external origin into the fiducial volume, and
events occurring outside the volume may be reconstructed
incorrectly inside.  Although the probability of such leakage
is very small, and the probability that such events will be above the analysis
energy threshold is also very small, the radioactivity levels outside the heavy
water volume are significantly higher than inside, and the leakage can
therefore be a non-negligible background to the neutrino signal.  

	For these backgrounds, neither the Monte Carlo nor analytic
models are likely to be good representations of the detector response,
for several reasons.  First is that the detector is
not nearly as well calibrated outside the fiducial volume as inside:
the optical and primary energy calibration sources can be deployed in a
much more limited number of places outside the heavy water than inside.
In addition, there is greater optical complexity in the outer regions
of the detector---the PMT angular response at high incidence needs
to be understood, the optical shadowing of the photocathodes by the
light concentrators becomes important, and the PMT-to-PMT variations in
efficiency are amplified as one gets nearer a particular area of the PMT
support structure.  Lastly, event leakage from this region into the inner
volume is caused by highly unusual circumstances, and the leakage fraction may
therefore be sensitive to detector artifacts such as electronic crosstalk,
miscalibrated PMT timing response, or coincidences between instrumental and
radioactive backgrounds.

	We therefore based the analysis of these `external' backgrounds primarily
on Th and U calibration source data, using the source data to create
radial profiles (pdfs in $R^3$) of the backgrounds and fitting these
profiles to the neutrino data.  To determine that the calibration
sources' $R^3$ profiles were reasonably insensitive to the specifics of the
source type and geometry, we compared the profile obtained using U calibration
source data to that obtained using Th calibration source data, and also
compared these radial profiles to those obtained with a set of shielded U and
Th calibration sources.  The shield blocked Cherenkov light created by the
$\beta$-decay in the sources' acrylic encapsulation.

To build the pdfs in $R^3$, we used data taken with the acrylic-encapsulated 
U and Th sources at many discrete locations within the H$_2$O.  
To create pdfs appropriate for the uniform distribution of radioactivity
expected in the neutrino data, we then weighted the source data by run time, 
and by volume by taking equal volumes around the source position.
Since the sources were untriggered, we subtracted the neutrino (and intrinsic
background) signals accumulated during the source run,  as well as 
photodisintegration neutrons (from $\gamma$s entering the heavy water region).

\begin{figure}
\centering
\includegraphics[height=0.3\textheight]{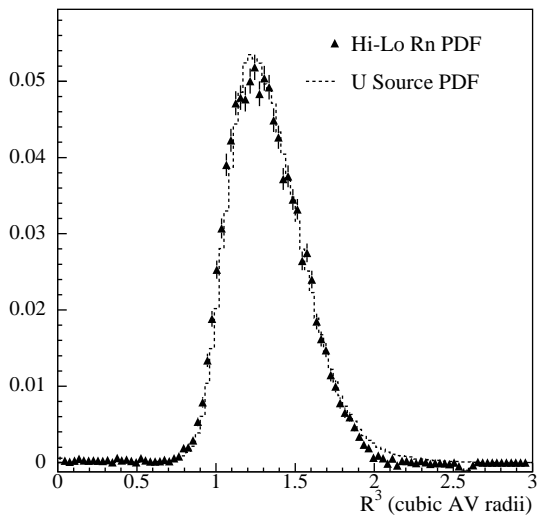}
\caption[Comparison between the PDF obtained from the high-low radon study and the H$_2$O PDF derived from encapsulated source data.]{Comparison between the PDF obtained from the high-low radon study and the H$_2$O pdf derived from acrylic source data. A value of $R^3=1$ corresponds to the radius of the AV. \protect\label{fig:hilopdfcomparison}}
\end{figure}

Figure~\ref{fig:hilopdfcomparison} shows the $R^3$ pdf
derived from U source data compared to a pdf created using neutrino data taken
during a period of high radon levels in the light water region.
We can see that the two agree well, despite the fact that one
is built from calibration data taken at discrete locations and the other 
from a distributed  source of Rn.  In Fig.~\ref{fig:hilopdfcomparison} and
in other $R^3$ distributions, we measure $R^3$ in units of cubic AV radii: $R^3
\equiv (R_{\rm fit}/R_{AV})^3$.  At the 600~cm AV radius, $R^3=1$.

To determine the relative contributions of each of the three sources of
background events (acrylic,
H$_2$O, and PSUP), we fit a linear combination of the three $R^3$ pdfs to 
the $R^3$ distribution of the data.
One problem with this approach is the lack of sufficient statistics in the
pdfs at the neutrino analysis threshold of $T_{\rm eff}$=5~MeV---even with very
hot calibration sources, it is difficult to get events above the
analysis threshold. To overcome this
problem, we performed the fit exclusively within the H$_2$O region, where
these backgrounds are highest (the radial range $1.02< R^3 <2.31$)
and with an energy selection of $T_{\rm eff}>$4~MeV.  The fit amplitudes for 
the three background components were then scaled to the intended kinetic
energy threshold of 5~MeV.  
The basic assumptions in this analysis are that there is no correlation
between $R^3$ and energy, and that the reconstruction does not get worse with
higher energy.  Several studies were performed to determine how various
effects (e.g. pile-up, crosstalk, or high noise rates) can affect
reconstruction.  None of the studied effects 
cause a significant effect in the reconstruction---the probability of
misreconstruction generally increases rather than decreases as energy is lowered.

\begin{figure}
\centering
\includegraphics[height=0.40\textheight]{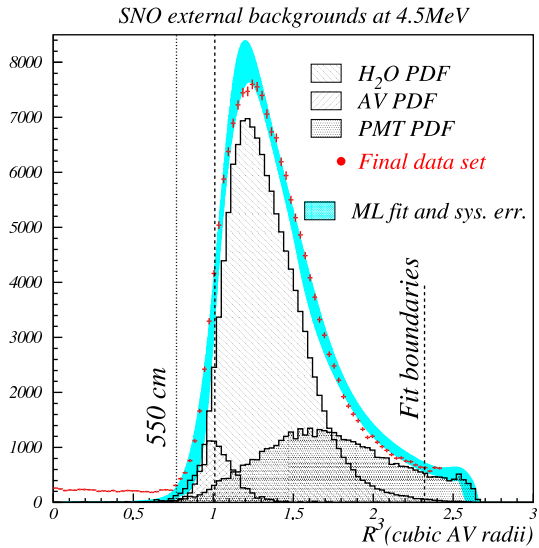}
\caption[$R^3$ fit at 4.5~MeV]{Fit of of $R^3$ pdfs created using calibration
source data to the neutrino data set, using an energy threshold of 
$T_{\rm eff}>4.0$~MeV.  The extended maximum likelihood method was used in the fit, and the band represents the systematic uncertainties. The $y$-axis is in units of Events/0.03 cubic AV radius.} \protect\label{fig:m}
\end{figure}

Figure~\ref{fig:m} shows the results of this $R^3$ fit.  The band shown in this
plot is the range of the systematic uncertainties.  For such a plot, we do not
necessarily expect the data points to be centered within the band, because the
systematic uncertainties are not normally
distributed.  Some of the sources of systematic uncertainty in this analysis
are similar to those for the neutrino analysis described in
Section~\ref{sec:sysunc}, such as vertex accuracy and energy scale.  In
addition, we evaluated uncertainties associated with the difference between the
different radioactive sources (U vs. Th) and the sensitivity of the
fit to the radial window chosen.  The overall 
uncertainties for the three sources of external $\beta$-$\gamma$ events are
$^{+31.7\%}_{-91.3\%}$ for events whose source was the acrylic vessel,
$^{+29.6\%}_{-9.1\%}$ for events whose source was the light water region, and
$^{+44.2\%}_{-11.1\%}$ for events from the PMT array.

Although the pdfs which were used in the fits above were binned, analytic
forms are used later in the data analysis for the extraction of the
CC, ES, and NC neutrino signals (see Section~\ref{sec:sigex}).
The analytic form of the $R^3$ profile for the PMT $\beta$-$\gamma$
backgrounds is an exponential (with $R^3$ in units of cubic AV radii): 
\begin{displaymath}
    f(R^3) \;=\;  \exp(-4.538+7.131 R^3) \,+\, 1.631.
\end{displaymath}
while the energy spectrum ($T_{\rm eff}>$4~MeV) of the PMT $\beta$-$\gamma$ background
can be approximated by a Gaussian distribution with $\mu$=1.416~MeV and
$\sigma$=0.960~MeV.  For the AV $\beta-\gamma$s, the $R^3$ distribution
can be approximated by a Gaussian with $\mu$=1.056 and $\sigma$=0.1267.
The energy spectrum ($T_{\rm eff}>$4~MeV) of the AV $\beta-\gamma$ background
can be approximated by a truncated Gaussian distribution with $\mu$=3.441~MeV and
$\sigma$=0.4617~MeV.

A number of consistency checks were done to ensure the validity of the
results.  A separate estimate of the external $\beta$-$\gamma$ tail background  was made
by calculating the ratio of the count rate within a monitoring window
around the acrylic-encapsulated calibration source ($\Omega_1$) to that in a window
within the fiducial volume ($\Omega_2$, $R_{\rm fit}<$550~cm).   By counting the number of
events in $\Omega_1$ during the neutrino runs, the $\beta$-$\gamma$ tail contribution in
$\Omega_2$ can then be estimated by scaling.  The scaled rate was found to be
consistent with the results obtained from the $R^3$ fit.  In addition, one can use
the Monte Carlo to predict, based on the radioactivity level of  each 
detector region, the number of tail events above the analysis threshold.  These
Monte Carlo results were also found to be in agreement with the results described
above.

Table~\ref{tbl:ctail_summary} is a summary of all estimated $\beta$-$\gamma$ 
backgrounds with $T_{\rm eff}>$5~MeV which reconstruct within the fiducial volume
($R_{\rm fit}<$550~cm.).  

\begin{table}
\caption[Summary of $\beta$-$\gamma$ background ($T_{\rm eff}>5$~MeV) in the fiducial volume ($R_{\rm fit}<550$~cm.) ]{Summary of $\beta$-$\gamma$ background ($T_{\rm eff}>5$~MeV) in the fiducial volume ($R_{\rm fit}<550$~cm.) for the Phase I data set.}
\protect\label{tbl:ctail_summary}
\begin{center}
\begin{tabular}{lr} \\ \hline \hline  
     Source    &  ~~~~Background Events \\ \hline
D$_2$O    & 20.4$^{+12.6}_{-5.5}$ \\  
AV      & 6.3$^{+2.9}_{-6.3}$ \\ 
H$_2$O   &2.8$^{+3.9}_{-2.8}$ \\ 
PMT   & 16.0$^{+10.5}_{-7.2}$ \\ \hline 
Total  & 45.5$^{+17.1}_{-11.4}$ \\ \hline \hline
\end{tabular}
\end{center}
\end{table}

\subsection{High Energy Gammas \label{sec:hegs}}

	In the ES-CC paper~\cite{snocc}, the high threshold of $T_{\rm
eff}>6.75$~MeV ensured that the number of $\beta$-$\gamma$ background events
from U and Th decays was negligibly small.  Thus the background estimates
discussed above in Sections~\ref{sec:certail} were not used.  There are,
however, sources of high energy $\gamma$ rays (HEGs) that get well above
2.6~MeV and could in principle wind up inside the fiducial volume and above
even the high 6.75~MeV threshold.  Thermal neutrons produced in $(\alpha,n)$
reactions from U and Th $\alpha$ emission can be captured on high density
materials such as the steel of the PSUP or the
cavity rock, and these captures can lead to HEGs. A second possible source is
direct $\gamma$ production through $(\alpha,p\gamma)$ reactions on light nuclei
such as Al in the PMT glass or concentrators.

	To estimate the number of these events above the $T_{\rm eff}=6.75$~MeV
threshold and inside 550~cm, we used a deployment of the $^{16}$N source out
near the PSUP and counted the number of events that made it into the fiducial
volume.  The count was normalized by the number of events in a small radial bin
just inward of the source location.  A count of the number of events in the
same radial bin was made for the neutrino data set, and was multiplied by the
ratio determined with the $^{16}$N source data. This `radial box' method
yielded a background estimate for HEGs inside the fiducial volume and above the
analysis energy threshold of $<0.8$\%.  For more details of this method, see
Ref.~\cite{the:msn}.

	For the NC~\cite{snonc} results, we found that the fit in $R^3$ 
for the external $\beta$-$\gamma$ background described above in
Section~\ref{sec:lowext} already accounted for the HEG background through the
pdf for the PMT $\beta$-$\gamma$s. We thus did
not include an independent estimate of these events for the NC paper, and the
third line of Table~\ref{tbl:ctail_summary} should be taken to
include the contribution from these high energy $\gamma$ rays.
	
\subsection{Additional Sources of Neutrons \label{sec:otherneutrons}}

In addition to the photo-disintegration backgrounds discussed in
Section~\ref{sec:pd_bgd}, there are other possible sources of neutrons:

\begin{itemize}
\item Spontaneous fission of $^{238}$U or $^{252}$Cf; 
\item Neutrons from cosmogenic sources; 
\item Deuteron breakup from alpha reactions; 
\item Neutron production from ($\alpha$,n) reactions; 
\item Neutrons produced by terrestrial and reactor antineutrinos; 
\end{itemize}

These will be discussed in the following sections.

\subsubsection{Spontaneous Fission}

Neutron backgrounds may arise from spontaneous fission of $^{238}$U or $^{252}$Cf.
Such fission events have unique characteristics, such as low energy gamma production
and the presence of multiple neutrons.  Many of these events are therefore removed
through the burst cuts discussed in Section~\ref{sec:dataproc} and
Appendix~\ref{sec:apdxa}, but here we estimate an upper bound on the number
remaining in the data set.  

The spontaneous fission half-life of $^{238}$U is $8.2 \pm 0.1 \times 10^{15}$
y~\cite{Holden}, corresponding to a branching ratio of $5.45 \times 10^{-7}$.  The
contribution of the spontaneous fission of $^{238}$U to neutron backgrounds can be
based on the measured concentration of $^{226}$Ra, but such an inference relies on
the assumption that the uranium decay chain is in equilibrium above radon.
Alternatively, we can use {\it ex-situ} measurements of $^{238}$U from the HTiO
adsorbent technique to determine the contribution from uranium directly.
Measurements of $^{238}$U from HTiO radioassays indicate less than
$10^{-15}$g$^{238}$U/g in the D$_2$O volume.  At such concentrations, the
contribution of spontaneous fission to the neutron background is much less than 1
event.

Spontaneous fission of $^{252}$Cf, introduced through the deployment of the
encapsulated neutron calibration source, can also occur.  Based upon our leach tests
of the deployed source, we estimate that it contributed much less than 1 event to
the final data set.

\subsubsection{Cosmogenic Sources \label{sec:cosmons}}

	SNO's great depth reduces the number of cosmic rays passing through
the detector to an extremely low rate---roughly three through-going
muons per hour within the PSUP enclosure. Nevertheless, cosmic rays---
which include muons as well as a low rate of atmospheric neutrino
interactions within the detector volume---are a potential source
of backgrounds.  Cosmic ray interactions may produce both radioactive
nuclides and neutrons.  As Section~\ref{sec:follcuts}
describes, we used two cuts to remove these events.  The first
cut, intended to remove both spallation nuclei and `follower neutrons',
eliminated all events that occurred within 20 s of a tagged muon event.
The second, intended to remove neutrons produced by untagged muons and
atmospheric neutrinos, removed all events in a 250~ms window following
any event with more than 60 hit PMTs.

	After these cuts, there are still potential sources of residual background
events; we address each of these in turn.

\begin{itemize}
\item Followers of External Muons

	One potential source of neutrons is muons passing outside
the detector volume, through the light water shield between the acrylic vessel
and the PSUP, between the PSUP and the rock,
or within the rock.  These high energy neutrons are typically produced through
photonuclear interactions between the muon and nuclei in the H$_2$O and through
secondary neutron production from subsequent interactions of the products of
the above reactions.  The high energy neutrons can penetrate through the water
shield surrounding the detector and contribute to the NC background.

To determine contamination from neutron events that passed through the light water
shield, we looked for follower events inside the fiducial volume
subsequent to events triggering the outward-looking (OWL) 
PMT array.  We found that the number of these follower events
was consistent with expectations from accidental coincidences alone, and
therefore the external muons are not a significant source of background
in the final neutrino candidate sample.

We also estimated the number of neutrons
produced from muon interactions in the rock, using both analytical models
and explicit Monte Carlo simulations.  Our estimate places the total
neutron event rate from muon-rock interactions below 0.18 neutrons/year,
not including losses due to reconstruction efficiencies.

\item Followers of Internal Muons

	Neutrons created by muons passing through the detector's fiducial volume are
removed through the muon follower cuts described above and in
Section~\ref{sec:follcuts}.  The efficiency of the cut is extremely high, as only a
small fraction of (the already small number of) muons originating outside the
detector and making it to the fiducial volume will be below Cherenkov threshold and
thus undetected.  Extrapolating from the number of followers we measure for detected
muons, we find that the number remaining in the data set after the muon follower
cuts is negligible.  The one exception are the muons created in association with
atmospheric neutrino interactions inside the detector volume, which are discussed
next.

\item Atmospheric Neutrinos 

The interactions of atmospheric neutrinos can produce primary events
within the fiducial volume of the detector, and the products of these
events---either neutrons or spallation nuclei---can contaminate the final data set.
Only a small subset of atmospheric neutrino interactions can mimic authentic solar
neutrino events.  Among these are neutral current events that release one or more
neutrons, neutral current events where a photon is released from the de-excitation
of $^{16}$O, or low energy charged current reactions.  These low energy atmospheric
neutrino interactions are often associated with a burst of events in the detector,
and are thus removed by the time-correlated cuts described in Section~\ref{sec:inst}
and Appendix~\ref{sec:apdxa}.  To estimate the background from the events that
remain after the cuts, we made use of a combination of analytic calculations of the
rates of atmospheric neutrino interactions and a full Monte Carlo simulation of the
propagation of their secondaries within the SNO detector.

We estimated the flux of atmospheric neutrinos using the calculations of 
Agrawal {\em et al.}~\cite{Bartol} for North American latitudes during solar 
maximum, and considered energy ranges from 50~MeV to 10~GeV.   
We included neutrino oscillations
assuming the measured $\nu_\mu \rightarrow \nu_\tau$ parameters from the
Super-Kamiokande Collaboration~\cite{SKatm}
($\Delta m^2 = 3\times10^{-3} {\rm~eV}^2$ and $\sin^2(2\theta) = 1$),
and path lengths through the Earth are taken into account in the angular
distribution of the flux. For these parameters, approximately 67\% of the
$\nu_{\mu}$ events remain after oscillation.

To calculate the interaction rates, we used the formalism of
Llewellyn-Smith~\cite{Smith}.  Since SNO possesses an isoscalar target
and the neutral current process does not distinguish between neutrino flavors, the
ratio of neutral current neutron interactions with muon charged current interactions
is $\approx 0.54 $.  By knowing the relative efficiencies of the
two types of events, we can normalize to the observed partially contained muons in
the detector.  

To estimate these efficiencies, the showers of particles produced in neutrino
interactions were propagated through the detector using SNO's Monte
Carlo simulation. We propagate the muons and hadrons through the SNO Monte Carlo
using the FLUKA hadron propagation code~\cite{fluka}. 

The systematic uncertainties associated with the atmospheric neutrino event rate in
SNO come mainly from uncertainties associated with the primary neutrino flux and
from the nuclear final state interactions.  Other errors that arise in the
calculation include uncertainties in the axial mass associated with the
quasi-elastic cross-section~\cite{Smith}, the application of Pauli suppression,
uncertainties in the oscillation parameters, and uncertainties in the pion-resonance
cross-section.  The total uncertainty due to the neutrino flux and cross-section
contributions is $\pm 30\%$.  Effects of final state interactions are dealt with in
the next section.

After the application of our
analysis cuts including fiducial volume and energy threshold (as described in
Section~\ref{sec:dataproc}) we find that 
the combined background from all atmospheric neutrino sources is 
$4 \pm 1$ events for the Phase I data set.  

\item Cosmogenic Production of $^{16}$N and Other Radioactivity

When a high-energy muon enters the SNO detector, several processes can
produce long-lived radioactive nuclei. The most common process is 
capture of a stopped muon on $^{16}$O, which produces a $^{16}$N nucleus.
Another process is muon-induced spallation, in which a muon
splits a nucleus into smaller fragments, which may be radioactive. These
radioactive nuclei can produce backgrounds to the neutral current solar neutrino
signal if they decay by the production of a neutron, or if they produce a gamma ray
with an energy above 2.2 MeV, which can photodisintegrate $^{2}$H. They
can also form a background to the charged current signal through the
Cherenkov light generated via $\beta$- or $\gamma$-decays.

Cosmic ray muons can also disintegrate $^{2}$H nuclei directly, producing prompt
neutrons.  The majority of these will capture and are removed by the muon
follower cut.  One must also consider muon capture on other nuclei which might
produce longer-lived nuclei, and (n,p), (n,$\alpha$), (p,n), and (p,$\alpha$)
reactions that produce long lived nuclei.

The dominant contribution of cosmogenic radioactivity to the background comes from
the production of $^{16}$N, which decays with a half-life of 7.13~s,  via either
muon capture or $(n,p)$ reactions.  An experimental measurement of this and other
spallation products is obtainable from the time dependence of muon followers, which
has been evaluated in several independent analyses.  The
presence of initial $^{16}$N is consistent with accidental background activity and
makes up a negligible portion of the total background rate. 

\end{itemize}

\subsubsection{Neutrons from ($\alpha$,n) Reactions}

Decays in the uranium and thorium chains produce alphas which in turn
can produce neutrons:

\begin{eqnarray*}
{\rm ^{ 2}H}& +\alpha \rightarrow & n + {\rm ^{ 1}H~-~2.223~MeV}, \\
{\rm ^{13}C}& +\alpha \rightarrow & n + {\rm ^{16}O~+~2.251~MeV}, \\
{\rm ^{17}O}& +\alpha \rightarrow & n + {\rm ^{20}Ne~+~5.871~MeV}, \\
{\rm ^{18}O}& +\alpha \rightarrow & n + {\rm ^{21}Ne~-~0.689~MeV}.
\end{eqnarray*}

The molecular targets of interest in SNO that could lead to the above reactions are
H$_2$O, D$_2$O, and acrylic - (C$_5$H$_8$O$_2$)$_n$.  The oxygen isotopes $^{17}$O
and $^{18}$O are somewhat enriched in D\(_2\)O;  natural oxygen is composed of
0.038\% $^{17}$O and 0.2\% $^{18}$O, while the fractional isotopic abundances in the
heavy water are 0.0485(5)\% and 0.320(3)\%, respectively~\cite{co2-h2o}.

All 14 alpha decays in the uranium and thorium chains are above threshold 
(6.6 MeV) for the
($\alpha$,n) reactions listed above.  The rates are $1.28 \times
10^{5}$ and $3.92 \times 10^{5}$ decays/year/$\mu$g of $^{232}$Th and $^{238}$U,
respectively.  In the heavy water, the main source of ($\alpha$,n) is 
$^{222}$Rn.    Contamination on the surface of the acrylic by radon
daughters, however, could yield more neutrons than expected from the U and Th
present in the heavy water.   Such neutrons would have a somewhat different
radial profile from neutrons generated in the D$_2$O volume.  As will be
discussed further in Section~\ref{sec:sigextns}, in our
Phase~I publications we performed a fit to the data using the expected radial
profile of external neutrons, allowing their number to float
in the fit.  We found in this fit that the total number of external neutrons
was consistent with our estimates for photodisintegration by external
radioactivity alone.  In a future publication, we will include updates to the
results here that explicitly fit for this potential source of neutrons.

The neutron yields from $^2$H($\alpha$,$\alpha$n)$^1$H and from
($\alpha$,n) reactions are summarized in Table \ref{tab:otherbkg}.

\subsubsection{Neutron Production From Terrestrial and Reactor Antineutrinos}

Antineutrino interactions with the light water, acrylic, and heavy water are an
additional source of background neutrons.  Such antineutrinos can be produced by
radioactive decays of uranium and thorium in the Earth's crust and
mantle, as well as by nearby fission reactors~\cite{Balantekin}.

Neutrons are produced in three antineutrino induced reactions:

\begin{eqnarray}
\overline{\nu}_{e} +p & \rightarrow & n+e^{+} -1.804\ {\rm MeV}\ \ (ccp) \\
\overline{\nu}_{e} +d & \rightarrow & n+n+e^{+} -4.03\ {\rm MeV}\ \ (ccd) \\
\overline{\nu}_{x} +d & \rightarrow & p+n+\overline{\nu}_{x} -2.223\ {\rm 
MeV}\ \ (ncd).
\end{eqnarray}

The charged current reaction on protons, ccp, has a threshold of 1.804~MeV.  There
are four beta-decays in the uranium and thorium decay chains that emit antineutrinos
above this threshold.  The charged current reaction on deuterium, ccd, has a larger
threshold of 4.03 MeV, so it need only be considered for reactor antineutrinos.  The
neutral-current reaction, ncd, has a threshold of 2.223 MeV, and thus must be
considered for reactor antineutrinos and for antineutrinos  from \(^{214}\)Bi  in
the uranium chain. There are two other decays, from \(^{212}\)Bi in the thorium
chain and \(^{234}\)Pa in the uranium chain, with antineutrino energies of 2.25 and
2.29 MeV, respectively, that are above the ncd reaction threshold.  The amount by
which they are over threshold, however, is small and their contribution is assumed
to be negligible.

In calculating the total contribution of antineutrinos to the background, effects
such as vacuum oscillations, reactor livetimes, and reactor efficiencies have been
taken into account. Table~\ref{tab:otherbkg} shows the results. The tabulated
numbers for the charged current include the fact that each interaction produces not
one but two neutrons per interaction.  These numbers are in agreement with the
background levels calculated for our limit on solar antineutrinos~\cite{antinu}.

\subsubsection{Summary of Other Neutron Backgrounds}

The neutron backgrounds from the sources discussed in this
section are summarized in Table \ref{tab:otherbkg}.
\begin{table}[h]
\caption{Neutron and gamma production and detection in the SNO detector
D$_2$O volume ($R<550$ cm).  The last column gives
the estimated background contribution to the data set for the pure D$_2$O
phase, after all analysis cuts have been applied.  Portions of the
neutron/gamma contribution errors are anti-correlated.}
\label{tab:otherbkg}
\medskip
\begin{center}
\begin{tabular}{lc}
\hline\hline
Source & Expected Number \\
       & of Detected Events \\
\hline
Fission [U,Cf] (neutrons) &  $\le 1$  \\
Fission [U,Cf] ($\gamma$s)  &  $\le 1$  \\
Atmospheric $\nu$ &  $4 \pm 1$  \\
\hline
$^2$H($\alpha$,$\alpha$n)$^1$H [Th] & $0.40 \pm 0.13$  \\
$^2$H($\alpha$,$\alpha$n)$^1$H [$^{222}$Rn] & $1.59\pm0.30$ \\
$^{17,18}$O($\alpha$,n)$^{20,21}$Ne [Th] & $\ll 1$  \\
$^{17,18}$O($\alpha$,n)$^{20,21}$Ne [$^{222}$Rn] & $\ll 1$  \\
$^{17,18}$O($\alpha$,n)$^{20,21}$Ne [$^{238}$U] & $\ll 1$ \\
\hline
$^{16}$N following muons &   $\le 1$  \\
Other spallation & $\leq 1$  \\
Muons follower neutrons & $\ll 1$   \\
Cosmogenic Rock neutrons & $0.18 \pm 0.01$  \\
Terrestrial and reactor $\bar{\nu}$ & $ 1^{+3}_{-1}$ \\
\hline
{\em Total Other Neutrons} &  $7^{+3}_{-1}$  \\
\hline
\hline
\end{tabular}
\end{center}
\end{table}

\subsection{Overall Background Summary \label{sec:bkdsumm}}

	Table~\ref{tbl:bkgd_summary} summarizes all sources of background
discussed in this section.  As will be discussed in Sections~\ref{sec:sigex}
and~\ref{sec:results}, the background numbers are subtraced off of the final,
fitted event totals.  In the case of the $\beta$-$\gamma$ backgrounds, the
numbers are used to fix the amplitudes of energy spectrum pdfs (the analytic
form of which were given in Sections~\ref{sec:lowint} and~\ref{sec:lowext})
during the signal extraction process.  For the internally-produced neutron
backgrounds, which look identical to neutrons produced by neutrino NC
reactions, the numbers are directly subtracted from the final fitted neutron
event total.  For external neutrons produced by radioactivity in the AV and
H$_2$O, a radial pdf is included in the signal extraction fit with its
amplitude fixed to the value given in the table.

\begingroup
\begin{table}[t]
\squeezetable
\caption{\label{tbl:bkgd_summary}Summary of estimated numbers of events 
for each source of background.}
\begin{ruledtabular}
\begin{tabular}{ll}
Source                                  & Events                        \\ \hline
Instrumental                            &  $<3$                         \\
D$_2$O photodisintegration              &  $44^{+8}_{-9}$               \\
H$_2$O + AV  photodisintegration        &  $27^{+8}_{-8}$               \\
Atmospheric $\nu$s     and             &                               \\
sub-Cherenkov threshold $\mu$s         &  $ 4 \pm 1$                   \\
Fission                                 &  $\ll1$                       \\
${}^{2}$H($\alpha,\alpha$)pn            &  $2 \pm 0.4$                  \\
${}^{17}$O($\alpha$,n)                  &  $\ll1$                       \\
Terrestrial and reactor $\bar{\nu}$s   &  $1^{+3}_{-1}$                \\
Cosmogenic neutrons from rock                       &  $\ll1$                       \\ \hline
Total neutron background                &  $78 \pm 12$                  \\ \hline
D$_2$O $\beta$-$\gamma$ &  $20^{+13}_{\phantom{1}-6}$   \\
H$_2$O $\beta$-$\gamma$ &  $3^{+4}_{-3}$                \\
AV $\beta$-$\gamma$                          &  $6^{+3}_{-6}$                \\
PMT $\beta$-$\gamma$ (+HEGs)                           &  $16^{+11}_{\phantom{1}-8}$   \\ \hline
Total $\beta$-$\gamma$ background              &  $45^{+18}_{-12}$
\end{tabular}
\end{ruledtabular}
\end{table}
\endgroup

\section{Signal Extraction Method \label{sec:sigex}}

We have described the analysis used to build accurate models of neutrino and
background signals in our detector, the processing of the data and the
measurement of the backgrounds.  After accomplishing those tasks we are in a
position to fit the data with the pdfs shown in Fig.~\ref{fig:pdfs}.  The fit
itself is an extended maximum likelihood method using binned pdfs.  We used
multiple sets of pdfs to verify our overall results. For example, we used both
pdfs based on the reconstructed kinetic energies as shown in
Fig.~\ref{fig:pdfs} and described in Section~\ref{sec:enecal}, and pdfs that
used only the total number of hits in each event (`$N_{\rm hit}$') as a measure
of the event energy.  These two approaches are identical other than in the
choice of energy variable.  As a further check, we fit the data using pdfs
constructed from an analytic model rather than from Monte Carlo simulation.

There are alternate approaches to fitting the energy spectra of the
data set.  In one method, we constrain the recoil electron spectra of
the CC and ES events to be that resulting from an undistorted $^8$B
neutrino spectrum.  This `constrained' fit is thus a test of the null
hypothesis that solar neutrinos do not oscillate, and is also
appropriate for the case of an energy-independent $\nu_e$ survival
probability, which is nearly correct for the LMA solution in this
energy region.  An alternate approach is to perform the fit without a
constraint on the CC energy spectral shape.  This may be done either
by excluding the energy variable from the signal extraction and so
using a pdf only in $R^3$ and $\cos \theta_\odot$, as was done in
our Phase I NC paper~\cite{snonc}, or else by fitting the CC energy spectrum
bin-by-bin while fixing the NC and background energy PDFs to their known shapes, as
in the Phase I ES-CC paper~\cite{snocc}.

We describe in this section the details of our signal extraction
method, and leave the presentation of the flux results to
Section~\ref{sec:results}.

\subsection{Extended Maximum Likelihood Method}

The basis of the signal extraction is to express the probability
distribution for neutrino events in the variables $E$, $R^3$, and
$\cos \theta_\odot$ with a linear superposition of pdfs corresponding to
different signals and backgrounds.  The total number of events
$\nu$ as a function of $E$, $R^3$, and $\cos \theta_\odot$ is
then
\begin{equation}
\nu(E,R^3,\cos \theta_\odot) = \sum_i N_i~f_i(E,R^3,\cos \theta_\odot).
\label{eq:sigex_decomposition}
\end{equation}
where $N_i$ is the number of events of type $i$ (eg. CC, ES, or NC), and $f_i$
is the probability distribution for events of that type,
normalized to unity.  The sum is taken over all signal types, and
over classes of background events for which pdfs may be constructed.  
In this section, we use $E$ to mean either $T_{\rm eff}$ or $N_{\rm hit}$; the
former for our primary signal extraction which uses the energy reconstructor, and the
latter for the verification signal extraction which uses the total light energy 
estimate.

The extended log likelihood then takes the form
\begin{equation}
\log L = -\sum_i N_i + \sum_j n_j \ln~ (\nu(E_j,R^3_j, \cos \theta_{\odot j}))
\end{equation}
where $j$ is a sum over all three-dimensional bins in the three
signal extraction parameters $E$, $R^3$, and $\cos \theta_\odot$, and
$n_j$ is the number of detected events in each bin.

In this analysis the numbers of CC, ES, and NC events are treated as
free parameters in the fit, while the numbers of background events of
each type are fixed, as described in
Section~\ref{sec:background_pdfs}.  The likelihood function is
maximized over the free parameters, and the best fit point yields the
number of CC, ES, and NC events along with a covariance matrix.

\subsection{Fitting with Monte Carlo pdfs}

Our reported results use the Monte Carlo simulation to generate pdfs for the 
neutrino signals over the three signal extraction variables: the effective
kinetic energy $T_{\rm eff}$ returned by the energy calibrator (see
Section~\ref{sec:enecal}), $R^3$, and $\cos \theta_{\odot}$.  
Generation of pdfs using the total number of hit PMTs ($N_{\rm hit}$) was done
similarly to what we describe here with the substitution of $N_{\rm hit}$ for
$T_{\rm eff}$.  (As mentioned in the previous section, here the variable $E$ will
denote either $T_{\rm eff}$ or $N_{\rm hit}$).

\subsubsection{Monte Carlo pdf Generation}
The pdfs were constructed by binning simulated events in these
three quantities, under the implicit
assumption that the full three dimensional pdf factorizes into separate
energy, radial, and angular components:
\begin{equation}
F(E, R^{3}, \cos \theta_{\odot}) = A(E)~B(R^3)~C(\cos \theta_\odot). 
\label{eq:facpdfs}
\end{equation}
The functions A, B, and C in Equation~\ref{eq:facpdfs} are shown in 
Fig.~\ref{fig:pdfs}. 

There are, in fact, modest correlations between energy and $R^3$ at
the few percent level, as well as a narrowing of the width of the
elastic scattering angular peak with increasing energy.  By testing
the signal extraction procedure on many sample Monte Carlo data sets,
we verified that these correlations introduced negligible bias in the
extracted fluxes and could therefore be ignored.

The Monte Carlo simulations used to create the pdfs were performed
run-by-run, matching the simulation inputs to the state of the
detector for each run as described in Section~\ref{sec:mcdaq}.  The
simulation for each run took account of the number of channels online,
threshold settings, the average PMT noise rate derived
from the 5~Hz pulsed trigger, and the measured livetime of the run.
The statistics for the Monte Carlo runs were 50 times the Standard
Solar Model prediction for each of the signals.

As described in Section~\ref{sec:enecal}, the mean energy response of the
detector varied slowly over the course of the data set.  This variation was
incorporated into the calculation of the energy for each event as a factor that
depended upon the time of the event relative to the start of the data set.
Monte Carlo simulations were done at a fixed energy scale.

For the analysis described in the ES-CC paper~\cite{snocc}, the energy component of
the pdfs was binned in 34 unequal bins between the lower and upper energy limits of
the analysis ($T_{\rm eff}=6.75-19.5$~MeV).  The first 33 bins were each 0.2574~MeV
wide, while the final bin was extended up to $T_{\rm eff}=19.5$~MeV.  For the
analysis in the Phase I NC paper~\cite{snonc}, the energy component of the pdfs was
binned in 42 bins between the lower and upper energy limits of the analysis
($T_{\rm eff}=5.0-19.5$~MeV kinetic energy).  Each of these bins was 0.25~MeV wide,
except for the last bin, which was extended to the upper energy limit.  In both the
ES-CC paper and the NC paper, the radial distribution was binned in 30 equal bins in
$R^3$ inside the 550~cm fiducial volume, and the angular pdfs were binned in 30
unequal bins of $\cos \theta_{\odot}$.  Fifteen equal bins spanned the region
between $-1 \le \cos \theta_\odot < 0.5$, and the remaining 15 bins spanned the
region between 0.5 and +1.  This unequal binning gives extra sensitivity to the
rapidly rising elastic scattering peak near $\cos \theta_\odot = 1$.  We binned the
data events in the same way.

\subsubsection{Fitting Procedure}

The pdfs for CC, ES, and NC events were generated
for a $^{8}$B spectrum.  The background pdfs described in
Section~\ref{sec:bkds} were used to subtract low energy backgrounds
(external neutrons, misreconstructed $\beta$-$\gamma$ events, etc.) by
fixing their amplitudes (see Section~\ref{sec:background_pdfs}) 
based on the measurements described in Section~\ref{sec:bkds}.

For signal extraction using the $^{8}$B spectral constraint, all three
signal pdfs are used. The `high level' cuts described in
Section~\ref{sec:cerbox} were not applied to the Monte Carlo simulated
events, but their efficiencies were included in the final flux
calculations (see Section~\ref{sec:hlcuteff}).  We used the Monte Carlo
generated SSM predictions for the expected number of events of each
signal type inside the fiducial volume and above the analysis energy
threshold to determine the acceptance of the detector. The extended
maximum likelihood fit returned the total number of extracted events
for each signal, the statistical uncertainty on the number of
extracted events, and a full statistical correlation matrix for the
extracted fluxes.  

The final flux values are determined by dividing the number of
extracted events by the predicted number of events from the Monte Carlo simulation,
and then correcting the flux for effects not modeled in the Monte
Carlo, including deadtime as described in Section~\ref{sec:livetime},
instrumental background cut acceptance, and high level (Cherenkov box)
cut acceptance.  Additional cross-section and scaling corrections were
applied, as described in detail in Section~\ref{sec:results}.
The result in each case is a `flux' for each interaction type in
units of neutrinos/cm$^2$/sec.  This 
is the equivalent total flux of $\nu_e$s from an undistorted $^8$B
energy spectrum that would yield the same number of interactions
inside the signal region as was observed for that signal type.

The signal extraction also calculates $\chi^2$ goodness of fit parameters
for the radial, angular, and energy projections of the data, as compared
to Monte Carlo predictions.

\subsubsection{Signal Extraction without a CC Energy Constraint
\label{sec:unconst}}

To extract a recoil electron energy spectrum, we must use the `unconstrained' 
approach in which the CC events are not assumed to have been created with a 
$^8$B neutrino
energy spectrum.  Two methods were used to implement this approach.
In the Phase I ES-CC paper~\cite{snocc}, the CC energy pdf was decomposed into a linear
sum of 11 components:
\begin{equation}
CC_{\rm pdf}(E) = \sum_{i=1}^{11} N_{CC,i} B_i(R^{3})~C_i(\cos \theta_{\odot})~\Delta_{i}(E)
\end{equation}
Here, $N_{CC,i}$ is the number of CC events in the $i$th bin of the CC energy
spectrum, and the radial and angular pdfs are binned separately in
each energy bin.  $\Delta_{i}(E)$ is defined to equal 1 if the
event energy $E$ lies in the $i$th energy bin, and equals zero otherwise.
This superposition corresponds to approximating the energy spectrum in
each bin by a step function.  The first 10 spectral bins covered the
range $T_{\rm eff}=6.75-11.9$~MeV, while the final bin extended
from $T_{\rm eff}=11.9$~MeV to $T_{\rm eff}=19.5$~MeV.

The 11 components of CC$_{\rm pdf}$ can then be treated as 11 independent CC
pdfs, along with the ES and NC pdfs.  The normalization of each pdf determines
the number of extracted CC events in that energy bin.  Only the CC spectrum is
so decomposed---we have fixed the ES energy pdf to be that created by an
undistorted $^{8}$B neutrino energy spectrum.  
Although it is technically
inconsistent to allow the CC shape to vary while the ES spectrum is kept fixed,
the flatness of the differential cross section ($d\sigma(E_{\nu})/dE_e$) for
the ES reaction,  the very low statistics of the ES electrons in SNO, and the fact
that the Super-Kamiokande Collaboration~\cite{SK} sees no distortion in the spectrum
of their ES electrons make this inconsistency a negligible effect on the analysis.
The NC (and background neutron) pdf need not be decomposed, because the  `energy'
spectrum is simply the response of the detector to the NC reaction's monoenergetic
6.25~MeV $\gamma$-ray, and holds no information about the incident neutrino energy.
The signal extraction proceeds as before with the $11+2$ signal pdfs (11 CC energy
pdfs, plus the ES and NC pdfs).  The extracted results give the fluxes and
uncertainties for each pdf, as well as a full correlation matrix.  This 13-parameter
fit was used to produce the CC energy spectrum in the Phase I ES-CC
paper~\cite{snocc}.

For results presented in the Phase I NC paper~\cite{snonc}, a simpler procedure was
used.  In this case the energy variable was not used in the signal extraction, and
instead two-dimensional pdfs in $R^3$ and $\cos \theta_\odot$ were constructed for
each signal.  The CC spectrum was not fit bin by bin, but rather the total numbers
of CC, ES, and NC events were determined from a 3-parameter fit.

\subsection{Background Subtraction During Signal Extraction}
\label{sec:background_pdfs}

	Because we fit for the three different signals,  we cannot simply
subtract the estimates of the backgrounds from the total event rate---we need
to decide how much each background contributes to each signal.  For
photodisintegration neutrons produced by radioactivity inside the D$_2$O
volume, this is relatively easy---these neutrons should look identical to the
NC signal.  For the $\beta$-$\gamma$ backgrounds from radioactivity inside and
outside the fiducial volume, we needed to use the energy pdfs described in
Sections~\ref{sec:lowint} and~\ref{sec:lowext}. For some backgrounds, like the
residual contamination from spallation products left after the follower cuts,
the number of events was too small to make using pdfs practical; they were
simply treated as upper limits with one-sided systematic uncertainties, applied
conservatively to each signal.  In the Phase I ES-CC paper~\cite{snocc}, which had a
higher analysis threshold, the same treatment was used.

	Backgrounds for which we used pdfs could in principle
be included as part of an overall fit for both the signals and backgrounds.
Nevertheless, because the most important information about these backgrounds comes
from events outside the signal region (either lower in energy or outside the
fiducial volume) we constrained the amplitudes of the backgrounds
based on the measurements described in Section~\ref{sec:bkds}.
Our signal extraction fit therefore included background pdfs of fixed 
amplitudes:
\begin{eqnarray}
f(E,R^3,\cos \theta) &= & N_{CC}f_{CC}(E,R^3,\cos \theta) \nonumber \\
&+ & N_{ES}f_{ES}(E,R^3,\cos \theta) \nonumber \\
&  + & N_{NC}f_{NC}(E,R^3,\cos \theta) \nonumber \\
 & + & \sum_i N_{bkgd,i}f_{bkgd,i}(E,R^3,\cos \theta_{\odot}) \nonumber \\
\end{eqnarray}
Here, $N_{CC}$, $N_{ES}$, and $N_{NC}$ are the fitted amplitudes of
the signal fluxes.  (As described in Section~\ref{sec:unconst}, for
the spectrally unconstrained fit in the ES-CC paper~\cite{snocc}, there is a CC
pdf for each CC spectral bin, giving additional free parameters.)  In
contrast, $N_{bkgd}$ is the fixed amplitude of the background pdf.  We
include a term in the sum for each source $i$ of background events.

To determine the effect of uncertainty in the amplitude of a
background $N_{bkgd,i}$, we vary $N_{bkgd,i}$ by its $\pm 1\sigma$ limits,
and repeat the signal extraction to determine the changes in the
extracted signal fluxes.  (That is, we change the assumed value of
$N_{bkgd,i}$ in the fit, but do not allow the value to float.)  

The backgrounds for which we included pdfs are the `external' neutrons
(those produced through photodisintegration by radioactivity outside the heavy water volume); the radioactivity from the
uranium and thorium chains originating inside the D$_2$O volume as
described in Section~\ref{sec:lowint}; and radioactivity from the
uranium and thorium chains originating inside the Acrylic Vessel
(including the AV `hot spot'), in the H$_2$O region, and in the PMTs as described in
Section~\ref{sec:lowext}.  As discussed in Section~\ref{sec:hegs}, we did not include a distinct pdf for high
energy ($>4$~MeV) $\gamma$ rays, because their number is included with the PMT
$\beta$-$\gamma$ events.  But because high energy gammas (HEG) 
have a different energy spectrum from PMT $\beta$-$\gamma$
events, there is an additional component of systematic uncertainty
on the total HEG+PMT~$\beta$-$\gamma$ number due to spectral
uncertainties.  
The sizes and uncertainties on the backgrounds were summarized in 
Table~\ref{tbl:bkgd_summary} of Section~\ref{sec:bkdsumm}.

\subsection{Fitting for the Neutrino Flavor Content}
\label{sec:fit_flavor}

	In addition to fitting for the three signal rates (CC, ES, and NC),
the SNO data allows us to also directly fit for the neutrino flavor content
by a straightforward change of variables:

\begin{eqnarray}
\label{eq.flavor_fit}
\phi_{CC} & = & \phi(\nu_e) \\
\phi_{ES} & = & \phi(\nu_e) + 0.1559 \phi(\nu_{\mu\tau}) \\
\phi_{NC} & = & \phi(\nu_e) + \phi(\nu_{\mu\tau}) 
\end{eqnarray}
Here the factor of 0.1559 is the ratio of the ES cross sections for
$\nu_{\mu\tau}$ and $\nu_e$ above $T_{\rm eff}=5.0$~MeV. 

Making this change of variables and fitting directly for the flavor content,
one reduces the task of doing a null hypothesis test of no flavor transformation to a single variable
test of $\phi(\nu_{\mu\tau})=0$.  By fitting directly for
$\phi(\nu_{\mu\tau})$, we automatically account for statistical and
systematic uncertainty covariances in the CC, ES, and NC
flux estimates. 
Note that this change of variables
implicitly assumes an energy-independent $\nu_e$ survival probability.

\subsection{Analytic Response Functions
\label{sec:anal_response}}

An alternative approach to signal extraction is to construct analytic
pdfs by convolving the expected signal distributions with SNO's
measured response functions.  In this technique, 
the same maximum likelihood fit is applied to a linear decomposition of
pdfs, but the pdfs in this case are calculated analytically
rather than by Monte Carlo simulation.  
The analytic approach works well because the detector is well-represented by simple
response functions in energy, position, and direction.  For others wishing to fit our
data set, these analytic response functions will be useful for creating pdfs.  
In the following, we describe the details of the pdf forms and analytic
response parameters.

We parameterized the energy response to electrons with a Gaussian functional form:
\begin{equation}
R(T_{\rm eff},T_{e})  = \frac{1}{\sqrt{(2 \pi)} \sigma_{T}(T_{e})}
\exp\left[ -  \frac{(T_{\rm eff} - T_{e} - \Delta_{T})^2}{2
    \sigma_{T}^2(T_{e})}  \right]  
\label{eresp}
\end{equation}
\noindent where $T_e$ is the true kinetic energy of the electron,
$T_{\rm eff}$ is the measured kinetic energy, $\sigma_{T}(T_{e})$ is the
energy resolution, and $\Delta_{T}$ is an energy offset that is
zero if the detector is correctly calibrated.  Table~\ref{response_pars} gives
the functional form of $\sigma_{T}(T_{e})$.

The energy spectral shape of the signal pdfs was modeled
by a convolution of the  solar neutrino spectra and cross sections with
the analytic response function.  For example, the charged current pdf is:
\begin{equation}
\frac{dN_{CC}}{dT_{\rm eff}} =
\int_{0}^{\infty} \int_{0}^{\infty}
\frac{d\sigma_{CC}}{dT_{e}}(E_{\nu}) \frac{d\Phi_e}{dE_\nu}
R(T_{e}, T_{\rm eff}) dT_{e} dE_{\nu}. 
\label{eq:ccspec}
\end{equation}
\noindent Here $dN_{CC}/dT_{\rm eff}$ is the number of charged current
interactions in the detector per target nucleus per unit MeV of measured electron
energy.  The CC cross section $d\sigma_{CC}/dT_e$ is given per MeV of true electron
energy, $d\Phi_e/dE_\nu$ is the $^8$B electron neutrino energy spectrum, and
$R(T_e,T_{\rm eff})$ is the energy response function given above in Eq.~\ref{eresp}.

The NC can be treated much more simply, because it represents the detector's
response to a monoenergetic $\gamma$ ray---we do not need to convolve an
analytic response function with an energy spectrum.  Instead, we used a Gaussian
to describe $dN_{NC}/dT_{\rm eff}$, with a fixed kinetic energy mean of 
$T_{\gamma} = 5.08$~MeV and a width $\sigma_{\gamma} = 1.11$~MeV:
\begin{equation}
  \frac{dN_{NC}}{dT_{\rm eff}} = \frac{1}{\sqrt{2\pi}\sigma_{\gamma}}
             \exp \left[
                  \frac{-(T_{\rm eff}-T_{\gamma})^2}{2\sigma_{\gamma}^2}
                  \right]
\end{equation}
The reduction in the effective energy mean $T_{\gamma}$ relative to the 6.25~MeV
total energy of the $\gamma$ ray itself is caused by the `loss' of energy to the 
Cherenkov threshold of each of the Compton-scattered electrons.

For the position resolution of the reconstruction method described in
Section~\ref{sec:recon}, we have a Gaussian distribution with exponential
tails. In one dimension (e.g., $x$), the position response is given by:
\begin{eqnarray}
R(x)& = & \frac{1-\alpha_{e}}{\sqrt{(2 \pi)}\sigma_P} \exp
\left[-\frac{1}{2} \left(\frac{x-\mu_P}{\sigma_P} \right)^2 \right] \nonumber \\
       & & + \frac{\alpha_{e}}{2 \tau_P} \exp \left[ - \frac{|x-\mu_P|}{\tau_P} \right] 
\label{presp}
\end{eqnarray}
where 
\begin{eqnarray}
\alpha_{e} & & \mbox{is the fractional exponential component}
\nonumber \\
\sigma_P     & & \mbox{is the Gaussian width} \nonumber \\
\mu_P        & & \mbox{is the Gaussian shift, and } \nonumber \\
\tau_P        && \mbox{ is the exponential slope.} \nonumber
\label{vertex_res_pars}
\end{eqnarray}
This analytic response function may be convolved with the true spatial 
distribution of events to estimate the fraction of events occurring inside the
fiducial volume.  Note that this expression is not accurate for misreconstructed
background events whose true position lies outside of the D$_2$O target.

For the angular response, we used the following functional form  for
the resolution function:

\begin{eqnarray}
P(\cos \theta) & = &  \alpha_{M} \frac{\beta_{M}\exp[\beta_{M}(\cos
 \theta - 1)]}{1-\exp(-2\beta_{M})} \nonumber \\
   & & + 
(1-\alpha_{M})\frac{\beta_{S}\exp[\beta_{S}(\cos \theta - 1)]}{1-\exp(-2\beta_{S})}
\label{aresp}
\end{eqnarray}
\noindent where $\cos \theta$ represents the angle between the reconstructed
(electron) event direction and the electron's initial direction. The expression has
two components: a main peak due to the true angular resolution of the detector, and a
broad tail due to multiple scattering of electrons.  The resolution function has
three parameters: the slopes of the two exponentials describing the main peak
($\beta_{S}$) and multiple scattering component ($\beta_{M}$), and the relative
fraction of these ($\alpha_{M}$).  This resolution function may be convolved with
the true distributions of $\cos \theta_\odot$ for CC and ES events to determine the
angular pdfs.

Table ~\ref{response_pars} shows the parameters and uncertainties
derived for all of the response functions given above.

\begin{table*}[th]
\caption{Analytic response functions for electrons, $\gamma$-rays and
  neutrons in SNO.  These parameters used in equations ~\ref{eresp},
  \ref{presp} and \ref{aresp} can be used to calculate the  three
  solar neutrino signal pdf's.\label{response_pars}}
\begin{center}
\begin{tabular}{lcll} \hline \hline
Component of pdf     &     Parameter        & Value                 & Uncertainty  \\ \hline
Energy response            &     $\Delta_{T}$        & 0
& 1.21\% $\times T_e$       \\
 (CC, ES)                     &     $\sigma_{T}$     & $ -0.0684 + 0.331\sqrt{T_{e}} + 0.0425T_{e}$ & 4.5\% for $T_{\rm eff}=5.0$~MeV, \\ 
               &              &      & 10\% for $T_{\rm eff}=18.7$~MeV (see Eq.~\ref{eq:sigTunc})  \\ \hline
Energy response      &     ${T_{\gamma}}$ & 5.08~MeV       & 1.21\% \\
 (NC)                     &     $\sigma_{\gamma}$ & 1.11~MeV & 4.5\% \\ \hline
Position             &     $\alpha_{e}$     &  0.55 (fixed)    &  0            \\
response             &     $\sigma_{P}$     &  13.3 cm    &     16 \%          \\
                     &     $\mu_{P}$        &  0 cm &   $ 0.01 \times R_{\rm fit}$ [cm]      \\
                     &     $\tau_{P}$       &  10.7 cm, 25 cm for $\gamma's$   & 16 \%             \\
Angular              &     $\alpha_{M}$     & 0.6 & fixed           \\
response             &     $\beta_{M}$      &  0.7495 + 0.5775 $E_{e}$ - 0.006262 $E_{e}^{2}$ & 5 \%                \\
                     &     $\beta_{S}$       & 4.815 +2.358 $E_{e}$ +
0.01208 $E_{e}^{2}$ & 14 \%        \\ \hline \hline

\end{tabular}
\end{center}
\end{table*}

\section{Flux Normalization \label{sec:norm}}

	The absolute normalization of the measured rates, and ultimately the
neutrino fluxes, depends upon careful accounting of detector livetime, 
efficiencies of all cuts applied to the data set, neutrino cross sections, and
the effective number of targets. In addition, for the neutral current reaction, we
need to know the overall neutron capture and detection efficiency. 
In this section we discuss our determination
of these normalization factors and their uncertainties.

\subsection{Livetime \label{sec:livetime}} 

SNO's primary clock is a 10 MHz oscillator disciplined to the Global
Position System's clock time, and is accurate to a few hundred nanoseconds.
Each event is stamped with the time measured by this clock. The
raw livetime for a run is determined from the elapsed counts of the
10 MHz clock between the first and last event in the run.  The elapsed
time between successive events is always less than 0.2 seconds due to the
presence in the data stream of events generated by a 5 Hz pulsed trigger, and so
the difference between the ``true'' livetime for a run and the elapsed time between
its first and last events is negligible.  An independent 50 MHz clock, which is the
master clock for the entire electronics system and defines the 5 Hz pulsed trigger
rate,  serves as an additional check of the livetime, and we find that the raw
livetimes estimated from the 10 MHz and 50 MHz clocks agree to within 0.006\%.
Finally, the raw livetime as measured by the 10 MHz clock is verified by counting
the number of pulsed triggers in the run and dividing by their 0.2 s period.

Time-based event cuts designed to remove ``bursts'' of instrumental backgrounds and
muon-induced spallation events reduce the effective detector livetime.  This
livetime correction is dominated by the ``muon follower short'' cut that removes all
events occurring within 20 seconds after a through-going muon.
Table~\ref{tab:LT_lt_corr} details the total livetime correction for each burst and
spallation cut.  The listed deadtime in the table for each cut is independent of the other
cuts, but the total includes the overlap between them and is thus smaller than the direct sum
of the numbers in the columns.

\begin{table}
\caption{The livetime correction imposed by the various cuts, together
with the combined correction, for the entire D$_2$O data set. The definitions
of the cut names can be found in Appendix~\ref{sec:apdxa}. The listed deadtimes for each cut
are independent of the other cuts, but the total includes the correlated overlaps between them
and is thus smaller than the direct sum of the columns.  \label{tab:LT_lt_corr}}
\begin{center}
\begin{tabular}{lcc}
\hline \hline
Cut                    &    Correction   &   Fractional Correction  \\
\hline
Retrigger              &     24.5 s      &      $9.1\times10^{-7}$       \\
Burst                  &     24.9 mins   &      $5.5\times10^{-5}$       \\
$N_{\rm hit}$ Burst             &      9.3 hours  &      0.0012              \\
Muon Follower Short    &    138.0 hours  &      0.0184              \\
Missed Muon Follower   &     21.5 hours  &      0.0029          \\
\hline
Combined Correction    &    156.9 hours  &      0.0213             \\
\hline \hline
\end{tabular}
\end{center}
\end{table}   

For the day-night asymmetry measurement described in
Section~\ref{sec:daynight}, we further divide the livetime into `day' and
`night' bins, where day livetime is defined as any time when the Sun is above
the horizon.  In an effort to reduce statistical
biases in the analysis, the data set for the day-night asymmetry measurement
was partitioned into two sets of approximately equal livetime.  Set 1 covered
the calendar period November 2, 1999-June 30, 2000, and Set 2 covered July 1,
2000-May 28, 2001.  Each set had substantial day and night components.
Analysis procedures were refined during the analysis of Set 1 and fixed before
Set 2 was analyzed. The latter thus served as an unbiased test.  This
open/blind separation was done in addition to the data division used in the
rest of the solar neutrino analysis and described in Section~\ref{sec:dataset}.

\begin{figure*}
\begin{center}
\includegraphics[height=0.3\textheight]{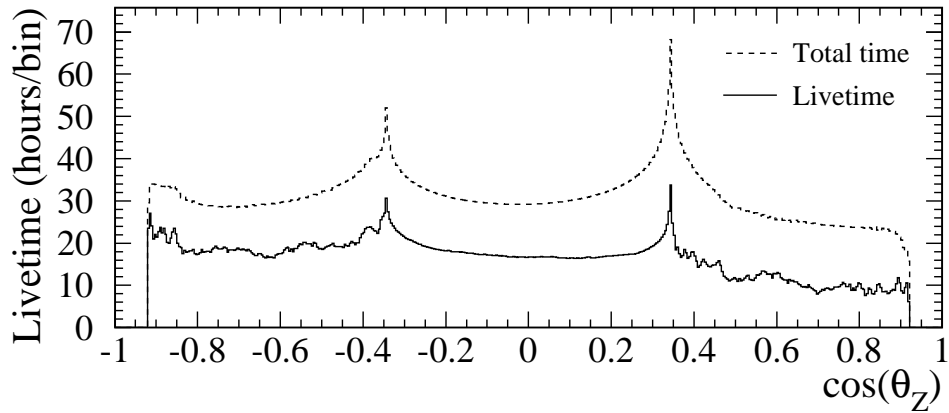}
\caption{The distribution of livetime and calendar time in 480
zenith angle bins for the D$_2$O data set. The dotted line labeled
`Total time' is the distribution of livetime that would
have resulted if the SNO detector were 100\% live during the entire
calendar time spanned by the full D$_2$O data set. The asymmetric structure in the figure corresponds to the effects of the Earth's axial tilt and the latitude of Sudbury.  The numerical values for the bin-by-bin livetimes are
given in Tables~\ref{tbl:coszenith1} and~\ref{tbl:coszenith2} of Appendix~\ref{sec:physint}.} 
\label{fig:LT_raw_10mhz_lt}
\end{center}
\end{figure*}

Table~\ref{tab:LT_summary} summarizes the final day and night livetimes.  The
combined data set has a day livetime of 128.5 days, and a night livetime of
177.9 days.  The livetime distribution in 480 zenith angle bins for the
entire data set is shown in Figure~\ref{fig:LT_raw_10mhz_lt}, and numerical
values for each bin are given in Tables~\ref{tbl:coszenith1}
and~\ref{tbl:coszenith2} of Appendix~\ref{sec:physint}. Also included in the
figure is the distribution of livetime that would have resulted if the SNO
detector were 100\% live during the entire calendar time spanned by the full
D$_2$O data set.

Maintenance work, detector
calibrations, and radiochemical assays are generally performed during
daylight hours.  Because data taken during these activities are not included
for solar neutrino analyses, the total day livetime is reduced relative
to the night livetime.  In addition, seasonal variations in the lengths
of day and night, when convolved with the SNO detector's exposure period,
introduce an additional difference in the day and night livetimes.

\begin{table} 
\caption{Summary of livetime results \label{tab:LT_summary}} 
\begin{center} 
\begin{tabular}{lcc} \hline \hline
Cut &    Day         &   Night      \\ \hline Raw Livetime           &
131.4 days    & 181.6 days   \\ Livetime Correction    &   68.4 hours   &
88.6 hours  \\ Corrected Livetime     &  128.5 days    & 177.9 days   \\
Open Data              &   64.4 days    &  92.9 days   \\ Blind Data
&   64.1 days    &  85.0 days   \\ \hline  \hline
\end{tabular} 
\end{center}
\end{table}

\subsection{Trigger Efficiency}

	We measured the trigger efficiency with the aid of a nearly isotropic
diffuse laser source, which was positioned at several places within the
detector volume, including the edge of the D$_2$O region~\cite{the:msn}.  The
trigger efficiency was measured by comparing an offline count of the number of
tubes firing in coincidence  with the trigger decision made by the detector
hardware.  The measurements showed that the efficiency was greater than 99.9\%
when 23 or more PMTs fired (roughly 3~MeV), and measurements made over a year
apart demonstrated the stability of the overall system.

\subsection{Reconstruction and Cut Efficiencies \label{sec:accept}}

	As described in Section~\ref{sec:dataproc}, we used several
cuts to remove backgrounds and to ensure that the fitted vertex and position
were consistent with light from a single Cherenkov electron.  In addition
to removing backgrounds, each cut also removes a small number of
neutrino signal events.  Given the large reduction in the raw data set,
we were particularly concerned that we demonstrate that the loss of acceptance
was small, robust, and stable.  We describe in this section our determination 
of the acceptance loss incurred by the cuts.
 
The cuts described in Section~\ref{sec:dataproc} fall into four
broad categories: 
\begin{itemize}
\item Time-correlated cuts (`burst cuts'): Removal of events based upon their time 
coincidence with each other and with certain special events such as muons.
\item Instrumental (`low level') cuts:
Removal of events before any reconstruction is done, based upon information such 
as PMT times and charges, event topology, or the presence of veto tubes.
\item Reconstruction quality cuts:
Removal of events in which the reconstruction algorithm either failed to
converge or for which the hypothesis of a single Cherenkov electron was
not satisfied.
\item Cherenkov Box (`high level') cuts.
Cuts which require an event to have a hit pattern and timing consistent
with Cherenkov light.
\end{itemize}
Cuts in the first category remove signal events through the deadtime they
create, as described in Section~\ref{sec:livetime}.

We examined the correlations between the cuts to understand whether we could
treat them separately.  Table~\ref{tab:SAC_n16corr_all} shows the number of
events which were removed by each cut suite using a sample of tagged 6.13~MeV
$\gamma$-rays from the $^{16}$N calibration source.
\begin{table*}
\caption{Number of events removed by
different sets of cuts for $^{16}$N calibration data inside the
solar neutrino analysis window.  Off-diagonal entries indicate the
number of events tagged by both sets of cuts.
\label{tab:SAC_n16corr_all}}
\begin{center}
\begin{tabular}{lcccc}
\hline \hline
                  &~Total Events~&~Instrumental Cuts~&~Reconstruction Quality Cuts~&~High Level Cuts\\
\hline 
Total Events      & 619362&        0        &    0 &   0           \\
Instrumental Cuts &  0    &     2657        &    1 &   63          \\
Reconstruction    &  0    &        1        &  258 &  258          \\
High Level Cuts   &  0    &       63        &  258 & 11245         \\
\hline \hline
\end{tabular}
\end{center}
\end{table*}
The correlations shown in the table between the instrumental cuts and the
reconstruction algorithm cuts, as well as that between the instrumental
cuts and the high level cuts, are weak enough that we can safely ignore
them.  The non-trivial correlation between the high level cuts and the
reconstruction cuts occurs because the high level cuts themselves use
information from the reconstructed vertex position.  We can also ignore this
correlation if we restrict the study of signal loss for the high level cuts to
events that have a good reconstructed vertex.

\subsubsection{The Acceptance of the Instrumental (`Low Level') Cuts
\label{sec:insteff}}

We measure the acceptance of the instrumental cuts using data from 
different calibration sources taken at different times.
The primary sources of data are scans using the tagged $^{16}$N source taken in
both the D$_2$O and H$_2$O regions. While these scans provide coverage
throughout the detector they are limited to the energy range of
the $^{16}$N source. We supplemented these data with the $^8$Li source,
which provides tagged electrons at higher energy, the diffuse laser source,
which provides optical photons of tunable intensity, and the pT source, which
provides untagged 19.8 MeV $\gamma$-rays.  Each source has its 
limitations but provides an important cross-check to the primary measurements
made with the $^{16}$N source.

\begin{figure}  
\begin{center}
\includegraphics[width=0.5\textwidth]{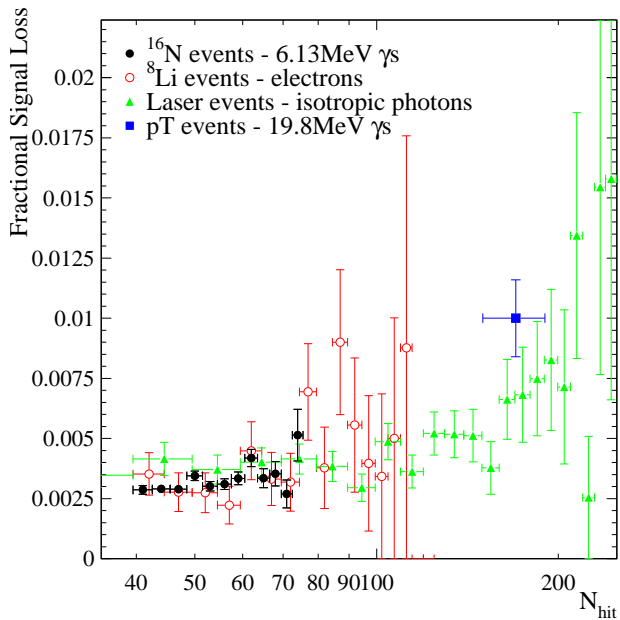}
\caption{The signal loss measured from various calibrations sources as a function of the number of hit PMTs.}
\label{fig:inst_sac}
\end{center}
\end{figure}

The signal loss measured using each of these sources as a function of
the number of hit PMTs ($N_{\rm hit}$) is shown in Figure~\ref{fig:inst_sac}.
The figure shows that the signal loss inside the range of interest for
measurements of events from $^8$B neutrinos (40-120 hits) is consistent with
being flat. The same is true as a function of position within the D$_2$O volume
and as a function of direction. The simplicity of these distributions allows
the signal loss to be easily calculated for the different classes of events.
Figure~\ref{fig:inst_sac} also shows good agreement between the
$^{16}$N and $^8$Li sources. This shows that the acceptance of the instrumental
cuts does not depend upon the particle type, and therefore we can use the
same acceptance for the electrons from the CC and ES reactions as we do for the
neutrons from the NC reaction. We obtain the central value for the signal loss
by fitting a flat distribution to the $^{16}$N and $^8$Li data. The best fit is
found to be $(0.311\pm0.007)\%$, where the uncertainty here is the statistical
uncertainty in the fit.

The systematic uncertainty in this measurement comes from a number of
sources. Uncertainties in calibrations of individual electronics channels were
checked by re-running the signal-loss measurement with the calibration quality
control flags turned off, leading to a 
one-sided systematic of $-0.021\%$. Deviations from the assumed flat distribution
provide a systematic uncertainty of $\pm 0.028\%$. The biggest
contributions to the systematic uncertainty arise from measurements of the
stability of the signal loss as a function of time. The increase in signal loss
due to faulty ADCs resulting in bad charge measurement on individual channels
was measured, resulting in a correction of $(+0.027\pm0.002)\%$.  The performance of
the instrumental cuts was monitored using periodic deployments of the 
$^{16}$N source. A systematic increase in
signal loss is observed over time, resulting in a one-sided systematic uncertainty
due to instability of $+0.11\%$. Combining these results in quadrature, the signal
loss of the instrumental cuts is found to be $(0.34^{+0.11}_{-0.03})\%$.

\subsubsection{The Acceptance of the Reconstruction Algorithm}

The reconstruction method described in Section~\ref{sec:recon} has three 
distinct failure modes: 
\begin{itemize} 
\item
The event may fail the figure-of-merit cuts, which test
how well the event fits the hypothesis of a single Cherenkov electron. The
figure-of-merit cuts act like high level cuts, and the acceptances of
the two sets of cuts are correlated. For the purpose of cut acceptance, we therefore
treat the figure-of-merit cuts together with the high level cuts.
\item
The reconstruction algorithm may not receive a good seed vertex. As described
in Section~\ref{sec:recon}, the algorithm uses a seed vertex
reconstructed using time information only. If the seed vertex lies
outside the detector, or no vertex is returned then the reconstruction
algorithm fails and the event is discarded.
\item 
The event may fail during (negative-log) likelihood function minimization. This
failure mode is relatively rare, being much less frequent than seed failure
mode.

\end{itemize}

It is difficult to know exactly how 
reconstruction acceptance varies as a function of position and
energy, because event location is, of course, not well known when reconstruction
fails.  Using scans taken with the $^{16}$N and $^8$Li sources, however,
we find the signal loss decreases with increasing energy and
increases sharply as events approach the acrylic vessel. The data from
these scans do not allow us to make a measurement of the signal
loss, but do allow us to place an upper limit of 0.3$\%$ for all
classes of signal within the fiducial volume.

\subsubsection{The Acceptance of the High Level Cuts \label{sec:hlcuteff}}

Unlike the instrumental cuts, the high level cuts rely upon timing and
hit pattern information only. For signal events these distributions
can be reproduced much more reliably by simulation than can
distributions such as the PMT charge distribution, and the
Monte Carlo simulation can be used to integrate the distributions of cut acceptance
for the high level cuts with the expected distributions for the three signals
observed in SNO.  Unlike the instrumental cuts and the reconstruction algorithm, the
high level cuts have different acceptances for each of the three signals (CC, ES and
NC).

The Monte Carlo simulation is used to calculate the acceptance for the high level
cuts because we have no electron calibration source that is unaffected by its own
hardware.  The $\theta_{ij}$ cut in particular is sensitive to the
amount of backward light. For electrons emitted by the $^8$Li source this backward
light is blocked, giving a distorted $\theta_{ij}$ distribution.  Furthermore,
events initiated by $\gamma$-rays have a different $\theta_{ij}$ distribution from
those initiated by electrons, because of the possibility that a second
Compton-scattered electron could contribute  light and produce a more isotropic hit
pattern.  Use of the Monte Carlo simulation allows these effects to be included. 

\begin{figure}  
\begin{center}
\includegraphics[width=0.5\textwidth]{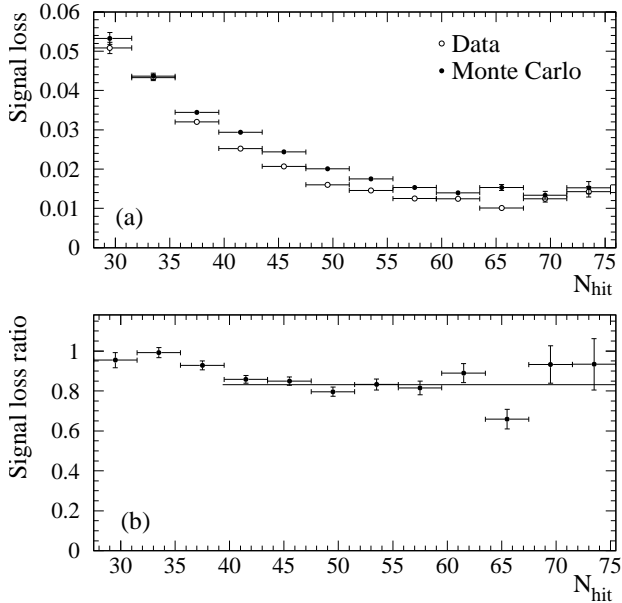}
\caption{The signal loss of the high level cuts for $^{16}$N $\gamma$-rays from data and Monte Carlo simulation (a), and the ratio of the two distributions (b).} 
\label{fig:sac_rat_hlc}
\end{center}
\end{figure}

Calibration data is not ignored, however. As
shown in Figure~\ref{fig:sac_rat_hlc}a, the
Monte Carlo simulation does not perfectly reproduce the measured signal loss for
$^{16}$N data. 
We see in the figure that the Monte Carlo simulation consistently overestimates the signal
loss, an effect which is mainly due to the non-perfect reproduction of the
$\theta_{ij}$ distribution by the simulation. To correct
for this effect, a signal loss scale factor is calculated from the ratio of the
data and Monte Carlo distributions (Fig.~\ref{fig:sac_rat_hlc}b). Above 40 hits the
scale factor is independent of the
number of hit PMTs. Below 40 hits the dominant contribution to the signal loss
changes from the $\theta_{ij}$ cut to the reconstruction figure-of-merit cuts,
resulting in a change of the scale factor. Very little of the signal data is
below 40 hits, however, and this effect can therefore be ignored. The scale factor
derived from a fit to the data above 40 hits is therefore used in the analysis.
Using the scale factor, the signal loss of electrons within the analysis region is
found to be $\sim$0.94$\%$ compared to 1.79$\%$ for neutrons.

There are three dominant contributions to the uncertainty of the
signal loss due to high level cuts. The systematic uncertainty in 
the scale factor and the statistical uncertainty in the Monte Carlo
simulated data sets contribute roughly equally to the uncertainty at the level of
0.05$\%$.  There is a much larger contribution to the uncertainty due to the
temporal stability of the cuts. Using the same $^{16}$N data set that was used
to monitor the instrumental cuts, this contribution to the uncertainty of the
signal loss is at the level of $^{+0.25}_{-0.11}\%$.

\subsubsection{Overall Cut Acceptance}
\label{sec:cutaccept}

As all three contributions to the signal loss (instrumental cuts, reconstruction
failures, and high level cuts) are small and
essentially uncorrelated, the combined signal loss can be found by
direct addition of the individual contributions. Calculation of the
uncertainty has a complication as the same data set was
used to measure many of the uncertainties due to stability. 
To account for this correlation these uncertainties are added linearly
and combined with the remaining uncertainties in quadrature. The
signal loss due to reconstruction was not measured; instead an upper
limit was placed at 0.3$\%$ and it was included as a contribution
of $0.15\pm0.15\%$ to the overall signal loss uncertainty. The final signal loss measurement for the three signals are therefore:
\begin{eqnarray*}
{\rm CC} & (1.43^{+0.39}_{-0.21})\% \\ 
{\rm ES} & (1.46^{+0.40}_{-0.21})\% \\
{\rm Neutrons} & (2.28^{+0.41}_{-0.23})\% \\ 
\end{eqnarray*}

\subsection{Target}
\subsubsection{Numbers of Deuterons}
\label{sec:nd2o}

The neutrino interaction rate depends on the number of targets within the
fiducial volume selected,  which, in turn, depends on the isotopic
enrichment and density.  The fiducial volume used, 
a 550-cm radius sphere, is defined by event
reconstruction supported by calibration.  A second volume is defined by the
acrylic vessel (AV) sphere itself, which provides both a geometrically
defined fiducial volume against which reconstruction can be checked, and
a precisely known volume of D$_2$O  that can be compared to the directly
weighed inventory.

We determined the density of the heavy water directly 
in a surface laboratory at the SNO site at temperatures in the range of
$17^\circ$ - $21^\circ$ C. We corrected to $11^\circ$ C using published
tabulations, as the actual operating temperature was $11.5^\circ$ C, at which
temperature the density differs negligibly.  Our measured density for the
heavy water is 1.10555(10) g cm$^{-3}$, and we add a
correction for compressibility which raises the value to 1.10563(10), as
the mean gauge pressure underground at SNO is 0.15 MPa.

The surveyed dimensions of the vessel, deviations from a spherical shape,
corrections for swelling and distortion, temperature, the measured D$_2$O
specific gravity and the compressibility give a calculated mass that may be
compared to the weighed inventory.  Table \ref{table:inventory} summarizes the
volumes, densities, and masses of the various components of the detector.  The
calculated mass of the D$_2$O has an uncertainty of roughly 0.3\%, dominated by
geometrical uncertainties. The volumetrically determined mass exceeds the inventory
mass by 828 kg, a discrepancy well within the estimated uncertainty. 

\begin{table}[h]
\caption{Heavy water inventory.}
\label{table:inventory}
\begin{center}
\begin{tabular}{lrl}\hline \hline
Quantity & & \\\hline
Temperature & $11.5^\circ$  & C \\
Density at 1 atm and $11.5^\circ$ C & 1.10555 & g cm$^{-3}$ \\
Mean gauge pressure in SNO & 0.15 & MPa \\
Isothermal Compressibility & 4.59 x 10$^{-4}$ & MPa$^{-1}$ \\
Corrected density & 1.10563 & g cm$^{-3}$ \\
Vessel radius as surveyed & 600.5(6) & cm \\
Vessel radius in service & 600.54(61) & cm \\
Calculated mass in sphere  & 1003049 & kg \\
Calculated mass in neck  & 8963 & kg \\ \hline \hline
\end{tabular}
\end{center}
\end{table}

The number of target deuterons also depends on  the isotopic abundance of
the heavy water.  Because neutron transport and detection are also sensitive to
the abundances, the isotopic mixture determines the characteristic radial
profile for capture events as well as the proportions of neutrons capturing
on each isotope.  The enrichment process also  affects the oxygen isotope
abundances.  The precise abundances of these isotopes are relevant for
corrections for the substantial neutron-capture cross section on $^{17}$O and
for neutrino charged-current interactions on $^{18}$O.

The hydrogen isotope mass fractions are determined by Fourier
Transform Infrared (FTIR) Spectroscopy on samples taken from the detector
volume recirculation path.  The mean measured isotopic abundance for deuterium
between October 1999 and March 2000 was 99.9176\% with
a standard deviation of 0.0023 based on 29 samples.  The corresponding
number fraction is 0.999084, which is the value we use in this article.
When measurements through November 2001 are included, the
resulting deuterium mass fraction is $99.9168\pm0.0021$\%.  The corresponding
number fraction is $99.9076\pm0.0021$\%. The largest uncertainty in the 
absolute isotopic measurement comes from
the accuracy of the standard, $\pm$0.01\%. 

The $^{16,17,18}$O isotope number
fractions were determined by analytic chemistry measurements made outside
the SNO collaboration.  Three independent techniques were applied to three
separate heavy-water samples.  Nuclear magnetic resonance~\cite{17nmr} as
well as infrared laser spectrometry was used to extract direct measurements
of the $^{17}$O and $^{18}$O number fractions~\cite{kerstel}. CO$_2$-water
equilibration was used to measure the $^{18}$O abundance~\cite{co2-h2o}.  
Recommended values are obtained by taking weighted averages
over the independent measurements.  All values are presented in
Table~\ref{tab:isotopes}.

The values given for the oxygen isotopes are very different from the ones
in~\cite{NIM} as a result of the new measurements.  A systematic error
associated with an ion-mass degeneracy is suspected to have influenced the
earlier measurement. 

The  number of molecules in a heavy water target
of mass $M$ is
\begin{widetext}
\begin{eqnarray}
N_M =  \frac{M}{2m_D f_D + 2m_H(1-f_D) + m_{17}f_{17} + m_{18}f_{18}+m_{16}(1-f_{17}-f_{18})},
\label{eq:nd}
\end{eqnarray}
\end{widetext}
\noindent where $f_D$, $f_{17}$, and $f_{18}$ are the atom-fraction
isotopic abundances of deuterium, $^{17}$O, and $^{18}$O, respectively,
and $m_i$ is the atomic mass of oxygen isotope $i$.  
There are 2$N_Mf_D$ deuterons in this target of mass $M$, so from Eq.~\ref{eq:nd} 
and the isotopic enrichment data there are$$
N_D = 6.0082(62) \times 10^{31}$$ deuterons in 1000 tonnes of SNO heavy water.
The error is from the uncertainty in the deuteron isotopic abundance 
(0.0023\% (stat.), 0.01\% (syst.)). For a given fiducial volume the 
error on the density of SNO heavy water (0.009\%) must be included.

The elastic scattering reaction similarly depends on the volume and
density. The number of electrons per mass $M$ of heavy water is 10$N_M$. 
There are thus $30.0684 \times 10^{31}$
electrons per 1000 tonnes. The dependence on composition is very weak.

\begin{table}[h]
\caption{SNO heavy water oxygen isotope number fractions. }
\label{tab:isotopes}
\begin{center}
\begin{tabular}{lcc} \hline\hline
Measurement Technique & $^{18}$O (\%) & $^{17}$O (\%) \\ \hline
IR Laser Spectrometry & 0.33$\pm$0.03 &  0.049$\pm$0.005 \\  
CO$_2$-water Equilibration &  0.320$\pm$0.006  &   0.0486$\pm$0.0009\\
$^{17}$O NMR &   0.311$\pm$0.004  &  0.0479$\pm$0.0006  \\ \hline
Recommended Values  &   0.320$\pm$0.003  &  0.0485$\pm$0.0005 \\ \hline\hline
\end{tabular}
\end{center}
\vspace{0.5cm}
\end{table}

\begin{table}[h]
\caption{Allowed nuclear matrix elements $|$M$|^2$ = BGT + BF and the resulting
cross sections, averaged over an undistorted $^8$B neutrino flux. Here, BGT stands
for the Gamow-Teller part of the matrix element, and BF the Fermi part.}
\label{table:BGT}
\begin{center}
\begin{tabular}{lllllc} \hline\hline
Target&$E_f$(MeV)& Q(MeV) & BGT&BF&$\sigma(^8$B)  \\ 
&&&&& (10$^{-42}$ cm$^2$) \\\hline 
$^{18}$O&0.0 & -1.655 &5.12&&4.14 \\
&1.04& -2.695 & &2.0&1.11 \\
&1.70& -3.355 & 0.21&&0.103 \\
\hline &Total &&&& 5.35(5) \\
\hline$^{17}$O& 0.0& -2.761 & 1.69&1.0&1.53(1) \\
$^{2}$H& 0.0 & -1.442 & & & 1.15(4) \\ \hline \hline
\end{tabular}
\end{center}
\end{table} 

\subsubsection{Other Isotopes \label{sec:iso}}

We include the terms for $^{17}$O and $^{18}$O in Equation~\ref{eq:nd} because
these rare isotopes of oxygen play a role similar to deuterium in their CC
interactions with $^{8}$B neutrinos.   Most of the cross section is due to a
super-allowed transition to the ground state, but, unlike deuterium, the final
states are narrow and stable to nucleon emission.  The interaction cross
sections have been calculated in~\cite{Haxton} and are summarized in Table~
\ref{table:BGT}.  Substituting the measured $^{17}$O and $^{18}$O abundances
gives the correction to the deuterium CC rate as 1.0078(10).  The main
uncertainty of 0.001 in this small correction factor comes from the variation
in Q-value from 1.4 to 3.3 MeV; the uncertainties in the isotopic abundances and the
matrix elements contribute very little (for pure Fermi and ground-state
Gamow-Teller matrix elements the uncertainties are 1\% or less).  The angular
distribution is slightly influenced.  For reactions on $^{18}$O it is
essentially flat while for $^{17}$O it is also weak but slightly forward 
peaked~\cite{Haxton}.  

\subsection{Neutron Capture and Detection Efficiency}
\label{sec:neutrons}
   
Several factors prevent the neutron detection efficiency from being unity.  
First, the finite D$_2$O volume means that some of the neutrons liberated from
deuterium can escape the heavy water and then capture on hydrogen in the acrylic
vessel or light water shield.  Second, free neutrons in the heavy water also have a
non-zero probability of being captured on nuclei other than deuterium, such as
hydrogen, $^{16}$O, $^{17}$O, and $^{18}$O.  
Lastly, our energy threshold and fiducial volume cuts remove a large fraction
of the 6.25 MeV capture $\gamma$-rays from the final data set.

We have measured the neutron capture efficiency by deploying a $^{252}$Cf
source at various positions throughout the heavy water volume. These
``point-source'' calibrations have been employed, together with Monte Carlo
simulation and an analytic diffusion model, to extract the capture efficiency
and its uncertainty relevant to a source of neutrons uniformly distributed
throughout the heavy water volume. As discussed in Section~\ref{sec:mcnp}, our
Monte Carlo simulation of neutron propagation and capture is based upon Los
Alamos National Laboratory's MCNP code.  
An analytic model for neutron transport in SNO has been
derived that relates the macroscopic quantities of interest such as absorption,
diffusion length, and lifetime, to the microscopic quantities such as isotopic
abundances and capture cross-sections.

\begin{figure}  
\begin{center}
\includegraphics[width=0.5\textwidth]{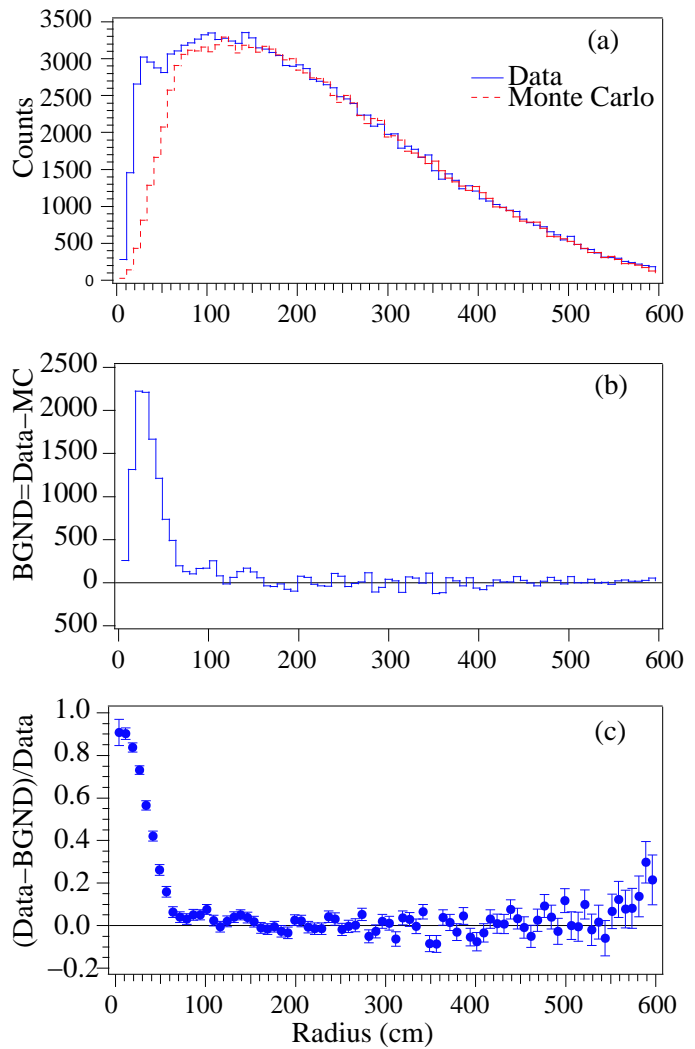}
\caption{Radial profile of neutron captures for $^{252}$Cf source deployed at
the center of the detector, compared to Monte Carlo simulation of the source.
In (a) we compare the raw data distribution of events to the Monte Carlo
simulation, and we see that the data has an excess due to the associated
$\gamma$s and $\beta$s produced by the source.  The difference between the two
curves is shown in (b), and the ratio of (b) to the data shown in (c).  The
drop off around 80~cm motivates the cut to remove the non-neutron events in
the calculation of the efficiency. \label{fig:radcf}} 
\end{center}
\end{figure}

\begin{figure}  
\begin{center}
\includegraphics[width=0.4\textwidth]{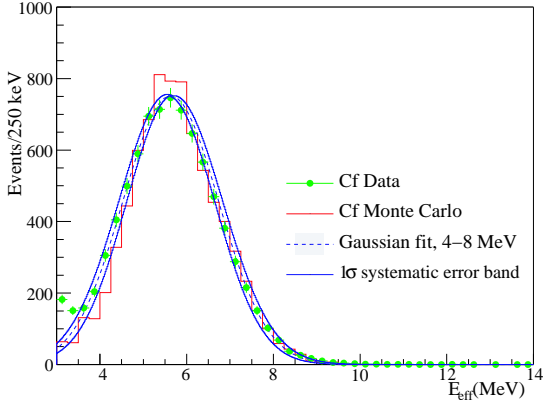}
\caption{Energy spectrum of $\gamma$-rays from neutron capture for deployment
of $^{252}$Cf source at the center of the detector. \label{fig:ngammas}}
\end{center}
\end{figure}

   The $^{252}$Cf source created fission $\gamma$s and $\beta$s as well as
neutrons.  These can contaminate the 6.25~MeV capture-gamma distribution of
interest. Since these backgrounds have a mean path length in D$_2$O that is
short in comparison to the mean neutron capture distance of about 120~cm, they
were avoided by requiring events to reconstruct more than 80~cm 
from the source. The loss of efficiency by invoking this cut is
determined via Monte Carlo simulation which accurately reproduces the radial profile of
neutron captures in the D$_2$O. An example is shown in Figure~\ref{fig:radcf}
for the radial profile obtained with the $^{252}$Cf source deployed near the
center of the heavy water volume. The associated 6.25~MeV gamma energy
distribution is shown in Figure~\ref{fig:ngammas}.

\begin{figure}  
\begin{center}
\includegraphics[width=0.4\textwidth]{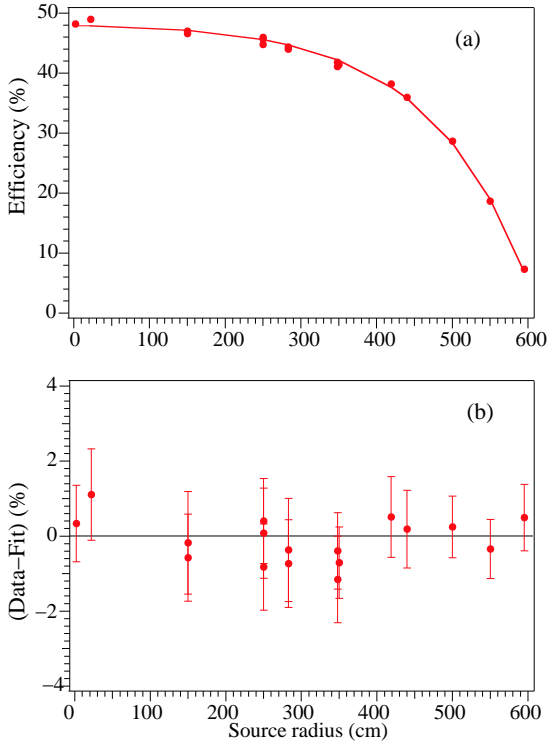}
\caption{Capture efficiency for neutrons from $^{252}$Cf source at several
locations throughout the detector volume. In (a) we compare the measurements
(filled dots) to the analytic calculation (solid curve), and in (b) the
difference between them. \label{fig:capeff}}
\end{center}
\end{figure}

   As can be seen in Fig.~\ref{fig:ngammas}, the 6.25~MeV gamma energy
distribution is well described by a Gaussian distribution. The number of
neutrons is determined from this distribution by fitting the centroid and width
to the calibration data above $T_{\rm eff}=5.0$~MeV and extrapolating the fit to zero energy.
In this way, correlated uncertainties associated with the absolute energy scale
and resolution are avoided. After correcting for the radial cut mentioned above
we obtain the total number of neutrons captured on deuterium for the $^{252}$Cf
at a given position in the detector. By knowing the livetime for a particular
calibration run and the absolute neutron yield of the $^{252}$Cf source we can
determine the capture efficiency for a point source deployed at a specific
location or radius in the detector.  Figure~\ref{fig:capeff} shows the results
from this exercise for a set of $^{252}$Cf calibration scans throughout the
detector.

   Fig.~\ref{fig:capeff} yields the absolute capture efficiency for
neutrons on deuterium that reconstruct within the D$_2$O volume when their
origin is a $^{252}$Cf source at a specific position in the detector. We need
to exploit this information to deduce the capture efficiency of interest,
namely the volume-weighted sum of neutrons captured from a source uniformly
distributed throughout the heavy water volume. To do so we require a
function to interpolate between the discrete calibration points which
can then be fed into the volume-weighted integral of interest. We have
developed an analytic neutron diffusion model that serves this purpose
well. The smooth curve in Fig.~\ref{fig:capeff} arises from a fit to the
calibration data using a two-parameter model predicting the radial profile for
point-neutron sources in the detector according to:

\begin{equation}
\epsilon(R) = \epsilon_0[1-F_{\rm escape}(R)]   
\label{eq:neutroneps}
\end{equation}
where 
\begin{eqnarray}
\label{eq:neutronesc}
F_{\rm escape}(R) & = & \frac{R_{AV}}{R}\frac{\sinh(\frac{R}{l})}{\sinh(\frac{R_e}{l})}[\cosh(\frac{R_e-R_{AV}}{l})\\ \nonumber 
& + & \frac{l}{R_{AV}}\sinh(\frac{R_e-R_{AV}}{l}) ] .
\end{eqnarray}

   In Equations~\ref{eq:neutroneps} and~\ref{eq:neutronesc}, $R$ is the
position of the point-source calibration data, measured in cm.  The leading
scale factor ($\epsilon_0$) in Eq.~\ref{eq:neutroneps} describes the capture
efficiency for the case where the SNO heavy water volume is infinite in extent.
In Eq.~\ref{eq:neutronesc}, $R_{AV}$ is the 600~cm radius of the heavy water
volume, and $R_e$ the radius at which a perfect absorber would need to be
placed to represent the effects of the acrylic and light water (roughly 15~cm
beyond the inner surface of the acrylic vessel). 

The escape of neutrons that arises due to the finite detector radius of $R_{AV}=600$~cm
and the non-zero diffusion length ($l$) explains the drop-off in efficiency for a
source closer to the AV.  A fit to the data yields:

\[ \epsilon_0=0.499 \pm 0.010 \] and 
\[ l=109.4 \pm 4.8~{\rm cm}. \]

   The same analytic diffusion model can be used to predict the capture
efficiency for a source of neutrons uniformly distributed throughout
the heavy water volume. It is described using the same two parameters
after integrating Eq.~\ref{eq:neutronesc} out to a fiducial volume of
radius $R_{f}$:

\begin{eqnarray*}
 F^{\rm NC}_{\rm escape} & = & \frac{1}{R_{AV}^3}[R_f^3 - 3l[R_f \cosh(\frac{R_f}{l})
-l\sinh(\frac{R_f}{l})] \\
& & \times \frac{[R\cosh(\frac{R_e-R_{AV}}{l}) +
l\sinh(\frac{R_e-R_{AV}}{l})]}{\sinh(\frac{R_e}{l})}] 
\end{eqnarray*}

   Using the parameters constrained in the fit to the point-source data
we deduce a NC neutron capture efficiency of:
$0.299\pm 0.011$.

   This efficiency corresponds to neutrons capturing on deuterium with an
effective detector energy threshold of zero and a full fiducial volume
of 600~cm. Monte Carlo simulation was used to determine the reduction
in efficiency relevant to our analysis threshold of $T_{\rm eff}$=5.0~MeV and 550~cm
fiducial volume.  In this case, the neutron detection efficiency relevant to our
analysis is $0.1438 \pm 0.0053$, with the breakdown of statistical and systematic
uncertainties outlined in Table~\ref{tbl:neff}.

\begin{table}
\caption{Statistical and systematic uncertainties on the neutron capture
measurement.}
\label{tbl:neff}
\begin{center}
\begin{tabular}{lc}
\hline \hline
Contribution &    Uncertainty (\%) \\
\hline
Energy Distribution &      1.74       \\
Source Standard &      2.20             \\
Source Exclusion &      0.86              \\
Source Position &      0.95          \\
\hline
Total Systematic Uncertainty &      3.09             \\
\hline
Statistics of $^{252}$Cf data   &      1.97       \\
\hline
Total Uncertainty &      3.68             \\
\hline \hline
\end{tabular}
\end{center}
\end{table}   

For verification of this `direct counting' method, we used a multiplicity analysis
that compared the number of neutrons detected per $^{252}$Cf decay to expectations
based on knowledge of the primary decay neutron multiplicity and Monte Carlo
simulation.  The results of the multiplicity analysis were in excellent agreement
with the direct counting method described above, albeit with somewhat larger
uncertainties.

\section{Final Flux Measurements \label{sec:results}}

     The cuts described in Section~\ref{sec:dataproc}, including the energy
threshold of $T_{\rm eff}=5.0$~MeV and the fiducial volume restriction of
$R_{\rm fit}<550$~cm, constitute our primary event selection criteria.  After application of
these cuts to the full data set,  2928 candidate neutrino events remain, and the
signal extraction fit is performed on this event sample.

     As a consistency check, the signal extraction fit was repeated using the total
number of hit tubes ($N_{\rm hit}$) as the estimate of event energy rather than the
prompt-time reconstructed energy described in Section~\ref{sec:enecal}.  For this
$N_{\rm hit}$-based analysis the energy threshold cut was replaced by a cut of
$N_{\rm hit} \ge 45$, chosen to give a total number of events in the final data
sample that matched the number using the cut on effective energy.  We further
explored the dependence on fiducial volume by performing fits to data that used both
tighter and looser radial cuts, including out into the H$_2$O volume.

	As discussed in Section~\ref{sec:sigex}, in our primary approach to
signal extraction, we used pdfs generated by the Monte Carlo.  For
verification, we also performed an extraction using pdfs generated by analytic
parameterizations of the response, as described in
Section~\ref{sec:anal_response}.  The analytic approach was also used for
our estimation of the neutrino mixing parameters, as discussed in
Appendix~\ref{sec:physint}.

This section will concentrate on the derivation of the flux results from the
Phase I NC paper~\cite{snonc}, but in Section~\ref{sec:ccprl_results} we will
comment on the high energy threshold analysis in the ES-CC paper~\cite{snocc}.
The Day-Night asymmetry measurement will be discussed in
Section~\ref{sec:daynight}.

\subsection{Spectrum-Constrained Fluxes}

The primary signal extraction was performed as described in Section~\ref{sec:sigex},
with three signal pdfs (plus background pdfs) in $T_{\rm eff}, R^3$, and $\cos \theta_\odot$,
with the CC and ES event energy spectra constrained to follow their expected shapes
for an undistorted $^{8}$B $\nu$ spectrum~\cite{ortiz}.  The raw numbers of
extracted signal events of each type are given in Table
\ref{tab.constrained_totals_rsp}.  The errors quoted here are symmetric parabolic
errors as calculated by MINUIT's HESSE routine~\cite{MINUIT}, and are very similar
to the MINOS asymmetric errors.  Table \ref{tab.constrained_corr_rsp} shows the full
correlation matrix for the signals obtained in the extraction process.

\begin{table}[t] 
\caption{Extracted numbers of CC, ES, and NC events in the full D$_2$O
  data set, with a $^{8}$B spectral constraint on the CC and ES
  spectra.  Errors are statistical only.  Note that the backgrounds discussed in
Section~\ref{sec:bkds} have been subtracted off in the manner discussed in
Section~\ref{sec:sigex}.
\label{tab.constrained_totals_rsp}
}
\begin{center}
\begin{tabular}{lr}
\hline \hline
Signal & Events \\
\hline
CC & $1967.71 \pm 61.36$ \\
ES & $263.64 \pm 25.68$ \\
NC & $576.53 \pm 48.82$ \\
\hline \hline
\end{tabular}
\end{center} 
\end{table}

\begin{table}[t] 
\caption{Statistical correlation matrix between CC, ES, and NC signals
  from the signal extraction with a $^8$B shape constraint.
\label{tab.constrained_corr_rsp}
}
\begin{center}
\begin{tabular}{lrrr}
\hline \hline
 & CC & ES & NC \\
\hline
CC & 1.000 & -0.162 & -0.520 \\
ES & -0.162 & 1.000 & -0.105 \\
NC & -0.520 & -0.105 & 1.000 \\
\hline \hline
\end{tabular}
\end{center} 
\end{table}

The raw number of extracted events of each signal type may be converted
to a flux through Equation~\ref{eq:fluxcalc}, which yields a flux in
units of $10^6$~neutrinos/cm$^2$/sec:
\begin{equation}
\phi_i = \frac{N_i}{N_{\rm MC}} \cdot L  \cdot
\frac{1}{\epsilon_{\rm cuts}} \cdot f_{{\rm O}}  \cdot {\cal E}
\cdot X 
\label{eq:fluxcalc}
\end{equation}

The various quantities are defined as:
\begin{description}
\item $N_i$:\\  Number of extracted events for a given signal type $i$,
as given in Table~\ref{tab.constrained_totals_rsp}
\item $N_{\rm MC}$:\\  Number of Monte Carlo events inside the signal region,
  for a total $^8$B flux of $1 \times 10^6$/cm$^2$/sec.  The number of
events we generated was 50 times the BPB2000 SSM prediction of 
$5.15\times 10^6 \nu~{\rm cm}^{-2}{\rm s}^{-1}$~\cite{bp2000}.
\item L:\\ Livetime correction factor.  This
  correction accounts for detector
  deadtime due to the imposition of time-correlated cuts (such as those that remove muon follower events).
\item $\epsilon_{\rm cuts}$:\\  Acceptance of low and high level cuts, as
described in Sections~\ref{sec:insteff} and~\ref{sec:hlcuteff}, that
are not applied to the Monte Carlo simulation. 
\item $f_{{\rm O}}$:\\ A correction to the CC flux due to CC neutrino
  interactions on $^{17}$O and $^{18}$O, as described in
Section~\ref{sec:iso}.  These interactions are not modeled in the
  Monte Carlo simulation.  This correction is applicable only to the CC flux.
\item ${\cal E}$:\\  Correction for eccentricity of the Earth's
orbit, which was not included in the Monte Carlo generation.  
\item $X$:  Minor corrections to the neutrino cross sections assumed in the 
Monte Carlo simulation.
For the CC and NC fluxes, this is a combination of the $g_A$
correction to the Butler, Chen, and Kong (BCK) cross section~\cite{BCK}, a downward revision of the NSGK cross section~\cite{NSGK}, and radiative corrections of
Kurylov~\etal~\cite{kmv}.  See
Section~\ref{sec:nuspecint} for further details. 
\end{description}

Table~\ref{tab.flux_correction_factors} contains the values of the
flux correction factors used for each signal.

\begin{table*}[t] 
\begin{center}
\caption[Flux correction factors for converting event totals to
fluxes.]
{Flux correction factors for converting event totals to fluxes.
The final entry (``Total'') is the product of all the corrections
which are applied to the ratio $N_i/MC$ to convert it into a flux 
in units of $10^6$ neutrino/cm$^2$/sec.
\label{tab.flux_correction_factors}
}
\begin{tabular}{lrccc}
\hline \hline
Correction & Symbol & CC & ES & NC \\
\hline
Livetime & Total/Corrected & 312.93/306.39 & 312.93/306.39 & 312.93/306.39 \\
Cut efficiency & $\epsilon_{\rm cuts}$ & $0.986^{+0.004}_{-0.002}$ & $0.985^{+0.004}_{-0.002}$ & $0.977^{+0.004}_{-0.002}$ \\
 $^{17}$O and $^{18}$O correction &  $f_{{\rm O}}$& 1/1.00793 & 1 & 1 \\
Eccentricity correction & ${\cal E}$ & 1/1.0069 & 1/1.0069 & 1/1.0069 \\
Cross section correction & X & 1/1.0162 & 1.02 & 1/1.0112 \\
\hline
Total Correction Factor &  & 1.0043 & 1.0500 & 1.0267 \\
\hline
Number of events  & $N$ & $1967.71 \pm 61.36$ & $263.64 \pm 25.68$ & $576.53 \pm 48.82$ \\
MC Prediction (for $10^6$ $\nu~{\rm cm}^{-2}{\rm s}^{-1}$) & $N_{\rm MC}$ & 1120.48 & 115.83 & 116.23 \\
\hline \hline
\end{tabular}
\end{center} 
\end{table*}

With all of these corrections applied, the extracted signal fluxes
are (statistical errors only):

\begin{center}
$\phi_{CC} = 1.76 ^{+0.06}_{-0.05} \times 10^{6}$~cm$^{-2}$~s$^{-1}$ \\
$\phi_{ES} = 2.39 ^{+0.24}_{-0.23} \times 10^{6}$~cm$^{-2}$~s$^{-1}$ \\
$\phi_{NC} = 5.09 ^{+0.44}_{-0.43} \times 10^{6}$~cm$^{-2}$~s$^{-1}$ \\
\end{center}

The physical interpretation of the ``flux'' for each interaction type
is that it is the equivalent flux of $^8$B $\nu_e$s produced from an
undistorted energy spectrum that would yield the same number of
events inside the signal region from that interaction as was seen in
the data set.

\begin{figure}
\includegraphics[width=3.7in]{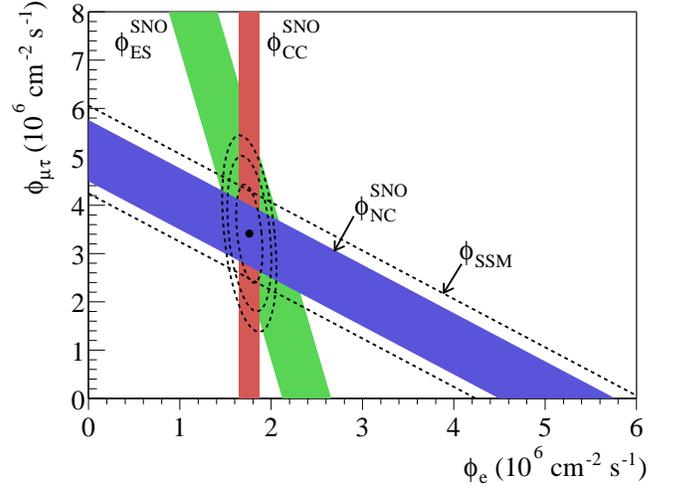}
\caption{\label{fig:hime_plot}Flux of ${}^{8}$B solar neutrinos which are
$\mu$ or $\tau$ flavor vs flux of electron neutrinos deduced from the
three neutrino reactions in SNO.  The diagonal bands show the total
${}^{8}$B flux as predicted by the BP2000 SSM~\cite{bp2000} (dashed lines)
and that measured with the NC reaction in SNO (solid band).  The
intercepts of these bands with the axes represent the $\pm 1\sigma$
errors.  The bands intersect at the fit values for $\phi_{e}$ and
$\phi_{\mu\tau}$, indicating that the combined flux results are
consistent with neutrino flavor transformation with no distortion
in the ${}^{8}$B neutrino energy spectrum.}
\end{figure}

The inequality of the CC, ES, and NC fluxes provides strong evidence
for a non-$\nu_e$ component to the $^8$B neutrino flux.
Figure~\ref{fig:hime_plot} shows the constraints on the flux of
$\nu_e$ versus the combined $\nu_{\mu}$ and $\nu_\tau$ fluxes derived from the CC, ES, and NC
rates.  Together the three rates are inconsistent with the
hypothesis that the $^8$B flux consists solely of $\nu_e$s, but are
consistent with an admixture consisting of $\sim 1/3~\nu_e$ and $2/3~
\nu_\mu$ and/or $\nu_\tau$.
\

\subsubsection{Goodness of Fit}
 
The signal extraction is done by a maximum likelihood fit which
does not readily yield an absolute goodness-of-fit parameter.  One
means of investigating the goodness of fit of the signal extraction is
to calculate the $\chi^2$ of the radial, energy, and angular marginal
distributions between the data and the best-fit sum of the weighted
pdfs.
This $\chi^2$ is defined as:
\begin{equation}
\chi^2 = \sum_{i=1}^{bins} (R_{\rm DATA}(i) - R_{\rm pdfs}(i))^{2}/R_{\rm DATA}
\end{equation}
Here, $R_{\rm DATA}(i)$ is the number of counts in the $i$th bin of the
data ($R$ may be a histogram in energy, angle, or radius).
$R_{\rm pdfs}(i)$ is the predicted number of counts in the $i$th bin,
found by weighting each signal pdf by the number of fitted events and
summing these renormalized pdfs.  This $\chi^2$ calculation does not
account for systematic uncertainties.
 
\begin{table}[t]
\caption[$\chi^2$ values between data and fit for the energy, radial,
and angular distributions, for a constrained fit]
{$\chi^2$ values between data and fit for the energy, radial,
and angular distributions, for the fit using the constraint that the 
effective kinetic energy spectrum results from an undistorted
$^8$B shape.
\label{tab.chi2_constrained}
}
\begin{center}
\begin{tabular}{lcr}
\hline \hline
Distribution & Number of Bins & $\chi^2$ \\
\hline
Energy & 42 & 34.58 \\
Radius & 30 & 39.28 \\
Angle & 30  & 19.85 \\
\hline \hline
\end{tabular}
\end{center}
\end{table}

Table~\ref{tab.chi2_constrained} shows the $\chi^2$ values for the fits using
the constraint that the effective kinetic energy spectrum results from an
undistorted $^8$B shape.  In each case the $\chi^2$ per degree of freedom is
close to one. One must be cautious in interpreting these results.  Although
the signal extraction fit has 3 free parameters, one should not subtract 3
degrees of freedom for each $\chi^2$, since the fit is a global fit to all
three distributions.  Furthermore, the actual signal extraction is a fit to the
three-dimensional data distribution, whereas the $\chi^2$s are calculated with
the marginal distributions.  These ``$\chi^2$'' values demonstrate that the
weighted sum of the signal pdfs provides a good match to the marginal energy,
radial, and angular distributions.

\begin{figure}
\includegraphics[width=3.73in]{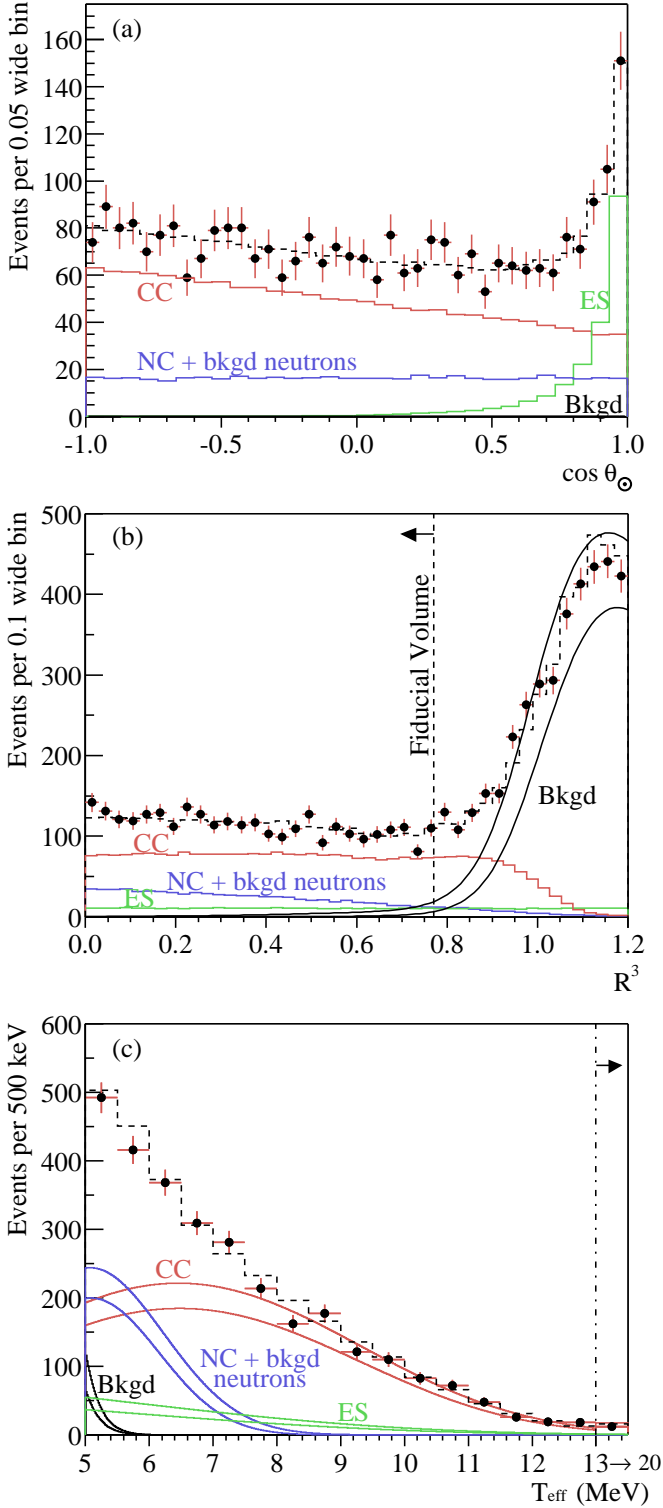}
\caption{\label{fig:r3_comp}(a) Distribution of $\cos\theta_{\odot}$ for
 $R_{\rm fit} \le 550$ cm. (b) Distribution of the radial
 variable $R^{3}=(R_{\rm fit}/R_{AV})^3$.  (c) Kinetic energy for $R_{\rm fit} \le 550$~cm.  Also
shown are the Monte Carlo predictions for CC, ES and NC + background neutron
events scaled to the fit results, and the calculated spectrum of
$\beta$-$\gamma$ background (Bkgd) events.  The dashed lines represent the
summed components, and the bands show $\pm 1\sigma$ statistical uncertainties
from the signal extraction fit.  All distributions are for events with $T_{\rm
eff}$$\geq$5 MeV. } 
\end{figure}

Figure~\ref{fig:r3_comp} shows the marginal radial, angular, and
energy distributions of the data along with Monte Carlo predictions
for CC, ES and NC + background neutron events, scaled by the fit
results.

\subsubsection{Results of Fitting for Flavor Content}

An alternative approach to doing a null hypothesis test for neutrino flavor
conversion, as discussed in Section~\ref{sec:fit_flavor}, is to fit for the fluxes
of $\nu_e$ and $\nu_{\mu\tau}$ directly.  This is a simple change of variables to
the standard signal extraction.  Fitting for the flavor content instead of the 3
signal fluxes, we find:

\begin{center}
$\phi(\nu_e) = 1.76 \pm 0.05 \times 10^{6}$~cm$^{-2}$~s$^{-1}$ \\
$\phi(\nu_{\mu\tau}) = 3.41 \pm 0.45 \times 10^{6}$~cm$^{-2}$~s$^{-1}$ \\
\end{center}

The statistical correlation coefficient between these values is $-0.678$.  
We will discuss the statistical significance of the non-zero $\phi(\nu_{\mu\tau})$
flux in Section~\ref{sec:finflux} where we include the systematic uncertainties.

\subsection{Sensitivity of Results to Choice of Energy Threshold,
  Fiducial Volume, and Energy Estimator}

To verify the stability of the extracted flux results, we repeated the
signal extraction and flux calculations with different choices of
fiducial volume, energy threshold, and energy estimator.  These
variations included restricting the fiducial volume to 500~cm and
450~cm, extending the fiducial volume to 620~cm and including external background pdfs, raising the energy threshold to $T_{\rm eff}> 5.5$~MeV,
and using the $N_{\rm hit}$ variable instead of the calibrated energy in MeV as
the energy variable in the signal extraction.  All of these variations
produced fluxes that agreed with the primary analysis within the
expected uncertainties.

\subsection{Inclusion of Additional $(\alpha,n)$ Neutrons from the Acrylic
\label{sec:sigextns}}

	As discussed in Section~\ref{sec:otherneutrons}, our estimates of the
contribution from neutrons produced by radioactivity external to the heavy
water volume are based on measurements of the U and Th content of the acrylic
and light water, and on expectations for the resultant number of
photodisintegration neutrons that pass our energy threshold and fiducial volume
cuts. In addition to these neutrons, $(\alpha,n)$ reactions on 
nuclei in the acrylic vessel are also a source of `external' neutrons, but  are
not included in our overall background estimates that lead to our neutrino
flux measurements.

To determine the effects of the inclusion of this background, subsequent to the
publication of the results in the NC paper~\cite{snonc}  we performed a signal
extraction fit in which we allowed the amplitude of the external neutron
background to float.  The results of this fit with a floating
external neutron background are consistent with the results with the background
level constrained to the value in Table~\ref{tbl:bkgd_summary} to within 
uncertainties.  Their inclusion would thus lead to a small increase in our
overall quoted systematic uncertainty. In a future publication we will include
updates to the flux measurements contained in this article that will explicitly
incorporate the (minor) effects of this background.

\subsection{Spectrum-Unconstrained Flux Results}

One can produce an ``unconstrained'' NC flux result, requiring
no assumptions about the CC energy spectrum, by doing an extraction
based only upon $R^3$ and $\cos \theta_{\odot}$.  This can be easily
implemented as a binned maximum likelihood fitter by setting the
number of energy bins in the pdfs to 1.  The resulting fit is 
equivalent to performing an unconstrained signal extraction with a single CC
spectral bin.  

Table~\ref{tab.unconstrained_nc_results} shows the results of the
unconstrained fit on the data set.
Table~\ref{tab.unconstrained_nc_corr_rsp} shows the correlation
matrix.  The anticorrelation between the CC and NC signals is nearly
-100\%.

\begin{table}[t] 
\begin{center}
\caption{Extracted numbers of events and fluxes for SNO's full D$_2$O
  data set, derived with no constraint on the shapes of the CC and ES
  energy spectra.Note that the backgrounds discussed in
Section~\ref{sec:bkds} have been subtracted off in the manner discussed in
Section~\ref{sec:sigex}.
\label{tab.unconstrained_nc_results}
}
\begin{tabular}{lrr}
\hline \hline
Signal & Events & Flux \\
\hline
CC & $1833.38 \pm 173.76$ & $1.64 \pm 0.16 \times 10^6$ \\
ES & $253.21 \pm 26.64$ & $2.30 \pm 0.24 \times 10^6$ \\
NC & $717.71 \pm 176.97$ & $6.42 \pm 1.57 \times 10^6$ \\
\hline \hline
\end{tabular}
\end{center} 
\end{table}

\begin{table}[t] 
\begin{center}
\caption[Correlation Matrix, Unconstrained Signal Extraction for NC
flux, $5 < T < 19.5$~MeV]
{Correlation Matrix, Unconstrained Signal Extraction for NC flux, $5
< T < 19.5$~MeV
\label{tab.unconstrained_nc_corr_rsp}
}
\begin{tabular}{lrrr}
\hline \hline
 & CC & ES & NC \\
\hline
CC & 1.000 & 0.208 & -0.950 \\
ES & 0.208 & 1.000 & -0.297 \\
NC & -0.950 & -0.297 & 1.000 \\
\hline \hline
\end{tabular}
\end{center} 
\end{table}

\subsection{Systematic Uncertainties}
\label{sec:sysunc_results}

	Three separate classes of systematic uncertainties need to be
propagated to the final flux calculation: uncertainties on the background
estimates, uncertainties that affect only the flux normalization, and 
uncertainties on the model used to generate the pdfs.  The last of these can 
affect both the pdf shapes and the overall normalization.  The handling of
background uncertainties is described in Section~\ref{sec:background_pdfs}, and
uncertainties on the backgrounds themselves are discussed in
Section~\ref{sec:bkds} and summarized in Table~\ref{tbl:bkgd_summary}.
Uncertainties that affect only the overall flux normalization---uncertainties
on acceptance loss of the applied cuts, on neutron capture efficiency, and
on the D$_2$O target---are applied directly to the final flux calculation.
Section~\ref{sec:norm} discussed these normalization uncertainties.

Systematic uncertainties that affect both the shapes of the pdfs and the
overall normalization are propagated to the final flux measurements by shifting
the radius, angle, or energy of the Monte Carlo events used to form the signal
pdfs, or, for the extraction using analytic pdfs described in
Section~\ref{sec:anal_response}, by varying the analytic detector response
parameters within their uncertainties.  These uncertainties are each discussed
in detail in Section~\ref{sec:sysunc}, and include uncertainties on the energy
scale, resolution, and non-linearity; vertex accuracy and resolution; and
angular resolution.

The effect of the shape-related systematic uncertainties is determined
by separately shifting the value of each affected parameter by its $\pm 1\sigma$
uncertainty, and then repeating the signal extraction and flux calculation with
the shifted pdfs.  For example, to model the $\pm 1.2\%$ systematic uncertainty
in the overall energy scale, the energies of all Monte Carlo events are first
shifted upward by 1.2\%, a set of perturbed pdfs is generated, and these
perturbed pdfs are used to perform a signal extraction and flux calculation.
Then a similar set of perturbed pdfs with the energies shifted by $-l$1.2\% is
generated and used.   For uncertainties affecting resolutions, the resolution
is `shifted' by convolving the pdfs with a Gaussian distribution. The Gaussian
convolution smears the pdfs, thus acting like a broadened resolution function.  

The perturbations to the pdf shapes are only applied to the signal pdfs, not to
background pdfs.  As described in Section~\ref{sec:background_pdfs}, the
amplitudes of the background pdfs are themselves varied between their $\pm
1\sigma$ limits, and these uncertainties are typically so large (30-50\%) that
they dominate over any shape-related uncertainty.  We have studied a number of
perturbations on the background pdf shapes themselves, such as varying their
radial profiles over wide ranges, from steeply sloped to almost flat, and have
seen negligible flux changes.  Generally speaking, the background pdfs fall so
rapidly in energy, that including them in the fit almost always tends to 
reduce the number of NC events in the lowest energy bin.

For the constrained fit in which one fits for the CC, ES, and NC
fluxes simultaneously, the systematic uncertainties are themselves
correlated between the different signals.  For the fit to the flavor
content ($\phi(\nu_e)$ and $\phi(\nu_{\mu\tau})$ described in
Section~\ref{sec:fit_flavor}, these correlations simplified---while there are
correlations between the electron and muon or tau neutrino fluxes, the null
hypothesis test is a simple one-variable test on $\phi(\nu_{\mu\tau})$.

\begin{center}
\begin{table*}
\caption{\label{tab.syserrors_constrained}Systematic uncertainties on
  the fluxes for the shape-constrained signal extraction. The relative
  ordering of the upper and lower  uncertainties  on the fluxes indicates that a systematic is
  correlated between two fluxes (same sign ordering) or anticorrelated
  (reverse sign ordering).  The `Experimental uncertainty' listed in the bottom
row refers to the contribution from systematic uncertainties propagated 
through the signal extraction process, but does not include normalization or
efficiency uncertainties, or theoretical uncertainties.
}
\begin{ruledtabular}
\begin{tabular}{llllll}
Source        &  CC Uncertainty.          & NC Uncertainty       & ES
                    Uncertainty & $\phi_e$ Uncertainty  &
                    $\phi_{\mu\tau}$ Uncertainty    \\ 
                    & (percent)         &(percent)
                    &(percent)&(percent)&(percent)\\ 
		    
Energy scale          & -4.2,+4.3 & +6.1,-6.2  & -3.1,+3
& +10.3,-10.4  \\
Energy resolution     & -0.9,+0.0 & +4.4,-0.0  & -0.4,+0.0 & -1.0,+0.0
& +6.8, -0.0  \\
Energy non-linearity  & -0.1,+0.1 & +0.4,-0.4 & 0.0        & -0.1,+0.1
& +0.6,-0.6 \\
Vertex resolution     &  0.0      & -0.1,+0.1 & 0.0        & 0.0 
& -0.2,+0.2    \\
Vertex accuracy       & -2.8,+2.9 & -1.8,+1.8  & -2.9,+2.9 & -2.8,+2.9
& -1.4,+1.4    \\
Angular resolution    & -0.2,+0.2 &  -0.3,+0.3 & +2.1,-2.0 & -0.1,+0.1
& +0.3,-0.3 \\
Internal source pd    & 0.0       &  -1.5,+1.6 & 0.0       & 0.0
& -2.0,+2.2  \\
External source pd    & -0.1,+0.1 &  -1.0,+1.0 & -0.1,+0.1 & -0.1,+0.1
& $\pm 1.4$    \\
D$_2$O $\beta$-$\gamma$      & -0.1,+0.2 & +1.2,-2.6  & +0.5,-0.2 & -0.1,+0.3
& +1.7,-3.7  \\
H$_2$O $\beta$-$\gamma$      & 0.0       &  -0.2,+0.4 & -0.1,+0.2 & 0.0
& -0.2,+0.6  \\
AV $\beta$-$\gamma$          & 0.0       &  -0.2,+0.2 & -0.1,+0.1 & 0.0
& -0.3,+0.3  \\
PMT $\beta$-$\gamma$         & -0.1,+0.1 &  +1.6,-2.1 & +0.1,-0.1 & -0.1,+0.1
& +2.2,-3.0  \\
Neutron capture       & 0.0       &  -4.0,+3.6 & 0.0       & -0.1,+0.1
& -5.8,+5.2  \\
Cut acceptance        & -0.2,+0.4 & -0.2,+0.4  & -0.2,+0.4 & -0.2,+0.4
& -0.2,+0.4  \\ \hline
Experimental uncertainty & -5.2,+5.2 &  -8.5,+9.1 & -4.8,+5.0 &
-5.3,+5.4 & -13.2,+14.1 \\ \hline
Cross section & $\pm 1.8$  & $\pm 1.3$ & --- & --- & $\pm 1.4$
\end{tabular}
\end{ruledtabular}
\end{table*}
\end{center}

Table~\ref{tab.syserrors_constrained} contains the systematic
uncertainties on the three signals and on the flavor-dependent fluxes.  Several
things should be noted.  First, separate positive and negative errors
are given for each systematic.  The ordering of signs on the systematic
uncertainties between columns indicates the sign of the correlation between
the signals in each column: same-sign ordering indicates correlations between
elements, while opposite sign ordering indicates anti-correlations.

Table \ref{tab.syserrors_unconstrained} gives the systematic errors
for the unconstrained analysis (fitting only with $R^3$ and $\cos
\theta_{\odot}$ between $5 < T_{\rm eff} < 19.5$~MeV).  The systematics must be
propagated separately for this fit, since the sensitivity to each
systematic has now changed.  For example, since the radial profile of
the signals is the dominant factor for separating CC from NC events,
systematics that affect the radial profiles, such as radial shift or
the amplitude of the AV $\beta-\gamma$ background (which has a steeply
changing radial profile), will have a much larger effect than they
have for the constrained signal extraction.

\begin{center}
\begin{table*}
\caption{\label{tab.syserrors_unconstrained}Systematic uncertainties on
  the fluxes for the shape-unconstrained signal extraction. The
  relative ordering of the upper and lower 
  uncertainties  on the fluxes indicates that a systematic is
  correlated between two fluxes (same sign ordering) or anticorrelated
  (reverse sign ordering). The `Experimental uncertainty' listed in the bottom
row refers to the contribution from systematic uncertainties propagated 
through the signal extraction process, but does not include normalization or
efficiency uncertainties, or theoretical uncertainties.
}
\begin{ruledtabular}
\begin{tabular}{llll}
Source        &  CC Uncertainty.          & NC Uncertainty       & ES
                    Uncertainty\\
                    & (percent)         &(percent)
                    &(percent) \\
		    
Energy scale          & -1.3,+1.4 & -3.7,+4.2 & -2.2,+2.3 \\
Energy resolution     & -0.0,+0.3 & -0.0,+0.2 & 0.0 \\
Energy non-linearity  & 0.0       & 0.0 & 0.0 \\
Vertex resolution     & -0.4,+0.5 & +0.9,-0.8 & -0.1,+0.1 \\
Vertex accuracy       & -0.8,+1.0 & -5.9,+5.6 & -2.3,+2.3 \\
Angular resolution    & -0.2,+0.3 & -1.2,+1.1 & +2.2,-2.0 \\
Internal source pd    & 0.0       & -1.1,+1.2 & 0.0 \\
External source pd    & -1.0,+1.0 & +1.1,-1.1 & -0.3,+0.3 \\
D$_2$O $\beta$-$\gamma$      & -0.6,+0.3 & -0.3,+0.1 & -0.2,+0.1 \\
H$_2$O $\beta$-$\gamma$      & -0.9,+2.1 & +1.6,-3.6 & -0.2,+0.7 \\
AV $\beta$-$\gamma$          & -1.2,+0.9 & +2.1,-1.6 & -0.4,+0.3 \\
PMT $\beta$-$\gamma$         & -1.0,+0.7 & +0.8,-0.6 & -0.3,+0.3 \\
Neutron capture       & -0.1,+0.1 & -3.6,+3.4 & 0.0 \\
Cut acceptance        & -0.2,+0.4 & -0.2,+0.4 & -0.2,+0.4 \\ \hline
Experimental uncertainty & -2.7,+3.2 & -9.1,+8.6 & -4.0,+4.2 \\
\hline
Cross section & $\pm 1.8$  & $\pm 1.3$ & --- \\
\end{tabular}
\end{ruledtabular}
\end{table*}
\end{center}

\subsection{Final Fluxes \label{sec:finflux}}

Combining the statistical, systematic, and theoretical uncertainties
our final extracted flux values for the constrained fit are: 

\begin{center}
$\phi_{CC} =
  1.76^{+0.06}_{-0.05}\mbox{(stat.)}^{+0.09}_{-0.09}~\mbox{(syst.)}
  \times 10^{6}$~cm$^{-2}$~s$^{-1}$ \\ 
$\phi_{ES} = 2.39^{+0.24}_{-0.23}\mbox{(stat.)}^{+0.12}_{-0.12}~\mbox{(syst.)}
\times 10^{6}$~cm$^{-2}$~s$^{-1}$ \\ 
$\phi_{NC} = 5.09^{+0.44}_{-0.43}\mbox{(stat.)}^{+0.46}_{-0.43}~\mbox{(syst.)}
\times 10^{6}$~cm$^{-2}$~s$^{-1}$ \\ 
\end{center}

\begin{center}
$\phi(\nu_e) =
  1.76^{+0.05}_{-0.05}\mbox{(stat.)}^{+0.09}_{-0.09}~\mbox{(syst.)}
  \times 10^{6}$~cm$^{-2}$~s$^{-1}$ \\ 
$\phi(\nu_{\mu\tau}) =
  3.41^{+0.45}_{-0.45}\mbox{(stat.)}^{+0.48}_{-0.45}~\mbox{(syst.)}
  \times 10^{6}$~cm$^{-2}$~s$^{-1}$ \\ 
\end{center}

Adding the statistical and systematic errors in quadrature, we find
that $\phi(\nu_{\mu\tau})$ is $5.3\sigma$ away from its null
hypothesis value of zero.

The `unconstrained NC flux', derived from fitting the data between
$5 < T_{\rm eff} < 19.5$~MeV only in
$R^3$ and $\cos \theta_{\odot}$, is:
\begin{center}
$\phi_{NC} =
  6.42^{+1.57}_{-1.57}\mbox{(stat.)}^{+0.55}_{-0.58}~\mbox{(syst.)} \times
10^{6}$~cm$^{-2}$~s$^{-1}$. \\ 
\end{center}

Both measurements of the total active fluxes $\phi_{NC}$, as well as the sum of 
$\phi(\nu_e)+\phi(\nu_{\mu\tau})$, are in good agreement with Standard Solar
Models~\cite{BP01,TC}.

\subsection{Verification with Analytic pdfs}
\label{sec:analpdf_ver}

As an independent check on the results of the previous sections, we
also fit the signals using pdfs generated with the analytically
parameterized detector responses as described in
Section~\ref{sec:anal_response}.  The propagation of systematic
uncertainties was also done analytically, by directly varying the
parameters in the analytical pdfs (rather than perturbing Monte Carlo
pdfs through smearing).  The analytic pdf method yielded results in
close agreement with the flux extraction using Monte Carlo pdfs.
Further details of this approach can be found in Ref.~\cite{the:mgb}.

\subsection{Results from Analysis with a High Energy Threshold
\label{sec:ccprl_results}}

The SNO collaboration's first physics publication, the ES-CC
paper~\cite{snocc},
presented the results of an analysis of the first 240.95 livedays of
SNO's D$_2$O data  using a high
kinetic energy threshold of 6.75~MeV.  Such a high energy threshold strongly
rejects low-energy background events from $\beta$-$\gamma$ decays and
reduced the need for a detailed characterization of all
backgrounds.  The high energy threshold also removes most
neutron events from the data set, so no attempt was made to produce a neutral
current measurement in that paper.  Instead, we chose to concentrate on
a CC flux result which, when combined with precise ES rate
measurements from Super-Kamiokande, provided the first direct evidence
that solar neutrinos change flavor.

The analysis in the ES-CC paper~\cite{snocc} is similar to that presented for the
full analysis of the complete D$_2$O data set described previously in
this section.  The only significant differences in the earlier analysis,
other than the different energy thresholds and the data set, are
\begin{itemize}
\item The high threshold analysis used only CC, ES, and neutron pdfs,
  with no background pdfs.  Limits on the number of background events
  were applied directly to the extracted numbers of CC and ES events.
\item No effort was made to determine the absolute neutron capture
  efficiency or the levels of uranium and thorium in the detector.
  Although the number of neutron events was extracted in the fit, we
  did not attempt to subtract neutron backgrounds or to
  convert this number into an NC flux.
\item An unconstrained CC energy spectrum was extracted from the data
  by fitting bin-by-bin for the number of CC events while constraining
  the NC and ES energy pdfs to have their nominal shapes, as described
  in Section~\ref{sec:sigex}.
\end{itemize}

\subsection{Analysis Verification Summary \label{sec:anal_ver}}

	As described in Section~\ref{sec:overview} and discussed throughout this
article,  for nearly every major analysis component we used one or more alternate
methods as a verification.  Table~\ref{tbl:analcomps} lists the multiple methods for
each component, as well as which one was used for the final flux numbers listed in
this section.  In some cases (such as the background estimates) the two methods were 
combined for the final measurements.
\begin{table*} 
\caption{Primary and secondary analysis methods used for verification.}
\label{tbl:analcomps}
\begin{center}
\begin{tabular}{llll}
\hline \hline
Component&    Primary approach & Verification approach & Section reference \\
\hline
Instrumental background cuts & Cut Set A & Cut Set B  & Section~\ref{sec:inst} and \\
                       &           &            & Appendix~\ref{sec:apdxa} \\ 
High Level (`Cherenkov Box') cuts & $\theta_{ij}$ vs. In-time ratio  & Two-pt. correlation function & Section~\ref{sec:cerbox} \\
                                  &                                  & vs. In-Time ratio \\ 
Vertex and direction reconstruction & Time+Angle fit  & Time-only fit &
Section~\ref{sec:recon} \\ 
Energy estimation & Energy reconstructor & $N_{\rm hit}$ & Section~\ref{sec:enecal} \\ 
Internal $\beta$-$\gamma$ backgrounds & Monte Carlo pdfs & Rn `spike' data &
Section~\ref{sec:certail}\\ 
External $\beta$-$\gamma$ backgrounds & Calibration source pdfs  & Monte Carlo model &
Section~\ref{sec:lowext} \\ 
Photodisintegration background & In-situ+Ex-situ & Ex-situ+In-situ &
Section~\ref{sec:dto_insitu} and \\ 
                              &                 &                 &
Section~\ref{sec:exsitu} \\
Neutron capture efficiency & Direct counting & Multiplicity analysis &
Section~\ref{sec:neutrons}\\ 
Livetime &  10~MHz+50~MHz clocks & Pulsed trigger events &
Section~\ref{sec:livetime}\\ 
Fiducial volume cut & 550~cm & Multiple volume cuts & Section~\ref{sec:fvolecut}\\ 
Signal Extraction pdfs & Monte Carlo  model & Analytic & Section~\ref{sec:sigex} \\ 
\hline \hline
\end{tabular} 
\end{center}
\end{table*}

\section{Day-Night Analysis \label{sec:daynight}}

\subsection{Introduction}

The favored explanation of neutrino flavor transformation in terms of
MSW-enhanced neutrino oscillations predicts, for some values of the mixing
parameters, observable spectral distortions and a measurable dependence on
solar zenith angle~\cite{bib:theo1,baltz87,bib:theo3}.  The latter might be
caused by interaction with matter in the Earth and would depend not only on
oscillation parameter values and neutrino energy, but also on the path length and
electron density through the Earth.  This `matter effect' can result in a difference
in the flavor content of the solar neutrino flux between night and day.  Observation
of a day-night asymmetry would be strong evidence that neutrino oscillations are the
correct explanation of the observed flavor transformation, as well as direct
evidence for a matter effect.

Day-night rate differences are customarily expressed in terms of an
asymmetry ratio, formed from the difference in the night ($N$) and day
($D$) event rates divided by their average:
\begin{equation}
A = \frac{N-D}{(N+D)/2}.
\label{eq:dn_asymm_def}
\end{equation}
This asymmetry ratio has the advantage that common systematics in $N$ and $D$
cancel, and can be neglected.  Although diurnal variations in systematics, and
certain other systematics, will not cancel, a day-night measurement
is in general limited by statistical and not systematic uncertainties.  

SNO's unique contribution to day-night measurements is its ability to
determine both the total neutrino flux and the electron neutrino flux.
Neutrino oscillation models with purely active neutrinos predict that
while the electron flux asymmetry $A_e$ will be in the range $\sim 0-0.15$,
the total flux asymmetry $A_{\rm tot}$ should be identically
zero.  Previous day-night measurements by the Super-Kamiokande
Collaboration have been only of the elastic scattering rate asymmetry
($A_{ES}$), which because of its neutral current sensitivity is a
linear combination of $A_e$ and $A_{\rm tot}$.  For SNO's measured CC/NC
ratio of $0.35:1$, one expects $A_{e} \approx 1.5 A_{ES}$.
Thus SNO has comparable day night sensitivity to the much larger
Super-Kamiokande detector, for equal livetimes and thresholds.

The day-night measurement is in principle simple, and builds
strongly upon the integral flux analysis.  At the most basic level,
one subdivides the data set into ``night'' and ``day''
portions, according to whether the Sun is below or above the horizon,
and then repeat the standard analysis on each individual data portion
separately.  The bulk of the work is in evaluating diurnal
systematic uncertainties in detector response and backgrounds, as well
as demonstrating the day-night stability of the detector. 

\subsection{Data Set}  

The day-night analysis is based on the same data set and cuts that were used
for the neutral current analysis (November 2, 1999 to May 28, 2001
UTC, with a livetime of 306.4 days.)  The data are divided into
``day'' and ``night'' portions based upon whether the Sun's elevation
is above or below the horizon.  Because the length of day is correlated
with the time of year, the eccentricity of the Earth's orbit
introduces a ``natural'' day-night rate difference due to $1/r^2$
variations in the Earth-Sun distance.  In the analysis the event
rates of the day and night data sets were corrected for the eccentricity.
The time-averaged inverse-square distance to the Sun
$\langle(\frac{1AU}{R})^2\rangle$ was 1.0002 and 1.0117 for the day and night
portions, respectively. Both values are greater than 1 because the detector had more
livetime during winter than summer for this data period.

	As described in Section~\ref{sec:livetime}, we also divided
the day/night data set into two sets of approximately equal livetime.  We used
one set of data to develop the analysis procedures and used the second as a
blind test of statistical bias.

\subsection{Determination Of Day-Night Systematic Uncertainties In
Detector Response} 

In an analysis of day-night differences using a ratio such as Equation
~\ref{eq:dn_asymm_def}, many systematic errors will cancel
and can be neglected.  Differential systematics
between day and night, such as a slight difference in energy scale,
can, however,  produce false day-night asymmetries.  Possible sources of diurnal differences in
detector response are the dominant systematic uncertainties in SNO's day-night
measurements.  Long-term variations in detector response can also lead to day-night
asymmetries through an ``aliasing'' effect.  Finally, directional dependencies in
detector response, convolved with the directional distributions of neutrino
events can also produce false day-night differences, particularly for the elastic
scattering (ES) signal.

A set of signals that are continuously present in the detector was
used to probe possible diurnal variations in detector response.
Further, a number of consistency tests that do
not yield better limits on systematics, but that provide additional
cross checks on detector stability, have been performed.  These checks
are described below.

\subsubsection{Long-term Energy Scale Drift}  
\label{sec:dn_energy_drift}
As described in Section~\ref{sec:ecalibrator} and shown in
Fig.~\ref{fig:edrift}, the SNO detector exhibited a slow long-term
decrease in detector gain, as measured by the mean $N_{\rm hit}$ for the $^{16}$N
calibration source.  The rate of this decrease ($\sim 2\%$/year) is so slow
that it does not directly produce a significant diurnal difference in energy
scale within a 24 hour period.  Nonetheless, because the length of day is
longer in summer than winter, such slow drifts in energy scale that are not
correctly measured and accounted for can cause a false day-night asymmetry.

The assigned energy of each event was corrected to account for the measured drift,
in principle eliminating this effect.  However, although
$^{16}$N calibration data were generally taken every 2-4 weeks, there were
gaps in the calibration schedule, and there is some uncertainty in
energy drift between calibration points.  

\begin{figure}  
\begin{center}
\includegraphics[height=0.25\textheight]{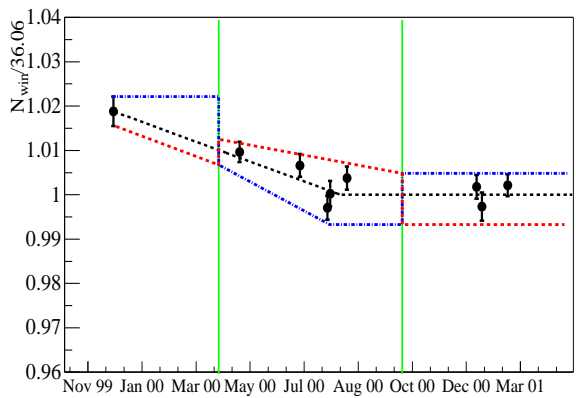}
\caption{Relative energy scale for $^{16}$N calibration data versus
calendar time.  The black curve is the measured energy drift.  The red
and blue curves represent ``worst case'' energy drift models designed
to maximize the relative energy scale difference between day and night data.
The energy estimate `$N_{\rm win}$' is discussed in Section~\ref{sec:enecal}.
}
\label{fig:worst_case_drifts}
\end{center}
\end{figure}

A conservative estimate of the effects of uncertainty in the time
dependence of the energy drift can be obtained by using ``worst case''
drift models, designed to exaggerate
the effects of an error.
In one extreme model, the energy drift is underestimated 
between the spring and fall equinoxes, when day is longer than
night, and is overestimated between the fall and spring equinoxes. (see Figure
\ref{fig:worst_case_drifts}).  A second extreme model has the
opposite error, overestimating the true drift in summer and
underestimating in the winter.  By systematically overestimating the
energy scale during one season and underestimating during the other, the difference
between the day and night energy scales due to long-term variations in energy drift
is maximized.  The worst case models are not meant to be realistic, but repeating
the analysis with the extreme models should yield bounds on the day-night
uncertainty from long-term energy scale drift.

\subsubsection{Diurnal Energy Scale} 
\label{sec:diurnalE}

Circadian variations in detector response could directly produce
diurnal variations in energy scale.  Numerous
sources of such variations can be imagined---diurnal ``sags'' in lab
power voltages, temperature variations in the lab, etc.  Regardless of
their source, the existence of such variations can be probed in
signals that are constantly present in the detector.

There is a solitary point of high background radioactivity, or `hot spot', on the
upper hemisphere of the acrylic vessel (see Fig.~\ref{fig:zvx}).  The origins of
this hot spot are unknown, but it is most likely a uranium or thorium contamination
inadvertently introduced during construction.  The event rate from the hot
spot is stable and sufficient to provide an excellent test of diurnal variations.
The hot spot radioactivity has a steeply falling energy spectrum, so that small
variations in energy scale translate into large variations in the number of counts
above an energy threshold.

Using the `hot spot' to measure diurnal variation in detector response 
requires that the intrinsic decay rate from the source is constant,
and that long-term variations in detector response, such as those
described in Section~\ref{sec:dn_energy_drift}, are corrected for.
The goal is to separate true diurnal variations from effects on
longer timescales that can ``alias'' into an apparent day-night
difference.  This is accomplished by dividing each data run
into ``day'' and ``night'' portions, and calculating a day-night
asymmetry for each run.  The vast majority of runs have durations less than 
24~hours, and so forming a day-night ratio on a run-by-run basis will
cancel detector variations at timescales much longer than a
day.

\begin{figure}  
\begin{center}
\includegraphics[height=0.26\textheight]{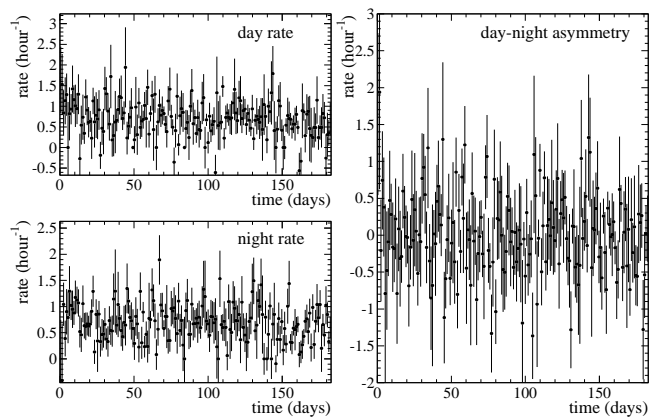}
\caption{Run-by-run day and night event rates from the acrylic hot
spot.  \label{fig:berkeley_blob}
} 
\end{center}
\end{figure}

Events from the acrylic hot spot are selected with a geometry cut.  The event rate
in regions of the same size on the acrylic vessel away from the hot spot is used to
estimate a background level.  Events are counted in a low energy monitoring window
between $27.3 < N_{\rm eff}^\prime < 40$, where $N_{\rm eff}^\prime$ is the $N_{\rm hit}$ of the
event corrected for long-term gain drifts and working tube checks (the
$N_{\rm eff}^\prime$ of Eq.~\ref{eq:neffprime} but with the drift correction,
$\epsilon_{\rm drift}$, included).  Figure~\ref{fig:berkeley_blob} shows the day and
night event rates from the hot spot for each run, as well as their difference.
The measured diurnal asymmetry in the hot spot event rate was $A = -1.8\% \pm
3.5\%$, consistent with zero.  The slope of the energy spectrum from the hot spot
radioactivity is found to be such that a 1\% shift in energy scale changes the event
rate above threshold by $10.3 \pm 2.4\%$.  Hence, the measured uncertainty on the
hot spot's rate asymmetry translates into a 0.3\% uncertainty in energy scale.
Examination of radioactivity event rates in monitoring  regions around the
PMTs and in the light water also show no diurnal rate variations, and yield
comparable limits on diurnal changes in energy scale.  

An interesting check on energy scale stability is provided by
the $^{252}$Cf neutron source.  This source was deployed overnight in
the detector, and substantial periods of day and night data were taken.
These data allow us to verify the energy scale stability for neutrons
during a single 24-hour period with high statistics.
Table~\ref{tab:DN_neut_checks} shows the total $N_{\rm hit}$ and the mean event
energy in MeV for these data.  No significant variations are seen in
the mean or width of the energy distribution between day and night.
Because these data cover only a single 24-hour period, they
do not probe all possible diurnal variations in response, but do
provide a reassuring complementary check on the studies of the hot
spot radioactivity.

\begin{table*}
\caption{The energy distribution of neutrons from the Cf source,
during the day and night.
\label{tab:DN_neut_checks}}
\begin{center}
\begin{tabular}{ l  c  c  c  c }
\hline \hline
\multicolumn{1}{c}{}   & \multicolumn{2}{c}{$N_{\rm hit}$} &\multicolumn{2}{c}{\rm Mean Event Energy (MeV)} \\ \hline
         & Mean     &  Width       & Mean       & Width          \\
\hline
Day      &$46.49\pm0.27$&$10.08\pm0.25$&$ 5.426\pm0.026$&$ 1.075\pm0.024$\\ 
Night    &$46.65\pm0.16$&$10.25\pm0.15$&$ 5.460\pm0.015$&$ 1.083\pm0.014$\\ 
Day-Night&$-0.16\pm0.31$&$-0.17\pm0.29$&$-0.034\pm0.030$&$-0.008\pm0.028$\\
\hline \hline
\end{tabular}
\end{center}
\end{table*}  

	Uncertainties associated with detector asymmetries---differences in
energy scale between the top and bottom of the detector, for example---were studied
by looking at $^{16}$N calibration source events and measuring the scale,
resolution, and other uncertainties as a function of direction and position within
the detector.  The effects on the asymmetries in the fluxes were then determined by
convolving the shifts due to these uncertainties with expected position and
direction distributions of neutrino events.

\subsection{Day-night Results}

\subsubsection{Day-night Integral Fluxes}

\begin{table*}
\caption{\label{tab:dn_sigex_results} The results of signal extraction, assuming
an undistorted $^8$B spectrum.  The systematic uncertainties (combined
set) include a component that cancels in the formation of the
$A$.  Except for the dimensionless $A$, the units
are $10^6$~cm$^{-2}$~s$^{-1}$.  Flux values have been rounded, but the
asymmetries were calculated with full precision.
}
\begin{ruledtabular}
\begin{tabular}{c|cc|cc|cc|c}
 & \multicolumn{2}{c}{Set 1} & \multicolumn{2}{c}{Set 2} &
\multicolumn{2}{c}{Combined} &   $A (\%)$ \\ 
signal & $\phi_{D}$ & $\phi_{N}$ & $\phi_{D}$ & $\phi_{N}$ &
$\phi_{D}$ & $\phi_{N}$ & \\  
\hline
CC & $1.53\pm0.12$ & $1.95\pm0.10$ & $1.69\pm0.12$ &$1.77\pm0.11$ &
$1.62\pm0.08\pm 0.08$  & $1.87\pm 0.07\pm 0.10$ & $+14.0 \pm~6.3
^{+1.5}_{-1.4}$ \\ 
ES & $2.91\pm0.52$ & $1.59\pm0.38$ & $2.35\pm0.51$ &$2.88\pm0.47$ &
$2.64\pm0.37\pm 0.12$ & $2.22 \pm0.30\pm 0.12$ &  $-17.4 \pm 19.5
^{+2.4}_{-2.2}$ \\ 
NC & $7.09\pm0.97$ & $3.95\pm0.75$ & $4.56\pm0.89$ &$5.33\pm0.84$ &
$5.69\pm0.66\pm 0.44$  & $4.63\pm0.57\pm 0.44$  &  $-20.4 \pm 16.9
^{+2.4}_{-2.5}$ \\ 
\end{tabular}
\end{ruledtabular}
\end{table*}

Table~\ref{tab:dn_sigex_results} contains extracted integral fluxes
for day and night data from the open data set (Set 1) and the blind
set (Set 2).  The fluxes have been normalized to an Earth-Sun distance
of 1~AU.  Due to the signal extraction process, the day-night
asymmetries for the individual signal rates are statistically
correlated.  For the combined data, $A_{CC}$ and $A_{NC}$ are strongly
anticorrelated, with a statistical correlation coefficient of
$\rho=-0.518$.  $A_{CC}$ and $A_{ES}$ have a correlation coefficient
of $\rho=-0.161$, while the coefficient between $A_{NC}$ and $A_{ES}$
is $\rho=-0.106$.  For the combined analysis, $A_{CC}$ is $+2.2\sigma$
from zero, while $A_{ES}$ and $A_{NC}$ are $-0.9\sigma$ and
$-1.2\sigma$ from zero, respectively.

\subsubsection{Day-night Energy Spectra}
\label{sec:dnspect}

\begin{figure}
\begin{center}
\includegraphics[width=3.6in]{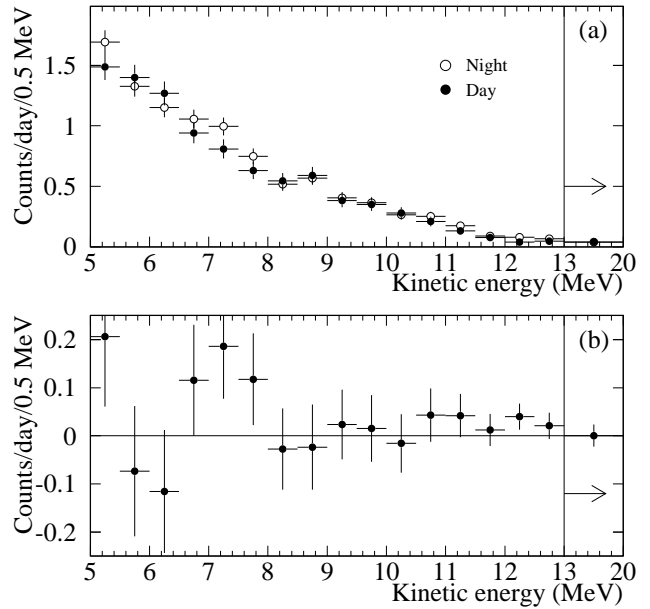}
\caption{\label{fig:DNspectra} (a) Energy spectra for day and night.
All signals and backgrounds contribute. The final bin extends from
13.0 to 20.0 MeV.  Numerical values for each bin are given in
Table~\ref{tab:DNspectra} in Appendix~\ref{sec:physint}. (b) Difference, \emph{night - day}, between the
spectra.  The day rate was $9.23 \pm 0.27$ events/day, and the night
rate was $9.79 \pm 0.24$ events/day.}
\end{center}
\end{figure}

Figure~\ref{fig:DNspectra} shows the day and night energy spectra for
all events (including the small contributions from radioactive
background). The integrated excess has a
significance of $1.55\sigma$.

\subsubsection{Integral Flux Asymmetries and Interpretation}

Table~\ref{tab:dn_sigex_results} shows the integral flux asymmetries
for the CC, ES, and NC signals.  Table~\ref{tab:dn_syserrs} gives the
systematic uncertainties on the asymmetry parameters.  All results are
derived under the assumption of a standard undistorted $^8$B energy
spectrum.

\begin{table}
\caption{\label{tab:dn_syserrs} Effect of systematic uncertainties on
$A~(\%)$.  For presentation, uncertainties have been
symmetrized and rounded.}
\begin{ruledtabular}
\begin{tabular}{c|c|c|c}
Systematic & $\delta A_{CC} $ & $\delta A_{ES} $ &
$\delta A_{NC} $ \\ 
\hline
Long-term energy scale drift       & 0.4 & 0.5 & 0.2 \\
Diurnal energy scale variation     & 1.2 & 0.7 & 1.6 \\
Directional energy scale var.      & 0.2 & 1.4 & 0.3 \\
Diurnal energy resolution var.     & 0.1 & 0.1 & 0.3 \\
Directional energy resolution var. & 0.0 & 0.1 & 0.0 \\
Diurnal vertex shift var.              & 0.5 & 0.6 & 0.7 \\
Directional vertex shift var. & 0.0 & 1.1 & 0.1 \\
Diurnal vertex resolution var.     & 0.2 & 0.7 & 0.5 \\
Directional angular recon. var.    & 0.0 & 0.1 & 0.1 \\
PMT $\beta$-$\gamma$ background    & 0.0 & 0.2 & 0.5 \\
AV+H$_2$O $\beta$-$\gamma$ bkgd.          & 0.0 & 0.6 & 0.2 \\
D$_2$O $\beta$-$\gamma$, neutrons bkgd.   & 0.1 & 0.4 & 1.2 \\
External neutrons bkgd.                 & 0.0 & 0.2 & 0.4 \\
Cut acceptance                          & 0.5 & 0.5 & 0.5 \\
\hline
Total                              & 1.5 & 2.4 & 2.4 \\
\end{tabular}
\end{ruledtabular}
\end{table}

\begin{table}
\caption{\label{tab:dn_electron} Measurement of the $\phi_e$ and
$\phi_{\rm tot}$ asymmetry for various constraints.  All analyses assume
an undistorted $^8$B spectrum.  }
\begin{ruledtabular}
\begin{tabular}{l|c}
Constraints & Asymmetry (\%)\\  
\hline
a) no additional constraint & $A_{CC} = 14.0 \pm 6.3
^{+1.5}_{-1.4}$\\ 
                            & $A_{NC} = -20.4 \pm 16.9
^{+2.4}_{-2.5}$\\ 
 & (see text for correlations)\\
\hline
b) $\phi_{ES} = (1-\epsilon) \phi_{e} + \epsilon \phi_{\rm tot}$ &
$A_{e} = 12.8 \pm 6.2 ^{+1.5}_{-1.4}$\\ 
&  $A_{\rm tot} = -24.2 \pm 16.1 ^{+2.4}_{-2.5}$\\ 
& correlation = -0.602\\
\hline
c) $\phi_{ES} = (1-\epsilon) \phi_{e} + \epsilon \phi_{\rm tot}$ & \\
$~~~A_{\rm tot} = 0$& $A_{e} = 7.0 \pm 4.9 ^{+1.3}_{-1.2}$\\
\hline
d) $\phi_{ES} = (1-\epsilon) \phi_{e} + \epsilon \phi_{\rm tot}$ &
$A_e(SK) = 5.3 \pm 3.7 ^{+2.0}_{-1.7}$\\ 
$~~~A_{\rm tot} = 0$& (derived from SK $A_{ES}$ \\
$~~~A_{ES}(SK) = 3.3\% \pm 2.2\% ^{+1.3}_{-1.2}\%$ & and SNO
total $^8$B flux)\\ 
\end{tabular}
\end{ruledtabular}
\end{table}

\begin{figure}
\begin{center}
\includegraphics[width=3.6in]{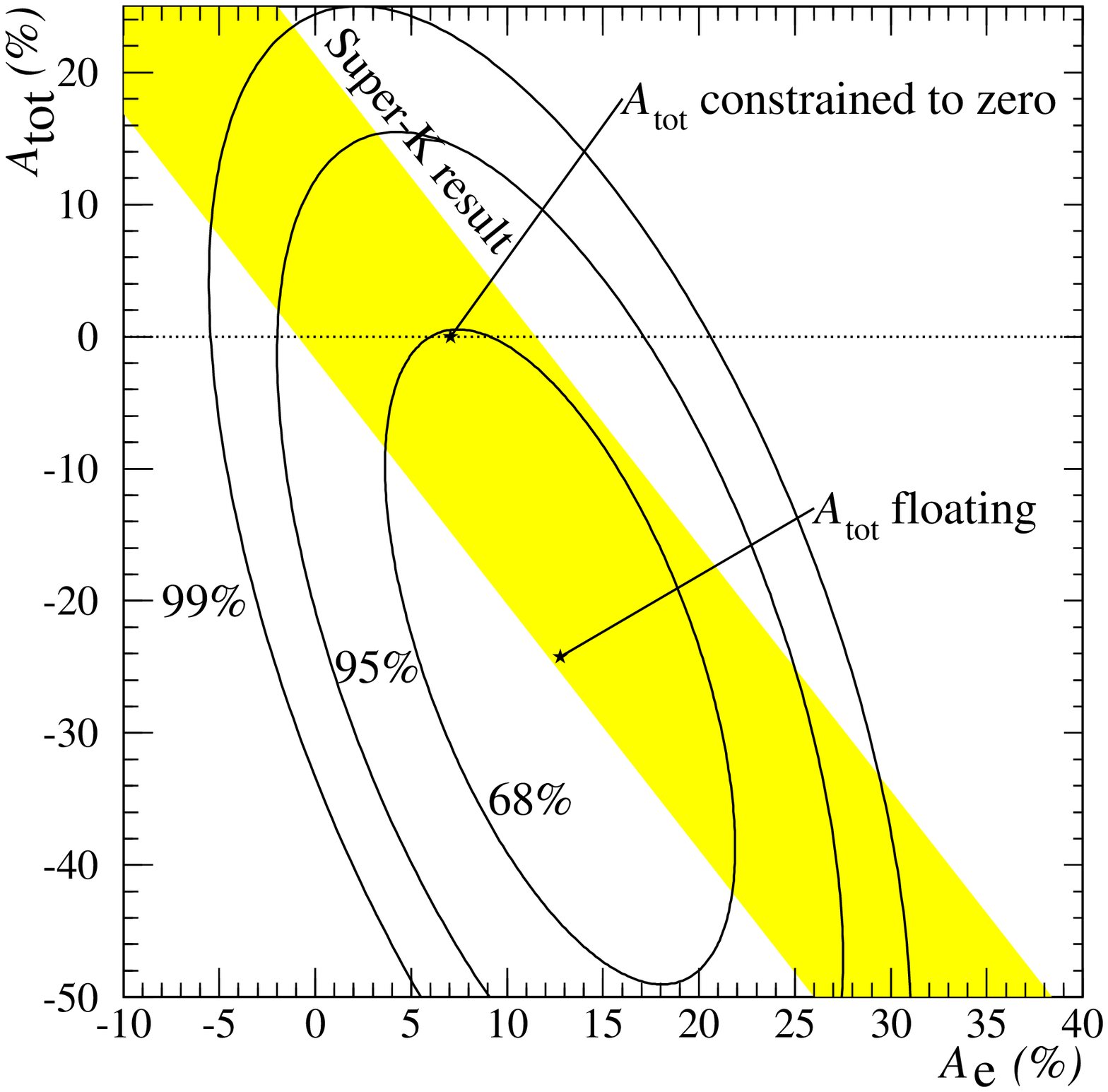}
\caption{\label{fig:DNcorr} Joint probability contours for
$A_{\rm tot}$ and $A_{e}$.  The points indicate the
results when $A_{\rm tot}$ is allowed to float and when it is
constrained to zero.  The diagonal band indicates the 68\% joint
contour for the Super-Kamiokande $A_{ES}$ measurement.  }
\end{center}
\end{figure}

The asymmetries on the individual neutrino reaction channels can be
recast as asymmetries on the neutrino flavor content.
Table \ref{tab:dn_electron} (a) shows the results for $A_e$
derived from the CC day and night rate measurements, i.e.,
$A_e = A_{CC}$.  However, the ES flux, when combined with the CC and
NC fluxes, contains additional information about the electron neutrino flux.
This information can be accounted for through a change of variables.  
Accordingly, the day and night flavor contents
were then extracted by changing variables to $\phi_{CC} = \phi_{e}$,
$\phi_{NC} = \phi_{\rm tot} = \phi_{e}+\phi_{\mu\tau}$ and $\phi_{ES} =
\phi_{e} + \epsilon \phi_{\mu\tau}$, where $\epsilon \equiv 1/6.48$ is
the ratio of the average ES cross sections above $T_{\rm eff}=5$~MeV for
$\nu_{\mu\tau}$ and $\nu_e$.  Table \ref{tab:dn_electron} (b) shows
the asymmetries of $\phi_e$ and $\phi_{\rm tot}$ with this additional
constraint from the ES rate measurements.  This analysis allowed for
an asymmetry in the total flux of $^8$B neutrinos (non-zero
$A_{\rm tot}$), with the measurements of $A_{e}$ and $A_{\rm tot}$ having a
strong anti-correlation.  Fig.~\ref{fig:DNcorr} shows the $A_e$
vs. $A_{\rm tot}$ joint probability contours.  Forcing $A_{\rm tot} = 0$, as
predicted by active-only models, yielded the result in 
Table~\ref{tab:dn_electron} (c) of $A_{e} = 7.0
\pm 4.9~\mathrm{(stat.)  ^{+1.3}_{-1.2}}\%~\mathrm{(sys.)}$.

Prior to SNO, the only day-night measurements of solar neutrinos were
those from the Super-Kamiokande experiment.  Because Super-Kamiokande
measures the elastic scattering rate, which is sensitive to a linear
combination of electron and non-electron neutrino rates, its
measurements alone cannot separately determine $A_e$ and $A_{\rm tot}$.
However, the SNO results can be used to break this covariance in the
Super-Kamiokande results.  The Super-Kamiokande (SK) collaboration
measured $A_{ES}(SK) = 3.3 \pm 2.2\%~\mathrm{(stat.)}^{+1.3}_{-1.2}\%~\mathrm{(sys.)}$~\cite{SK}.
The ES measurement includes a neutral current component, which reduces
the asymmetry for this reaction relative to $A_{e}$~\cite{Robustsig}.
$A_{ES}(SK)$ may be converted to an equivalent electron flavor
asymmetry using the total neutrino flux measured by SNO, yielding
$A_{e}(SK) = 5.3 \pm 3.7 ^{+2.0}_{-1.7}$ (Table
\ref{tab:dn_electron} (d)).  This value is in good agreement with SNO's
direct measurement of $A_e$, as seen in Fig.~\ref{fig:DNcorr}.  Taking
a weighted average of the SNO and Super-Kamiokande measurements of
$A_e$ yields an asymmetry of $A_e = 6.0\% \pm 3.2\%$.

\section{Summary and Conclusions}

	We have detailed here the results from the Sudbury Neutrino
Observatory's Phase I data set.  The Phase I data were taken with an integral
exposure to solar $^8$B neutrinos of 0.65 kt-years. Heavy water, without
any additives, was both the target and detection medium.  The heavy water
provided us with three neutrino detection reactions: a charged current reaction
exclusive to $\nu_e$'s, a neutral current reaction sensitive to all flavors,
and an elastic scattering reaction that is primarily sensitive to $\nu_e$ but
has a small sensitivity to other flavors.  Under the assumption that the solar
$^8$B flux is composed entirely of $\nu_e$ and that its spectrum is undistorted,
we find that the measured fluxes using each of the three reactions are:
\begin{eqnarray*}
\phi_{CC} & = &
  1.76^{+0.06}_{-0.05}\mbox{(stat.)}^{+0.09}_{-0.09}~\mbox{(syst.)}
  \times 10^{6}~{\rm cm}^{-2}~{\rm s}^{-1} \\
\phi_{ES}  &= & 2.39^{+0.24}_{-0.23}\mbox{(stat.)}^{+0.12}_{-0.12}~\mbox{(syst.)}
\times 10^{6}~{\rm cm}^{-2}~{\rm s}^{-1} \\
\phi_{NC}  &=  &5.09^{+0.44}_{-0.43}\mbox{(stat.)}^{+0.46}_{-0.43}~\mbox{(syst.)} \times 10^{6}~{\rm cm}^{-2}~{\rm s}^{-1}. 
\end{eqnarray*}
The flux of neutrinos measured by $\phi_{NC}$ is significantly larger
than that measured by $\phi_{CC}$, thus leading to the conclusion that
neutrinos of flavors other than $\nu_e$ must be a substantial component of the
solar flux.  Resolving the fluxes above directly into flavor components yields
\begin{eqnarray*}
\phi(\nu_e) & = &
  1.76^{+0.05}_{-0.05}\mbox{(stat.)}^{+0.09}_{-0.09}~\mbox{(syst.)}
  \times 10^{6}~{\rm cm}^{-2}~{\rm s}^{-1} \\
\phi(\nu_{\mu\tau}) & = &
  3.41^{+0.45}_{-0.45}\mbox{(stat.)}^{+0.48}_{-0.45}~\mbox{(syst.)}
  \times 10^{6}~{\rm cm}^{-2}~{\rm s}^{-1} 
\end{eqnarray*}
showing that $\phi(\nu_{\mu\tau})$ is 5.3$\sigma$ away from zero.  The total
flux of $^8$B neutrinos, as measured by $\phi_{NC}$, is in excellent agreement
with the predictions of Standard Solar Models.

	We have also looked for an asymmetry in the day and night neutrino
fluxes, as would be expected for neutrino oscillations driven by the MSW
effect.  We find that the day-night asymmetry in the electron neutrino flux is
\[ A_{e} = 7.0
\pm 4.9~\rm{(stat.)  ^{+1.3}_{-1.2}}\%~\rm{(syst.)} \]
when we constrain the day-night asymmetry in the total flux to be zero.

These results collectively represent the first solar-model independent
measurements of the solar $^8$B neutrino flux, and the first inclusive
appearance measurement of neutrino oscillations.  In addition, they provide
the first direct confirmation of the predictions of the Standard Solar Model,
and have thus solved the long-standing solar neutrino problem.

\section{Acknowledgments}
 This research was supported by: Canada:
Natural Sciences and Engineering Research Council, Industry Canada,
National Research Council, Northern Ontario Heritage Fund, Atomic
Energy of Canada, Ltd., Ontario Power Generation, High Performance
Computing Virtual Laboratory, Canada Foundation for Innovation; US:
Department of Energy, National Energy Research Scientific Computing
Center, Alfred P. Sloan Foundation; UK: Particle Physics and Astronomy
Research Council. We thank the SNO technical staff for their strong
contributions.  We thank Inco, Ltd. for hosting this project.

\section{Approach to Estimation of Mixing Parameters for Two-Neutrino Oscillations}
\label{sec:physint}

In Section~\ref{sec:results}, the measurements of the rates of the three event
types---CC, NC, and ES---were made under the assumption that the $^8$B energy
spectrum is undistorted.  These measurements thus provide a null hypothesis test
that neutrinos from the Sun change flavor on their way to detectors on Earth. As
shown in Section~\ref{sec:results}, this null hypothesis was rejected at
5.3$\sigma$.   To derive constraints on mixing parameters, however, we must
explicitly take into account the oscillation model which may alter the shape
of the neutrino spectra.

	In our Phase I Day-Night paper~\cite{snodn}, we reported our first constraints on the mixing
parameters including data from SNO and other solar neutrino experiments.  For
that analysis, we start from the day and night energy spectra reported here in
Section~\ref{sec:daynight}, rather than using the null hypothesis results of
Section~\ref{sec:results} or the asymmetry reported in
Section~\ref{sec:daynight}.  In this section, we describe the methods used in
the Day-Night paper~\cite{snodn} to extract these bounds.

\subsection{Outline of Method}

To generate MSW contours using the data presented in this article,  we use a
`forward fitting' technique~\cite{saltphys}. We make predictions for the CC, ES, and NC spectra by
convolving a given theoretical model (e.g. a particular point in MSW parameter
space) with SNO's response functions.  Adding these together, and then adding the
energy spectra expected for the low energy backgrounds, we obtain a prediction for
the total energy spectrum that SNO should see for all events.  We then compare
this prediction to the measured SNO day and night energy spectra shown in
Fig.~\ref{fig:DNspectra} (and given here in Table~\ref{tab:DNspectra}), and
calculate a goodness-of-fit parameter.  The day and night energy spectra contain the
sum of CC, ES, NC, and background events; they therefore include all of the flux and
shape information needed to test a given oscillation hypothesis.  Unlike the signal
extraction procedure described in Sections~\ref{sec:sigex} and~\ref{sec:results},
the estimation of mixing parameters described here relies solely upon the
information contained in the energy spectra of the three signals---the radial and
$\cos \theta_{\odot}$ distributions are not used.  The approach discussed here also
differs from the primary signal extraction procedure described in
Section~\ref{sec:sigex} in that it does not use Monte Carlo generated pdfs as the
model for the energy spectra, but rather more closely follows the analytic pdf
approach discussed in Sections~\ref{sec:anal_response} and~\ref{sec:anal_ver}.
\begin{table}
\caption{Bin-by-bin contents of day and night energy spectra shown in
Fig.~\ref{fig:DNspectra}.  These are the numbers used in the SNO mixing parameter
analysis as described in the text.  The second and third columns give the 
boundaries of each energy bin. These data can be obtained
from~\cite{snoweb}. \label{tab:DNspectra}}
\begin{center}
\begin{tabular}{rcccc}
\hline \hline
Bin & $T_{\rm min}$ & $T_{\rm max}$ & $N_{\rm day}$ & $N_{\rm night}$ \\
    &    (MeV)      &   (MeV)       &               &                 \\ 
\hline
   1&    5.0&    5.5&     191&    301 \\
   2&    5.5&    6.0&     180&    236 \\
   3&    6.0&    6.5&     163&    205 \\
   4&    6.5&    7.0&     121&    188 \\
   5&    7.0&    7.5&     104&    177 \\
   6&    7.5&    8.0&      81&    133 \\
   7&    8.0&    8.5&      70&     92 \\
   8&    8.5&    9.0&      76&    101 \\
   9&    9.0&    9.5&      49&     72 \\
  10&    9.5&   10.0&      45&     65 \\
  11&   10.0&   10.5&      36&     47 \\
  12&   10.5&   11.0&      27&     45 \\
  13&   11.0&   11.5&      17&     31 \\
  14&   11.5&   12.0&      10&     16 \\
  15&   12.0&   12.5&       5&     14 \\
  16&   12.5&   13.0&       6&     12 \\
  17&   13.0&   20.0&       5&      7 \\
\hline \hline
\end{tabular}
\end{center}
\end{table}

The following outline gives the basic steps in our mixing parameter analysis:
\begin{enumerate}
\item We start with a particular theoretical model for which we want to
calculate a goodness of fit.  For example, the model might be 2-$\nu$
oscillations with $\tan^2 \theta = 0.4$ and $\Delta m^2 = 2 \times
10^{-5}$.
\item We calculate the electron neutrino survival probability as a
function of energy, and the probability that the neutrino interacts as
$\nu_{\mu,\tau}$.  (For active-only oscillations, these add to 1.)
\item We convolve the $^8$B neutrino energy spectrum, modified by the survival
probability for the hypothesized mixing parameters, with the
differential CC cross-section and the SNO energy response function to
yield a prediction for the shape of the CC energy spectrum SNO should
detect.  We normalize the amplitude by SNO's livetime, the number of
targets, etc.  If the $^8$B flux is allowed to float in the fit, an
additional scale parameter is included on the normalization of the pdf.
\item We do the same thing for the ES reaction, remembering to include
the contribution from $\nu_{\mu,\tau}$ with the appropriate relative
cross-sections.
\item Neutral current interactions generate a Gaussian pdf in energy,
as described in Section~\ref{sec:anal_response}.  In other words, the
shape of the energy spectrum from neutron captures is independent of 
the neutrino energy.  We therefore use the theoretical model only to make a
prediction for how many neutron capture events on deuterium SNO should
have seen.  We then normalize the neutron energy pdf by this amount.
\item SNO's energy spectrum contains small numbers of events from
radioactive backgrounds.  These include background neutrons from
sources such as photodisintegration, and Cherenkov tail events from
$\beta-\gamma$ decays in the detector.  The shapes and amplitudes of
the background pdf are given on the SNO web site, along with their
uncertainties.  The amplitudes are fixed by the SNO analysis, and
so are not allowed to float as free parameters in the fit.
\item We sum the energy spectra for the CC, ES, NC, and background
contributions.  We then compare the resulting shape to the total
energy spectrum from SNO.  We evaluate a goodness of fit  
(e.g. $\chi^2$ between the model spectrum and the data).  We then
repeat the procedure for other solutions or point in parameter space,
and form $\Delta \chi^2$ contours.  The spectra for Day and Night are treated
separately, both added as terms in the overall $\chi^2$.
\end{enumerate}

  In this approach, the SNO energy spectrum gets assigned only
statistical uncertainties.  The systematic uncertainties in the SNO
response functions (energy scale, neutron capture efficiency, etc) are
treated as uncertainties on the model prediction for the energy
spectrum.  Similarly, uncertainties in the background amplitudes
become systematic uncertainties on the model to which the SNO data
gets compared.  The systematic errors are of course correlated from
bin-to-bin, but can be treated by standard covariance matrix
techniques.  

\subsection{Neutrino Flux and Survival Probability}
\label{sec:psurv}

Neutrino production was calculated starting with the fluxes given in
BP2000~\cite{bp2000}. We used both $^8$B and hep fluxes in our analysis, allowing
the $^8$B flux to float in some of the fits.  The spectral shape of the hep
neutrinos was taken from Bahcall~\cite{jnb_na,bahcallweb}. 
The $^{8}$B spectral
shape was from Ortiz {\em et al.}~\cite{ortiz}. Zeros were added at
both ends of the Ortiz table to improve interpolation.

MSW survival probabilities for electron neutrinos to reach the Earth's
surface were calculated using the solar neutrino production regions
and electron density profile given in
BP2000~\cite{bp2000}. Calculations of vacuum oscillation survival
probabilities were not averaged over the production regions in the Sun
but were averaged over the annual variation of the Earth-Sun distance.
Survival probabilities for the quasivacuum oscillation region between
the vacuum and MSW regimes were calculated using the analytic
procedure of Lisi {\em et~al.}~\cite{lisi01}.
Survival probabilities for electron neutrinos traveling
through the Earth were calculated using the electron density profile
taken from~\cite{earthref}.

A number of comparisons and checks were carried out to ensure our
prescription was consistent with others found in the literature.  For
example, 
our calculations
suggest Earth regeneration effects should be strong in SNO for 10~MeV
neutrinos when $\delta m^{2} \approx 10^{-5} {\rm eV}^{2}$, similar
to the result found in ~\cite{baltz87}, and, as also
found in~\cite{bks01},
we find no significant Earth regeneration
occurs for $\delta m^{2}/E < 10^{-8}{\rm eV}^{2}/$~MeV.

\subsection{Interaction Cross Sections}
\label{sec:xsec}

For interactions in SNO, neutrino-deuteron CC and NC interaction cross
sections were taken from the effective field theory calculations of
Butler {\em et~al.}~\cite{BCK}. A value of $5.6~{\rm fm}^{3}$ was
adopted for the $L_{1A}$ counter term in these calculations to provide
good agreement with the potential model calculations of Nakamura {\em
et~al.}~\cite{nsa}.  Neutrino-electron elastic scattering cross
sections were calculated using the formulae given in
Bahcall~\cite{jnb_na}.  In addition, neutrino cross sections on
chlorine and gallium needed for global fits were those of
Bahcall~\cite{bahcallweb}. A point was added to the table for
chlorine at 0.861 MeV ($2.67 \times 10^{-42} {\mathrm cm}^{2}$, taken
from~\cite{jnb_na}), to help get the correct contribution from
$^{7}$Be neutrinos to the chlorine experiment.

\subsection{Calculation of CC and ES Electron Spectra}

The prediction for the measured energy spectra for the recoil electrons from the CC 
reaction is given by Eq.~\ref{eq:ccspec}, integrated over the detector livetime and
multiplied by the number of targets $N_D$:
\begin{widetext}
\begin{equation}
\frac{dN_{\mathrm CC}}{dT_{\rm eff}} = N_{D} \int_{\rm livetime} dt \int_{0}^{\infty} dE_{\nu} \frac{d\Phi_e}{dE_\nu} \int_{0}^{\infty}
dT_{e} \frac{d\sigma_{\mathrm CC}(E_{\nu})}{dT_{e}}
R(T_{e}, T_{\rm eff}) 
\label{eq:dnccdt}
\end{equation}
\end{widetext}

where $d\Phi_{e}/dE_{\nu}$ is the differential flux (the energy spectrum) of
electron neutrinos at the detector calculated as described above in
Section~A\ref{sec:psurv} (it includes the survival probability for the hypothesized
mixing parameters),
$d\sigma_{\mathrm CC}/dT_{e}$ is the CC $\nu-d$ differential cross section with respect to
the true recoil electron kinetic energy $T_{e}$ (discussed above in
Section~A\ref{sec:xsec} and further in Sections~\ref{sec:nuspecint}
and~\ref{sec:results}), and $R(T_{e},T)$ is the detector response function to
electrons given in Eq.~\ref{eresp} and Table~\ref{response_pars} of
Section~\ref{sec:anal_response}.  The number of deuteron targets, $N_{D}$, is given
in Section~\ref{sec:nd2o} for 1000 tonnes of SNO heavy water, and must be multiplied
by the fraction of the volume within SNO's 550~cm radial cut.  To calculate electron
recoil spectra for comparison with the day and night energy spectra, we integrate
$dN_{\mathrm CC}/dT_{\rm eff}$ over limits corresponding to the boundaries of each spectral bin.  

Except for a contribution from non-electron neutrinos, the prediction for the 
measured electron spectrum for the ES reaction is similar:
\begin{widetext}
\begin{equation}
\frac{dN_{\mathrm ES}}{dT_{\rm eff}} = n_e \int_{\rm livetime} dt \int_{0}^{\infty} dE_{\nu}
\left[ \frac{d\Phi_e}{dE_\nu} \int_{0}^{\infty} dT_{e} \frac{d\sigma^{e}_{\mathrm ES}(E_{\nu})}{dT_{e}} R(T_{e},T_{\rm eff})+ \frac{d\Phi_{\mu,\tau}}{dE_\nu} \int_{0}^{\infty} dT_e \frac{d\sigma^{\mu,\tau}_{\mathrm ES}(E_{\nu})}{dT_{e}}(E_{\nu}) R(T_{e}, T_{\rm eff}) \right]
\end{equation}
\end{widetext}

where $d\Phi_{\mu,\tau}/dE_{\nu}$ is the energy spectrum of muon and tau neutrinos
at the detector. Assuming only active neutrinos,
\begin{equation}
\frac{d\Phi_{SSM}}{dE_{\nu}} = \frac{d\Phi_{e}}{dE_{\nu}} +
                                      \frac{d\Phi_{\mu,\tau}}{E_{\nu}}. 
\end{equation}
$n_e$ is the total number of target electrons,
given in Section~\ref{sec:nd2o} for 1000 tonnes of heavy water.  It, too, must be
scaled by the fraction of the volume inside 550~cm.  Like the differential CC
rate, we integrate $dN_{\mathrm ES}/dT_{\rm eff}$ to provide estimates of the number of events
in each recoil-electron energy bin.

\subsection{Calculation of Detected NC Neutron Rate}
\label{sec:ncrate_physint}

	As described in Section~\ref{sec:anal_response}, the `spectrum' of events
from the NC reaction is just the response of the detector to the monoenergetic
$\gamma$ rays, and we have used a Gaussian to characterize the shape of this
effective kinetic energy spectrum with a mean of $T_{\gamma}=5.08$~MeV and a width
$\sigma_{\gamma}=1.11$~MeV.  We therefore only need to calculate the absolute
normalization of this distribution, which depends in part on the number of 
neutrons produced $N^{\rm prod}_{NC}$ through the NC reaction:
\begin{widetext}
\begin{equation}
 N^{\rm prod}_{NC} =   N_{D} \int_{\rm livetime} dt \int_{0}^{\infty} dE_{\nu}  \frac{d\Phi_{SSM}}{dE_{\nu}}
    \sigma_{NC}(E_{\nu}) 
\end{equation}
\end{widetext}
where $\sigma_{NC}(E_{\nu})$ is the total NC cross section as a function of
neutrino energy.  

	To convert the number of neutrons produced by the NC reaction, $N^{\rm
prod}_{NC}$, to the number actually detected, $N^{\rm det}_{NC}$, we need to
multiply by the neutron capture and detection efficiencies. As detailed in
Section~\ref{sec:neutrons},  the probability that a neutron generated at a random
location inside the 600~cm-radius acrylic vessel will capture on a deuteron 
is $29.9\% \pm 1.1\%$.  Not all of the captured neutrons will be inside SNO's
fiducial volume and above SNO's energy threshold.  As also described in
Section~\ref{sec:neutrons}, for neutrons generated throughout the acrylic vessel,
$14.4\% \pm 0.53\%$ will be detected inside 550~cm and above $T=5$~MeV.
We therefore have
\[ N^{\rm det}_{NC} = 0.144\times N^{\rm prod}_{NC} \]
Note this differs from the calculation for the CC and ES events, for which we
multiplied the total number produced in each effective kinetic energy bin by just
the ratio of the 550~cm fiducial volume to the total 600~cm acrylic vessel
volume.

For the purposes of constructing predicted energy spectra for comparison to our
measured day and night spectra, we 
break the total detection probability of $14.4\%$ into an energy
component and a radial component, to allow easier application of systematic errors
on energy scale and radial reconstruction.
With this separation, we find that:
\begin{itemize}
\item $27.01\% \pm 0.99\%$ of NC neutrons capture on deuterons inside
550~cm, producing events with detectable Cherenkov light
\item 53.2\% of all neutrons have reconstructed effective kinetic energies 
above $T_{\rm eff}>5$~MeV, as can be determined using the energy spectrum for 6.25~MeV
$\gamma$ rays given above and in Section~\ref{sec:anal_response}.
\end{itemize}
We can now recalculate a new neutron
detection efficiency for varying shifts in (for example) energy scale by
reevaluating what fraction of the neutron energy spectrum is above the threshold,
then multiplying by the ``radial'' part 27.01\%.

\subsection{Livetime}

	As discussed in Section~\ref{sec:norm} and summarized in Table~\ref{tab:LT_summary}, the `day'
livetime for the SNO Phase I data set measured 128.5 days and the `night' livetime 177.9 days.
Fig.~\ref{fig:LT_raw_10mhz_lt} showed the distribution of this livetime over 480 bins in
the zenith angle $\cos\theta_{Z}$.  The values in the figure are given here in 
Tables~\ref{tbl:coszenith1} and~\ref{tbl:coszenith2}.
\begin{table*}
\caption{SNO Phase I livetime as a function of zenith angle $\cos\theta_{Z}$. The 
table shows the first 240 bins of Fig.~\ref{fig:LT_raw_10mhz_lt} of Section~\ref{sec:norm}, corresponding to an
even division of the region $-1< \cos\theta_Z < 0$. These data can be 
obtained from~\cite{snoweb}. \label{tbl:coszenith1}} 
\begin{center}
\begin{tabular}{cccccccccc} \hline \hline 
~Bin~ & ~~Time (s)~~ & ~Bin~ & ~~Time (s)~~ & ~Bin~ & ~~Time (s)~~ & ~Bin~ & ~~Time (s)~~ & ~Bin~ & ~~Time (s)~~  \\ \hline 
1 &         0.     & 49 &   62557.04     & 97 &   62982.45     & 145 &   81155.52     & 193 &   65399.10     \\ 
2 &         0.     & 50 &   63826.41     & 98 &   63787.66     & 146 &   84341.83     & 194 &   65637.09     \\ 
3 &         0.     & 51 &   66604.71     & 99 &   66201.56     & 147 &   85427.77     & 195 &   65446.03     \\ 
4 &         0.     & 52 &   64183.09     & 100 &   67866.33     & 148 &   85537.79     & 196 &   64818.23     \\ 
5 &         0.     & 53 &   64576.03     & 101 &   68584.96     & 149 &   86001.10     & 197 &   64797.89     \\ 
6 &         0.     & 54 &   65123.99     & 102 &   65068.79     & 150 &   84170.44     & 198 &   64800.28     \\ 
7 &         0.     & 55 &   64859.73     & 103 &   67233.30     & 151 &   82621.84     & 199 &   64601.40     \\ 
8 &         0.     & 56 &   64058.89     & 104 &   64686.10     & 152 &   82383.84     & 200 &   64476.81     \\ 
9 &         0.     & 57 &   66499.73     & 105 &   67244.77     & 153 &   80350.16     & 201 &   64349.98     \\ 
10 &         0.     & 58 &   68018.73     & 106 &   67248.45     & 154 &   84034.05     & 202 &   64489.91     \\ 
11 &         0.     & 59 &   69503.30     & 107 &   69622.96     & 155 &   91947.32     & 203 &   64226.22     \\ 
12 &         0.     & 60 &   67884.50     & 108 &   71553.39     & 156 &   94835.76     & 204 &   63956.47     \\ 
13 &         0.     & 61 &   66940.01     & 109 &   70630.96     & 157 &   97317.92     & 205 &   63513.82     \\ 
14 &         0.     & 62 &   66881.68     & 110 &   72455.70     & 158 &   110378.3     & 206 &   63373.80     \\ 
15 &         0.     & 63 &   67582.49     & 111 &   72594.23     & 159 &   98593.37     & 207 &   63495.21     \\ 
16 &         0.     & 64 &   68785.34     & 112 &   71415.52     & 160 &   92173.38     & 208 &   63340.18     \\ 
17 &         0.     & 65 &   67764.55     & 113 &   71318.52     & 161 &   88951.77     & 209 &   63090.16     \\ 
18 &         0.     & 66 &   64389.93     & 114 &   71358.01     & 162 &   86595.58     & 210 &   62792.86     \\ 
19 &         0.     & 67 &   66387.95     & 115 &   69312.26     & 163 &   84660.95     & 211 &   62684.35     \\ 
20 &   84533.33     & 68 &   66300.30     & 116 &   72729.52     & 164 &   82793.54     & 212 &   62228.04     \\ 
21 &   97593.33     & 69 &   66037.54     & 117 &   71753.02     & 165 &   81456.33     & 213 &   62354.96     \\ 
22 &   86614.20     & 70 &   65925.23     & 118 &   68760.13     & 166 &   80262.84     & 214 &   62319.11     \\ 
23 &   75566.42     & 71 &   66750.92     & 119 &   67004.47     & 167 &   79041.59     & 215 &   62454.68     \\ 
24 &   79532.30     & 72 &   68012.68     & 120 &   67529.37     & 168 &   77823.67     & 216 &   62330.83     \\ 
25 &   77045.07     & 73 &   66688.96     & 121 &   66701.97     & 169 &   76724.69     & 217 &   61784.18     \\ 
26 &   81880.93     & 74 &   66405.37     & 122 &   67285.90     & 170 &   75893.26     & 218 &   61767.18     \\ 
27 &   85593.41     & 75 &   67102.45     & 123 &   68590.23     & 171 &   75053.70     & 219 &   61874.20     \\ 
28 &   80641.81     & 76 &   66336.70     & 124 &   69383.94     & 172 &   74714.66     & 220 &   61549.47     \\ 
29 &   84675.29     & 77 &   65337.44     & 125 &   69906.62     & 173 &   74178.74     & 221 &   61513.18     \\ 
30 &   79656.66     & 78 &   66537.74     & 126 &   70594.98     & 174 &   73096.82     & 222 &   61637.83     \\ 
31 &   72727.22     & 79 &   65395.91     & 127 &   72027.86     & 175 &   72504.97     & 223 &   61620.56     \\ 
32 &   72032.57     & 80 &   63817.60     & 128 &   69610.27     & 176 &   71872.47     & 224 &   61179.25     \\ 
33 &   76068.43     & 81 &   62694.98     & 129 &   68918.78     & 177 &   71173.97     & 225 &   61342.76     \\ 
34 &   81562.41     & 82 &   61596.10     & 130 &   68994.07     & 178 &   70925.52     & 226 &   61209.52     \\ 
35 &   84678.54     & 83 &   61388.55     & 131 &   69004.28     & 179 &   70039.62     & 227 &   60897.05     \\ 
36 &   80041.36     & 84 &   63142.58     & 132 &   68985.91     & 180 &   69598.54     & 228 &   60820.14     \\ 
37 &   71661.50     & 85 &   63019.42     & 133 &   68357.59     & 181 &   68997.95     & 229 &   60109.34     \\ 
38 &   68868.23     & 86 &   61704.64     & 134 &   71152.03     & 182 &   68646.09     & 230 &   60264.41     \\ 
39 &   68064.79     & 87 &   60329.91     & 135 &   69895.10     & 183 &   68173.02     & 231 &   60365.65     \\ 
40 &   63073.40     & 88 &   59583.88     & 136 &   69267.49     & 184 &   68019.66     & 232 &   60484.48     \\ 
41 &   65690.94     & 89 &   60845.42     & 137 &   71743.50     & 185 &   67635.93     & 233 &   60201.59     \\ 
42 &   64269.88     & 90 &   59964.68     & 138 &   73458.68     & 186 &   67184.90     & 234 &   59894.00     \\ 
43 &   64671.51     & 91 &   60300.15     & 139 &   72150.24     & 187 &   66659.48     & 235 &   60171.55     \\ 
44 &   65497.02     & 92 &   59803.61     & 140 &   70592.44     & 188 &   66500.74     & 236 &   60055.25     \\ 
45 &   65543.83     & 93 &   59537.41     & 141 &   72060.95     & 189 &   66303.93     & 237 &   60159.95     \\ 
46 &   66562.79     & 94 &   61625.99     & 142 &   72246.81     & 190 &   65949.48     & 238 &   60156.89     \\ 
47 &   62863.50     & 95 &   62733.15     & 143 &   76672.17     & 191 &   65782.26     & 239 &   59953.44     \\ 
48 &   62216.66     & 96 &   62959.43     & 144 &   79998.69     & 192 &   65350.60     & 240 &   59716.41     \\ 
\hline \hline 
\end{tabular}
\end{center}
\end{table*}

\begin{table*}
\caption{SNO Phase I livetime as a function of zenith angle $\cos\theta_{Z}$. The 
table shows the second 240 bins of Fig.~\ref{fig:LT_raw_10mhz_lt} of Section~\ref{sec:norm}, corresponding to an
even division of the region $0< \cos\theta_Z < 1$. These data can be obtained
from~\cite{snoweb}. \label{tbl:coszenith2}} 
\begin{center}
\begin{tabular}{cccccccccc} \hline \hline 
~Bin~ & ~~Time (s)~~ & ~Bin~ & ~~Time (s)~~ & ~Bin~ & ~~Time (s)~~ & ~Bin~ & ~~Time (s)~~ & ~Bin~ & ~~Time (s)~~  \\ \hline 
241 &   60159.48     & 289 &   61038.79     & 337 &   61164.14     & 385 &   46882.55     & 433 &   37335.41     \\ 
242 &   60157.90     & 290 &   61508.05     & 338 &   56770.42     & 386 &   46182.11     & 434 &   33368.73     \\ 
243 &   60234.48     & 291 &   61546.99     & 339 &   51308.69     & 387 &   42744.34     & 435 &   32445.89     \\ 
244 &   59981.35     & 292 &   61547.62     & 340 &   53072.36     & 388 &   41460.28     & 436 &   27420.13     \\ 
245 &   60192.39     & 293 &   61706.14     & 341 &   56323.44     & 389 &   40766.97     & 437 &   28998.35     \\ 
246 &   60250.41     & 294 &   61701.27     & 342 &   58190.27     & 390 &   40664.09     & 438 &   33648.66     \\ 
247 &   60303.48     & 295 &   61624.77     & 343 &   56005.26     & 391 &   37794.84     & 439 &   32459.73     \\ 
248 &   60448.31     & 296 &   62144.61     & 344 &   53765.05     & 392 &   40818.96     & 440 &   29045.29     \\ 
249 &   60374.68     & 297 &   62326.85     & 345 &   52351.17     & 393 &   39825.29     & 441 &   28530.63     \\ 
250 &   60324.01     & 298 &   62056.93     & 346 &   51606.84     & 394 &   40105.02     & 442 &   31042.25     \\ 
251 &   60375.73     & 299 &   62596.74     & 347 &   52845.12     & 395 &   39035.45     & 443 &   36630.90     \\ 
252 &   60272.48     & 300 &   62648.16     & 348 &   53221.47     & 396 &   39014.06     & 444 &   37523.19     \\ 
253 &   60155.11     & 301 &   62655.70     & 349 &   53412.16     & 397 &   37545.49     & 445 &   34152.77     \\ 
254 &   60531.04     & 302 &   62806.52     & 350 &   55313.84     & 398 &   36553.44     & 446 &   30558.32     \\ 
255 &   60394.99     & 303 &   63160.08     & 351 &   56830.81     & 399 &   35125.43     & 447 &   30646.48     \\ 
256 &   60195.88     & 304 &   63317.90     & 352 &   53013.16     & 400 &   34513.82     & 448 &   30777.36     \\ 
257 &   59620.40     & 305 &   64031.32     & 353 &   47838.07     & 401 &   34882.41     & 449 &   29097.11     \\ 
258 &   59352.50     & 306 &   64792.95     & 354 &   43675.86     & 402 &   34156.55     & 450 &   29732.82     \\ 
259 &   59370.32     & 307 &   65207.07     & 355 &   42474.98     & 403 &   36379.56     & 451 &   31460.36     \\ 
260 &   59585.25     & 308 &   65752.78     & 356 &   41186.11     & 404 &   34642.27     & 452 &   32689.98     \\ 
261 &   59575.30     & 309 &   66737.20     & 357 &   39666.04     & 405 &   33300.05     & 453 &   31884.92     \\ 
262 &   59367.36     & 310 &   67193.61     & 358 &   40520.35     & 406 &   32057.59     & 454 &   37457.38     \\ 
263 &   59073.72     & 311 &   68265.95     & 359 &   41378.50     & 407 &   31470.17     & 455 &   42695.23     \\ 
264 &   58882.82     & 312 &   69167.70     & 360 &   42517.52     & 408 &   28722.06     & 456 &   38338.34     \\ 
265 &   59243.31     & 313 &   70034.46     & 361 &   41348.18     & 409 &   29699.14     & 457 &   31178.37     \\ 
266 &   59288.02     & 314 &   71069.70     & 362 &   39566.09     & 410 &   31005.69     & 458 &   29230.07     \\ 
267 &   58948.56     & 315 &   72152.85     & 363 &   41592.60     & 411 &   32534.68     & 459 &   34298.59     \\ 
268 &   58954.50     & 316 &   73329.67     & 364 &   39813.68     & 412 &   31686.04     & 460 &   38157.72     \\ 
269 &   59257.32     & 317 &   74909.52     & 365 &   40729.54     & 413 &   33545.51     & 461 &   22054.96     \\ 
270 &   59034.95     & 318 &   77241.68     & 366 &   42001.31     & 414 &   33590.75     & 462 &         0.     \\ 
271 &   59083.38     & 319 &   79539.92     & 367 &   43734.07     & 415 &   32907.48     & 463 &         0.     \\ 
272 &   59281.79     & 320 &   82756.59     & 368 &   44741.57     & 416 &   31964.73     & 464 &         0.     \\ 
273 &   59547.59     & 321 &   88017.92     & 369 &   40951.43     & 417 &   32842.52     & 465 &         0.     \\ 
274 &   59799.04     & 322 &   99021.12     & 370 &   40466.54     & 418 &   34557.49     & 466 &         0.     \\ 
275 &   60162.81     & 323 &   121739.2     & 371 &   40200.36     & 419 &   34609.77     & 467 &         0.     \\ 
276 &   60114.32     & 324 &   89277.84     & 372 &   42005.69     & 420 &   34484.57     & 468 &         0.     \\ 
277 &   59583.07     & 325 &   78515.95     & 373 &   45166.04     & 421 &   34879.89     & 469 &         0.     \\ 
278 &   59605.48     & 326 &   66065.66     & 374 &   44578.41     & 422 &   36527.20     & 470 &         0.     \\ 
279 &   59491.66     & 327 &   68601.35     & 375 &   45959.60     & 423 &   38488.16     & 471 &         0.     \\ 
280 &   59935.23     & 328 &   62846.25     & 376 &   48257.65     & 424 &   37494.09     & 472 &         0.     \\ 
281 &   59986.53     & 329 &   59989.86     & 377 &   48027.89     & 425 &   35050.48     & 473 &         0.     \\ 
282 &   60121.65     & 330 &   61566.98     & 378 &   47269.43     & 426 &   35148.81     & 474 &         0.     \\ 
283 &   60357.52     & 331 &   66459.38     & 379 &   45325.18     & 427 &   36982.53     & 475 &         0.     \\ 
284 &   60465.19     & 332 &   63723.37     & 380 &   45823.05     & 428 &   31821.25     & 476 &         0.     \\ 
285 &   60541.61     & 333 &   59033.32     & 381 &   45573.78     & 429 &   33763.16     & 477 &         0.     \\ 
286 &   60905.05     & 334 &   60657.42     & 382 &   42281.97     & 430 &   31914.01     & 478 &         0.     \\ 
287 &   61118.30     & 335 &   64283.69     & 383 &   46222.83     & 431 &   35033.18     & 479 &         0.     \\ 
288 &   61107.27     & 336 &   64612.13     & 384 &   45803.84     & 432 &   38348.48     & 480 &         0.     \\ 
\hline \hline 
\end{tabular}
\end{center}
\end{table*}

\subsection{Corrections to the SNO Calculations}

The number of events calculated above need to be corrected for the signal loss
incurred by the application  of the cuts described in Section~\ref{sec:dataproc}.
The loss for each signal, and the uncertainties on these losses, is listed in
Section~\ref{sec:cutaccept}.  The same losses apply to both the day and night
spectrum, and are treated as uncorrelated.

\subsection{Backgrounds}

As discussed in Section~\ref{sec:bkds}, there are two primary sources of
backgrounds to the neutrino signals: neutrons from photodisintegration and other
processes, and low energy `Cherenkov' backgrounds from radioactivity inside and
outside the fiducial volume.  The two sources of background are essentially
independent of one another.  The overall summary of the backgrounds and 
uncertainties are listed in Table~\ref{tbl:bkgd_summary}.  In the fits for the 
mixing parameters, the asymmetric error bars for the low energy backgrounds were
symmetrized by taking their average.  The background event numbers given in
Table~\ref{tbl:bkgd_summary} represent the total number of detected events, and they
therefore do not need any further correction for the cut losses described in the
previous section, livetime, energy threshold, or fiducial volume.
Table~\ref{tbl:bkdspec} shows the bin-by-bin background event numbers divided
between day and night, with their uncertainties.  These are the numbers used in the
calculation of the mixing parameters discussed here.
\begin{table*}
\caption{Bin-by-bin contents of day and night energy spectra for neutron ($n$) and 
low energy Cherenkov (${Ch}$) backgrounds.  Columns labeled with $\sigma$ indicate
the (symmetrized) uncertainty on the background numbers. The overall summary of the 
integral numbers of background events, listed by source, can be found in 
Table~\ref{tbl:bkgd_summary}.  The second and third columns give the  
boundaries of each energy bin. These data can be obtained
from~\cite{snoweb}. \label{tbl:bkdspec}}
\begin{center}
\begin{tabular}{rcccccccccc}
\hline \hline
Bin & $T_{\rm min}$ & $T_{\rm max}$ & $N_{n}$ & $\sigma_{n}$ & $N_{Ch}$ & $\sigma_{Ch}$ &  $N_{n}$ & $\sigma_{n}$ & $N_{Ch}$ & $\sigma_{Ch}$ \\
    &    (MeV)      &   (MeV)       &  Day & Day & Day & Day & Night & Night & Night & Night \\ 
\hline
    1&   5.0&   5.5&  10.3928&   1.6092&  16.7125&   5.5747&  15.9916&   2.4809&  26.2490&   8.5824 \\
    2&   5.5&   6.0&   8.7606&   1.3565&   0.9377&   0.3584&  13.4801&   2.0913&   1.4727&   0.5413 \\
    3&   6.0&   6.5&   6.0286&   0.9335&   0.0479&   0.0207&   9.2763&   1.4391&   0.0752&   0.0300 \\
    4&   6.5&   7.0&   3.3867&   0.5244&   0.0019&   0.0009&   5.2112&   0.8084&   0.0030&   0.0013 \\
    5&   7.0&   7.5&   1.5532&   0.2405&   0.0001&   0.0000&   2.3899&   0.3708&   0.0001&   0.0001 \\
    6&   7.5&   8.0&   0.5815&   0.0900&   0.0000&   0.0000&   0.8947&   0.1388&   0.0000&   0.0000 \\
    7&   8.0&   8.5&   0.1777&   0.0275&   0.0000&   0.0000&   0.2735&   0.0424&   0.0000&   0.0000 \\
    8&   8.5&   9.0&   0.0443&   0.0069&   0.0000&   0.0000&   0.0682&   0.0106&   0.0000&   0.0000 \\
    9&   9.0&   9.5&   0.0090&   0.0014&   0.0000&   0.0000&   0.0139&   0.0022&   0.0000&   0.0000 \\
   10&   9.5&  10.0&   0.0015&   0.0002&   0.0000&   0.0000&   0.0023&   0.0004&   0.0000&   0.0000 \\
   11&  10.0&  10.5&   0.0002&   0.0000&   0.0000&   0.0000&   0.0003&   0.0000&   0.0000&   0.0000 \\
   12&  10.5&  11.0&   0.0000&   0.0000&   0.0000&   0.0000&   0.0000&   0.0000&   0.0000&   0.0000 \\
   13&  11.0&  11.5&   0.0000&   0.0000&   0.0000&   0.0000&   0.0000&   0.0000&   0.0000&   0.0000 \\
   14&  11.5&  12.0&   0.0000&   0.0000&   0.0000&   0.0000&   0.0000&   0.0000&   0.0000&   0.0000 \\
   15&  12.0&  12.5&   0.0000&   0.0000&   0.0000&   0.0000&   0.0000&   0.0000&   0.0000&   0.0000 \\
   16&  12.5&  13.0&   0.0000&   0.0000&   0.0000&   0.0000&   0.0000&   0.0000&   0.0000&   0.0000 \\
   17&  13.0&  20.0&   0.0000&   0.0000&   0.0000&   0.0000&   0.0000&   0.0000&   0.0000&   0.0000 \\
\hline \hline
\end{tabular}
\end{center}
\end{table*}

\subsection{Incorporation of Systematic Uncertainties}

	In addition to the statistical uncertainties for each bin in the SNO
spectra, there are systematic uncertainties on the detector response functions.
In our forward fitting technique, these result in systematic uncertainties on the
model prediction for the total energy spectrum.

   As was presented in Section~\ref{sec:sysunc_results}, we have here also
uncertainties on the amplitudes of the backgrounds, uncertainties on the overall
normalization of the signals, and uncertainties on the model we use to create our
predictions for the signal energy spectra. Unlike the primary method described in
Section~\ref{sec:sysunc_results}, here we incorporate the uncertainties on the
model not by shifting and `smearing' Monte Carlo-generated pdfs, but by directly
varying the parameters in the analytic response functions.  For example, we
characterized the energy response to electrons with the Gaussian shown in
Eq.~\ref{eresp},
\[ R(T_{\rm eff},T_{e})  = \frac{1}{\sqrt{(2 \pi)} \sigma_{T}(T_{e})}
\exp\left[ -  \frac{(T_{\rm eff} - T_{e} - \Delta_{T})^2}{2 \sigma_{T}^2(T_{e})}  \right],  \]
in which both the energy resolution $\sigma_{T}$ and the energy scale offset 
$\Delta_T$ are parameters for which we have measured systematic uncertainties.
To propagate these uncertainties, we vary these parameters by the $\pm 1\sigma$
uncertainties and re-calculate the predicted energy spectra through the 
convolution of Equation~\ref{eq:dnccdt}.  Of course, a change in the shape of the
energy spectrum also leads to a change in the number of events above threshold for
each signal, and this number is also re-calculated when varying the response
parameters.
	
Table~\ref{response_pars} of Section~\ref{sec:sigex} lists all the parameter
uncertainties used in creating analytic pdfs.  (Because the estimation of mixing
parameters we do here does not use solar direction information, the uncertainties
on the angular resolution listed in the table are not needed).  

Uncertainties that affect only the overall normalization---the cut acceptances,
neutron capture efficiency, target volume, etc.---are given in
Section~\ref{sec:norm}.  The neutron capture efficiency is treated as discussed
above in Section~A\ref{sec:ncrate_physint}.

The uncertainties on the amplitudes of the backgrounds are given in
Section~\ref{sec:bkds}.  As Section~\ref{sec:sysunc_results} explained, no
additional shape-related uncertainties were propagated for the backgrounds---their
pdf shapes were taken as those given in Section~\ref{sec:bkds} and only the
amplitudes allowed to vary within $\pm 1\sigma$.  The uncertainties on these
amplitudes were symmetrized as described above.

The systematic uncertainties will generally affect all bins of the energy spectrum
in a correlated way.  We therefore construct $N \times N$ covariance matrices for
the systematics, which are then added to the statistical uncertainties on the
spectral data (a diagonal matrix) to get a total uncertainty matrix that is used
to form the SNO $\chi^2$.

\subsection{Inclusion of Other Data Sets}

In our Phase I Day-Night paper~\cite{snodn}, we published MSW exclusion plots using only the SNO day and 
night spectral information, as well as in combination with the results of other
solar neutrino experiments.
Those analyses were based on a chi-squared statistic in the usual
way, {\em viz.}
\[ \chi^{2} = \sum_{j_{1},j_{2}=1}^{N} (O_{j_{1}} - O_{j_{1}}^{exp})
              [\sigma_{j_{1}j_{2}}^{2}(tot)]^{-1}
              (O_{j_{2}} - O_{j_{2}}^{exp})            \]
where $O_{j}^{exp}$ and $O_{j}$ are the experimental value and
the theoretical prediction, respectively, for each observable 
(rate measurement or spectral bin), and $[\sigma^2_{j{1}{j2}}(tot)]^{-1}$
is the inverse of the covariance matrix for the observables.

In the case of global fits using other data sets, the error matrix was taken to be
a summation of contributions from all the considered data, 
\[ \sigma^{2}(tot)=\sigma^{2}(exp)+\sigma_{R}^{2}+\sigma_{S}^{2}(SNO)+ \sigma_{S}^{2}(SK), \] 
where $\sigma^{2}(exp)$ contains both the statistical and
systematic errors from the rate measurements and statistical errors for the
spectral measurements. Correlations among the rate measurements,
$\sigma_{R}^{2}$, were handled according to the prescription of Fogli and
Lisi~\cite{fogli95}.  It was implicitly assumed that there were no correlations
between the rate and spectral measurements or between the SNO and Super-Kamiokande
spectral measurements.

The Super-Kamiokande spectral data was taken from Ref.~\cite{SK}, which were quoted as
as fractions relative to the  BP2000 value of
$\phi_{^{8}B} = 5.15 \times 10^{-6} {\mathrm cm}^{-2}s^{-1}$.  
The errors used for these numbers were obtained by combining in quadrature the positive
statistical errors with the positive uncorrelated systematic errors given in~\cite{SK}.

Super-Kamiokande's energy response to electrons was taken to be a Gaussian with a
resolution whose width was 1.5 MeV for a 10 MeV electron and which scaled as
$\sqrt{T_{e}}$~\cite{fogli01}.  

The chlorine and gallium experiments do not have any spectral information
associated with their data.  Theoretical yields for these experiments are
therefore simple integrations over the flux and cross section: 
\begin{equation}
R_{X}=\int_{0}^{\infty}dE_{\nu}\phi_{\nu_{e}}(E_{\nu})\sigma_{X}(E_{\nu}) 
\end{equation}
where $X$ is chlorine or gallium, $\phi_{\nu_{e}}$ is the sum of all solar 
fluxes, and the units are SNUs ($1~{\rm SNU} = 10^{-36} s^{-1}$).

Neutrino production was calculated starting with the fluxes given in
BP2000~\cite{bp2000} for the eight neutrino producing reactions that
occur in the pp and CNO chains. The shapes for the hep, pp
and CNO neutrinos were taken from Bahcall~\cite{jnb_na,bahcallweb}.

The neutrino cross sections on chlorine and gallium fits were taken from
Ref.~\cite{bahcallweb}. A point was added to the table for chlorine at 0.861 MeV
($2.67 \times 10^{-42} {\mathrm cm}^{2}$, taken from~\cite{jnb_na}), to help get the correct
contribution from $^{7}$Be neutrinos to the chlorine experiment.

Combining our SNO analysis with the data and theoretical yield calculations for
the gallium and chlorine experiments and the energy spectral data and calculated
predictions for Super-Kamiokande, gives a best fit $\Delta m^2=5.0\times 10^{-5}$eV$^2$ and
$\tan^2\theta=0.34$.

\section{Instrumental Background Cuts \label{sec:apdxa}}

	We created two independently developed sets of cuts designed to remove
instrumental backgrounds.  The cuts were developed using data collected primarily
during the first four months of production (November 1999 - February
2000) and the SNO commissioning data. A small set of data was hand-scanned
after the application of the cuts, to look for additional instrumental
backgrounds. There were four design goals: residual background
contamination after application of the cuts should be less than 1\%, the
acceptance for genuine neutrino events should be greater than 99\% for events
produced inside a 7~m radius,  the bias in the cut acceptance should be small, and
the cuts should be insensitive to bad PMT calibrations.  The two sets of cuts were
benchmarked against each other, with good agreement.  Cut Set A was used for the
final analysis.

\subsection{Cut Set A}

Cut Set A used a set of sixteen cuts:   
\begin{itemize}

\item Analog Measurement Board  (AMB)

	The AMB monitors the analog trigger signals, producing a measurement of the
integral and peak of the `energy sum' trigger signal (a signal that was
proportional to the amount of charge detected by each PMT, see
Section~\ref{sec:dataset}).  It thus provides a measurement of the total charge
deposited in the event that is independent of the channel-by-channel
digital measurements.  For each event, the measured integral and peak of
the energy sum trigger signal are compared to the expectations for each event based
upon the number of hit PMT's.  If an event has too much or too little charge (over
4$\sigma$ away from expectation for either the integral or peak) then it is rejected.
The expectations come from calibration data with the $^{16}$N source.  Of all the
cuts, this cut removes the largest fraction of the instrumental backgrounds.

\item  QCluster (Charge with Hit Cluster)

As described in Section~\ref{sec:inst} PMT `flasher events' deposit a very high
charge in a single PMT, which causes many nearby hits through crosstalk in the
cables and electronics. The QCluster cut identifies such events by finding clusters
of channel hits surrounding a high charge hit.

\item  QvT  (Charge versus Time)

In a flasher event the high charge tube appears early because the remaining
hits are due to emitted light detected on the opposite side of the
detector. The QvT cut removes an event if the highest charge PMT is above a charge
threshold and is more than 60~ns earlier than the median time of the remaining hits.

\item  Q/$N_{\rm hit}$  (Charge over $N_{\rm hit}$)

The Q/$N_{\rm hit}$ cut uses a measurement of the charge averaged over all the hits in an
event.  It is similar to the Analog Measurement Board cut described above,
except that the digitally measured average charge of the PMTs is
used rather than the analog energy sum. To provide immunity to bad
channel calibrations, the 10\% with the highest charge
are rejected from the calculation. As a consequence of this
filtering, the cut used is one-sided and used only to remove the low end of the
charge distribution, thus eliminating events due to electronic pickup.

\item Outward-Looking Tube 

Any event with three or more hits in the outward-looking PMTs on the outside of
the phototube support structure.

\item  Neck 

The neck cut uses two of the PMTs deployed in the neck of the acrylic vessel
to remove events from light created at the acrylic-water boundary and around
calibration hardware. An event is removed the two neck tubes
fire, or if only one neck tube fires, and is early in time and above a charge
threshold.

\item  In-Time Cut

	Solar neutrino events in SNO produce Cherenkov light which has
a very narrow time distribution---much less than 1~ns.
Many instrumental backgrounds produce light distributed over
many nanoseconds, and thus can removed.  The simplest approach is to
require a large fraction of the PMT hits to occur in a short window
of time.  Because the instrumental background cuts are applied to the data
well before event reconstruction, the time window used
is very wide (as opposed to that used in the post-reconstruction
`in-time ratio' cut described in Section~\ref{sec:cerbox}).  Regardless of
where an event occurs, the Cherenkov light should reach a PMT within no more than
the $\sim $85~ns light transit time across the detector. The in-time cut uses the
ratio of the number of hits within a 93~ns window to the total number of hits to
reject events.

\item Fitterless Time Spread 
	
	Although the `in-time cut' removes sources of events with very
wide timing distributions (anything which produces steady light, such as
a glowing PMT base), flasher events do not have such a wide timing distribution.
Although the vast majority of the flasher PMT's are removed by cuts based
on the presence of a high charge tube, in cases where the tube's signal path
is broken neither the high charge tube itself nor its associated crosstalk
hits may be in the event, and timing information becomes the only handle.
To remove these `blind flasher' events before reconstruction we look at
the distribution of PMT hit times for adjacent tubes, which are expected to
be close in time if the light originates from a point source.  The median
of the time differences between PMT pairs is then used as a cut
parameter.  The cut rejects roughly 50$\%$ of the flasher events where the cluster
and high charge tube have been removed in software.

\item Crate isotropy 

Internal pickup events have distinct electronic channel hit patterns, as typically
PMTs connected to two adjacent cards (or the cards on the edge) of a crate will
fire without any others. The crate isotropy cut removes events with more than
a given fraction of hits on two adjacent cards.

\item Flasher Geometry Cut

Events in which the flasher tube itself is missing, but its
associated crosstalk hits are present, can be removed by looking
for a cluster of hits on the side of the detector opposite from the majority
of the hits.  The flasher geometry cut searches for
all possible clusters of a given size and computes the mean distance
from each such cluster to the remaining hits. Events with a cluster separated  
by more than 12~m from the remaining tubes are eliminated.

\item Retriggers 

	Large events can cause the trigger system
to retrigger immediately after the end of its lockout period. The retrigger
may be due to optical photons continuing
to bounce around inside the detector, or because PMT afterpulses can
fire microseconds later. Flasher events have very high light levels originating
from a single tube, and therefore the tube often produces very large 
afterpulses.  To remove these, all  events that occur within 
5$\mu$s of a preceding event are cut from the data set.

\item Bursts (Short Window and `$N_{\rm hit}$-Burst')

Two burst cuts are used by Cut Set A; one which cuts any events which occur
within a very short time window, and one which cuts high $N_{\rm hit}$ events which
occur within a wider window.  For the first cut, if more than 
three events occur within 1~ms, the entire burst is removed. For the
second cut, only events with more than  40 hits are considered, and if six or
more of these occur within 4 second the entire burst is removed.

\item Trigger Bits

As a backup to the other cuts, two cuts operate based upon the energy sum
triggers. One cut removes events that have only 
the low-gain energy-sum trigger bit set, and another cut does the same for
events which have the outward-looking (veto) tube energy sum bit set. \\

\item Data Acquisition Artifacts

	Event data are occasionally not properly collected by the data acquisition 
system.  This can happen at very high data rates which cause the event buffers 
to flush early, or  because a channel's trigger ID is incorrect and no
corresponding event header can be found.  Other rare problems are the presence
in an event of two hits from the same channel.  Such events are all removed
using the data acquisition tags and information.

\end{itemize}

\subsection{Cut Set B}

Cut Set B used 17 cuts in total.  Among the major differences between this set and Set A were the fact that the cuts were designed to be robust to errors in low-level (electronics) calibrations, by relying either upon raw ADC values or on quantities which did not require any calibration.  In addition, a database of channels with frequent high-charge hits  was generated on a run-by-run basis prior to the application of cuts.  This ``high-charge cut frequency" (HQCF) database is used in the identification of instrumental background events in Cut Set B.

\begin{itemize}

\item Burst 

	The Cut Set B burst cut removes any event which occurs within 1$\mu$s
of a previous event, thus removing any event caused by a retriggering of the data
acquisition system (whose minimum time between triggers is $\sim$440~ns).  PMT
flasher events as well as high voltage breakdown in the PMT bases, connectors, or
cables often produce such retriggers. 

\item Trigger Bit Cuts 

	Several types of events whose sources are not Cherenkov light within
the detector are tagged by the trigger system.  Cut Set B uses these trigger
bit tags to remove pulsed trigger events, software-triggered events, and events
associated with the GPS timing system.  Events that were not tagged as
resulting from a 93~ns hit-coincidence trigger are also removed. 

\item QBC (Charge Bad Channel)

Flashing PMTs typically have anomalously high deposited charge.  Poorly
operating electronics channels may also fire with high charges in coincidence
with a Cherenkov event.  These channels are identified and stored in the HQCF
database on a run-by-run basis.  The QBC cut searches for high-charge hits and
removes the event if the offending channels are not in the HQCF database.  

\item QTC (Charge-Time-Cluster)

The charge-time-cluster (QTC) cut uses the geometric clustering of PMTs in electronics space and 
the timing of hits with anomalously low or high charge to identify flasher and
electronic noise events.  This cut is similar to a sequential application of
the QvT and Qcluster cuts of Cut Set A.   An event with an early anomalous hit,
whose charge is significantly different from the median charge of all hits, is
removed if the associated channel is not in the HQCF database.    This
criterion reduces data loss when a misbehaving channel is firing at high
frequency during a run, as such hits may be in accidental coincidence with a
Cherenkov light event.   Events with hit channels clustering around a channel
with an anomalously high charge are also removed.

\item PMT Timing RMS and Kurtosis

Some instrumental background events exhibit a much larger spread in the PMT hit
times than Cherenkov light events.  Events resulting from high voltage
breakdown in the PMT bases or connectors, which can produce long ($\sim$ ms)
pulses of light, often have raw time distributions that are flat across the
event window.  The root mean square of the raw PMT timing distribution ($t_{\rm
rms}$) alone is not sufficient to distinguish these background events
unambiguously, as a small fraction of the Cherenkov events also have a large
$t_{\rm rms}$.  However, the PMT timing distribution for Cherenkov events are
leptokurtic, as opposed to the platykurtic nature of these instrumental
background events.  By employing $t_{\rm rms}$ and the kurtosis of the PMT
timing distribution in a two-dimensional cut, instrumental backgrounds with
anomalously wide and flat distributions of PMT times are effectively removed.

\item Neck 

This cut is similar to the corresponding neck cut in Cut Set A.  An event is
removed if at least two of the four PMTs deployed in the neck of the acrylic
fire.  An event is also removed when a neck-deployed PMT fires with a very high
charge.  

\item FGC (Flasher Geometry Cut)

This cut is the predecessor to the cut of Cut Set A with the same name.  The
primary difference between them is in cluster identification; this cut requires
more PMT hits in a cluster.

\item QQP (Two-Charge Cut)

This is a two-dimensional cut that uses the total channel charges (from summing
the charges in all hit channels) and the integral charge from the AMB, both
averaged over all PMT hits, as the cut parameters.  This cut is effective
against electrical noise, which normally have very low integral charges.
Electrical  discharge events, which have very high deposited charge, would
saturate the AMB integral charge channel because of its limited dynamic range.
In such instances, this cut identifies such discharge events by imposing
additional cut criteria on the pulse height measured by the AMB.  

\item Correlated Channel Count Rates

	Electronics boards in which several adjacent channels had high count
rates are flagged so that events created by pickup from the data acquisition
readout can be removed.  Electrical noise pickup events may have a
disproportional number of hits in a single crate.  Events with such
concentration of hits in a crate are also removed.

\item Veto Tube Cut

	Muons and muon-related events, as well as any event which produces light
external to the phototube support sphere, are cut using a combination of the
veto PMTs in the neck region of the acrylic vessel, the outward-looking PMTs
installed on the outside of the phototube support sphere, and a set of 23 PMTs 
suspended between the phototube support sphere and the rock of the cavity.

\end{itemize}

\bibliographystyle{apsrev}

\begin{thebibliography}{999}
\bibitem{egsnote}
\bibinfo{note}{Subsequent to the results
published in Refs.~\cite{snocc,snonc,snodn}, we found that EGS4 needed a small 
correction to allow for the neglect of Mott terms in the electron scattering
corss-section and other approximations in the treatment of multiple
scattering~\cite{nsp}. The effects of the Mott terms on the analyses presented here
were found to be negligible.}
\bibitem{saltphys}
\bibinfo{note}{In our more recent publications~\cite{nsp,salt1}, based on data from SNO Phase
II in which NaCl was dissolved within the heavy water, we were able to use the
derived fluxes and spectra from the signal extraction process, rather than the
summed spectra as discussed in this article.  This was possible because of the
better discrimination between neutrons and electrons provided by the $\gamma$
cascade from neutron capture on Cl.}
\end{thebibliography}

\end{document}